\documentclass[rmp,aps,twocolumn,nofootinbib]{revtex4-1}

\usepackage{natbib}
\usepackage{graphicx}
\usepackage[dvipsnames]{xcolor}
\usepackage{amsmath,amsthm,amssymb}
\usepackage{xfrac}
\usepackage{microtype}
\usepackage{comment}
\usepackage{ulem}

\usepackage[hidelinks]{hyperref}

\usepackage{cleveref}
\crefname{equation}{Eq.}{Eqs.}
\Crefname{equation}{Equation}{Equations}
\crefname{figure}{Fig.}{Figs.}
\Crefname{figure}{Figure}{Figures}
\crefname{section}{Sec.}{Secs.}
\Crefname{section}{Section}{Sections}
\crefname{appendix}{Appendix}{Apps.}
\Crefname{appendix}{Appendix}{Apps.}
\crefname{paragraph}{Sec.}{Secs.}
\crefname{table}{Table}{Tables}

\def\hX{\hat{X}}
\def\hP{\hat{P}}

\newcommand{\eq}[1]{Eq.~\eqref{#1}}
\newcommand{\secref}[1]{Sec.~\ref{#1}}

\newcommand{\figref}[1]{Fig.~\ref{#1}}

\newcommand{\rhoS}{\rho}

\newcommand{\wa}{\omega_{q}}

\newcommand{\twc}{\tilde\omega_r}
\newcommand{\wc}{\omega_r}
\newcommand{\wdrive}{\omega_\mathrm{d}}
\newcommand{\wmod}{\omega_m}

\newcommand{\zc}{Z_r}

\newcommand{\wif}{\omega_\mathrm{IF}}
\newcommand{\wlo}{\omega_\mathrm{LO}}

\newcommand{\wrf}{\omega_\mathrm{RF}}
\newcommand{\philo}{\phi_\mathrm{LO}}
\newcommand{\phidrive}{\phi_\mathrm{d}}

\newcommand{\gphi}{\gamma_\varphi}

\newcommand{\SNR}{\mathrm{SNR}}
\newcommand{\tmeas}{\tau_m}
\newcommand{\Emeas}{E_m}
\newcommand{\Fmeas}{F_m}

\newcommand{\ket}[1]{| #1 \rangle}
\newcommand{\bra}[1]{\langle #1 |}

\newcommand{\me}[3]{\bra{#1}#2\ket{#3}}
\newcommand{\av}[1]{\langle #1 \rangle}
\newcommand{\braket}[1]{\langle #1 \rangle}

\newcommand{\sz}[1]{\hat\sigma_{z#1}}
\newcommand{\sx}[1]{\hat\sigma_{x#1}}
\newcommand{\sy}[1]{\hat\sigma_{y#1}}
\newcommand{\smm}[1]{\hat\sigma_{-#1}}
\newcommand{\spp}[1]{\hat\sigma_{+#1}}

\newcommand{\aop}{\hat a}

\newcommand{\ad}{\hat a^\dag}
\newcommand{\ada}{\hat a^\dag \hat a}

\newcommand{\zp}{\mathrm{zpf}}
\newcommand{\Fzp}{\Phi_\zp}
\newcommand{\Qzp}{Q_\zp}

\newcommand{\bop}{\hat b}
\newcommand{\bd}{\hat b^\dag}
\newcommand{\bdb}{\hat b^\dag \hat b}

\newcommand{\aamp}{\hat a_\mathrm{amp}}
\newcommand{\aampd}{\hat a^\dag_\mathrm{amp}}
\newcommand{\aopf}{\hat a_f} 
\newcommand{\adf}{\hat a^\dag_f} 
\newcommand{\bout}{\hat b_\mathrm{out}}

\newcommand{\boutd}{\hat b^\dag_\mathrm{out}}
\newcommand{\bin}{\hat b_\mathrm{in}}

\newcommand{\bind}{\hat b^\dag_\mathrm{in}}

\newcommand{\dg}{^\dagger}
\newcommand{\Dop}{\hat D}

\newcommand{\hH}{\hat H}
\newcommand{\h}[1]{\hat{#1}}
\newcommand{\hHL}{\hat H_\mathrm{L}}
\newcommand{\hHNL}{\hat H_\mathrm{NL}}

\newcommand{\bdw}{\hat b^\dag_\omega}
\newcommand{\bw}{\hat b_\omega}

\newcommand{\hQ}{\hat Q}
\newcommand{\hF}{\hat \Phi}
\newcommand{\hV}{\hat V}
\newcommand{\hn}{\hat n}
\newcommand{\hvphi}{\hat \varphi}
\newcommand{\Fnot}{\Phi_0}

\newcommand{\ncrit}{n_\mathrm{crit}}

\newcommand{\tr}{{\mathrm{Tr}}}

\renewcommand{\Im}{\text{Im}}

\newcommand{\cC}{\mathcal{C}}

\newcommand{\cZeff}{\mathcal{Z}^\text{eff}}

\newcommand{\half}{\frac{1}{2}}
\newcommand{\hc}{\text{H.c.}}
\def\*#1{\mathbf{#1}}
\newcommand{\tml}{\text{tml}}

\begin{document}

\title{Circuit Quantum Electrodynamics}

\author{Alexandre Blais}\email[Send comments and feedback to ]{circuitqed.review@gmail.com}
\affiliation{Institut quantique and D\'epartement de Physique, Universit\'e de Sherbrooke, Sherbrooke J1K 2R1 QC, Canada}
\affiliation{Canadian Institute for Advanced Research, Toronto, ON, Canada}
\author{Arne L. Grimsmo}
\affiliation{Centre for Engineered Quantum Systems, School of Physics, The University of Sydney, Sydney, NSW 2006, Australia}
\author{S. M. Girvin}
\affiliation{Yale Quantum Institute, PO Box 208 334, New Haven, CT 06520-8263 USA}
\author{Andreas Wallraff}
\affiliation{Department of Physics, ETH Zurich, CH-8093, Zurich, Switzerland.}

\begin{abstract}
  Quantum mechanical effects at the macroscopic level were first explored in Josephson junction-based superconducting circuits in the 1980's. In the last twenty years, the emergence of quantum information science has intensified research toward using these circuits as qubits in quantum information processors. The realization that superconducting qubits can be made to strongly and controllably interact with microwave photons, the quantized electromagnetic fields stored in superconducting circuits, led to the creation of the field of circuit quantum electrodynamics (QED), the topic of this review. While atomic cavity QED inspired many of the early developments of circuit QED, the latter has now become an independent and thriving field of research in its own right. Circuit QED allows the study and control of light-matter interaction at the quantum level in unprecedented detail. It also plays an essential role in all current approaches to quantum information processing with superconducting circuits. In addition, circuit QED enables the study of hybrid quantum systems, such as quantum dots, magnons, Rydberg atoms, surface acoustic waves, and mechanical systems, interacting with microwave photons.
  Here, we review the coherent coupling of superconducting qubits to microwave photons in high-quality oscillators focussing on the physics of the Jaynes-Cummings model, its dispersive limit, and the different regimes of light-matter interaction in this system. We discuss coupling of superconducting circuits to their environment, which is necessary for coherent control and measurements in circuit QED, but which also invariably leads to decoherence. Dispersive qubit readout, a central ingredient in almost all circuit QED experiments, is also described. Following an introduction to these fundamental concepts that are at the heart of circuit QED, we discuss important use cases of these ideas in quantum information processing and in quantum optics. Circuit QED realizes a broad set of concepts that open up new possibilities for the study of quantum physics at the macro scale with superconducting circuits and applications to quantum information science in the widest sense.
\end{abstract}

\maketitle
\today

\clearpage
\tableofcontents

\section{Introduction}

Circuit quantum electrodynamics (QED) is the study of the interaction of nonlinear superconducting circuits, acting as artificial atoms or as qubits for quantum information processing, with quantized electromagnetic fields in the microwave frequency domain. Inspired by cavity QED \cite{Haroche2006,Kimble1998}, a field of research originating from atomic physics and quantum optics, circuit QED has led to advances in the fundamental study of light-matter interaction, in the development of quantum information processing technology \cite{Kjaergaard2019,Krantz2019,Wendin2017,Clarke2008,Blais2020}, and in the exploration of novel hybrid quantum systems \cite{Xiang2013a,Clerk2020}.

First steps toward exploring the quantum physics of superconducting circuits were made in the mid-1980's. At that time, the question arose whether quantum phenomena, such as quantum tunneling or energy level quantization, could be observed in macroscopic systems of any kind \cite{Leggett1980,Leggett1984}. One example of such a macroscopic system is the Josehphson tunnel junction \cite{Josephson1962,Tinkham2004} formed by a thin insulating barrier at the interface  between two superconductors and in which macroscopic quantities such as the current flowing through the junction or the voltage developed across it are governed by the dynamics of a macroscopic order parameter \cite{Eckern1984}. This macroscopic order parameter relates to the density and common phase of Bose-condensed Cooper pairs of electrons. The first experimental evidences for quantum effects in these circuits \cite{Clarke1988} were the observation of quantum tunneling of the phase degree of freedom of a Josephson junction \cite{Devoret1985}, rapidly followed by the measurement of quantized energy levels of the same degree of freedom \cite{Martinis1985}.

While the possibility of observation of coherent quantum phenomena in Josephson junction-based circuits, such as coherent oscillations between two quantum states of the junction and the preparation of quantum superpositions was already envisaged in the 1980's \cite{Tesche1987}, the prospect of realizing superconducting qubits for quantum computation revived interest in the pursuit of this goal \cite{Bouchiat1998,Shnirman1997,Bocko1997,Makhlin1999,Makhlin2001}. In a groundbreaking experiment, time-resolved coherent oscillations with a superconducting qubit were observed in 1999 \cite{Nakamura1999}. Further progress resulted in the observation of coherent oscillations in coupled superconducting qubits \cite{Pashkin2003,Yamamoto2003} and in significant improvements of the  coherence times of these devices by exploiting symmetries in the Hamiltonian underlying the description of the circuits \cite{Vion2002}.

In parallel to these advances, in atomic physics and quantum optics, cavity QED developed into an excellent setting for the study of the coherent interactions between individual atoms and quantum radiation fields \cite{Rempe1987,Thompson1992,Brune1996,Haroche1989}, and its application to quantum communication \cite{Kimble2008} and quantum computation  \cite{Kimble1998,Haroche2006}. In the early 2000's, the concept of realizing the physics of cavity QED with superconducting circuits emerged with proposals to coherently couple superconducting qubits to microwave photons in open 3D cavities \cite{AlSaidi2001,Yang2003, You2003}, in discrete LC oscillators \cite{Makhlin2001,Buisson2001}, and in large Josephson junctions \cite{Marquardt2001,Plastina2003, Blais2003a}. The prospect of realizing strong coupling of superconducting qubits to photons stored in high-quality coplanar waveguide resonators, together with suggestions to use this approach to protect qubits from decoherence, to read out their state, and to couple them to each other in a quantum computer architecture advanced the study of cavity QED with superconducting circuits \cite{Blais2004}. This possibility of exploring both the foundations of light-matter interaction and advancing quantum information processing technology with superconducting circuits motivated the rapid advance in experimental research, culminating in the first experimental realization of a circuit QED system achieving the strong coupling regime of light-matter interaction where the coupling overwhelms damping \cite{Wallraff2004,Chiorescu2004}.

Circuit QED combines the theoretical and experimental tools of atomic physics, quantum optics and the physics of mesoscopic superconducting circuits not only to further explore the physics of cavity QED and quantum optics in novel parameter regimes, but also to allow the realization of engineered quantum devices with technological applications. Indeed, after 15 years of continuous development, circuit QED is now a leading architecture for quantum computation. Simple quantum algorithms have been implemented \cite{DiCarlo2009}, cloud-based devices are accessible, demonstrations of quantum-error correction have approached or reached the so-called break-even point \cite{Ofek2016,Hu2019}, and devices with several tens of qubits have been operated with claims of quantum supremacy \cite{Arute2019}. 

More generally, circuit QED is opening new research directions. These include the development of quantum-limited amplifiers and single-microwave photon detectors with applications ranging from quantum information processing to the search for dark matter axions, to hybrid quantum systems \cite{Clerk2020} where different physical systems such as NV centers \cite{Kubo2010}, mechanical oscillators \cite{Aspelmeyer2013}, semiconducting quantum dots \cite{Burkard2020}, or collective spin excitations in ferromagnetic crystals \cite{Lachance-Quirion2019} are interfaced with superconducting quantum circuits. 

In this review, we start in \cref{sec:SupQuantumCircuits} by introducing the two main actors of circuit QED: high-quality superconducting oscillators and superconducting artificial atoms. The latter are also known as superconducting qubits in the context of quantum information processing. There are many types of superconducting qubits and we choose to focus on the transmon \cite{Koch2007}. This choice is made because the transmon is not only the most widely used qubit but also because this allows us to present the main ideas of circuit QED without having to delve into the very rich physics of the different types of superconducting qubits. Most of the material presented in this review applies to other qubits without much modification. \Cref{sec:LightMatter} is devoted to light-matter coupling in circuit QED including a discussion of the Jaynes-Cummings model and its dispersive limit. Different methods to obtain approximate effective Hamiltonians valid in the dispersive regime are presented. \Cref{sec:environment} addresses the coupling of superconducting quantum circuits to their electromagnetic environment, considering both dissipation and coherent control. In \cref{sec:readout}, we turn to measurements in circuit QED with an emphasis on dispersive qubit readout. Building on this discussion, \cref{sec:CouplingRegimes} presents the different regimes of light-matter coupling which are reached in circuit QED and their experimental signatures. In the last sections, we turn to two applications of circuit QED: quantum computing in \cref{sec:quantumcomputing} and quantum optics in \cref{sec:QuantumOptics}. 

Our objective with this review is to give the reader a solid background on the foundations of circuit QED rather than showcasing the very latest developments of the field. We hope that this introductory text will allow the reader to understand the recent advances of the field and to become an active participant in its development.

\section{\label{sec:SupQuantumCircuits}Superconducting quantum circuits}

Circuit components with spatial dimensions that are small compared to the relevant wavelength can be treated as lumped elements~\cite{Devoret1997}, and we start this section with a particularly simple lumped-element circuit: the quantum LC oscillator. We subsequently discuss the closely related two- and three-dimensional microwave resonators that play a central role in circuit QED experiments and which can be thought of as distributed versions of the LC oscillator with a set of harmonic frequencies. Finally, we move on to nonlinear quantum circuits with Josephson junctions as the source of nonlinearity, and discuss how such circuits can behave as artificial atoms. We put special emphasis on the transmon qubit~\cite{Koch2007}, which is the most widely used artificial atom design in current circuit QED experiments.

\subsection{\label{sec:HO}The quantum LC resonator}

\begin{figure}
  \centering
  \includegraphics{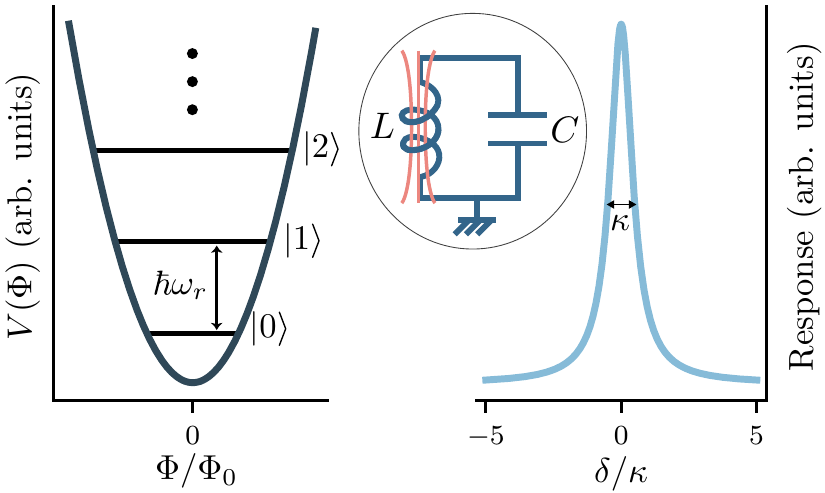}
  \caption{\label{fig:LCpotential}
  (Left) Harmonic potential versus flux of the LC circuit with $\Phi_0 = h/2e$ the flux quantum. (Right) Response of the oscillator to an external perturbation as a function of the detuning $\delta$ of the perturbation from the oscillator frequency. Here $\kappa = \wc/Q$, with $Q$ the oscillator's quality factor, is the full width at half maximum (FWHM) of the oscillator response. Equivalently, $1/\kappa$ is the average lifetime of the single-photon state $\ket 1$ before it decays to $\ket 0$. (Inset) Lumped-element LC oscillator of inductance $L$ and capacitace $C$.
  }
  \end{figure}

An LC oscillator is characterized by its inductance $L$ and capacitance $C$ or, equivalently, by its angular frequency $\wc = 1/\sqrt{LC}$ and characteristic impedance $Z_r=\sqrt{L/C}$. The total energy of this oscillator is given by the sum of its charging and inductive energy
\begin{equation}\label{eq:HLC}
H_{LC} = \frac{Q^2}{2C} + \frac{\Phi^2}{2L},
\end{equation}
where $Q$ is the charge on the capacitor and $\Phi$ the flux threading the inductor, see \cref{fig:LCpotential}. 
Charge is related to current, $I$, from charge conservation by $Q(t) = \int_{t_0}^t dt'\, I(t')$, and flux to voltage from Faraday's induction law by $\Phi(t) = \int_{t_0}^t dt'\, V(t')$, where we have assumed that the voltage and current are zero at an initial time $t_0$, often taken to be in the distant past \cite{Vool2017}.
 
It is instructive to rewrite $H_{LC}$ as
\begin{equation}\label{eq:HLCAnalogy}
H_{LC} = \frac{Q^2}{2C} + \frac{1}{2}C\wc^2\Phi^2.
\end{equation}
This form emphasizes the analogy of the LC oscillator with a mechanical oscillator of coordinate $\Phi$, conjugate momentum $Q$, and mass $C$. With this analogy in mind, quantization proceeds in a manner that should be well known to the reader: The charge and flux variables are promoted to non-commuting observables satisfying the commutation relation
\begin{equation}
[\hF,\hQ] = i\hbar.
\end{equation}
It is further useful to introduce the standard annihilation $\aop$ and creation $\ad$ operators of the harmonic oscillator. With the above mechanical analogy in mind, we choose these operators as
\begin{align}\label{eq:HatPhiQ}
\hF = \Fzp(\ad+\aop),
\qquad
\hQ = i\Qzp (\ad-\aop),
\end{align}
with $\Fzp = \sqrt{\hbar/2\wc C} = \sqrt{\hbar Z_r/2}$ and $\Qzp = \sqrt{\hbar\wc C/2} = \sqrt{\hbar/2 Z_r}$ the characteristic magnitude of the zero-point fluctuations of the flux and the charge, respectively. With these definitions, the above Hamiltonian takes the usual form
\begin{equation}
\hH_{LC} = \hbar\wc (\ada+1/2),
\end{equation}
with eigenstates that satisfy $\ada \ket n = n \ket n$ for $n=0,1,2,\dots$ In the rest of this review, we follow the convention of dropping from the Hamiltonian the factor of $\sfrac12$ corresponding to zero-point energy. The action of $\ad = \sqrt{1/2\hbar Z_r}(\hF-iZ_r\hQ)$ is to create a quantized excitation of the flux and charge degrees of freedom of the oscillator or, equivalently of the magnetic and electric fields. In other words, $\ad$ creates a photon of frequency $\wc$ stored in the circuit.

While formally correct, one can wonder if this quantization procedure is relevant in practice. In other words, is it possible to operate LC oscillators in a regime where quantum effects are important? For this to be the case, at least two conditions must be satisfied. First, the oscillator should be sufficiently well decoupled from uncontrolled degrees of freedom such that its energy levels are considerably less broad than their separation. In other words, we require the oscillator's quality factor $Q = \wc/\kappa$, with $\kappa$ the oscillator linewidth or equivalently the photon loss rate, to be large. An approach to treat the environment of a quantum system is described  in~\secref{sec:environment}. Because losses are at the origin of level broadening, superconductors are ideal to reach the quantum regime. In practice, most circuit QED devices are made of thin aluminum films evaporated on low-loss dielectric substrates such as sapphire or high-resistivity silicon wafers. Mainly for its larger superconducting gap, niobium is sometimes used in place of aluminum. In addition to internal losses in the metals forming the LC circuit, care must also be taken to minimize the effect of coupling to the external circuitry that is essential for operating the oscillator. As will be discussed below, large quality factors ranging from $Q \sim 10^{3}$ to $10^{8}$ can be obtained in the laboratory~\cite{Frunzio2005,Bruno2015,Reagor2016}.

Given that your microwave oven has a quality factor approaching $10^4$~\cite{Vollmer2004}, it should not come as a surprise that large Q-factor oscillators can be realized in state-of-the-art laboratories. Relation to kitchen appliances, however, stops here with the second condition requiring that the energy separation $\hbar\wc$ between adjacent eigenstates be larger than thermal energy $k_BT$. Since $1\, \mathrm{GHz} \times h/k_B \sim 50$~mK, the condition $\hbar\wc\gg k_BT$ can be satisfied easily with microwave frequency circuits operated  at $\sim 10$~mK in a dilution refrigerator. These circuits are therefore operated at temperatures far below the critical temperature ($\sim 1-10$ K) of the superconducting films from which they are made.

With these two requirements satisfied, an oscillator with a frequency in the microwave range can be operated in the quantum regime. This means that the circuit can be prepared in its quantum-mechanical ground state $\ket{n=0}$ simply by waiting for a time of the order of a few photon lifetimes $T_\kappa = 1/\kappa$. It is also crucial to note that the vacuum fluctuations in such an oscillator can be made large. For example, taking reasonable values $L\sim0.8$~nH and $C\sim0.4$~pF, corresponding to $\wc/2\pi\sim 8$ GHz and $Z_r\sim 50~\Omega$, the ground state is characterized by vacuum fluctuations of the voltage of variance as large as 
$\Delta V_0 = [\av{\hV^2} - \av{\hat V}^2]^{1/2} = \sqrt{\hbar\wc/2C} \sim 1~\mu$V, with $\hV = \hQ/C$. As will be made clear later, this leads to large electric field fluctuations and therefore to large electric-dipole interactions when coupling to an artificial atom.

\subsection{\label{sec:2D}2D resonators}

Quantum harmonic oscillators come in many shapes and sizes, the	 LC oscillator being just one example. Other types of harmonic oscillators that feature centrally in circuit QED are microwave resonators where the electromagnetic field is confined either in a planar, essentially two-dimensional structure (2D resonators) or in a three-dimensional volume (3D resonators). The boundary conditions imposed by the geometry of these different resonators lead to a discretization of the electromagnetic field into a set of modes with distinct frequencies, where each mode can be thought of as an independent harmonic oscillator. Conversely (especially for the 2D case) one can think of these modes as nearly dissipationless acoustic plasma modes of superconductors.

\begin{figure}[t]
  \centering
  \includegraphics{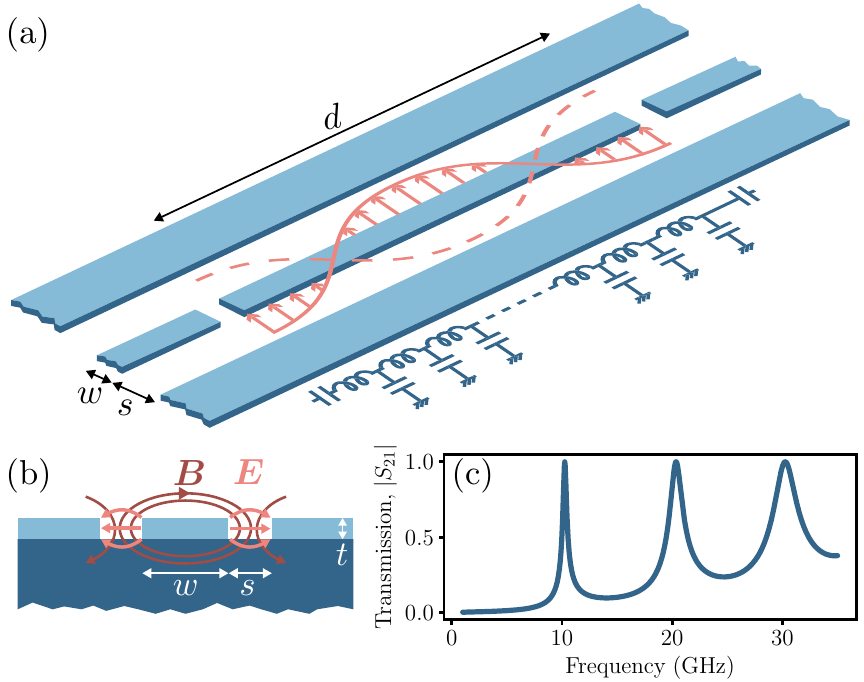}
  \caption{(a) Schematic layout of a $\lambda/2$ coplanar waveguide resonator of length $d$, center conductor width $w$, ground plane separation $s$, together with its capacitively coupled input and output ports. The cosine shape of the second mode function ($m=1$) is illustrated with pink arrows. Also shown is the equivalent lumped element circuit model. Adapted from \textcite{Blais2004}. (b) Cross-section cut of the coplanar waveguide resonator showing the substrate (dark blue), the two ground planes and the center conductor (light blue) as well as schematic representations of the E and B field distributions. (c)~Transmission versus frequency for an overcoupled resonator. The first three resonances of frequencies $f_m = (m+1)f_0$ are illustrated with $f_0 = v_0/2d \sim 10$ GHz and linewidth $\kappa_m/2\pi = f_m/Q$. 
  }  \label{fig:coplanaracrhitecture}
\end{figure}

Early experiments in circuit QED were motivated by the observation of large quality factors in coplanar waveguide resonators in experiments for radiation detectors~\cite{Day2003} and by the understanding of the importance of presenting a clean electromagnetic environment to the qubits. Early circuit QED experiments were performed with these 2D coplanar waveguide resonators \cite{Wallraff2004}, which remains one of the most commonly used architectures today.

A coplanar waveguide resonator consist of a coplanar waveguide of finite length formed by a center conductor of width $w$ and thickness $t$, separated on both sides by a distance $s$ from a ground plane of the same thickness, see Fig.~\ref{fig:coplanaracrhitecture}(a)~\cite{Pozar2012, Simons2001}. Both conductors are typically deposited on a low-loss dielectric substrate 
of permittivity $\varepsilon$ and thickness much larger than the dimensions $w, s, t$. This planar structure acts as a transmission line along which signals are transmitted in a way analogous to a conventional coaxial cable. As in a coaxial cable, the coplanar waveguide confines the electromagnetic field to a small volume between its center conductor and the ground, see Fig.~\ref{fig:coplanaracrhitecture}(b).
The dimensions of the center conductor, the gaps, and the thickness of the dielectric are chosen such that the field is concentrated between the center conductor and ground, and radiation in other directions is minimized. This structure supports a quasi-TEM mode~\cite{Wen1969}, with the electromagnetic field partly in the dielectric substrate and in the vacuum (or other dielectric) above the substrate, and with the largest concentration in the gaps between the center conductor and the ground planes. In practice, the coplanar waveguide can be treated as an essentially dispersion-free, linear dielectric medium. Ideally, the loss of coplanar waveguides is only limited by the conductivity of the center conductor and the ground plane, and by the loss tangent of the dielectric. To minimize losses, superconducting metals such as aluminum, niobium or niobium titanium nitride (NbTiN), are used in combination with dielectrics of low loss tangent, such as sapphire or high-resistivity silicon.

Similarly to the lumped LC oscillator, the electromagnetic properties of a coplanar waveguide resonator are described by its characteristic impedance $\zc = \sqrt{l_0/c_0}$ and the speed of light in the waveguide $v_0 = 1/\sqrt{l_0c_0}$, where we have introduced the capacitance to ground $c_0$ and inductance $l_0$ per unit length~\cite{Simons2001}.
Typical values of these parameters are $\zc \sim 50\,\Omega$ and $v_0 \sim 1.3\times 10^8$ m/s, or about a third of the speed of light in vacuum~\cite{Goppl2008}.
For a given substrate, metal thickness and center conductor width, the characteristic impedance can be adjusted by varying the parameters $w$, $s$ and $t$ of the waveguide \cite{Simons2001}. In the coplanar waveguide geometry, transmission lines of constant impedance $\zc$ can therefore be realized for varying center conductor width $w$ by keeping the ratio of $w/s$ close to a constant \cite{Simons2001}. This allows the experimenter to fabricate a device with large $w$ at the edges for convenient interfacing, and small $w$ away from the edges to minimize the mode volume or simply for miniaturization.

A resonator is formed from a coplanar waveguide by imposing boundary conditions of either zero current or zero voltage  at the two endpoints separated by a distance $d$. Zero current is achieved by micro-fabricating a gap in the center conductor (open boundary), while zero voltage can be achieved by grounding an end point (shorted boundary). A resonator with open boundary conditions at both ends, as illustrated in Fig.~\ref{fig:coplanaracrhitecture}(a), has a fundamental frequency $f_0 = v_0/2d$ with harmonics at $f_m = (m+1) f_0$, and is known as a $\lambda/2$ resonator. On the other hand, $\lambda/4$ resonators with fundamental frequency $f_0 = v_0/4d$ are obtained with one open end and one grounded end. 

A typical example is a $\lambda/2$ resonator of length $1.0$ cm and speed of light $1.3\times 10^8$ m/s corresponding to a fundamental frequency of $6.5$ GHz. This coplanar waveguide geometry is very flexible and a large range of frequencies can be achieved. In practice, however, the useful frequency range is restricted from above by the superconducting gap of the metal from which the resonator is made (82 GHz for aluminum). Above this energy, losses due to quasiparticles increase dramatically. Low frequency resonators can be made by using long, meandering, coplanar waveguides. For example, in \textcite{Sundaresan15}, a resonator was realized with a length of $0.68$ m and a fundamental frequency of $f_0 = 92$ MHz. With this frequency corresponding to a temperature of 4.4 mK, the low frequency modes of such long resonators are, however, not in the vacuum state. Indeed, according to the Bose-Einstein distribution, the thermal occupation of the fundamental mode frequency at 10 mK is $\bar n_\kappa = 1/(e^{h f_0/k_B T}-1) \sim 1.8$. Typical circuit QED experiments rather work with resonators in the range of 5--15 GHz where, conveniently, microwave electronics is well developed.

As already mentioned, entering the quantum regime for a given mode $m$ requires more than $\hbar \omega_m\gg k_B T$. It is also important that the linewidth $\kappa_m$ be small compared to the mode frequency $\omega_m$. As for the LC oscillator, the linewidth can be expressed in terms of the quality factor $Q_m$ of the resonator mode as $\kappa_m = \omega_m/Q_m$. An expression for the linewidth in terms of circuit parameters is given in~\cref{sec:environment}. There are multiple sources of losses and it is common to distinguish between internal losses due to coupling to uncontrolled degrees of freedom (dielectric and conductor losses at the surfaces and interfaces, substrate dielectric losses, non-equilibrium quasiparticles, vortices, two-level fluctuators\ldots) and external losses due to coupling to the input and output ports used to couple signals in and out of the resonator \cite{Goppl2008}. In terms of these two contributions, the total dissipation rate of mode $m$ is $\kappa_m = \kappa_{\mathrm{ext},m} + \kappa_{\mathrm{int},m}$ and the total, or loaded, quality factor of the resonator is therefore $Q_{\mathrm{L},m} = (Q^{-1}_{\mathrm{ext},m} + Q^{-1}_{\mathrm{int},m})^{-1}$. It is always advantageous to maximize the internal quality factor and much effort have been invested in improving resonator fabrication such that values of $ Q_\mathrm{int} \sim 10^5$ are routinely achieved. A dominant source of internal losses in superconducting resonators at low power are believed to be two-level systems (TLSs) that reside in the bulk dielectric, in the metal substrate, and in the metal-vacuum and substrate-vacuum interfaces where the electric field is large~\cite{Sage:2011,Wang2015n}. Internal quality factors over $10^6$ can be achieved by careful fabrication minimizing the occurrence of TLSs and by etching techniques to avoid substrate-vacuum interfaces in regions of high electric fields~\cite{Vissers2010b,Megrant2012,Bruno2015,Calusine2018}. 

On the other hand, the external quality factor can be adjusted by designing the capacitive coupling at the open ends of the resonator to input/output transmission lines. In coplanar waveguide resonators, these input and output coupling capacitors are frequently chosen either as a simple gap of a defined width in the center conductor, as illustrated in Fig.~\ref{fig:coplanaracrhitecture}(a), or formed by interdigitated capacitors  \cite{Goppl2008}. The choice $Q_\mathrm{ext} \ll Q_\mathrm{int}$ corresponding to an `overcoupled' resonator is ideal for fast qubit measurement, which is discussed in more detail in \cref{sec:readout}. On the other hand, undercoupled resonators, $Q_\mathrm{ext} \gg Q_\mathrm{int}$,  where dissipation is only limited by internal losses which are kept as small as possible, can serve as quantum memories to store microwave photons for long times. Using different modes of the same resonator \cite{Leek2010}, or combinations of resonators \cite{Johnson2010, Kirchmair2013}, both regimes of high and low external losses can also be combined in the same circuit QED device. A general approach to describe losses in quantum systems is described in \cref{sec:environment}.

Finally, the magnitude of the vacuum fluctuations of the electric field in coplanar waveguide resonators is related to the mode volume. While the longitudinal dimension of the mode is limited by the length of the resonator, which also sets the fundamental frequency $d \sim \lambda/2$, the transverse dimension can be adjusted over a broad range. Commonly chosen transverse dimensions are on the order of $w \sim 10 \, \rm{\mu m}$ and $s \sim 5\,\rm{\mu m}$ \cite{Wallraff2004}. If desired, the transverse dimension of the center conductor may be reduced to the sub-micron scale, up to a limit set by the penetration depth of the  superconducting thin films which is typically of the order of 100 to 200 nm. When combining the typical separation $s \sim 5\,\rm{\mu m}$ with the magnitude of the voltage fluctuations $\Delta V_0 \sim 1~\mu$V already expected from the discussion of the LC circuit, we find that the zero-point electric field in coplanar resonator can be as large as $\Delta E_0 = \Delta V_0/s \sim 0.2$ V/m. This is at least two orders of magnitude larger than the typical amplitude of $\Delta E_0$ in the 3D cavities used in cavity QED~\cite{Haroche2006}. As will become clear later, together with the large size of superconducting artificial atoms, this will lead to the very large light-matter coupling strengths which are characteristic of circuit QED.

\subsubsection{\label{sec:telegrapher}Quantized modes of the transmission line resonator}

While only a single mode of the transmission line resonator is often considered, there are many circuit QED experiments where the multimode structure of the device plays an important role. In this section, we present the standard approach to finding the normal modes of a distributed resonator, first using a classical description of the circuit.

\begin{figure}[t]
  \centering
  \includegraphics{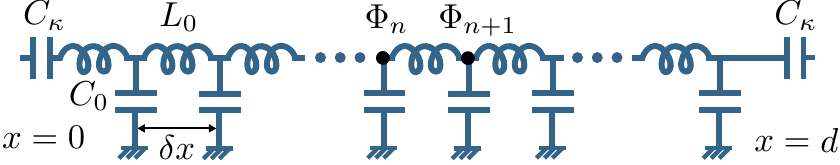}
  \caption{Telegrapher model of an open-ended transmission line resonator of length $d$. $L_0$ and $C_0$ are, respectively, the inductance and capacitance associated to each node $n$ of flux $\Phi_n$. The resonator is coupled to external transmission lines (not shown) at its input and output ports via the capacitors $C_\kappa$.}
\label{fig:TelegrapherResonator}
\end{figure}

The electromagnetic properties along the $x$-direction of a coplanar waveguide resonator of length $d$ can be modeled using a linear, dispersion-free one-dimensional medium. \Cref{fig:TelegrapherResonator} shows the telegrapher model for such a system where the distributed inductance of the resonator's center conductor is represented by the series of lumped elements inductances and the capacitance to ground by a parallel combination of capacitances~\cite{Pozar2011}. Using the flux and charge variables introduced in the description of the LC oscillator, the energy associated to each capacitance is $Q_n^2/2C_0$ while the energy associated to each inductance is $(\Phi_{n+1}-\Phi_n)^2/2L_0$. In these expressions, $\Phi_n$ is the flux variable associated with the $n$th node and $Q_n$ the conjugate variable which is the charge on that node. Using the standard approach~\cite{Devoret1997}, we can thus write the classical Hamiltonian corresponding to \Cref{fig:TelegrapherResonator} as
\begin{equation}
    H = \sum_{n=0}^{N-1}\left[ \frac{1}{2C_0} Q_n^2 + \frac{1}{2L_0}(\Phi_{n+1}-\Phi_n)^2 \right].
\end{equation}
It is useful to consider a continuum limit of this Hamiltonian where the size of a unit cell $\delta x$ is taken to zero. For this purpose, we write $C_0 = \delta x \, c_0$ and $L_0 = \delta x\, l_0$, with $c_0$ and $l_0$  the capacitance and inductance per unit length, respectively. Moreover, we define a continuum flux field via $\Phi(x_n) = \Phi_n$ and charge density field $Q(x_n) = Q_n/\delta x$.
We can subsequently take the continuum limit $\delta x \to 0$ while keeping $d=N\Delta x$ constant to find
\begin{equation}\label{eq:HamiltonianResonatorContinous}
H = \int_{0}^{d} dx\,
\left\{
  \frac{1}{2 c_0} Q(x)^2 + \frac{1}{2l_0}\left[\partial_x \Phi(x)\right]^2
\right\},
\end{equation}
where we have used that $\partial_x \Phi(x_n) = \lim_{\delta x \to 0}\, (\Phi_{n+1}-\Phi_n)/\delta x$. In this expression, the charge $Q(x,t) = c_0 \partial_t \Phi(x,t)$ is the canonical momentum to the generalized flux $\Phi(x,t) = \int_{-\infty}^t dt'\, V(x,t')$, with $V(x,t)$ the voltage to ground on the center conductor.

Using Hamilton's equations together with~\cref{eq:HamiltonianResonatorContinous}, we find that the propagation along the transmission line is described by the wave equation 
\begin{equation}\label{eq:WaveEq}
v_0^2 \frac{\partial^2\Phi(x,t)}{\partial x^2} - \frac{\partial^2\Phi(x,t)}{\partial t^2} = 0,
\end{equation}
with $v_0=1/\sqrt{l_0c_0}$ the speed of light in the medium.
The solution to~\eq{eq:WaveEq} can be expressed in terms of normal modes
\begin{equation}\label{eq:ResonatorModeDecomp}
\Phi(x,t) = \sum_{m=0}^\infty u_m(x) \Phi_m(t),
\end{equation}
with $\ddot\Phi_m = -\omega_m^2\Phi_m$ a function of time oscillating at the mode frequency $\omega_m$ and
\begin{equation}\label{eq:ResonatorModeFunc}
u_m(x) = A_m \cos\left[ k_m x + \varphi_m\right],
\end{equation}
being the spatial profile of the mode with amplitude $A_m$. The  wavevector $k_m = \omega_m/v_0$ and the phase $\varphi_m$ are set by the boundary conditions. For an open-ended $\lambda/2$-resonator these are
\begin{equation}\label{eq:ZeroCurrentBoundary}
I(x=0,d) = -\frac{1}{l_0}\left.\frac{\partial \Phi(x,t)}{\partial x}\right|_{x=0,d} = 0,
\end{equation}
corresponding to the fact that the current vanishes at the two extremities. A $\lambda/4$-resonator is modeled by requiring that the voltage $V(x,t) = \partial_t \Phi(x,t)$ vanishes at the grounded boundary. Asking for~\eq{eq:ZeroCurrentBoundary} to be satisfied for every mode implies that $\varphi_m = 0$ and that the wavevector is discrete with $k_m = m\pi/d$. Finally, it is useful to choose the normalization constant $A_m$ such that
\begin{equation}
\frac 1d \int_{0}^{d} dx\, u_m(x)u_{m'}(x) = \delta_{mm'},
\end{equation}
resulting in $A_m = \sqrt{2}$. This normalization implies that the amplitude of the modes in a 1D resonator goes down with the square root of the length $d$.

Using this normal mode decomposition in~\eq{eq:HamiltonianResonatorContinous}, the Hamiltonian can now be expressed in the simpler form
\begin{equation}\label{eq:NormalModeHResonator}
H = \sum_{m=0}^\infty
\left[
\frac{Q_m^2}{2C_\mathrm{r}} + \frac{1}{2} C_\mathrm{r} \omega_m^2 \Phi_m^2
\right],
\end{equation}
where $C_\mathrm{r} = dc_0$ is the total capacitance of the resonator and $Q_m = C_\mathrm{r} \dot\Phi_m$  the charge  conjugate to $\Phi_m$.
We immediately recognize this Hamiltonian to be a sum over independent harmonic oscillators, cf.~\eq{eq:HLC}.

Following once more the quantization procedure of Sec.~\ref{sec:HO}, the two conjugate variables $\Phi_m$ and $Q_m$ are promoted to non-commuting operators
\begin{align}
  \hF_m &= 
  \sqrt{\frac{\hbar Z_m}{2}}
 (\ad_m+\aop_m),\label{eq:ResonatorFlux}\\
  \hQ_m &= i
  \sqrt{\frac{\hbar}{2 Z_m}}
  (\ad_m-\aop_m),\label{eq:ResonatorCharge}
\end{align}
with $Z_m = \sqrt{L_m/C_\mathrm{r}}$
the characteristic impedance of mode $m$ and $L_m^{-1} \equiv C_r\omega_m^2$. Using these expressions in \cref{eq:NormalModeHResonator} immediately leads to the final result
\begin{equation}\label{eq:H_cavity}
\hH = \sum_{m=0}^\infty \hbar\omega_m \ad_m \aop_m,
\end{equation}
with $\omega_m = (m+1) \omega_0$ the mode frequency and $\omega_0/2\pi = v_0 /2d$ the fundamental frequency of the $\lambda/2$ transmission-line resonator.

To simplify the discussion, we have assumed here that the medium forming the resonator is homogenous. In particular, we have ignored the presence of the input and output port capacitors in the boundary condition of \cref{eq:ZeroCurrentBoundary}. In addition to lowering the external quality factor $Q_\mathrm{ext}$, these capacitances modify the amplitude and phase of the mode functions, as well as shift the mode frequencies. In some contexts, it can also be useful to introduce one or several Josephson junctions directly in the center conductor of the resonator. As discussed in \cref{sec:transmon}, this leads to a Kerr nonlinearity, $-K\hat{a}^{\dag2}\aop^2/2$, of the oscillator. Kerr nonlinear coplanar waveguide resonators have been used, for example, as near quantum-limited linear amplifiers~\cite{Castellanos2008}, bifurcation amplifiers~\cite{Metcalfe2007}, and to study phenomena such as quantum heating~\cite{Ong2013}. We discuss the importance of quantum-limited amplification for qubit measurement in \cref{sec:readout} and applications of the Kerr nonlinearity in the context of quantum optics in \cref{sec:QuantumOptics}. A theoretical treatment of the resonator mode functions, frequencies and Kerr nonlinearity in the presence of resonator inhomogeneities, including embedded junctions, can be found in~\textcite{Bourassa2012} and is discussed in~\secref{sec:bbq}.

\subsection{\label{sec:3D}3D resonators}

Although their physical origin is not yet fully understood, dielectric losses at interfaces and surfaces are important limiting factors to the internal quality factor of coplanar transmission line resonators and lumped element LC oscillators, see~\textcite{Oliver2013b} for a review. An approach to mitigate the effect of these losses is to lower the ratio of the field energy stored at interfaces and surfaces to the energy stored in vacuum. Indeed, it has been observed that planar resonators with larger feature sizes ($s$ and $w$), and hence weaker electric fields near the interfaces and surfaces, typically have larger internal quality factors~\cite{Sage:2011}.

This approach can be pushed further by using three-dimensional microwave cavities rather than planar circuits~\cite{Paik2011}. In 3D resonators formed by a metallic cavity, a larger fraction of the field energy is stored in the vacuum inside the cavity rather than close to the surface. As a result, the surface participation ratio can be as small as $10^{-7}$ in 3D cavities, in comparison to $10^{-5}$ for typical planar geometries \cite{Reagor2015}. Another potential advantage is the absence of dielectric substrate. In practice, however, this does not lead to a major gain in quality factor since, while coplanar resonators can have air-substrate participation ratio as large as 0.9, the bulk loss tangent of sapphire and silicon substrate is significantly smaller than that of the interface oxides and does not appear to be the limiting factor~\cite{Wang2015n}.

In practice, three-dimensional resonators come in many different shapes and sizes, and can reach higher quality factors than lumped-element oscillators and 1D resonators. Quality factors has high as $4.2\times 10^{10}$ have been reported at 51 GHz and 0.8 K with Fabry-P\'erot cavities formed by two highly-polished cooper mirrors coated with niobium~\cite{Kuhr2007}. Corresponding to single microwave photon lifetimes of 130 ms, these cavities have been used in landmark cavity QED experiments~\cite{Haroche2006}. Similar quality factors have also been achieved with niobium micromaser cavities at 22~GHz and 0.15 K~\cite{Varcoe2000}. In the context of circuit QED, commonly used geometries include rectangular~\cite{Paik2011,Rigetti2012} and coaxial $\lambda/4$ cavities~\cite{Reagor2016}. The latter have important practical advantages in that no current flows near any seams created in the assembly of the device.

\begin{figure}
  \centering
  \includegraphics[width=0.9\columnwidth]{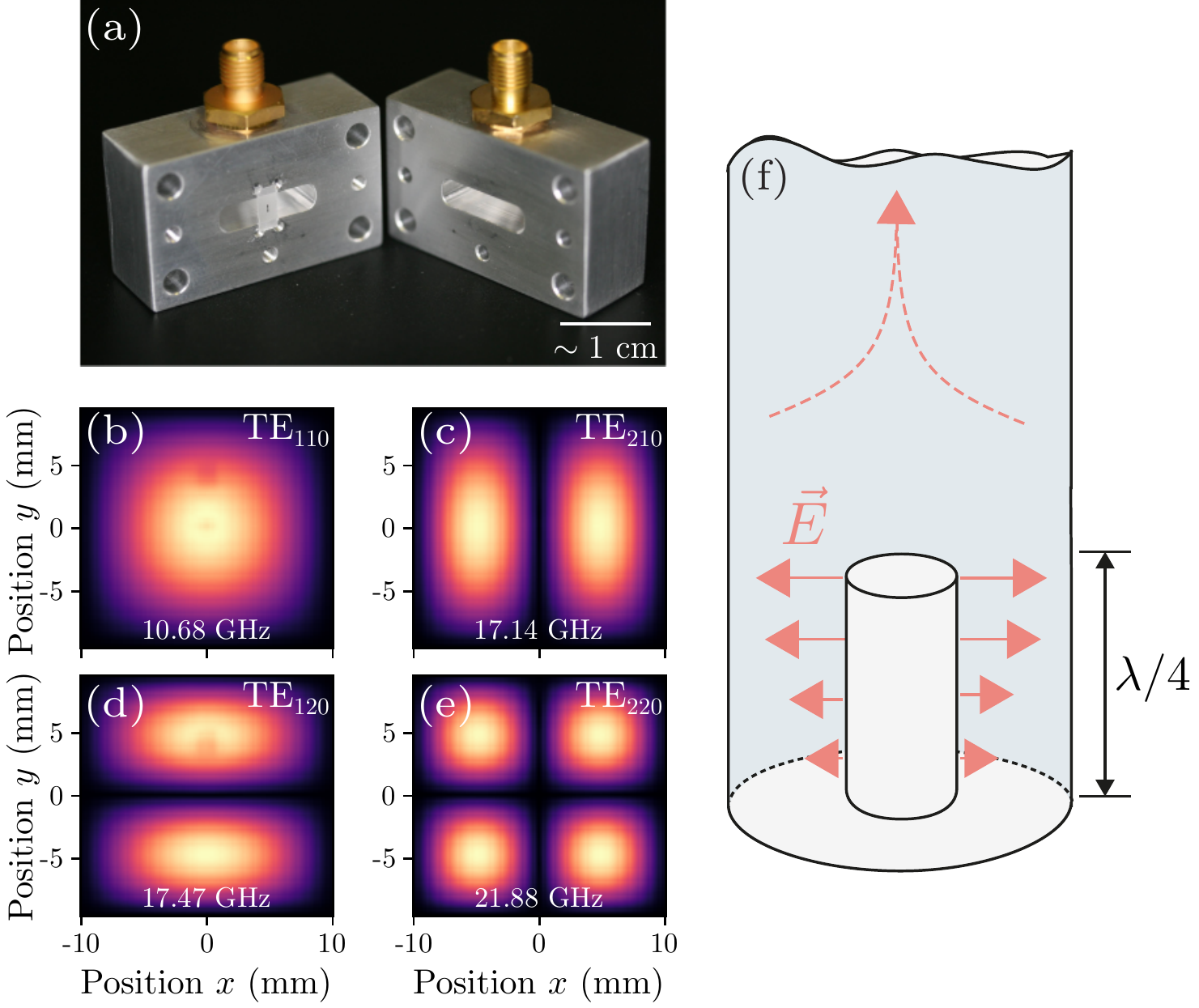}
 \caption{\label{fig:3Dmodes} (a) Photograph of a 3D rectangular superconducting cavity showing interior volume of the waveguide enclosure housing a sapphire chip and transmon qubit, with two symmetric coaxial connectors for coupling signals in and out. Credit: IBM. (b-e) First four $\mathrm{TE}_{mnl}$ modes of a 3D rectangular superconducting cavity obtained from COMSOL\texttrademark. Credit: Dany Lachance-Quirion. (f) Schematic representation of a coaxial $\lambda/4$ cavity with electric field (full line) pointing from the inner conductor to the sidewalls and  evanescent field (dashed line) rapidly decaying from the top of the inner conductor. Adapted from \textcite{Reagor2016}.}
\end{figure}

As illustrated in \cref{fig:3Dmodes}(a) and in close analogy with the coplanar waveguide resonator, rectangular cavities are formed by a finite section of a rectangular waveguide terminated by two metal walls acting as shorts. This three-dimensional resonator is thus simply vacuum surrounded on all sides by metal, typically aluminum to maximize the internal quality factor or copper if magnetic field tuning of components placed inside the cavity is required. The metallic walls impose boundary conditions on the electromagnetic field in the cavity, leading to a discrete set of TE and TM cavity modes of frequency~\cite{Pozar2012}
\begin{equation}
\omega_{mnl} = c\sqrt{
\left(\frac{m\pi}{a}\right)^2
+
\left(\frac{n\pi}{b}\right)^2
+
\left(\frac{l\pi}{d}\right)^2
},
\end{equation}
labelled by the three integers $(l,m,n)$ and where $c$ is the speed of light, and $a$, $b$ and $d$ are the cavity dimensions. Dimensions of the order of a centimeter lead to resonance frequencies in the GHz range for the first modes. The TE modes, to which superconducting artificial atoms couple, are illustrated in \cref{fig:3Dmodes}(b-e). Because these modes are independent, once quantized, the cavity Hamiltonian again takes the form of \cref{eq:H_cavity} corresponding to a sum of independent harmonic oscillators. We return to the question of quantizing the electromagnetic field in arbitrary geometries in \secref{sec:bbq}.

As already mentioned, a major advantage of 3D cavities compared to their 1D or lumped-element analogs is their high quality factor or, equivalently, long photon lifetime. A typical internal Q factor for rectangular aluminum cavities is $5\times 10^6$, corresponding to a photon lifetime above 50~$\mu$s~\cite{Paik2011}. These numbers are even higher for coaxial cavities where $Q_\mathrm{int} = 7 \times 10^7$, or above a millisecond of photon storage time, has been reported~\cite{Reagor2016}. Moreover, this latter type of cavity is more robust against imperfections which arise when integrating 3D resonators with Josephson-junction-based circuits. Lifetimes up to 2 s have also been reported in niobium cavities that were initially developed for accelerators \cite{Romanenko2020}. At such long photon lifetimes, microwave cavities are longer-lived quantum memories than the transmon qubit which we will introduce in the next section. This has led to a new paradigm for quantum information processing where information is stored in a cavity with the role of the qubit limited to providing the essential nonlinearity~\cite{Mirrahimi2014}. We come back to these ideas in \cref{sec:CatCodes}.

\subsection{\label{sec:transmon}The transmon artificial atom}

Although the oscillators discussed in the previous section can be prepared in their quantum mechanical ground state, it is challenging to observe clear quantum behavior with such linear systems. Indeed, harmonic oscillators are always in the correspondence limit and some degree of nonlinearity is therefore essential to encode and manipulate quantum information in these systems \cite{Leggett1984a}. Fortunately, superconductivity allows to introduce nonlinearity in quantum electrical circuits while avoiding losses. 
Indeed, the Josephson junction is a nonlinear circuit element that is compatible with the requirements for very high quality factors and operation at millikelvin temperatures. 
The physics of these junctions was first understood in 1962 by Brian Josephson, then a 22-year-old PhD candidate~\cite{Josephson1962,McDonald2001}. 

Contrary to expectations~\cite{Bardeen1962}, Josephson showed that a dissipationless current, i.e.~a supercurrent, could flow between two superconducting electrodes separated by a thin insulating barrier. More precisely, he showed that this supercurrent is given by $I = I_c \sin\varphi$, where $I_c$ is the junction's critical current and $\varphi$ the phase difference between the superconducting condensates on either sides of the junction~\cite{Tinkham1996}. The critical current, whose magnitude is determined by the junction size and material parameters, is the maximum current that can be supported before Cooper pairs are broken. Once this happens, dissipation kicks in and a finite voltage develops across the junction accompanied by a resistive current.
Clearly, operation in the quantum regime requires currents well below this critical current. Josephson also showed that the time dependence of the phase difference $\varphi$ is related to the voltage across the junction according to $d\varphi/dt = 2\pi V /\Phi_0$, with $\Phi_0 = h/2e$ the flux quantum. It is useful to write this expression as $\varphi(t) = 2\pi \Phi(t)/\Phi_0~(\mathrm{mod}~2\pi)= 2\pi \int dt'\, V(t')/\Phi_0~(\mathrm{mod}~2\pi)$, with $\Phi(t)$ the flux variable already introduced in \secref{sec:HO}. The mod $2\pi$ in the previous equalities takes into account the fact that the superconducting phase $\varphi$ is a compact variable on the unit circle, $\varphi = \varphi+2\pi$, while $\Phi$ can take arbitrary real values.

Taken together, the two Josephson relations make it clear that a Josephson junction relates current $I$ to flux $\Phi$. This is akin to a geometric inductance whose constitutive relation $\Phi = L I$ also links these two quantities. For this reason, it is useful to define the Josephson inductance
\begin{equation}\label{eq:LJ}
L_J (\Phi)
= \left(\frac{\partial I}{\partial \Phi}\right)^{-1}
= \frac{\Phi_0}{2\pi I_c} \frac{1}{\cos(2\pi\Phi/\Phi_0)}.
\end{equation}
In contrast to geometric inductances, $L_J$ depends on the flux.
As a result, when operated below the critical current, the Josephson junction can be thought of as a nonlinear inductor. 

\begin{figure}
\centering
\includegraphics{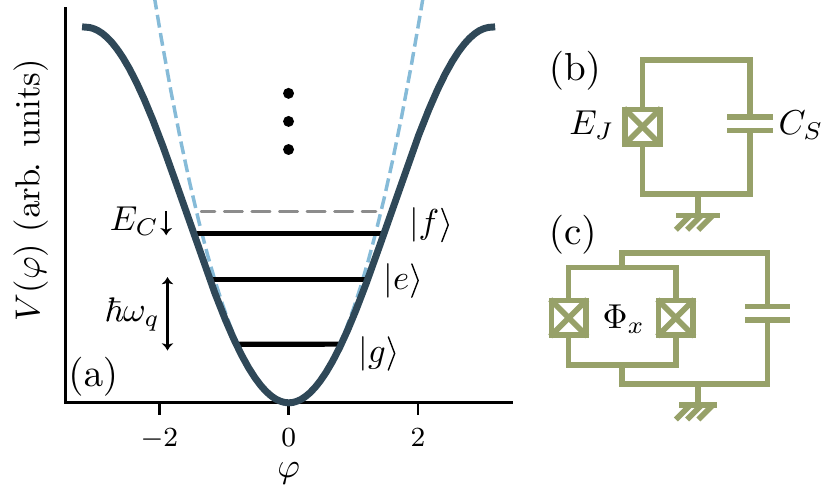}
\caption{\label{fig:transmonpotential}
(a) Cosine potential well of the transmon qubit (full line) compared to the quadratic potential of the LC oscillator (dashed lines). The spectrum  of the former as eigenstates labelled $\{\ket{g},\ket{e},\ket{f},\ket{h}\ldots\}$ and is characterized by an anharmonicity $-E_C$. (b) Circuit for the fixed-frequency transmon qubit. The square with a cross represents a Josephson junction with Josephson energy $E_J$ and junction capacitance $C_J$. (c) By using a SQUID rather than a single junction, the  frequency of the transmon qubit becomes flux tunable.
}
\end{figure}

Replacing the geometric inductance $L$ of the LC oscillator discussed in \secref{sec:HO} by a Josephson junction, as in \cref{fig:transmonpotential}(b), therefore renders the circuit nonlinear. In this situation, the energy levels of the circuit are no longer equidistant. If the nonlinearity and the quality factor of the junction are large enough, the energy spectrum resembles that of an atom, with well-resolved and nonuniformly spread spectral lines. We therefore often refer to this circuit as an \emph{artificial atom}~\cite{Clarke1988}. In many situations, and as is the focus of much of this review, we can furthermore restrict our attention to only two energy levels, typically the ground and first excited states, forming a qubit.

To make this discussion more precise, it is useful to see how the Hamiltonian of the circuit of \cref{fig:transmonpotential}(b) is modified by the presence of the Josephson junction taking the place of the linear inductor. While the energy stored in a linear inductor is $E = \int dt\, V(t) I(t) = \int dt\, (d\Phi/dt) I = \Phi^2/2L$, where we have used $\Phi = L I$ in the last equality, the energy of the nonlinear inductance rather takes the form
\begin{equation}
E  = I_c \int dt\, \left(\frac{d\Phi}{dt}\right) \sin\left(\frac{2\pi}{\Phi_0}\Phi\right)  = - E_J \cos\left(\frac{2\pi}{\Phi_0}\Phi\right),
\end{equation}
with $E_J = \Phi_0I_c/2\pi$ the Josephson energy. This quantity is proportional the rate of tunnelling of Cooper pairs across the junction. Taking into account this contribution, the quantized Hamiltonian of the capacitively shunted Josephson junction therefore reads
(see \cref{sec:AppendixTRcoupling})
\begin{equation}\label{eq:Hsj}
\begin{split}
\hH_\mathrm{T}
&= \frac{(\hQ-Q_g)^2}{2C_\Sigma} - E_J \cos\left(\frac{2\pi}{\Phi_0}\hF\right) \\ 
& = 4E_C (\hn-n_g)^2 - E_J \cos\hvphi.
\end{split}
\end{equation}
In this expression, $C_\Sigma=C_J+C_S$ is the total capacitance, including the junction's capacitance $C_J$ and the shunt capacitance $C_S$. In the second line, we have defined the charge number operator $\hn = \hQ/2e$, the phase operator $\hvphi = (2\pi/\Phi_0)\hat\Phi$ (mod $2\pi$) and the charging energy $E_C = e^2/2C_\Sigma$.
We have also included a possible offset charge $n_g = Q_g/2e$ due to capacitive coupling of the transmon to external charges. The offset charge can arise from spurious unwanted degrees of freedom in the transmon's environment or from an external gate voltage $V_g = Q_g/C_g$. As we show below, the choice of $E_J$ and $E_C$ is crucial in determining the system's sensitivity to the offset charge.

\begin{figure}
\centering
\includegraphics{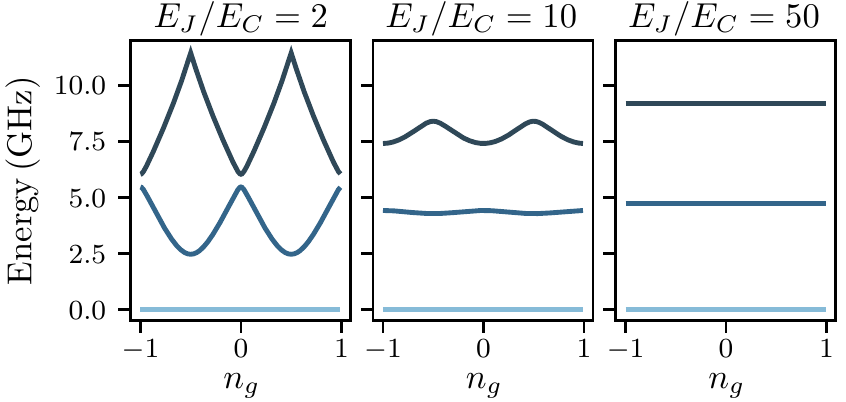}
\caption{\label{fig:transmonregime}Frequency difference $\omega_j-\omega_0$ for the first three energy levels of the transmon Hamiltonian obtained from numerical diagonalization of~\cref{eq:Hsj} expressed in the charge basis $\{\ket{n}\}$ for different $E_J/E_C$ ratios and a fixed plasma frequency $\omega_p/2\pi = 5$ GHz. For large values of $E_J/E_C$ the energy levels become insensitive to the offset charge $n_g$. }
\end{figure}

The spectrum of $\hat H_T$ is controlled by the ratio $E_J/E_C$, with different values of this ratio corresponding to different types of superconducting qubits; see for example the reviews~\cite{Makhlin2001,Zagoskin2007,Clarke2008,Kjaergaard2019}. Regardless of the parameter regime, one can always express the Hamiltonian in the diagonal form 
 $\hH = \sum_j \hbar \omega_j \ket j \bra j$
in terms of its eigenfrequencies $\omega_j$ and eigenstates $\ket j$.
In the literature, two notations are commonly used to label these eigenstates: $\{\ket{g},\ket{e},\ket{f},\ket{h}\ldots\}$ and, when there is not risk of confusion with resonator Fock states, $\{\ket{0},\ket{1},\ket{2}\ldots\}$. Depending on the context, we will use both notation in this review. 
\Cref{fig:transmonregime} shows the energy difference $\omega_j-\omega_0$ for the three lowest energy levels for different ratios $E_J/E_C$ as obtained from numerical diagonalization of \cref{eq:Hsj}. If the charging energy dominates, $E_J/E_C < 1$, the eigenstates of the Hamiltonian are approximately given by eigenstates of the charge operator, $\ket j \simeq \ket n$, with $\hat n\ket n = n\ket n$. In this situation, a change in gate charge $n_g$ has a large impact on the transition frequency of the device. As a result, unavoidable charge fluctuations in the circuit's environment lead to corresponding fluctuations in the qubit transition frequency and consequently to dephasing.

To mitigate this problem, a solution is to work in the \emph{transmon regime} where the ratio $E_J/E_C$ is large~\cite{Koch2007,Schreier2008}. Typical values are  $E_J/E_C \sim 20 - 80$. In this situation, the charge degree of freedom is highly delocalized due to the large Josephson energy. For this reason, and as clearly visible in \cref{fig:transmonregime}(c), the first energy levels of the device become essentially independent of the the gate charge. It can in fact be shown that the charge dispersion, which describes the variation of the energy levels with gate charge, decreases exponentially with $E_J/E_C$ in the transmon regime~\cite{Koch2007}. The net result is that the coherence time of the device is much larger than at small $E_J/E_C$. However, as is also clear from \cref{fig:transmonregime}, the price to pay for this increased coherence is the reduced anharmonicity of the transmon, anharmonicity that is required to control the qubit without causing unwanted transitions to higher excited states. 
Fortunately, while charge dispersion is exponentially small with $E_J/E_C$, the loss of anharmonicity has a much weaker dependence on this ratio given by $\sim (E_J/E_C)^{-1/2}$.
As will be discussed in more detail in \cref{sec:quantumcomputing}, because of the gain in coherence, the reduction in anharmonicity is not an impediment to controlling the transmon state with high fidelity.

While the variance of the charge degree of freedom is large when $E_J/E_C\gg 1$, the variance of its conjugate variable $\hvphi$ is correspondingly small, with $\Delta\hvphi = \sqrt{\braket{\hvphi^2}-\braket{\hvphi}^2} \ll1$. In this situation, it is instructive to rewrite \eq{eq:Hsj} as 
\begin{equation}\label{eq:HsjLinNonLin}
\begin{split}
\hH_\mathrm{T}  = 4E_C \hn^2 +  \frac{1}{2}E_J \hvphi^2 - E_J\left(\cos\hvphi + \frac{1}{2}\hvphi^2\right),
\end{split}
\end{equation}
the first two terms corresponding to an LC circuit of capacitance $C_\Sigma$ and inductance $E_J^{-1}(\Fnot/2\pi)^2$, the linear part of the Josephson inductance \eq{eq:LJ}.
We have dropped the offset charge $n_g$ in~\cref{eq:HsjLinNonLin} on the basis that the frequency of the relevant low-lying energy levels is insensitive to this parameter. Importantly, although these energies are not sensitive to variations in $n_g$, it is still possible to use an external oscillating voltage source to cause transition between the transmon states. We come back to this later. The last term of \cref{eq:HsjLinNonLin} is the nonlinear correction to this harmonic potential which, for $E_J/E_C\gg 1$ and therefore $\Delta\hvphi\ll1$ can be truncated to its first nonlinear correction
\begin{equation}\label{eq:HsjPhi4}
\begin{split}
\hH_\mathrm{q}  = 4E_C \hn^2 + \frac{1}{2}E_J \hvphi^2 - \frac{1}{4!}E_J\hvphi^4.
\end{split}
\end{equation}
As expected from the above discussion, the transmon is thus a weakly anharmonic oscillator.  Note that in this approximation, the phase $\hvphi$ is treated as a continuous variable with eigenvalues extending to arbitrary real values, rather than enforcing $2\pi$-periodicity. This is allowed as long as the device is in a regime where $\hvphi$ is sufficiently localized, which holds for low-lying energy eigenstates in the transmon regime with $E_J/E_C \gg 1$.

Following the previous section, it is then useful to introduce creation and annihilation operators chosen to diagonalize the first two terms of \eq{eq:HsjPhi4}. Denoting these operators $\bd$ and $\bop$, in analogy to \eq{eq:HatPhiQ} we have
\begin{align}
\hvphi &= \left(\frac{2 E_C}{E_J}\right)^{1/4}(\bd+\bop),\label{eq:phiTransmon}\\
\quad
\hn &= \frac{i}{2}\left(\frac{E_J}{2 E_C}\right)^{1/4}(\bd-\bop).\label{eq:nTransmon}
\end{align} 
This form makes it quite clear that fluctuations of the phase $\hvphi$ decrease with $E_J/E_C$, while the reverse is true for the conjugate charge $\hn$. Using these expressions in \eq{eq:HsjPhi4} finally leads to\footnote{The approximate Hamiltonian \cref{eq:HsjPhi4b} is not bounded from below -- an artefact of the truncation of the cosine operator. Care should therefore be taken when using this form, and it should strictly speaking only be used in a truncated subspace of the original Hilbert space.}
\begin{equation}\label{eq:HsjPhi4b}
\begin{split}
\hH_\mathrm{q}
&= \sqrt{8E_CE_J}\bdb - \frac{E_C}{12}(\bd+\bop)^4\\
&\approx \hbar\omega_q \bdb - \frac{E_C}{2}\bd\bd\bop\bop,
\end{split}
\end{equation}
where $\hbar\omega_q = \sqrt{8E_CE_J}-E_C$. In the second line,  we have kept only terms that have the same number of creation and annihilation operators. This is reasonable because, in a frame rotating at $\omega_q$, any terms with an unequal number of $\bop$ and $\bd$ will be oscillating. If the frequency of these oscillations is larger than the prefactor of the oscillating term, then this term rapidly averages out and can be neglected~\cite{Cohen-Tannoudji1977}. This rotating-wave approximation (RWA) is valid here if $\hbar\omega_q \gg E_C/4$, an inequality that is easily satisfied in the transmon regime.

The particular combination $\omega_p = \sqrt{8E_CE_J}/\hbar$ is known as the Josephson plasma frequency and corresponds to the frequency of small oscillations of the effective particle of mass $C$ at the bottom of a well of the cosine potential of the Josephson junction. In the transmon regime, this frequency is renormalized by a `Lamb shift' equal to the charging energy $E_C$ such that $\omega_q = \omega_p-E_C/\hbar$ is the transition frequency between ground and first excited state. 
Finally, the last term of \eq{eq:HsjPhi4b} is a Kerr nonlinearity, with $E_C/\hbar$ playing the role of Kerr frequency shift per excitation of the nonlinear oscillator~\cite{Walls2008}. To see this even more clearly, it can be useful to rewrite \eq{eq:HsjPhi4b} as $H_\mathrm{q} = \hbar\tilde \omega_q(\bdb)\bdb$, where the frequency $\tilde \omega_q(\bdb) = \omega_q-E_C(\bdb-1)/2\hbar$ of the oscillator is a decreasing function of the excitation number $\bdb$. Considering only the first few levels of the transmon, this simply means that the $e$--$f$ transition frequency is smaller by $E_C$ than the $g$--$e$ transition frequency, see \cref{fig:transmonpotential}(a). In other words, in the regime of validity of the approximation made to obtain \cref{eq:HsjPhi4}, the anharmonicity of the transmon is $-E_C$ with a typical value of $E_C/h \sim 100$--$400$~MHz~\cite{Koch2007}. Corrections to the anharmonicity from $-E_C$ can be obtained numerically or by keeping higher-order terms in the expansion of \cref{eq:HsjPhi4}.

While the nonlinearity $E_C/\hbar$ is small with respect to the oscillator frequency $\omega_q$, it is in practice much larger than the spectral linewidth that can routinely be obtained for these artificial atoms and can therefore easily be spectrally resolved. As a result, and in contrast to more traditional realizations of Kerr nonlinearities in quantum optics, it is possible with superconducting quantum circuits to have a large Kerr nonlinearity even at the single-photon level. Some of the many implications of this observation will be discussed further in this review. For quantum information processing, the presence of this nonlinearity is necessary to address only the ground and first excited state without unwanted transition to other states. In this case, the transmon acts as a two-level system, or qubit. However, it is important to keep in mind that the transmon is a multilevel system and that it is often necessary to include higher levels in the  description of the device to quantitatively explain experimental observations. These higher levels can also be used to considerable advantage in some cases \cite{Rosenblum2018b,Reinhold2019,Elder2020,Ma2019a}.

\subsection{\label{sec:tunabletransmon}Flux-tunable transmons}

A useful variant of the transmon artificial atom is the flux-tunable transmon, where the single Josephson junction is replaced with two parallel junctions forming a superconducting quantum interference device (SQUID), see \cref{fig:transmonpotential}(c)~\cite{Koch2007}. The transmon Hamiltonian then  reads
\begin{equation}
\begin{split}
\hH_\mathrm{T}
& = 4E_C \hn^2 - E_{J1} \cos\hvphi_1 - E_{J2} \cos\hvphi_2,
\end{split}
\end{equation}
where $E_{Ji}$ is the Josephson energy of junction $i$, and $\hvphi_i$ the phase difference across that junction.
In the presence of an external flux $\Phi_x$ threading the SQUID loop and in the limit of negligible geometric inductance of the loop, flux quantization requires that $\hvphi_1-\hvphi_2 = 2\pi\Phi_x/\Phi_0~(\bmod\,2\pi)$~\cite{Tinkham1996}. Defining the average phase difference $\hvphi = (\hvphi_1+\hvphi_2)/2$, the Hamiltonian can then be rewritten as~\cite{Koch2007,Tinkham1996}
\begin{equation}\label{eq:HTransmonFluxTunable}
  \hH_\mathrm{T} = 4E_C \hn^2 - E_J(\Phi_x)\cos(\hvphi-\varphi_0),
\end{equation}
where
\begin{equation}
  \begin{aligned}
  E_J(\Phi_x) ={}
  E_{J\Sigma}
  \cos\left(\frac{\pi\Phi_x}{\Phi_0}\right)
  \sqrt{1+d^2\tan^2\left(\frac{\pi\Phi_x}{\Phi_0}\right)},
  \end{aligned}
\end{equation}
with $E_{J\Sigma} = E_{J2}+E_{J1}$ and $d=(E_{J2}-E_{J1})/E_{J\Sigma}$ the junction asymmetry. The phase $\varphi_0 = d \tan(\pi\Phi_x/\Phi_0)$ can be ignored for a time-independent flux~\cite{Koch2007}. According to \cref{eq:HTransmonFluxTunable}, replacing the single junction with a SQUID loop yields an effective flux-tunable Josephson energy $E_J(\Phi_x)$. In turn, this results in a flux tunable transmon frequency $\omega_q(\Phi_x) = \sqrt{8E_C|E_J(\Phi_x)|}-E_C/\hbar$.~\footnote{The absolute value arises because when expanding the Hamiltonian in powers of $\hvphi$ in \cref{eq:HsjPhi4}, the potential energy term must always be expanded around a minimum. This discussion also assumes that the ratio $|E_J(\Phi_x)|/E_C$ is in the transmon range for all relevant $\Phi_x$.} In practice, the transmon frequency can be tuned by as much as one GHz in as little as 10--20 ns \cite{Rol2019a,Rol2020}. Dynamic range can also be traded for faster flux excursions by increasing the bandwidth of the flux bias lines. This possibility is exploited in several applications, including for quantum logical gates as discussed in more detail in~\cref{sec:quantumcomputing}.

As will become clear later, this additional control knob can lead to dephasing due to noise in the flux threading the SQUID loop. With this in mind, it is worth noticing that a larger asymmetry $d$ leads to a smaller range of tunability and thus also to less susceptibility to flux noise~\cite{Hutchings2017}. Finally, first steps towards realizing voltage tunable transmons where a semiconducting nanowire takes the place of the SQUID loop have been demonstrated~\cite{Casparis2018,Luthi2018}.

\subsection{Other superconducting qubits}

While the transmon is currently the most extensively used and studied superconducting qubit, many other types of superconducting artificial atoms are used in the context of circuit QED. In addition to working with different ratios of $E_J/E_C$, these other qubits vary in the number of Josephson junctions and the topology of the circuit in which these junctions are embedded. This includes charge qubits~\cite{Bouchiat1998, Shnirman1997,Nakamura1999}, flux qubits~\cite{Orlando1999,Mooij1999} including variations with a large shunting capacitance~\cite{You2007,Yan2016a}, phase qubits~\cite{Martinis2002}, the quantronium~\cite{Vion02}, the fluxonium~\cite{Manucharyan2009} and the $0-\pi$ qubit~\cite{Gyenis2019a,Brooks2013}, amongst others. For more details about these different qubits, the reader is referred to reviews on the topic~\cite{Makhlin2001,Zagoskin2007,Clarke2008,Krantz2019,Kjaergaard2019}.

\section{\label{sec:LightMatter} Light-matter interaction in circuit QED}

\subsection{Exchange interaction between a transmon and an oscillator}
\label{sec:TransmonOscillator}

\begin{figure}
  \centering
  \includegraphics{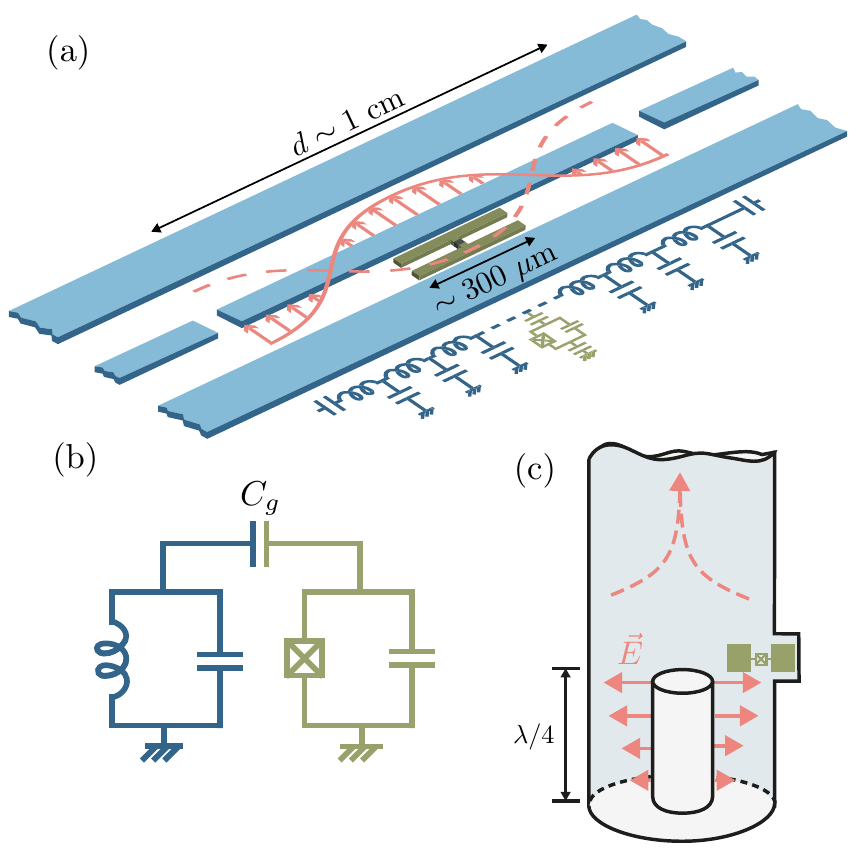}
  \caption{Schematic representation of a transmon qubit (green) coupled to (a) a 1D transmission-line resonator, (b) a lumped-element LC circuit and (c) a 3D coaxial cavity. Panel (a) is adapted from \textcite{Blais2004} and panel (c) from \textcite{Reagor2016}.
  } \label{fig:transmonCQED}
\end{figure}

Having introduced the two main characters of this review, the quantum harmonic oscillator and the transmon artificial atom, we are now ready to consider their interaction. Because of their large size coming from the requirement of having a low charging energy (large capacitance), transmon qubits can very naturally be capacitively coupled to microwave resonators, see \cref{fig:transmonCQED} for schematic representations. With the resonator taking the place of the classical voltage source $V_g$, capacitive coupling to a resonator can be introduced in the transmon Hamiltonian~\cref{eq:Hsj} with a quantized gate voltage $n_g \to -\hn_r$,  representing the charge bias of the transmon due to the resonator
(the choice of sign is simply a common convention in the literature that we will adopt here, see \cref{sec:AppendixTRcoupling}). The Hamiltonian of the combined system is therefore \cite{Blais2004}
\begin{equation}\label{eq:HTransmonResonator}
\hH = 4E_C (\hn+\hn_r)^2 - E_J \cos\hvphi - \sum_m\hbar\omega_m \hat a_m^\dagger \hat a_m,
\end{equation}
where
$\hn_r = \sum_m \hn_m$ with $\hat n_m = (C_g/C_m) \hat Q_m/2e$ the contribution to the charge bias due to the $m$th resonator mode. Here, $C_g$ is the gate capacitance and $C_m$ the associated resonator mode capacitance. To simplify these expressions, we have assumed here that $C_g \ll C_\Sigma, C_m$. A derivation of the Hamiltonian of \cref{eq:HTransmonResonator} that goes beyond the simple replacement of $n_g$ by $-\hn_r$ and without the above assumption can be found in~\cref{sec:AppendixTRcoupling} for the case of a single LC oscillator coupled to the transmon.

Assuming that the transmon frequency is much closer to one of the resonator modes than all the other modes, say $|\omega_0-\omega_q| \ll |\omega_m-\omega_q|$ for $m\ge 1$, we truncate the sum over $m$ in \cref{eq:HTransmonResonator} to a single term. In this single-mode approximation, the Hamiltonian reduces to a single oscillator of frequency denoted $\wc$ coupled to a transmon. It is interesting to note that, regardless of the physical nature of the oscillator---for example a single mode of a 2D or 3D resonator---it is possible to represent this Hamiltonian with an equivalent circuit where the transmon is capacitively coupled to an LC oscillator as illustrated in \figref{fig:transmonCQED}(b). This type of formal representation of complex geometries in terms of equivalent lumped element circuits is generally known as ``black-box quantization''~\cite{Nigg2012}, and is explored in more detail in \secref{sec:bbq}. As will be discussed in \cref{sec:Purcell}, there are many situations of experimental relevance where ignoring the multi-mode nature of the resonator leads to inaccurate predictions.

Using the creation and annihilation operators introduced in the previous section, in the single-mode approximation \eq{eq:HTransmonResonator} reduces to
\footnote{One might worry about the term $\hn_r^2$ arising from~\cref{eq:HTransmonResonator}. However, this term can be absorbed in the charging energy term of the resonator mode, see \cref{eq:HLC}, and therefore leads to a renormalization of the resonator frequency which we omit here for simplicity. See~\cref{eq:app:cq:HTransmonResonatorNoApprox,eq:app:cq:HTransmonResonator} for further details.}
\begin{equation}\label{eq:HTransmonRabi}
\begin{aligned}
  \hH \approx{}& \hbar\wc \ada
  + \hbar\omega_q \bdb - \frac{E_C}{2}\bd\bd\bop\bop\\
  &- \hbar g (\bd-\bop)(\ad-\aop),
\end{aligned}
\end{equation}
where $\wc$ is the frequency of the mode of interest and 
\begin{equation}\label{eq:gTransmon}
  \begin{aligned}
    g 
  ={}& \wc \frac{C_g}{C_\Sigma} \left(\frac{E_J}{2E_C}\right)^{1/4} \sqrt{\frac{\pi \zc}{R_K}},
  \end{aligned}
\end{equation}
the oscillator-transmon, or light-matter, coupling constant. Here, $\zc$ is the characteristic impedance of the resonator mode and $R_K = h/e^2 \sim 25.8$~k$\Omega$ the resistance quantum. The above Hamiltonian can be simplified further in the experimentally relevant situation where the coupling constant is much smaller than the system frequencies, $|g| \ll \wc,\, \omega_q$. Invoking the rotating-wave approximation, it simplifies to
\begin{equation}\label{eq:HTransmonJC}
\begin{aligned}
  \hH \approx{}& \hbar\wc \ada
  + \hbar\omega_q \bdb - \frac{E_C}{2}\bd\bd\bop\bop\\
  &+ \hbar g (\bd\aop+\bop\ad).
\end{aligned}
\end{equation}

As can be seen from \eq{eq:nTransmon}, the prefactor $(E_J/2E_C)^{1/4}$ in \eq{eq:gTransmon} is linked to the size of charge fluctuations in the transmon. By introducing a length scale $l$ corresponding to the distance a Cooper pair travels when tunneling across the transmon's junction, it is tempting to interpret \eq{eq:gTransmon} as $\hbar g = d_0\mathcal{E}_0$ with $d_0 = 2e l (E_J/32 E_C)^{1/4}$ the dipole moment of the transmon and $\mathcal{E}_0= (\wc/l)(C_g/C_\Sigma)\sqrt{\hbar \zc/2}$ the resonator's  zero-point electric field as seen by the transmon. Since these two factors can be made large, especially so in the transmon regime where $d_0 \gg 2e l$, the electric-dipole interaction strength $g$ can be made very large, much more so than with natural atoms in cavity QED. It is also instructive to express \cref{eq:gTransmon} as 
\begin{equation}\label{eq:g_alpha}
    g = \wc \frac{C_g}{C_\Sigma} \left(\frac{E_J}{2E_C}\right)^{1/4} \sqrt{\frac{\zc}{Z_\mathrm{vac}}}\sqrt{2\pi\alpha},
\end{equation}
where $\alpha = Z_\mathrm{vac}/2R_K$ is the fine-structure constant and $Z_\mathrm{vac}= \sqrt{\mu_0/\epsilon_0} \sim 377~\Omega$ the impedance of vacuum with $\epsilon_0$ the vacuum permittivity and $\mu_0$ the vacuum permeability \cite{Devoret2007}. To find $\alpha$ here should not be surprising because this quantity characterizes the interaction between the electromagnetic field and charged particles. Here, this interaction is reduced by the fact that both $\zc/Z_\mathrm{vac}$ and $C_g/C_\Sigma$ are smaller than unity. Very large couplings can nevertheless be achieved by working with large values of $E_J/E_C$ or, in other words, in the transmon regime. Large $g$ is therefore obtained at the expense of reducing the transmon's relative anharmonicity $-E_C/\hbar\omega_q \simeq -\sqrt{E_C/8E_J}$. We note that the coupling can be increased by boosting the resonator's impedance, something that can be realized, for example, by replacing the resonator's center conductor with a junction array \cite{Andersen2017,Stockklauser2017}.

Apart from a change in the details of the expression of the coupling $g$, the above discussion holds for transmons coupled to lumped, 2D or 3D resonators. Importantly, by going from 2D to 3D, the resonator mode volume is made significantly larger leading to an important reduction in the vacuum fluctuations of the electric field. As first demonstrated in \textcite{Paik2011}, this can be made without change in the magnitude of $g$ simply by making the transmon larger thereby increasing its dipole moment. As illustrated in Fig.~\ref{fig:transmonCQED}(c), the transmon then essentially becomes an antenna that is optimally placed within the 3D resonator to strongly couple to one of the resonator  modes.

To strengthen the analogy with cavity QED even further, it is useful to restrict the description of the transmon to its first two levels. This corresponds to making the replacements $\bd \rightarrow\spp{} = \ket{e}\bra{g}$ and $\bop \rightarrow\smm{} = \ket{g}\bra{e}$ in \eq{eq:HTransmonRabi} to obtain the well-known Jaynes-Cummings Hamiltonian~\cite{Haroche2006,Blais2004}
\begin{equation}\label{eq:HJC}
\hH_\mathrm{JC}
=
\hbar\wc \ada
+ \frac{\hbar\omega_q}{2}\sz{}
+ \hbar g (\ad\smm{} + \aop\spp{}),
\end{equation}
where we use the convention $\sz{} = \ket e\bra e - \ket g \bra g$.
The last term of this Hamiltonian describes the coherent exchange of a single quantum between light and matter, here realized as a photon in the oscillator or an excitation of the transmon.

\subsection{The Jaynes-Cummings spectrum}\label{sec:JaynesCummings}

The Jaynes-Cummings Hamiltonian is an exactly solvable model which very accurately describes many situations in which an atom, artificial or natural, can be considered as a two-level system in interaction with a single mode of the electromagnetic field. This model can yield qualitative agreement with experiments in situations where only the first two levels of the transmon, $\ket{\sigma = \{g, e\}}$, play an important role. It is often the case, however, that quantitative agreement between theoretical predictions and experiments is obtained only when accounting for higher transmon energy levels and the multimode nature of the field. Nevertheless, since a great deal of insight can be gained from this, in this section we consider the Jaynes-Cummings model more closely.

\begin{figure}
  \centering
  \includegraphics{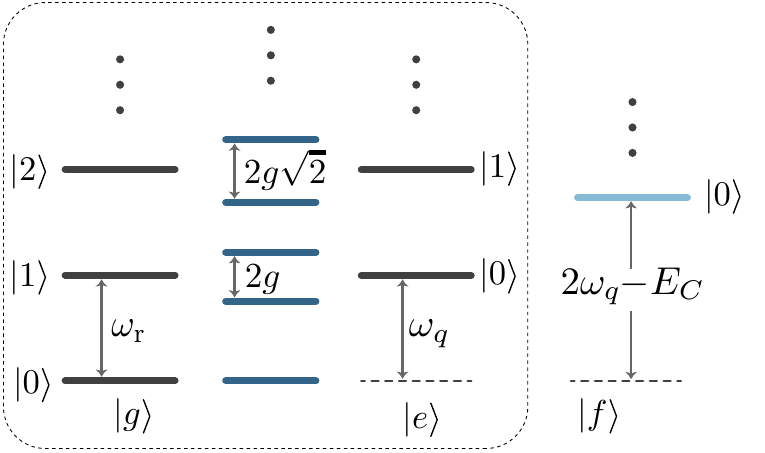}
  \caption{Box: Energy spectrum of the uncoupled (gray lines) and dressed (blue lines) states of the Jaynes-Cummings Hamiltonian at zero detuning, $\Delta = \wa-\wc = 0$. 
Transmon states are labelled $\{\ket{g},\ket{e}\}$ while photon number in the cavity are labelled $\ket{n=0,1,2\ldots}$ and are plotted vertically.  The degeneracy of the two-dimensional manifolds of states with $n$ quanta is lifted by $2g\sqrt{n}$ by the electric-dipole coupling. The light blue line outside of the main box represents the third excited state of the transmon, labelled $\ket{f}$. Although this is not illustrated here, the presence of this level shifts the dressed states in the manifolds with $n\ge2$ quanta~\cite{Fink2008}.
 \label{fig:JCSpectrum}}
\end{figure}

In the absence of coupling $g$, the \emph{bare} states of the qubit-field system are labelled $\ket{\sigma,n}$ with $\sigma$ defined above and $n$ the photon number. The \emph{dressed} eigenstates of the Jaynes-Cummings Hamiltonian, $\ket{\overline{\sigma, n}} = \hat U^\dagger \ket{\sigma,n}$, can be obtained from these bare states using the Bogoliubov-like unitary transformation~\cite{Boissonneault2009,Carbonaro1979}
\begin{equation}
	\hat U = \exp \left[\Lambda(\hat N_T) (\ad\smm{} - \aop\spp{})\right],
	\label{eqn:JCdiag_transform}
\end{equation}
where we have defined
\begin{equation}\label{eq:LambdaNT}
	\Lambda(\hat N_T) = \frac{\arctan\left(2\lambda\sqrt{\hat N_T}\right)}{2\sqrt{\hat N_T}}.
\end{equation}
Here, $\hat N_T = \ad\aop + \spp{}\smm{}$ is the operator associated with the total number of excitations, which commutes with $\hH_\mathrm{JC}$, and $\lambda = g/\Delta$ with $\Delta = \wa-\wc$ the qubit-resonator detuning.
Under this transformation, $\hH_\mathrm{JC}$ takes the diagonal form
\begin{equation}\label{eq:HJCdiagonal}
\begin{split}
  &\hH_D
		 = \hat U^\dag \hH_\mathrm{JC} \hat U\\
		& = \hbar\wc\ada + \frac{\hbar\wa}{2}\sz{} - \frac{\hbar\Delta}{2} \left( 1 - \sqrt{1+4\lambda^2\hat N_T} \right) \sz{}.
\end{split}
\end{equation}
The dressed-state energies can be read directly from this expression and, as illustrated in Fig.~\ref{fig:JCSpectrum}, the Jaynes-Cummings spectrum consists of doublets $\{\ket{\overline{g,n}},\ket{\overline{e,n-1}}\}$ of fixed excitation number\footnote{To arrive at these expressions, we have added $\hbar \wc/2$ to $\hH_D$. This global energy shift is without consequences.}
\begin{align}\label{eq:JCEnergies}
  \begin{split}
E_{\overline{g,n}} = \hbar n \wc - \frac\hbar 2 \sqrt{\Delta^2 + 4g^2n},\\
E_{\overline{e,n-1}} = \hbar n \wc + \frac\hbar 2 \sqrt{\Delta^2 + 4g^2n},
  \end{split}
\end{align}
and of the ground state $\ket{\overline{g,0}}=\ket{g,0}$ of energy $E_{\overline{g,0}} = -\hbar\wa/2$. The excited dressed states are 
\begin{equation}\label{eq:JCEigenstates}
\begin{split}
\ket{\overline{g, n}} &= \cos(\theta_n/2)\ket{g,n}-\sin(\theta_n/2)\ket{e,n-1},\\
\ket{\overline{e, n-1}} &= \sin(\theta_n/2)\ket{g,n} + \cos(\theta_n/2)\ket{e,n-1},
\end{split}
\end{equation}
with $\theta_n = \arctan(2g\sqrt{n}/\Delta)$. 

A crucial feature of the energy spectrum of~\eq{eq:JCEnergies} is the scaling with the photon number $n$. In particular, for zero detuning, $\Delta = 0$, the energy levels $E_{\overline{g,n}}$ and $E_{\overline{e,n-1}}$ are split by $2g\sqrt{n}$, in contrast to two coupled harmonic oscillators where the energy splitting is independent of $n$. Experimentally probing this spectrum thus constitutes a way to assess the quantum nature of the coupled system~\cite{Carmichael1996,Fink2008}. We return to this and related features of the spectrum in~\secref{sec:resonant}.

\subsection{\label{sec:dispersive}Dispersive regime}

On resonance, $\Delta = 0$, the dressed-states \cref{eq:JCEigenstates} are maximally entangled qubit-resonator states implying that the qubit is, by itself, never in a well-defined state, i.e.~the reduced state of the qubit found by tracing over the resonator is not pure. For quantum information processing, it is therefore more practical to work in the dispersive regime where the qubit-resonator detuning is large with respect to the coupling strength, $|\lambda| = |g/\Delta|\ll 1$. In this case, the coherent exchange of a quanta between the two systems described by the last term of $\hH_\mathrm{JC}$ is not resonant, and interactions take place only via virtual photon processes. Qubit and resonator are therefore only weakly entangled and a simplified model obtained below from second-order perturbation theory is often an excellent approximation. As the virtual processes can involve higher energy levels of the transmon, it is however crucial to account for its multi-level nature. For this reason, our starting point will be the Hamiltonian of \cref{eq:HTransmonJC} and not its two-level approximation \cref{eq:HJC}.

\subsubsection{\label{sec:dispersive:SW}Schrieffer-Wolff approach}

\begin{figure}
  \centering
  \includegraphics{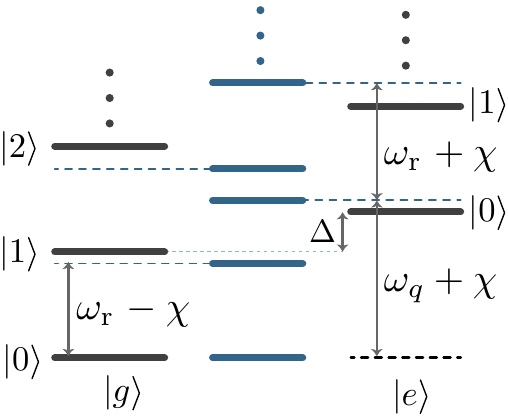}
  \caption{Energy spectrum of the uncoupled (gray lines) and dressed states in the dispersive regime (blue lines). The two lowest transmon states are labelled $\{\ket{g},\ket{e}\}$ while photon numbers in the cavity are labelled $\ket{n=0,1,2\ldots}$ and are plotted vertically. In the dispersive regime, the $g-e$ transition frequency of the qubit in the absence of resonator photons is Lamb shifted and takes value $\wa+\chi$. Moreover, the cavity frequency is pulled by its interaction with the qubit and takes the qubit-state dependent value $\wc\pm\chi$.
\label{fig:DispersiveSpectrum}}
\end{figure}

To find an approximation to~\cref{eq:HTransmonJC} valid in the dispersive regime, we perform a Schrieffer-Wolff transformation to second order~\cite{Koch2007}. As shown in~\cref{sec:unitarytransforms}, as long as the interaction term in~\cref{eq:HTransmonJC} is sufficiently small, the resulting effective Hamiltonian is well approximated by
\begin{equation}\label{eq:HTransmonDispersiveSW}
  \begin{aligned}
  \hH_\mathrm{disp} \simeq{}& \hbar \wc \ada 
  + \hbar\omega_q \bdb - \frac{E_C}{2}\bd\bd\bop\bop\\
  +& \sum_{j=0}^\infty \hbar \left(\Lambda_{j} + \chi_{j}\ad\aop\right) \ket{j} \bra{j},
  \end{aligned}
\end{equation}
where $\ket{j}$ label the eigenstates of the transmon which, under the approximation used to obtain~\cref{eq:HsjPhi4b}, are just the eigenstates of the number operator $\bd\bop$. Moreover, we have defined
\begin{subequations}
\begin{align}
    \Lambda_{j} ={}& \chi_{j-1,j},\quad
    \chi_{j} = \chi_{j-1,j} - \chi_{j,j+1},\\
    \chi_{j-1,j} ={}& \frac{j g^2}{\Delta - (j - 1)E_C/\hbar}.
\end{align}
for $j>0$ and $\Lambda_0 = 0$, $\chi_0 = -g^2/\Delta$.
Here $\chi_j$ are known as dispersive shifts, while $\Lambda_j$ are Lamb shifts and are signatures of vacuum fluctuations~\cite{Lamb1947,Bethe1947,Fragner2008}.

\end{subequations}
Truncating \eq{eq:HTransmonDispersiveSW} to the first two levels of the transmon leads to the more standard form of the dispersive Hamiltonian~\cite{Blais2004}
\begin{equation}\label{eq:HQubitDispersive}
  \begin{aligned}
    \hH_\text{disp} \approx{}& \hbar \wc' \ada + \frac{\hbar\omega_q'}{2} \sz{}
    + \hbar \chi \ada \sz{},
  \end{aligned}
\end{equation}
where $\chi$ is the qubit state-dependent dispersive cavity-shift
with~\cite{Koch2007}
\begin{equation}\label{eq:HQubitDispersiveParametersSW}
\begin{aligned}
\wc' ={}& \wc - \frac{g^2}{\Delta-E_C/\hbar},\quad
\omega_q' = \omega_q + \frac{g^2}{\Delta},\\
\chi ={}& -\frac{g^2E_C/\hbar}{\Delta(\Delta - E_C/\hbar)}.
\end{aligned}
\end{equation}
These dressed frequencies are what are measured experimentally in the dispersive regime. It is  important to emphasize that the frequencies entering the right-hand-sides of \cref{eq:HQubitDispersiveParametersSW} are the \emph{bare} qubit and resonator frequencies.
The spectrum of this two-level dispersive Hamiltonian is illustrated in \cref{fig:DispersiveSpectrum}. Much of this review is devoted to the consequences of this dispersive Hamiltonian for qubit readout and quantum information processing. We note that the Scrieffer-Wolff transformation also gives rise to resonator and qubit self-Kerr nonlinearities at fourth order~\cite{Zhu2012}. 

As already mentioned, these perturbative results are valid when the interaction term in~\cref{eq:HTransmonJC} is sufficiently small compared to the energy splitting of the bare transmon-oscillator energy levels, $|\lambda|=|g/\Delta|\ll1$. Because the matrix elements of the operators involved in the interaction term scale with the number of photons in the resonator and the number of qubit excitations, a more precise bound on the validity of \cref{eq:HTransmonDispersiveSW} needs to take into account these quantities. As discussed in~\cref{sec:SW:transmon}, we find that for the above second order perturbative results to be a good approximation, the oscillator photon number $\bar n$ should be much smaller than a critical photon number $n_\text{crit}$ 
\begin{equation}\label{eq:ncrit}
   \bar n \ll n_\text{crit} \equiv \frac{1}{2j + 1} \left(\frac{|\Delta-j E_C/\hbar|^2}{4 g^2} - j\right),
\end{equation}
where $j=0,1,\dots$ refers to the qubit state as before.
For $j=0$, this yields the familiar value $n_\text{crit}=(\Delta/2g)^2$ for the critical photon number expected from the Jaynes-Cummings model~\cite{Blais2004}, while setting $j=1$ gives a more conservative bound. In either case, this gives only a rough estimate for when to expect higher-order effects to become important. 

It is worth contrasting~\cref{eq:HQubitDispersiveParametersSW} to the results expected from performing a dispersive approximation to the Jaynes-Cummings model~\cref{eq:HJC}, which leads to $\chi=g^2/\Delta$ (see~\cref{sec:AppendixDispersiveTLS}, \textcite{Blais2004,Boissonneault2010}). This agrees with the above result in the limit of very large $E_C$ but, since $E_C/\hbar$ is typically rather small compared to $\Delta$ in most transmon experiments, the two-level system Jaynes-Cummings model gives a poor prediction for the dispersive shift $\chi$ in practice. The intuition here is that $E_C$ determines the anharmonicity of the transmon.  Two coupled harmonic oscillators can shift each other's frequencies, but only in a state-independent manner.  Thus the dispersive shift must vanish in the limit of $E_C$ going to zero.

\subsubsection{\label{sec:dispersive:bb}Bogoliubov approach}

We now present an approach to arrive at~\cref{eq:HQubitDispersive} that can be simpler than performing a Schrieffer-Wolff transformation and which is often used in the circuit QED literature.

To proceed, it is convenient to write~\cref{eq:HTransmonJC} as the sum of a linear and a nonlinear part, $\hH = \hHL + \hHNL$, where
\begin{align}
  \hHL ={}& \hbar\wc \ada + \hbar\omega_q \bdb + \hbar g (\bd\aop+\bop\ad),\label{eq:HTransmonDispersiveL}\\
  \hHNL ={}&- \frac{E_C}{2}\bd\bd\bop\bop\label{eq:HTransmonDispersiveNL}.
\end{align}
The linear part $\hHL$ can be diagonalized exactly using the Bogoliubov transformation
\begin{equation}\label{eq:UDispersive}
\hat U_\mathrm{disp} = \exp\left[\Lambda (\hat a^\dagger \hat b - \hat a \hat b^\dagger) \right].
\end{equation}
Under this unitary, the annihilation operators transform as $\hat U_\mathrm{disp}^\dagger \aop \hat U_\mathrm{disp} = \cos(\Lambda)\aop + \sin(\Lambda)\hat b$ and $\hat U_\mathrm{disp}^\dagger \hat b \hat U_\mathrm{disp} = \cos(\Lambda)\hat b - \sin(\Lambda)\aop$. With the choice $\Lambda = \half \arctan(2\lambda)$,
this results in the diagonal form
\begin{equation}\label{eq:HDiagonalLinear}
\begin{aligned}
  \hat U_\mathrm{disp}^\dagger \hHL \hat U_\mathrm{disp} 
  = \hbar\twc  \ad \aop + \hbar\tilde\omega_q \hat b^\dagger \hat b,
\end{aligned}
\end{equation}
with the dressed frequencies 
\begin{subequations}\label{eq:DressedFrequenciesDispersive}
\begin{align}
  \twc  ={}& \half\left(\wc + \omega_q - \sqrt{\Delta^2+4g^2}\right),\\
  \tilde\omega_q ={}& \half\left(\wc + \omega_q + \sqrt{\Delta^2+4g^2}\right).
\end{align}
\end{subequations}

Applying the same transformation to $\hHNL$ and, in the dispersive regime, expanding the result in orders of $\lambda$ leads to the dispersive Hamiltonian (see \cref{sec:AppendixDispersiveTransformation})	
\begin{equation}\label{eq:HTransmonDispersive}
  \begin{aligned}
    \hH_\text{disp} 
    ={}& \hat U_\mathrm{disp}^\dagger \hat H \hat U_\mathrm{disp} \\
    \simeq{}& \hbar\twc\ada + \hbar\tilde\omega_q \bdb \\
  +&\frac{\hbar K_a}{2} \hat a^\dagger \hat a^\dagger \aop \aop + \frac{\hbar K_b}{2} \bd\bd\bop\bop+ \hbar \chi_{ab} \ada \bdb,
  \end{aligned}
\end{equation}
where we have introduced 
\begin{equation}\label{eq:HTransmonDispersiveParameters}
  \begin{aligned}
    K_a &\simeq -\frac{E_C}{2\hbar }\left(\frac{g}{\Delta}\right)^4,\;
    K_b \simeq -E_C/\hbar,\\
    \chi_{ab} &\simeq -2 \frac{g^2 E_C/\hbar}{\Delta(\Delta-E_C/\hbar)}.
  \end{aligned}
\end{equation}
The first two of these quantities are self-Kerr nonlinearities, while the third is a cross-Kerr interaction. All are negative in the dispersive regime. As discussed in \cref{sec:AppendixDispersiveTransformation}, the above expression for $\chi_{ab}$ is obtained after performing a Schrieffer-Wolff transformation to eliminate a term of the form $\hat b^\dagger \hat b\,\hat a^\dagger \hat b + \hc$ that results from applying $U_\mathrm{disp}$ on $H_\mathrm{NL}$. Higher-order terms in $\lambda$ and other terms rotating at frequency $\Delta$ or faster have been dropped to arrive at \cref{eq:HTransmonDispersive}.
These terms are given in~\cref{eq:app:dispersive:H_NL_lambda}.

Truncating \eq{eq:HTransmonDispersive} to the first two levels of the transmon correctly leads to~\cref{eq:HQubitDispersive,eq:HQubitDispersiveParametersSW}. Importantly, these expressions are not valid if the excitation number of the resonator or the transmon is too large or if $|\Delta| \sim E_C/\hbar$. The regime $0 < \Delta < E_C$, known as the straddling regime, is qualitatively different from the usual dispersive regime. It is characterized by positive self-Kerr and cross-Kerr nonlinearities, $K_a, \chi_{ab} > 0$, and is better addressed by exact numerical diagonalization of~\cref{eq:HTransmonResonator}~\cite{Koch2007}.

A remarkable feature of circuit QED is the large nonlinearities that are achievable in the dispersive regime. Dispersive shifts larger than the resonator or qubit linewidth, $\chi > \kappa,\, \gamma$, are readily realized in experiments, a regime referred to as strong dispersive coupling~\cite{Gambetta2006,Schuster2007a}. Some of the consequences of this regime are discussed in Sec.~\ref{sec:DispersiveConsequences}. It is also possible to achieve large self-Kerr nonlinearities for the resonator, $K_a > \kappa$.\footnote{Of course, the transmon is itself an oscillator with a very large self-Kerr given by $\hbar K_b = -E_C$.} These nonlinearities can be enhanced by  embedding Josephson junctions in the center conductor of the resonator~\cite{Bourassa2012,Ong2013}, an approach which is used for example in quantum-limited parametric amplifiers~\cite{Castellanos2008} or for the preparation of quantum states of the microwave electromagnetic field~\cite{Kirchmair2013,Holland2015,Puri2017}.

\subsection{\label{sec:bbq}Josephson junctions embedded in multimode electromagnetic environments}

So far, we have focussed on the capacitive coupling of a transmon to a single mode of an oscillator. For many situations of experimental relevance it is, however, necessary to consider the transmon, or even multiple transmons, embedded in an electromagnetic environment with a possibly complex geometry, such as a 3D cavity.

\begin{figure}
  \centering
  \includegraphics[width=\columnwidth]{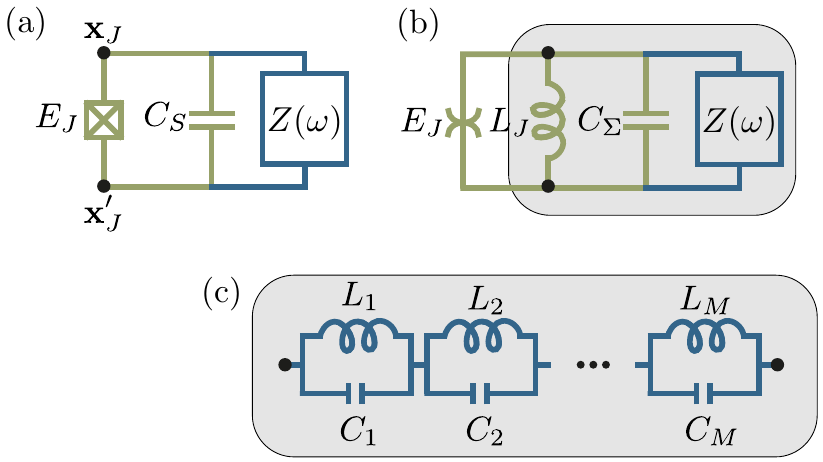}
  \caption{(a) Transmon qubit coupled to an arbitrary impedance, such as that realized by a 3D microwave cavity.  (b) The Josephson junction can be represented as a capacitive element $C_J$ and a linear inductive element $L_J$ in parallel with a purely nonlinear element indicated here by the spiderlike symbol. Here, $C_\Sigma = C_S+C_J$ is the parallel combinaison of the Josephson capacitance and the shunting capacitance of the transmon. (c) Normal mode decomposition of the parallel combination of the impedance $Z(\omega)$ together with $L_J$ and $C_\Sigma$ represented  by effective $LC$ circuits.
 \label{fig:bbq}}
\end{figure}

Consider the situation depicted in~\figref{fig:bbq}(a) where a capacitively shunted Josephson junction is embedded in some electromagnetic environment represented by the impedance $Z(\omega)$. To keep the discussion simple, we consider here a single junction but the procedure can easily be extended to multiple junctions. As discussed in \secref{sec:transmon}, the Hamiltonian of the shunted junction \cref{eq:HsjLinNonLin} 
can be decomposed into a linear term of capacitance $C_\Sigma = C_S + C_J$ and linear inductance $L_J = E_J^{-1}(\Fnot/2\pi)^2$, and a purely nonlinear element. This decomposition is illustrated in~\figref{fig:bbq}(b), where the spider symbol represents the nonlinear element~\cite{Manucharyan2007,Bourassa2012}. 

We assume that the electromagnetic environment is linear, nonmagnetic and has no free charges and currents. Since $C_\Sigma$ and $L_J$ are themselves linear elements, we might as well consider them part of the electromagnetic environment too, something that is illustrated by the box in \cref{fig:bbq}(b). Combining all linear contributions, we write a Hamiltonian for the entire system, junction plus the surrounding electromagnetic environment, as $\hat H = \hHL + \hHNL$ with
\begin{equation}
    \hHNL = -E_J\left(\cos\hat\varphi + \half \hat\varphi^2\right)
\end{equation}
the nonlinear part of the transmon Hamiltonian already introduced in \cref{eq:HsjLinNonLin}.
 A good strategy is to first diagonalize the linear part, $\hHL$, which can in principle be done exactly much like was done in \secref{sec:dispersive}. Subsequently, the phase difference $\hat\varphi$ across the junction can be expressed as a linear combination of the eigenmodes of $\hHL$, a decomposition which is then used in $\hHNL$.

A convenient choice of canonical fields for the electromagnetic environment are the electric displacement field $\hat{\*D}(\*x)$ and the magnetic field $\hat{\*B}(\*x)$, which can be expressed in terms of bosonic creation and annihilation operators~\cite{Bhat2006}
\begin{subequations}\label{eq:bbq:DB}
\begin{align}
  \hat{\*D}(\*x) ={}& \sum_m \left[\*D_m(\*x)\hat a_m + \hc \right], \label{eq:bbq:D}\\
  \hat{\*B}(\*x) ={}& \sum_m \left[\*B_m(\*x)\hat a_m + \hc \right], \label{eq:bbq:B}
\end{align}
\end{subequations}
where $[\hat a_m, \hat a_n^\dagger] = \delta_{mn}$.
The more commonly used electric field is related to the displacement field through $\hat{\*D}(\*x) = \varepsilon_0\hat{\*E}(\*x)  + \hat{\*P}(\*x)$, where $\hat{\*P}(\*x)$ is the polarization of the medium. Moreover, the mode functions $\*D_m(\*x)$ and $\*B_m(x)$ can be chosen to satisfy orthogonality and normalization conditions such that
\begin{equation}\label{eq:bbq:H_L}
  \hHL = \sum_m \hbar\omega_m \hat a_m^\dagger \hat a_m.
\end{equation}
In \cref{eq:bbq:DB,eq:bbq:H_L}, we have implicitly assumed that the eigenmodes form a discrete set. If some part of the spectrum is continuous, which is the case for infinite systems such as open waveguides, the sums must be replaced by integrals over the relevant frequency ranges. The result is very general, holds for arbitrary geometries, and can include inhomogeneities such as partially reflecting mirrors, and materials with dispersion~\cite{Bhat2006}. We will, however, restrict ourselves to discrete spectra in the following. 

Diagonalizing $\hHL$ amounts to determining the mode functions $\{\hat{\*D}_m(\*x), \hat{\*B}_m(\*x)\}$, which is essentially a classical electromagnetism problem that can, e.g., be approached using numerical software such as finite element solvers~\cite{minev2019catching}.
Assuming that the mode functions have been found,
we now turn to $\hHNL$ for which we relate $\hat \varphi$ to the bosonic operators $\hat a_m$. This can be done by noting again that $\hat\varphi(t) = 2\pi \int dt'\, \hat V(t')/\Phi_0$, where the voltage is simply the line integral of the electric field $\hat V(t) = \int d\*l \cdot \hat{\*E}(\*x) = \int d\*l \cdot \hat{\*D}(\*x)/\varepsilon$ across the junction~\cite{Vool2016}. Consequently, the phase variable can be expressed as
\begin{equation}\label{eq:bbq:phase}
  \hat \varphi = \sum_m \left[\varphi_m \hat a_m + \hc \right],
\end{equation}
where $\varphi_m = i(2\pi/\Phi_0) \int_{\*x_J'}^{\*x_J} d \*l \cdot \*D_m(\*x)/(\omega_m\varepsilon)$ 
is the dimensionless magnitude of the zero-point fluctuations of the $m$th mode as seen by the junction and the boundary conditions defined as in \cref{fig:bbq}(a).

Using~\cref{eq:bbq:phase} in $\hHNL$ we  expand the cosine to fourth order in analogy with~\cref{eq:HsjPhi4}. This means that we are assuming that the capacitively shunted junction is well in the transmon regime, with a small anharmonicity relative to Josephson energy.
Focusing on the dispersive regime where all eigenfrequencies $\omega_m$ are sufficiently well separated, and neglecting fast-rotating terms in analogy with \secref{sec:dispersive}, leads to
\begin{equation}\label{eq:blackbox:HNL}
  \begin{aligned}
    \hHNL \simeq{}& \sum_m \hbar \Delta_m \hat a_m^\dagger \hat a_m
    + \half \sum_{m}\hbar K_m (\hat a_m^\dagger)^2 \hat a_m^2, \\
    &+ \sum_{m > n}\hbar \chi_{m,n}\hat a_m^\dagger \hat a_m \hat a_n^\dagger \hat a_n,
  \end{aligned}
\end{equation}
where $\Delta_m = \half \sum_n \chi_{m,n}$, $K_m = \frac{\chi_{m,m}}{2}$ and
\begin{align}
  \hbar \chi_{m,n} = -E_J \varphi_m^2\varphi_n^2.
\end{align}
It is also useful to introduce the energy participation ratio $p_m$, defined to be the fraction of the total inductive energy of mode $m$ that is stored in the junction $p_m = (2E_J/\hbar\omega_m)\varphi_m^2$~\cite{minev2019catching} such that we can write
\begin{equation}
  \chi_{m,n} = -\frac{\hbar\omega_m\omega_n}{4 E_J} p_m p_n.
\end{equation}

As is clear from the above discussion, finding the nonlinear Hamiltonian can be reduced to finding the eigenmodes of the system and the zero-point fluctuations of each mode across the junction. Of course, finding the mode  frequencies $\omega_m$ and zero-point fluctuations $\varphi_m$, or alternatively the energy participation ratios $p_m$, can be complicated for a complex geometry. As already mentioned this is, however, an entirely classical electromagnetism problem~\cite{Bhat2006,minev2019catching}.

An alternative approach is to represent the linear electromagnetic environment seen by the purely nonlinear element as an impedance $Z(\omega)$, as illustrated in~\figref{fig:bbq}(c). Neglecting loss, any such impedance can be represented by an equivalent circuit of possibly infinitely many LC oscillators connected in series. The eigenfrequencies $\hbar \omega_m = 1/\sqrt{L_m C_m}$, can be determined by the real parts of the zeros of the admittance $Y(\omega) = Z^{-1}(\omega)$, and the effective impedance of the $m$'th mode as seen by the junction can be found from $\cZeff_m = 2/[\omega_m \Im\,Y'(\omega_m)]$~\cite{Nigg2012,Solgun2014}. The effective impedance is related to the zero-point fluctuations used above as $\cZeff_m = 2(\Phi_0/2\pi)^2\varphi_m^2/\hbar = R_K\varphi_m^2/(4\pi)$. From this point of view, the quantization procedure thus reduces to the task of determining the impedance $Z(\omega)$ as a function of frequency.

\subsection{Beyond the transmon: multilevel artificial atom}

In the preceding sections, we have relied on a perturbative expansion of the cosine potential of the transmon under the assumption $E_J/E_C \gg 1$. To go beyond this regime one can instead resort to exact diagonalization of the transmon Hamiltonian. Returning to the full transmon-resonator Hamiltonian~\cref{eq:HTransmonResonator}, we write \cite{Koch2007} 
\begin{equation}\label{eq:HMultileveltResonator}
\begin{aligned}
\hH ={}& 4E_C \hn - E_J \cos\hvphi + \hbar\wc \hat a^\dagger \hat a
+ 8E_C \hat n\hat n_r\\
={}& \sum_j \hbar \omega_j \ket j \bra j + \hbar\wc \hat a^\dagger \hat a
+ i \sum_{ij} \hbar g_{ij} \ket i\bra j(\hat a\dg - \hat a),
\end{aligned}
\end{equation}
where $\ket j$ are now the eigenstates of the bare transmon Hamiltonian $\hH_T = 4E_C \hn - E_J\cos\hvphi$ obtained from numerical diagonalization and we have defined
\begin{equation}
    \hbar g_{ij} = 2e\frac{C_g}{CC_\Sigma} \Qzp \braket{i|\hn|j}.
\end{equation}
The eigenfrequencies $\omega_j$ and the matrix elements $\braket{i|\hat n|j}$ can be computed numerically in the charge basis. Alternatively, they can be determined by taking advantage of the fact that, in the phase basis, \eq{eq:Hsj} takes the form of a Mathieu equation whose exact solution is known~\cite{Cottet2002a,Koch2007}.

The second form of~\cref{eq:HMultileveltResonator} written in terms of energy eigenstates $\ket j$ is a very general Hamiltonian that can describe an arbitrary multilevel artificial atom capacitively coupled to a resonator.
Similarly to the discussion of \cref{sec:dispersive}, in the dispersive regime where $|g_{ij}| \sqrt{n+1} \ll |\omega_i-\omega_j-\omega_r|$ for all relevant atomic transitions $i \leftrightarrow j$ and with $n$ the oscillator photon number, it is possible to use a Schrieffer-Wolff transformation to approximately diagonalize~\cref{eq:HMultileveltResonator}. As discussed in~\cref{sec:SW:multilevel}, to second order one finds~\cite{Zhu2012}
\begin{equation}
  \begin{aligned}
  \hH \simeq{}& \sum_j \hbar (\omega_j + \Lambda_j) \ket j \bra j + \hbar \wc \aop\dg \aop \\
  &+ \sum_j \hbar \chi_j \aop\dg \aop \ket j \bra j,
  \end{aligned}
\end{equation}
where
\begin{subequations}
\begin{align}
\Lambda_j ={}& \sum_i \frac{|g_{ij}|^2}{\omega_j-\omega_i-\omega_r},\\
\chi_j ={}& \sum_i \left(\frac{|g_{ij}|^2}{\omega_j-\omega_i-\omega_r} - \frac{|g_{ij}|^2}{\omega_i-\omega_j-\omega_r} \right).
\end{align}
\end{subequations}
This result is, as already stated, very general, and can be used with a variety of artificial atoms coupled to a resonator in the dispersive limit. Higher order expressions can be found in~\cite{Boissonneault2010,Zhu2012}.

\subsection{Alternative coupling schemes}

Coupling the electric dipole moment of a qubit to the zero-point electric field of an oscillator is the most common approach to light-matter coupling in a circuit but it is not the only possibility. Another approach is  to take advantage of the mutual inductance between a flux qubit and the center conductor of a resonator to couple the qubit's magnetic dipole to the resonator's magnetic field. Stronger interaction can be obtained by galvanically connecting the flux qubit to the center conductor of a transmission-line resonator \cite{Bourassa2009}. In such a situation, the coupling can be engineered to be as large, or even larger, than the system frequencies allowing to reach what is known at the ultrastrong coupling regime, see \cref{sec:ultrastrong}.

Yet another approach is to couple the qubit to the oscillator in such a way that the resonator field does not result in qubit transitions but only shifts the qubit's frequency. This is known as longitudinal coupling and is represented by the Hamiltonian \cite{Didier2015c,Kerman2013,Billangeon2015,Billangeon2015a,Richer2016,Richer2017a}
\begin{equation}\label{eq:H_Longitudinal}
	\hH_\mathrm{z}
	=
	\hbar\wc \ada
	+ \frac{\hbar\omega_q}{2}\sz{}
	+ \hbar g_z (\ad+\aop)\sz{}.
\end{equation}
Because light-matter interaction in $\hH_\mathrm{z}$ is proportional to $\sz{}$ rather than $\sx{}$ , the longitudinal interaction does not lead to dressing of the qubit by the resonator field of the form discussed in \cref{sec:JaynesCummings}. Some of the consequences of this observation, in particular for qubit readout, are discussed in \cref{sec:ReadoutOtherApproaches}.

\section{\label{sec:environment}Coupling to the outside world: The role of the environment}

So far we have dealt with isolated quantum systems. A complete description of quantum electrical circuits, however, must also take into account a description of how these systems couple to their environment, including any measurement apparatus and control circuitry. In fact, the environment plays a dual role in quantum technology: Not only is a description of quantum systems as perfectly isolated unrealistic, as coupling to unwanted environmental degrees of freedom is unavoidable, but a perfectly isolated system would also not be very useful since we would have no means of controlling or observing it. For these reasons, in this section we consider quantum systems coupled to external transmission lines. We also introduce the input-output theory which is of central importance in understanding qubit readout in circuit QED in the next section.

\subsection{\label{sec:wiringup} Wiring up quantum systems with transmission lines}

\begin{figure}[t]
  \centering
  \includegraphics{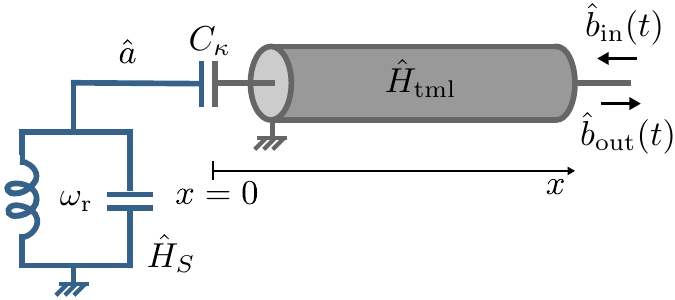}
  \caption{
  LC circuit capacitively coupled to a semi-infinite transmission line used to model both damping and driving of the system. Here, $\bin(t)$ and $\bout(t)$ are the oscillators input and output fields, respectively. 
  \label{fig:inout}
  }
  \end{figure}

We start the discussion by considering transmission lines coupled to individual quantum systems, which are a model for losses and can be used to apply and receive quantum signals for control and measurement. To be specific, we consider a semi-infinite coplanar waveguide transmission line capacitively coupled at one end to an oscillator, see \cref{fig:inout}. The semi-infinite transmission line can be considered as a limit of the coplanar waveguide resonator of finite length already discussed in~\secref{sec:telegrapher} where one of the boundaries is now pushed to infinity, $d\to\infty$. Increasing the length of the transmission line leads to a densely packed frequency spectrum, which in its infinite limit must be treated as a continuum. In analogy with \eq{eq:H_cavity}, the Hamiltonian of the transmission line is consequently
\begin{equation}\label{eq:Html}
  \hH_\tml = \int_0^\infty d\omega\, \hbar\omega \bdw\bw,
\end{equation}
where the mode operators now satisfy $[\hat b_\omega,\hat b_{\omega'}^\dagger]=\delta(\omega-\omega')$. Similarly, the position-dependent flux and charge operators of the transmission line are, in analogy with \cref{eq:ResonatorModeDecomp,eq:ResonatorModeFunc,eq:ResonatorFlux,eq:ResonatorCharge}, given in the continuum limit by~\cite{Yurke2004}
\begin{subequations}
\begin{align}
  \hat\Phi_\tml(x) ={}& \int_0^\infty d\omega\, \sqrt{\frac{\hbar}{\pi\omega cv}} \cos\left(\frac{\omega x}{v}\right)(\hat b_{\omega}^\dagger + \hat b_{\omega}), \label{eq:Phi_tml}\\
  \hat Q_\tml(x) ={}& i\int_0^\infty d\omega\, \sqrt{\frac{\hbar\omega c}{ \pi v}} \cos\left(\frac{\omega x}{v}\right) (\hat b_{\omega}^\dagger - \hat b_{\omega}).\label{eq:Q_tml}
\end{align}
\end{subequations}
These are the canonical fields of the transmission line and in the Heisenberg picture under~\cref{eq:Html} are related through $\hat Q_\tml(x,t) = c\dot{\hat \Phi}_\tml(x,t)$. In these expressions, $v=1/\sqrt{lc}$ is the speed of light in the transmission line, with $c$ and $l$ the capacitance and inductance per unit length, respectively.

Considering capacitive coupling of the line to the oscillator at $x=0$, the total Hamiltonian takes the form
\begin{equation}\label{eq:HSBCavityNonMarkov}
  \hH = \hat H_S + \hat H_\tml - \hbar\int_{0}^\infty d\omega\, \lambda(\omega)(\hat b_{\omega}^\dagger-\hat b_{\omega})(\ad-\aop),
\end{equation}
where $\hat H_S = \hbar\wc \hat a^\dagger \hat a$ is the oscillator Hamiltonian.  Moreover, $\lambda(\omega) = (C_\kappa/\sqrt{c C_r})\,\sqrt{\wc \omega/2 \pi v}$ is the frequency-dependent coupling strength, with $C_\kappa$ the coupling capacitance and $C_r$ the resonator capacitance. These expressions neglect small renormalizations of the capacitances due to $C_\kappa$, as discussed in~\cref{sec:AppendixTRcoupling}.

In the following, $\lambda(\omega)$ is assumed sufficiently small compared to $\wc$, such that the interaction can be treated as a perturbation. In this situation, the system's $Q$ factor is large and the oscillator only responds in a small bandwidth around $\wc$. It is therefore reasonable to take $\lambda(\omega)  \simeq \lambda(\wc)$ in \cref{eq:HSBCavityNonMarkov}. Dropping rapidly oscillating terms finally leads to~\cite{Gardiner1999}
\begin{equation}\label{eq:HSBCavityMarkov}
  \hH \simeq \hat H_S + \hat H_\tml + \hbar\int_{0}^\infty d\omega\, \lambda(\wc)(\aop\hat b_{\omega}^\dagger-\ad\hat b_{\omega}).
  \end{equation}
Under the well-established Born-Markov approximations, \eq{eq:HSBCavityMarkov} leads to a Lindblad-form Markovian master equation for the system's density matrix $\rhoS$~\cite{Gardiner1999,Carmichael2002,Breuer2002}
\begin{equation}\label{eq:ME_harmonic}
\dot\rhoS = -i[\hat H_S,\rhoS] + \kappa (\bar n_\kappa + 1) \mathcal{D}[\aop]\rhoS + \kappa \bar n_\kappa \mathcal{D}[\ad]\rhoS,
\end{equation}
where
$\kappa = 2\pi \lambda(\wc)^2 = Z_\tml \wc^2 C_\kappa^2 /C_\mathrm{r}$
is the photon decay rate, or linewidth, of the oscillator introduced earlier. Moreover, $\bar n_\kappa = \bar n_\kappa(\wc)$ is the number of thermal photons of the transmission line as given by the Bose-Einstein distribution, $\braket{\hat b_\omega^\dagger \hat b_{\omega'}} = \bar n_\kappa(\omega)\delta(\omega-\omega')$, at the system frequency $\wc$ and environment temperature $T$. The symbol $\mathcal{D}[\hat O]\bullet$ represents the dissipator
\begin{equation}\label{eq:dissipator}
\mathcal{D}[\hat O]\bullet = \hat O \bullet \hat O^\dag - \frac{1}{2} \left\{\hat O^\dag \hat O, \bullet \right\}, 
\end{equation}
with $\{\cdot, \cdot \}$ the anticommutator. Focussing on the second term of \eq{eq:ME_harmonic}, the role of this superoperator can be understood intuitively by noting that the term $\hat O \rho \hat O^\dag$ with $\hat O = \aop$
in \eq{eq:dissipator} acts on the Fock state $\ket n$ as $\aop\ket n\bra n\ad = n\ket{n-1}\bra{n-1}$. The second term of \eq{eq:ME_harmonic} therefore corresponds to photon loss at rate $\kappa$. Finite temperature stimulates photon emission, boosting the loss rate to $\kappa (\bar n_\kappa + 1)$. On the other hand, the last term of \eq{eq:ME_harmonic} corresponds to absorption of thermal photons by the system. Because $\hbar\wc\gg k_BT$ at dilution refrigerator temperatures, it is often assumed that $\bar n_\kappa\rightarrow 0$. Deviations from this expected behavior are, however, common in practice due to residual thermal radiation propagating along control lines connecting to room temperature equipment and to uncontrolled sources of heating. Approaches to mitigate this problem using absorptive components are being developed \cite{Corcoles2011,Wang2019}.

\subsection{\label{sec:inout}Input-output theory in electrical networks}

While the master equation describes the system's damped dynamics, it provides no information on the fields radiated by the system. Since radiated signals are what are measured experimentally, it is of practical importance to include those in our model.

This is known as the input-output theory for which two standard approaches exist. The first is to work directly with~\cref{eq:HSBCavityMarkov} and consider Heisenberg picture equations of motion for the system and field annihilation operators $\hat a$ and $\hat b_\omega$. This is the route taken by Gardiner and Collett, which is widely used in the quantum optics literature~\cite{Collett1984,Gardiner1985}. 

An alternative approach is to introduce a decomposition of the transmission line modes in terms of left- and right-moving fields, linked by a boundary condition at the position of the oscillator which we take to be $x=0$ with the transmission line at $x\ge0$~\cite{Yurke1984}. The advantage of this approch is that the oscillator's input and output fields are then defined in terms of easily identifiable left- ($\hat b_{L\omega}$) and right-moving ($\hat b_{R\omega}$) radiation field components propagating along the transmission line. 
To achieve this, we replace the modes $\cos(\omega x/v) \hat b_\omega$ in Eqs.~(\ref{eq:Phi_tml}) and (\ref{eq:Q_tml}) by $(\hat b_{R\omega}e^{i \omega x /v} + \hat b_{L\omega}e^{-i \omega x /v} )/2$. Since the number of degrees of freedom of the transmission line has seemingly doubled, the modes $\hat b_{L/R\omega}$ cannot be independent. Indeed, the dynamics of one set of modes is fully determined by the other set through a boundary condition linking the left- and right-movers at $x=0$.

To see this, it is useful to first decompose the voltage $\hat V(x,t) = \dot{\hat \Phi}_\tml(x,t)$ at $x=0$ into left-moving (input) and right-moving (output) contributions as $\hat V(t) = \hat V(x=0,t) = \hat V_\text{in}(t) + \hat V_\text{out}(t)$, where
\begin{equation}
  \begin{aligned}
    \hat V_\text{in/out}(t) ={}
    i\int_0^\infty d\omega\, \sqrt{\frac{\hbar\omega}{4 \pi cv}} e^{i\omega t} \hat b_{\text{L/R}\omega}\dg 
  + \hc
  \end{aligned}
\end{equation}
The boundary condition at $x=0$ follows from Kirchhoff's current law
\begin{equation}\label{eq:voltage_inout}
 \hat I(t) =   \frac{\hat V_\text{out}(t) - \hat V_\text{in}(t)}{Z_\tml},
\end{equation}
where the left-hand side $\hat I(t) = (C_\kappa/C_r)\dot{\hat Q}_r(t)$ is the current injected by the sample, with $\hat Q_r$ the oscillator charge (see~\cref{app:inout} for a derivation), while the right-hand side is the transmission line voltage difference at $x=0$.\footnote{Note that if instead we have a boundary condition of zero current at $x=0$, it would follow that $\hat V_\text{in}(t) = \hat V_\text{out}(t)$, i.e.~the endpoint simply serves as a mirror reflecting the input signal.} A mode expansions of the operators involved in~\eq{eq:voltage_inout} leads to the standard input-output relation (see~\cref{app:inout} for details)
\begin{equation}\label{eq:inoutrel}
   \bout(t) - \bin(t) = \sqrt{\kappa} \hat a(t),
\end{equation}
where the input and output fields are defined as
\begin{align}
  \bin(t) ={}& \frac{-i}{\sqrt{2\pi}} \int_{-\infty}^\infty d\omega\,\hat b_{L\omega}e^{-i(\omega-\wc) t},\label{eq:bin}\\
  \bout(t) ={}& \frac{-i}{\sqrt{2\pi}} \int_{-\infty}^\infty d\omega\,\hat b_{R\omega}e^{-i(\omega-\wc) t}\label{eq:bout}
\end{align}
and satisfy the commutation relations $[\bin(t),\bind(t')]=[\bout(t),\boutd(t')]=\delta(t-t')$. To arrive at~\cref{eq:inoutrel}, terms rotating at $\omega + \wc$ have been dropped based on the already mentioned assumption that the system only responds to frequencies $\omega\simeq \wc$ such that these terms are fast rotating~\cite{Yurke2004}. In turn, this  approximation allows to extend the range of integration from $(0,\infty)$ to $(-\infty,\infty)$ in \cref{eq:bin,eq:bout}. We have also approximated $\lambda(\omega) \simeq \lambda(\wc)$ over the relevant frequency range. These approximations are compatible with those used to arrive at the Lindblad-form Markovian master equation of \cref{eq:ME_harmonic}.

The same expressions and approximations can be used to obtain the equation of motion for the resonator field $\hat a(t)$ in the Heisenberg picture, which takes the form (see~\cref{app:inout} for details)
\begin{equation}\label{eq:inouteom}
  \dot{\hat a}(t) = i[\hat H_S, \hat a(t)] - \frac{\kappa}{2}\hat a(t) + \sqrt{\kappa}\bin(t).
\end{equation}
This expression shows that the resonator dynamics is determined by the input field (in practice, noise or drive), while~\eq{eq:inoutrel} shows how the output can, in turn, be found from the input and the system dynamics. The output field thus holds information about the system's response to the input and which can be measured to, indirectly, give us access to information about the dynamics of the system. As will be discussed in more detail in \cref{sec:readout}, this can be done, for example, by measuring the voltage at some $x>0$ away from the oscillator. Under the approximations used above, this voltage can be expressed as
\begin{equation}\label{eq:TLvoltage}
  \begin{aligned}
    \hat V(x, t) 
    \simeq{}& \sqrt{\frac{\hbar \wc Z_\tml}{2}} \bigg[ e^{i\wc x/v - i\wc t} \bout(t) \\
    &+ e^{-i \wc x/v - i\wc t} \bin(t) + \hc\bigg].
  \end{aligned}
\end{equation}
Note that this approximate expression assumes that all relevant frequencies are near $\wc$ and furthermore neglects all non-Markovian time-delay effects.

In this section we have considered a particularly simple setup: A single quantum system connected to the endpoint of a semi-infinite transmission line. More generally, quantum systems can be made to interact by coupling them to a common transmission line, and multiple transmission lines can be used to form quantum networks. These more complex setups can be treated using the SLH formalism, which generalizes the results in this section~\cite{Gough2009,Combes2016a}.

\subsection{\label{sec:qubitnoise}Qubit relaxation and dephasing}

\begin{figure}
\centering
\includegraphics{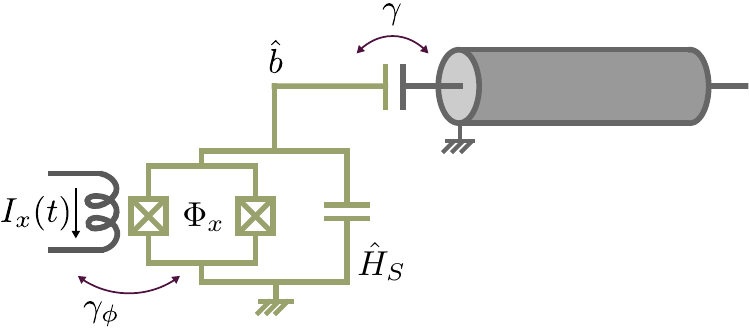}
\caption{\label{fig:qubitnoise}
Transmon qubit capacitively coupled to a semi-infinite transmission line and inductively to a flux line. These ports are used to control the qubit state and to change its transition frequency. They also lead to qubit decay into the transmission line and to dephasing due to flux noise.}
\end{figure}

The master equation~\eq{eq:ME_harmonic} was derived for an oscillator coupled to a transmission line, but this form of the master equation is quite general. In fact,~\eq{eq:HSBCavityNonMarkov} is itself a very generic system-bath Hamiltonian that can be used to model dissipation due to a variety of different noise sources~\cite{Caldeira81}. To model damping of an arbitrary quantum system, for example a transmon qubit or a coupled resonator-transmon system, the operator $\hat a$ in~\eq{eq:HSBCavityNonMarkov} is simply replaced with the relevant system operator that couples to the transmission line (or, more generally, the bath).

For the case of a transmon, see \cref{fig:qubitnoise}, $\hat H_S$ in~\cref{eq:ME_harmonic} is replaced with the Hamiltonian $\hH_\mathrm{q}$ of~\cref{eq:HsjPhi4b} together with the additional replacements $\mathcal D[\hat a]\bullet \to \mathcal{D}[\hat b]\bullet$, $\mathcal D[\hat a\dg]\bullet \to \mathcal D[\hat b\dg]\bullet$, and $\kappa \rightarrow \gamma$. Here, $\gamma = 2\pi\lambda(\wa)^2$ is the relaxation rate of the artificial atom which is related to the qubit-environment coupling strength evaluated at the qubit frequency.
This immediately leads to the master equation
\begin{equation}\label{eq:MEQubitBs}
    \dot\rhoS = {-}i[\hH_\mathrm{q},\rhoS]
    + \gamma (\bar n_{\gamma} + 1) \mathcal{D}[\bop]\rhoS + \gamma \bar n_{\gamma} \mathcal{D}[\bd]\rhoS,
\end{equation}
where $\rhoS$ now refers to the transmon state and $\bar n_{\gamma}$ is the thermal photon number of the transmon's environment. It is often assumed that $\bar n_{\gamma} \rightarrow 0$ but, just like for the oscillator, a residual thermal population is often observed in practice~\cite{Corcoles2011,Wang2019}.

Superconducting quantum circuits can also suffer from dephasing caused, for example, by fluctuations of parameters controlling their transition frequency and by dispersive coupling to other degrees of freedom in their environment. For a transmon, a phenomenological model for dephasing can be introduced by adding the following term to the master equation~\cite{Carmichael2002}
\begin{equation}\label{eq:MEDephasingTransmon}
2\gphi\mathcal{D}[\hat b\dg \hat b]\rhoS,
\end{equation}
with $\gphi$ the pure dephasing rate. Because of its insensitivity to charge noise (see \cref{fig:transmonregime}), $\gphi$ is often very small for the 0-1 transition of transmon qubits~\cite{Koch2007}. Given that charge dispersion increases exponentially with level number, dephasing due to charge noise can, however, be apparent on higher transmon levels, see for example \textcite{Egger2019}. Another source of dephasing for the transmon is the residual thermal photon population of a resonator to which the transmon is dispersively coupled. This can be understood from the form of the interaction in the  dispersive regime, $\chi_{ab} \ada \bdb$, where fluctuations of the photon number lead to a fluctuations in the qubit frequency and therefore to dephasing \cite{Bertet2005,Schuster2005,Gambetta2006,Rigetti2012}. We note that a term of the form of \cref{eq:MEDephasingTransmon} but with $\bdb$ replaced by $\ada$ can be added to the master equation of the oscillator to model this aspect. Oscillator dephasing rates are, however, typically small and this contribution is often neglected \cite{Reagor2016}. Other sources of relaxation and dephasing include two-level systems within the materials and interfaces of the devices \cite{Mueller2019}, quasiparticles \cite{Glazman2020} generated by a number of phenomenon including infrared radiation \cite{Corcoles2011,Barends2011} and even ionizing radiation \cite{Vepsalainen2020}.

Combining the above results, the master equation for a transmon subject to relaxation and dephasing assuming $\bar n_{\gamma}\to 0$ is 
\begin{equation}\label{eq:ME_transmon}
  \begin{aligned}
    \dot\rhoS ={}& -i[\hH_\mathrm{q},\rhoS] + \gamma \mathcal{D}[\bop]\rhoS + 2\gamma_\varphi\mathcal D[\hat b\dg \hat b]\rhoS.
  \end{aligned}
\end{equation}
It is common to express this master equation in the two-level approximation of the transmon, something that is obtained simply by taking $\hH_\mathrm{q} \to \hbar\omega_a \sz{}/2$, $\hat b\dg \hat b \to \left(\sz{} + 1\right)/2$, $\hat b \to \smm{}$ and $\hat b\dg \to \spp{}$.

Note that the rates $\gamma$ and $\gamma_\varphi$ appearing in the above expressions are related to the characteristic $T_1$ relaxation time and $T_2$ coherence time of the artificial atom which are defined as~\cite{schoelkopf:2003a}
\begin{subequations}
\begin{align}\label{eq:T1}
  T_1 ={}& \frac{1}{\gamma_1} = \frac{1}{\gamma_\downarrow + \gamma_\uparrow}
  \simeq \frac{1}{\gamma},\\
  T_2 ={}& \frac{1}{\gamma_2} = \left( \frac{\gamma_1}{2} + \gamma_\varphi \right)^{-1},
  \label{eq:T2_definition}
\end{align}
\end{subequations}
where $\gamma_\downarrow = (\bar n_\gamma + 1)\gamma$, $\gamma_\uparrow = \bar n_\gamma \gamma$. The approximation in \cref{eq:T1} is for $\bar n_{\gamma}\rightarrow 0$.
At zero temperature, $T_1$ is the characteristic time for the artificial atom to relax from its first excited state to the ground state. On the other hand, $T_2$ is the dephasing time, which quantifies the characteristic lifetime of coherent superpositions, and includes both a contribution from pure dephasing ($\gamma_\varphi)$ and relaxation ($\gamma_1$). Current best values for the $T_1$ and $T_2$ time of transmon qubits is in the 50 to 120~$\mu$s range for aluminum-based transmons \cite{Wei2019,Nersisyan2019,Devoret2013,Kjaergaard2019}. Relaxation times above 300~$\mu$s have been reported in transmon qubits where the transmon pads have been made with tantalum rather than aluminum, but the Josephson junction still made from aluminum and aluminum oxide \cite{Place2020}. Other superconducting qubits also show large relaxation and coherence times. Examples are $T_1,\,T_2\sim300~\mu$s for heavy-fluxonium qubits \cite{Zhang2020}, and $T_1 \sim 1.6$~ms and $T_2\sim25~\mu$s for the $0-\pi$ qubit \cite{Gyenis2019a}.

Qubit relaxation and incoherent excitation occur due to uncontrolled exchange of GHz frequency photons between the qubit and its environment. These processes are observed to be well described by the Markovian master equation  of \cref{eq:ME_transmon}. In contrast, the dynamics leading to dephasing are typically non-Markovian, happening at low-frequencies (i.e.~slow time scales set by the phase coherence time itself). As a result, these processes cannot very accurately be described by a Markovian master equation such as~\cref{eq:ME_transmon}. This equation thus represents a somewhat crude approximation to dephasing in superconducting qubits. That being said, in practice, the Markovian theory is still useful in particular because it correctly predicts the results of experiments probing the steady-state response of the system.

\subsection{Dissipation in the dispersive regime}\label{sec:DissipationDispersive}

\begin{figure}
\centering
\includegraphics{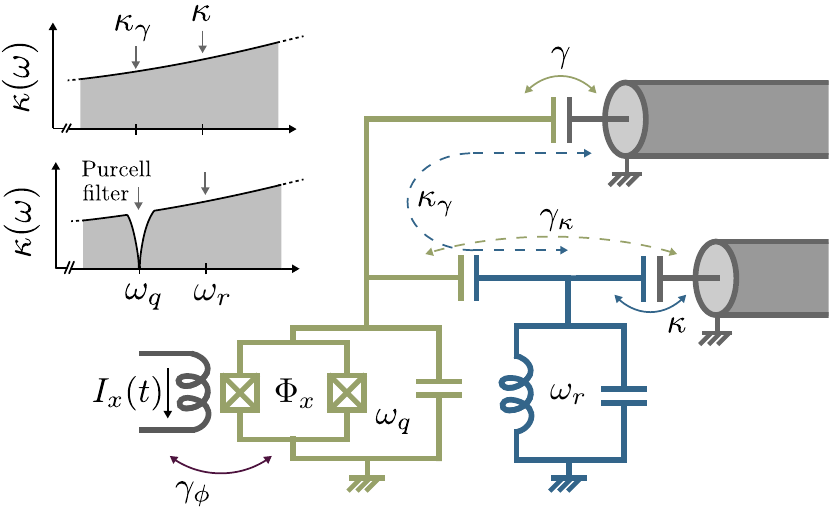}
\caption{\label{fig:dispersivenoise}
Because the dressed states in the dispersive regime are entangled qubit-cavity states, cavity damping at the rate $\kappa$ leads to qubit relaxation at the Purcell rate $\gamma_\kappa$. Conversely, qubit relaxation leads to cavity decay at the rate inverse Purcell rate $\kappa_\gamma$. Adding a Purcell filter (not shown) reduces the cavity density of states at the qubit frequency and therefore suppresses Purcell decay.}
\end{figure}

We now turn to a description of dissipation for the coupled transmon-resonator system of~\cref{sec:LightMatter}. Assuming that the transmon and the resonator are coupled to independent baths as illustrated in \cref{fig:dispersivenoise}, the master equation for this composite system is (taking $\bar n_{\kappa,\gamma}\rightarrow0$ for simplicity)
\begin{equation}\label{eq:JCMasterEquation}
  \begin{aligned}
    \dot\rhoS ={}& -i[\hat H,\rhoS] + \kappa \mathcal{D}[\aop]\rhoS + \gamma \mathcal{D}[\bop]\rhoS \\
    &+ 2\gamma_\varphi\mathcal D[\hat b\dg \hat b]\rhoS,
  \end{aligned}
\end{equation}
where $\rhoS$ is now a density matrix for the total system, and $\hat H$ describes the coupled system as in~\cref{eq:HTransmonJC}. Importantly, the above expression is only valid at small values of $g/(\wc,\wa)$. This is because energy decay occurs via transitions between system eigenstates while the above expression describes transitions between the uncoupled bare states. A derivation of the master equation valid at arbitrary $g$ can be found, for example, in \textcite{Beaudoin2011}.

More important to the present discussion is the fact that, at first glance, \cref{eq:JCMasterEquation} gives the impression that dissipative processes influence the transmon and the resonator in completely independent manners. However, because $\hat H$ entangles the two systems, the loss, for example, of a resonator photon can lead to qubit relaxation.  Moving to the dispersive regime, a more complete picture of dissipation emerges after applying the unitary transformation \eq{eq:UDispersive} not only on the Hamiltonian but also on the above master equation.
Neglecting fast rotating terms and considering corrections to second order in $\lambda$ (which is consistent if $\kappa,\,\gamma,\,\gamma_\varphi = \mathcal{O}(E_C g^2/\Delta^2)$), leads to the dispersive master equation~\cite{Boissonneault2009}
\begin{equation}\label{eq:DispersiveMasterEquation}
  \begin{aligned}
    \dot\rho_\mathrm{disp} ={}& -i[\hat H_\text{disp},\rho_\mathrm{disp}] \\
    & + (\kappa+\kappa_\gamma) \mathcal{D}[\aop]\rho_\mathrm{disp} + (\gamma+\gamma_\kappa) \mathcal{D}[\bop]\rho_\mathrm{disp}\\
    &+ 2\gamma_\varphi\mathcal D[\hat b\dg \hat b]\rho_\mathrm{disp}\\
    &+ \gamma_\Delta \mathcal D[\hat a\dg \hat b]\rho_\mathrm{disp}
    + \gamma_\Delta \mathcal D[\hat b\dg \hat a]\rho_\mathrm{disp},
  \end{aligned}
\end{equation}
where
\begin{align}\label{eq:DispersiveRates}
  \gamma_\kappa ={}& \left(\frac{g}{\Delta}\right)^2\kappa,\;\;
  \kappa_\gamma = \left(\frac{g}{\Delta}\right)^2\gamma,\;\;
  \gamma_{\Delta} = 2\left(\frac{g}{\Delta}\right)^2\gamma_\varphi,
\end{align}
and $\rho_\mathrm{disp} = \hat U_\mathrm{disp}^\dag \rhoS \hat U_\mathrm{disp}$ is the density matrix in the dispersive frame. This expression has three new rates, the first of which is known as the Purcell decay rate $\gamma_\kappa$~\cite{Purcell1946}. This rate captures the fact that the qubit can relax by emission of a resonator photon. It can be understood simply following \eq{eq:JCEigenstates} from the form of the dressed eigenstate $\ket{\overline{e, 0}} \sim \ket{e,0} + (g/\Delta)\ket{g,1}$ which is closest to a qubit excitation $\ket{e}$. This state is the superposition of the qubit first excited state with no photon and, with probability $(g/\Delta)^2$, the qubit ground state with a photon in the resonator. The latter component can decay at the rate $\kappa$ taking the dressed excited qubit to the ground state $\ket{g, 0}$ with a rate $\gamma_\kappa$. A similar intuition also applies to $ \kappa_\gamma$, now associated with a resonator photon loss through a qubit decay event.

The situation is more subtle for the last line of \eq{eq:DispersiveMasterEquation}. Following \textcite{Boissonneault2008,Boissonneault2009}, an effective master equation for the transmon only can be obtained from \eq{eq:DispersiveMasterEquation} by approximately eliminating the resonator degrees of freedom. This results in transmon relaxation and excitation rates given approximately by $\bar n \gamma_{\Delta}$, with $\bar n$ the average photon number in the resonator. Commonly known as dressed-dephasing, this leads to spurious transitions during qubit measurement and can be interpreted as originating from dephasing noise at the detuning frequency $\Delta$ that is up- or down-converted by readout photons to cause spurious qubit state transitions.

Because we have taken the shortcut of applying the dispersive transformation on the master equation, the above discussion neglects the frequency dependence of the various decay rates. In a more careful derivation, the dispersive transformation is applied on the system plus bath Hamiltonian, and only then is the master equation derived~\cite{Boissonneault2009}. The result has the same form as \eq{eq:DispersiveMasterEquation}, but with different expressions for the rates. Indeed, it is useful to write $\kappa = \kappa(\wc)$ and $\gamma = \gamma(\wa)$ to recognize that, while photon relaxation is probing the environment at the resonator frequency $\wc$, qubit relaxation is probing the environment at $\wa$. With this notation, the first two rates of \eq{eq:DispersiveRates} become in the more careful derivation $\gamma_\kappa = (g/\Delta)^2 \kappa(\wa)$ and $\kappa_\gamma = (g/\Delta)^2 \gamma(\wc)$. In other words, Purcell decay occurs by emitting a photon at the qubit frequency and not at the resonator frequency as suggested by the completely white noise model used to derive \eq{eq:DispersiveRates}. In the same way, it is useful to write the dephasing rate as $\gamma_\varphi = \gamma_\varphi(\omega \to 0)$ to recognize the importance of low-frequency noise. Using this notation, the rates in the last two terms of \cref{eq:DispersiveMasterEquation} become, respectively,  $\gamma_\Delta = 2(g/\Delta)^2\gamma_\varphi(\Delta)$ and $\gamma_{-\Delta} = 2(g/\Delta)^2\gamma_\varphi(-\Delta)$~\cite{Boissonneault2009}. In short, dressed dephasing probes the noise responsible for dephasing at the transmon-resonator detuning frequency $\Delta$. This observation was used to probe this noise at GHz frequencies by \textcite{Slichter2012}.

It is important to note that the observations in this section result from the qubit-oscillator dressing that occurs under the Jaynes-Cummings Hamiltonian. For this reason, the situation is very different if the electric-dipole interaction leading to the Jaynes-Cummings Hamiltonian is replaced by a longitudinal interaction of the form of \cref{eq:H_Longitudinal}. In this case, there is no light-matter dressing and consequently no Purcell decay or dressed-dephasing \cite{Kerman2013,Billangeon2015}. This is one of the advantages of this alternative light-matter coupling.

\subsection{Multi-mode Purcell effect and Purcell filters}\label{sec:Purcell}

Up to now we have considered dissipation for a qubit dispersively coupled to a single-mode oscillator. Replacing the latter with a multi-mode resonator leads to dressing of the qubit by all of the resonator modes and therefore to a modification of the Purcell decay rate. Following the above discussion, one may then expect the contributions to add up, leading to the modified rate $\sum_{m=0}^\infty (g_m/\Delta_m)^2\kappa_m$, with $m$ the mode index. However, when accounting for the frequency dependence of $\kappa_m$, $g_m$ and $\Delta_m$, this expression diverges~\cite{Houck2008}. It is possible to cure this problem using a more refined model including the finite size of the transmon and the frequency dependence of the impedance of the resonator's input and output capacitors~\cite{Bourassa2012b,Malekakhlagh2017}.

Given that damping rates in quantum electrical circuits are set by classical system parameters~\cite{Leggett1984}, a simpler approach to compute the Purcell rate exists. It can indeed be shown that $\gamma_\kappa =  \mathrm{Re}[Y(\wa)]/C_\Sigma$, with $Y(\omega) = 1/Z(\omega)$ the admittance of the electromagnetic environment seen by the transmon~\cite{Esteve1986,Houck2008}. This expression again makes it clear that relaxation probes the environment (here represented by the admittance) at the system frequency. It also suggests that engineering the admittance $Y(\omega)$ such that it is purely reactive at $\wa$ can cancel Purcell decay (see the inset of \cref{fig:dispersivenoise}). This can be done, for example, by adding a transmission-line stub of appropriate length and terminated in an open circuit at the output of the resonator, something which is known as a Purcell filter~\cite{Reed2010}.  Because of the increased freedom in optimizing the system parameters (essentially decoupling the choice of $\kappa$ from the qubit relaxation rate), various types of Purcell filters are commonly used experimentally~\cite{Jeffrey2014,Bronn2015b,Walter2017}.

\subsection{\label{sec:drives}Controlling quantum systems with microwave drives}

While connecting a quantum system to external transmission lines leads to losses, such connections are nevertheless necessary to control and measure the system. Consider a continuous microwave tone of frequency $\wdrive$ and phase $\phidrive$ applied to the input port of the resonator. A simple approach to model this drive is based on the input-output approach of \cref{sec:inout}. Indeed, the drive can be taken into account by replacing the input field $\bin(t)$ in~\cref{eq:inouteom} with $\bin(t) \to \bin(t) + \beta(t)$, where $\beta(t) = A(t) \exp(-i\wdrive t -i\phidrive)$ is the coherent classical part of the input field of amplitude $A(t)$. The resulting term $\sqrt{\kappa}\beta(t)$ in the Langevin equation can be absorbed in the system Hamiltonian with the replacement $\hat H_S \to \hat H_S + \hat H_\text{d}$ where
\begin{equation}\label{eq:DriveOnCavity}
  \hat H_\text{d} = \hbar \left[\varepsilon(t) \ad e^{-i\wdrive t - i \phidrive} + \varepsilon^*(t)\aop e^{i\wdrive t + i \phidrive}\right]),
\end{equation}
with $\varepsilon(t) = i\sqrt{\kappa}A(t)$ the possibly time-dependent amplitude of the drive as seen by the resonator mode. Generalizing to multiple drives on the resonator and/or drives on the transmon is straightforward.

Moreover, the Hamiltonian $\hat H_\text{d}$ is the generator of displacement in phase space of the resonator. As a result, by choosing appropriate parameters for the drive, evolution under $\hH_\text{d}$ will bring the intra-resonator state from vacuum to an arbitrary coherent state~\cite{Gardiner1999,Carmichael2002}
\begin{equation}\label{eq:CoherentState}
  \ket\alpha = \Dop(\alpha)\ket0 = e^{-|\alpha|^2/2}\sum_{n=0}^\infty \frac{\alpha^n}{\sqrt{n!}}\ket{n},
\end{equation}
where $\Dop(\alpha)$ is known as the displacement operator and takes the form
\begin{equation}\label{eq:DisplacementOp}
  \Dop(\alpha) = e^{\alpha \ad - \alpha^* \aop}.
\end{equation}
As discussed in the next section, coherent states play an important role in qubit readout in circuit QED.

It is important to note that $\hat H_\text{d}$ derives from \cref{eq:inouteom} which is itself the result of a rotating-wave approximation. As can be understood from \cref{eq:Hsj}, before this approximation, the drive rather takes the form
$i\hbar \varepsilon(t) \cos(\wdrive t + \phidrive)(\ad  - \aop)$. Although $\hat H_\text{d}$ is sufficient in most cases of practical interest, departures from the predictions of \cref{eq:DriveOnCavity} can been seen at large drive amplitudes \cite{Pietikainen2017,Verney2019}.

\section{\label{sec:readout}Measurements in circuit QED} 

Before the development of circuit QED, the quantum state of superconducting qubits was measured by fabricating and operating a measurement device, such as a single-electron transistor, in close proximity to the qubit~\cite{Makhlin2001}. A challenge with such an approach is that the readout circuitry must be strongly coupled to the qubit during measurement so as to extract information on a time scale much smaller than $T_1$, while being well decoupled from the qubit when the measurement is turned off to avoid unwanted back-action. Especially given that measurement necessarily involves dissipation~\cite{Landauer1991}, simultaneously satisfying these two requirements is challenging. Circuit QED, however, has several advantages to offer over the previous approaches. Indeed, as discussed in further detail in this section, qubit readout in this architecture is realized by measuring scattering of a probe tone off an oscillator coupled to the qubit. This approach first leads to an excellent measurement on/off ratio since qubit readout only occurs in the presence of the probe tone. A second advantage is that the necessary dissipation now occurs away from the qubit, essentially at a voltage meter located at room temperature, rather than in a device fabricated in close proximity to the qubit. Unwanted energy exchange is moreover inhibited when working in the dispersive regime where the effective qubit-resonator interaction \eq{eq:HQubitDispersive} is such that even the probe-tone photons are not absorbed by the qubit. As a result, the 
backaction on the qubit is to a large extent limited to the essential dephasing that quantum measurements must impart on the measured system leading, in principle, to a quantum non-demolition (QND) qubit readout.

Because of the small energy of microwave photons with respect to optical photons, single-photon detectors in the microwave frequency regime are still being developed, see \cref{sec:SinglePhotonDetector}. Therefore, measurements in circuit QED rely on amplification of weak microwave signals followed by detection of field quadratures using heterodyne detection. Before discussing qubit readout, the objective of the next subsection is to explain these terms and go over the main challenges related to such measurements in the quantum regime.

\subsection{Microwave field detection}
\label{sec:FieldDetection}

\begin{figure}
  \centering
  \includegraphics[width=\columnwidth]{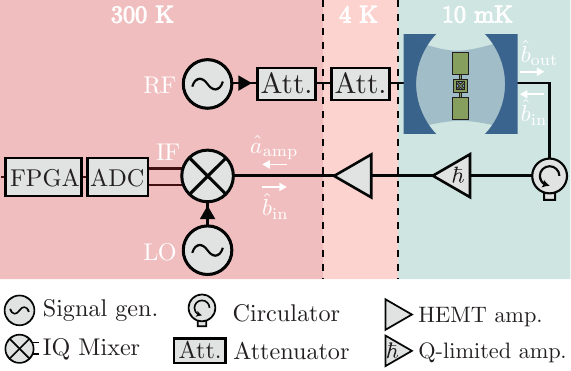}
  \caption{\label{fig:MeasurementChain}
    Schematic representation of the microwave measurement chain for field detection in circuit QED, with the resonator depicted as a Fabry-Perot cavity. The signal (RF) from a microwave source is applied to the input port of the resonator first passing through attenuators to reduce the level of  thermal radiation. After passing through a circulator, the resonator's output field is first amplified by a quantum-limited amplifier, such as a JPA or a JTWPA, and then by an HEMT amplifier. The signal is then mixed with a local oscillator (LO). The signal at the output of the mixer is digitized with an analog-to-digital converter (ADC) and can be further processed by a field-programmable gate array (FPGA). The two lines at the output of the mixer correspond to the two quadratures of the field. The temperature at which the different components are operated is indicated.
  }
\end{figure}

\Cref{fig:MeasurementChain} illustrates a typical measurement chain in circuit QED. The signal of a microwave source is directed to the input port of the resonator first going through a series of attenuators thermally anchored at different stages of the dilution refrigerator. The role of these attenuators is to absorb the room-temperature thermal noise propagating towards the sample. The field transmitted by the resonator is first amplified, then mixed with a reference signal, converted from analog to digital, and finally processed with an FPGA or recorded. Circulators are inserted before the amplification stage to prevent noise generated by the amplifier from reaching the resonator. Circulators are directional devices that transmit signals in the forward direction while strongly attenuating signals propagating in the reverse direction (here coming from the amplifier)~\cite{Pozar2011}. In practice, circulators are bulky off-chip devices relying on permanent magnets that are not compatible with the requirement for integration with superconducting quantum circuits. They also introduce additional losses, for example due to insertion losses and off-chip cable losses. Significant effort is currently being devoted to developing compact, on-chip, superconducting circuit-based circulators~\cite{Kamal2011,Chapman2017,Lecocq2017,Abdo2019}.

In practice, the different components and cables of the measurement chain have a finite bandwidth which we will assume to be larger than the bandwidth of the signal of interest $\bout(t)$ at the output of the resonator. To account for the finite bandwidth of the measurement chain and to simplify the following discussion, it is useful to consider the filtered output field
\begin{equation}\label{eq:bout_filtered}
  \begin{aligned}
  \aopf(t) ={}&
  (f \star \bout)(t) \\
  ={}& \int_{-\infty}^\infty d\tau f(t-\tau) \bout(\tau)\\
  ={}& \int_{-\infty}^\infty d\tau f(t-\tau) \left[\sqrt{\kappa}\aop(\tau) + \bin(\tau)\right],
  \end{aligned}
\end{equation}
which is linked to the intra-cavity field $\aop$ via the input-ouput boundary condition \cref{eq:inoutrel} which we have used in the last line. In this expression, the filter function $f(t)$ is normalized to $\int_{-\infty}^\infty dt |f(t)|^2 = 1$ such that $[\aopf(t),\aopf\dg(t)]=1$. As will be discussed later in the context of qubit readout, in addition to representing the measurement bandwidth, filter functions are used to optimize the distinguishability between the qubit states.

Ignoring the presence of the circulator and assuming that a phase-preserving amplifier (i.e.~an amplifier that amplifies both signal quadratures equally) is used, in the first stage of the measurement chain the signal is transformed according to~\cite{Caves1982,Clerk2010} 
\begin{equation}\label{eq:amplifier_inout}
  \aamp = \sqrt G \aopf + \sqrt{G-1} \hat h^\dag,
\end{equation}
where $G$ is the power gain and $\hat{h}^\dag$ accounts for noise added by the amplifier. The presence of this  added noise is required for the amplified signal to obey the bosonic commutation relation, $[\aamp,\aampd]=1$. Equivalently, the noise must be present because the two quadratures of the signal are canonically conjugate.  Amplification of both quadratures without added noise would allow us to violate the Heisenberg uncertainty relation between the two quadratures.  

In a standard parametric amplifier, $\aopf$ in \cref{eq:amplifier_inout} represents the amplitude of the signal mode and $h$ represents the amplitude of a second mode called the idler.  The physical interpretation of \cref{eq:amplifier_inout} is that an ideal amplifier performs a Bogoljubov transformation on the signal and idler modes.  The signal mode is amplified, but the requirement that the transformation be canonical implies that the (phase conjugated and amplified) quantum noise from the idler port must appear in the signal output port. Ideally, the input to the idler is vacuum with $\av{\hat{h}^\dag\hat{h}} = 0$ and $\av{\hat{h}\hat{h}^\dag} = 1$, so the amplifier only adds quantum noise. Near quantum-limited amplifiers with $\sim 20$ dB power gain approaching this ideal behavior are now routinely used in circuit QED experiments. These Josephson junction-based devices, as well as the distinction between phase-preserving and phase-sensitive amplification, are discussed further in~\cref{sec:QO:AmplificationSqueezing}.

\begin{figure}[t]
  \centering
  \includegraphics[width=0.7\columnwidth]{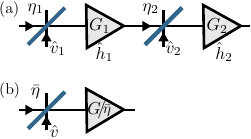}
  \caption{\label{fig:QuantumEfficiency}
    (a) Amplification chain with amplifiers of gain $G_{1,2}$ and noise mode $\hat h_{1,2}$ with attenuation modeled by beam splitters of transmitivity $\eta_{1,2}$. The beam splitters each have a vacuum port with vacuum mode $\hat v_{1,2}$ such that $\av{\hat v^\dag_{1,2}\hat v_{1,2}}=0$. The quantum efficiency derived from this model is $\eta = 1/(N_T+1)\le 1$, with $N_T=\av{\hat h_T^\dag \hat h_T}$ the total added noise number given in \cref{eq:NoiseNumber}. (b) Alternative model where a noisy amplifier is modeled by a noiseless amplifier preceded by a beam splitter of transmitivity $\bar\eta$. The quantum efficiency derived from this model is $\bar \eta = 1/(2\mathcal{A}+1) \le 1/2$ with $\mathcal{A}$ the added noise given in \cref{eq:AddedNoise}.
  }
\end{figure}

To measure the very weak signals that are typical in circuit QED,  the output of the first near-quantum limited amplifier is further amplified by a low-noise high-electron-mobility transistor (HEMT) amplifier. The latter acts on the signal again following \cref{eq:amplifier_inout}, now with a larger power gain $\sim 30-40$ dB but also larger added noise photon number. 
The very best cryogenic HEMT amplifiers in the 4-8 GHz band have noise figures as low as $\av{\hat{h}^\dag\hat{h}} \sim 5 - 10$. However, the effect of attenuation due to cabling up to the previous element of the amplification chain, i.e.~a quantum-limited amplifier or the sample of interest itself, can degrade this figure significantly. A more complete understanding of the added noise in this situation can be derived from \cref{fig:QuantumEfficiency}(a). There, beam splitters of transmissivity $\eta_{1,2}$ model the attenuation leading to the two amplifiers of gain labelled $G_1$ and $G_2$. Taking into account  vacuum noise $\hat v_{1,2}$ at the beam splitters, the input-output expression of this chain can be cast under the form of \cref{eq:amplifier_inout} with a total gain $G_T = \eta_1\eta_2G_1G_2$ and noise mode $\hat h_T^\dag$ corresponding to the total added noise number 
\begin{equation}\label{eq:NoiseNumber}
  \begin{split}
    N_T
    &= \frac{1}{G_T-1}
    \Big[
    \eta_1(G_1-1)G_2(N_1+1) \\
     &\quad\quad\quad\quad +(G_2-1)(N_2+1)  
    \Big] -1\\
    &\approx \frac{1}{\eta_1}\left[1+N_1+\frac{N_2}{\eta_2 G_1}\right]-1,
  \end{split}
\end{equation}
with $N_i = \av{\hat h^\dag_i \hat h_i}$ with $i=$ 1, 2, $T$. The last expression corresponds to the large gain limit. If the gain $G_1$ of the first amplifier is large, the noise of the chain is dominated by the noise $N_1$ of the first amplifier. This emphasizes the importance of using near quantum-limited amplifiers with low noise in the first stage of the chain. In the literature, the quantum efficiency $\eta = 1/(N_T + 1)$ is often used to characterize the measurement chain, with $\eta = 1$ in the ideal case $N_T = 0$.

It is worthwhile to note that another definition of the quantum efficiency can often be found in the literature. This alternative definition is based on \cref{fig:QuantumEfficiency}(b) where a noisy amplifier of gain $G$ is replaced by a noiseless amplifier of gain $G/\bar\eta$ preceded by a fictitious beam splitter of transmittivity $\bar\eta$ adding vacuum noise to the amplifier's input \cite{Leonhardt1993}. The quantum efficiency corresponds, here, to the transmittivity $\bar\eta$ of the fictitious beam splitter. The input-ouput relation of the network of \cref{fig:QuantumEfficiency}(b) with its noiseless phase-preserving amplifier reads $\aamp = \sqrt{G/\bar\eta}(\sqrt{\bar\eta}\aopf+\sqrt{1-\bar\eta}\hat v)$, something which can be expressed as
\begin{equation}\label{eq:amplifier_noise1}
  \av{|\aamp|^2} = \frac{G}{\bar\eta}\left[(1-\bar\eta)\frac{1}{2}+\bar\eta\av{|\aopf|^2}\right],
\end{equation}
with $\av{|\hat O|^2} = \av{\{\hat O^\dag,\hat O\}}/2$ the symmetrized fluctuations. The first term of the above expression corresponds to the noise added by the amplifier, here represented by vacuum noise added to the signal before amplification, while the second term corresponds to noise in the signal at the input of the amplifier. 
On the other hand, \Cref{eq:amplifier_inout} for a noisy amplifier can also be cast in the form of \cref{eq:amplifier_noise1} with
\begin{equation}\label{eq:amplifier_noise2}
   \av{|\aamp|^2} = G(\mathcal{A}+\av{|\aopf|^2}),
\end{equation}
where we have introduced the added noise
\begin{equation}\label{eq:AddedNoise}
  \mathcal{A}  = \frac{(G-1)}{G}\left(\av{\hat h^\dag\hat h}+\frac12\right). 
\end{equation}
In the limit of low amplifier noise $\av{\hat h^\dag\hat h} \rightarrow 0$ and large gain, the added noise is found to be bounded by $\mathcal{A} \ge (1-G^{-1})/2 \simeq 1/2$ corresponding to half a photon of noise \cite{Caves1982}. Using \cref{eq:amplifier_noise1,eq:amplifier_noise2}, the quantum efficiency of a phase-preserving amplifier can therefore be written as $\bar\eta = 1/(2\mathcal{A}+1) \le 1/2$ and is found to be bounded by 1/2 in the ideal case. Importantly, the concept of quantum efficiency is not limited to amplification, and can be applied to the whole measurement chain illustrated in \cref{fig:MeasurementChain}.

Using \cref{eq:TLvoltage,eq:amplifier_inout}, the voltage after amplification can be expressed as 
\begin{equation}\label{eq:Vamp}
  \hat V_\mathrm{amp}(t) \simeq{} \sqrt{\frac{\hbar \wrf Z_\tml}{2}} \left[ e^{- i\wrf t} \aamp + \hc\right],
\end{equation}
where $\wrf$ is the signal frequency. To simplify the expressions, we have dropped the phase associated to the finite cable length. We have also dropped the contribution from the input field $\bin(t)$ moving towards the amplifier in the opposite direction at this point, cf.~\cref{fig:MeasurementChain}, because this field is not amplified and therefore gives a very small contribution compared to the amplified output field. Recall, however, the contribution of this field to the filtered signal~\cref{eq:bout_filtered}.

\begin{figure}[t]
  \centering \includegraphics[width=0.5\columnwidth]{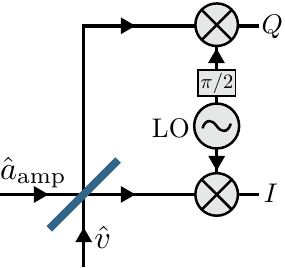}\\
  \caption{Schematic representation of an IQ mixer. The RF signal $\aamp$ is split into two parts at a power divider, here illustrated as a beam splitter to account for added noise due to internal modes. Ideally, only vacuum noise $\hat v$ is introduced at that stage. The two outputs are combined with a local oscillator (LO) at mixers. By phase shifting the LO by $\pi/2$ in one of the two arms, it is possible to simultaneously measure the two quadratures of the field.
}\label{fig:IQMixer}
\end{figure}

Different strategies can be used to extract information from the amplified signal, and here we take the next stage of the chain to be an IQ-mixer. As schematically illustrated in \cref{fig:IQMixer}, in this microwave device the signal first encounters a power divider, illustrated here as a beam splitter to account for added noise due to internal modes, followed in each branch by mixers with local oscillators (LO) that are offset in phase by $\pi/2$. The LO consists in a reference signal of well-defined amplitude $A_\mathrm{LO}$, frequency $\wlo$ and phase $\philo$:
\begin{equation}
V_\mathrm{LO}(t) = A_\mathrm{LO}\cos(\wlo t - \philo).
\end{equation}
Mixers use nonlinearity to down-convert the input signal to a lower frequency referred to as the intermediate frequency (IF) signal. Describing first the signal as a classical voltage $V_\mathrm{RF}(t)=A_\mathrm{RF}\cos(\wrf t+ \phi_\mathrm{RF})$, the output at one of these mixers is~\cite{Pozar2011}
\begin{equation}\label{eq:Mixer}
\begin{split}
V_\mathrm{mixer}(t)
&= K V_\mathrm{RF}(t)  V_\mathrm{LO}(t) \\
&= \frac{1}{2}K A_\mathrm{LO}A_\mathrm{RF} \left\{
\cos[(\wlo-\wrf)t-\philo]
\right.\\&\left.\qquad
+\cos[(\wlo+\wrf)t-\philo]
\right\},
\end{split}
\end{equation}
where $K$ accounts for voltage conversion losses. According to the above expression, mixing with the LO results in two sidebands at frequencies $\wlo\pm\wrf$. The high frequency component is filtered out with a low-pass filter (not shown) leaving only the lower sideband of frequency $\wif = \wlo-\wrf$. The choice  $\wif \neq0$ is known as heterodyne detection. Taking the LO frequency such that $\wif$ is in the range of few tens to a few hundreds of MHz, the signal can be digitized using an analog to digital converter (ADC) with a sampling rate chosen in accordance with the bandwidth requirements of the signal to be recorded. This bandwidth is set by the choice of IF frequency and the signal bandwidth, typically a few MHz to a few tens of MHz if set by the  bandwidth $\kappa/2\pi$ of the cavity chosen for the specific circuit QED application such as qubit readout. The recorded signal can then be averaged, or analyzed in more complex ways, using real-time field-programmable gate array (FPGA) electronics or processed offline. A detailed discussion of digital signal processing in the context of circuit QED can be found in \textcite{Salathe2018}.

Going back to a quantum mechanical description of the signal by combining \cref{eq:Vamp,eq:Mixer}, the IF signals at the $I$ and $Q$ ports of the IQ-mixer read
\begin{subequations}\label{eq:V_IF}
  \begin{align}
  \begin{split}
    \hat V_\mathrm{I}(t) &= 
    V_\mathrm{IF}
    \left[\hX_f(t) \cos(\wif t) - \hP_f(t) \sin(\wif t)\right]\\
    &\quad+ \hat V_\mathrm{noise,I}(t),
  \end{split}\\
  \begin{split}
    \hat V_\mathrm{Q}(t) &= 
    -V_\mathrm{IF}
    \left[\hP_f(t) \cos(\wif t) + \hX_f(t) \sin(\wif t)\right]\\
    &\quad+ \hat V_\mathrm{noise,Q}(t),
  \end{split}
  \end{align}
\end{subequations}
where we have taken $\philo = 0$ in the $I$ arm of the IQ-mixer, and $\philo = \pi/2$ in the $Q$ arm. We have defined $V_\mathrm{IF} = K A_\mathrm{LO} \sqrt{\kappa G Z_\tml \hbar \wrf/2}$, and $\hat V_\mathrm{noise,I/Q}$ as the contributions from the amplifier noise and any other added noise. We have also introduced the quadratures
\begin{equation}
\hX_f=\frac{\adf+\aopf}{2}, \qquad  \hP_f=\frac{i(\adf-\aopf)}{2},
\end{equation}
the dimensionless position and momentum operators of the simple harmonic oscillator, here defined such that $[\hX_f,\hP_f] = i/2$. Taken together, $\hat V_\mathrm{I}(t)$ and $\hat V_\mathrm{Q}(t)$ trace a circle in the $x_f-p_f$ plane and contain information about the two quadratures at all times. It is therefore possible to digitally transform the signals by going to a frame where they are stationary using the rotation matrix 
\begin{equation}
  R(t) = 
  \begin{pmatrix}
    \cos(\wif t) & -\sin(\wif t)\\
    \sin(\wif t) & \cos(\wif t)
  \end{pmatrix}
\end{equation}
to extract $\hX_f(t)$ and $\hP_f(t)$.

We note that the case $\wif = 0$ is generally known as homodyne detection~\cite{Leonhardt1997,Gardiner1999,Wiseman2010,Pozar2011}. In this situation, 
the IF signal in one of the arms of the IQ-mixer is  proportional to
\begin{equation}\label{eq:X_homodyne}
\begin{split}
\hX_{f,\philo}
&= \frac{\ad_f e^{i\philo} + \aop_f e^{-i\philo}}{2}\\
&= \hX_f\cos\philo + \hP_f\sin\philo.
\end{split}
\end{equation}
While this is in appereance simpler and therefore advantageous, this approach is succeptible to $1/f$ noise and drift because the homodyne signal is at DC. It is also worthwhile to note that homodyne detection as realized with the approach described here differs from optical homodyne detection which can be performed in a noiseless fashion (in the present case, noise is added at the very least by the phase-preserving amplifiers and the noise port of the IQ mixer).  The reader is referred to \textcite{Schuster2005} and \textcite{Krantz2019} for more detailed discussions of the different field measurement techniques and their applications in the context of circuit QED.

\subsection{Phase-space representations and their relation to field detection}
\label{sec:PhaseSpace}

In the context of field detection, it is particularly useful to represent the quantum state of the electromagnetic field using phase-space representations. There exists several such representations and here we focus on the Wigner function and the Husimi-$Q$ distribution~\cite{Carmichael2002,Haroche2006}. This discussion applies equality well to the intra-cavity field $\aop$ as to the filtered output field $\aopf$.

The Wigner function is a quasiprobability distribution given by the Fourier transform
\begin{equation}\label{eq:Wigner}
W_\rho(x,p) = \frac{1}{\pi^2}\iint_{-\infty}^\infty dx' dp' C_\rho(x',p') e^{2i (px'  -  xp')}
\end{equation}
of the characteristic function
\begin{equation}
C_\rho(x,p) = \mathrm{Tr}\left\{\rho \,e^{2i(p\hX-x\hP)}\right\}.
\end{equation}
With $\rho$ the state of the electromagnetic field, $C_\rho(x,p)$ can be understood as the expectation value of the displacement operator
\begin{equation}
\Dop(\alpha) = e^{2i(p\hX-x\hP)} = e^{\alpha \ad - \alpha^* \aop},
\end{equation}
with $\alpha = x + i p$, see \cref{eq:DisplacementOp}.

\begin{figure}[t]
  \centering
  \includegraphics[width=0.8\columnwidth]{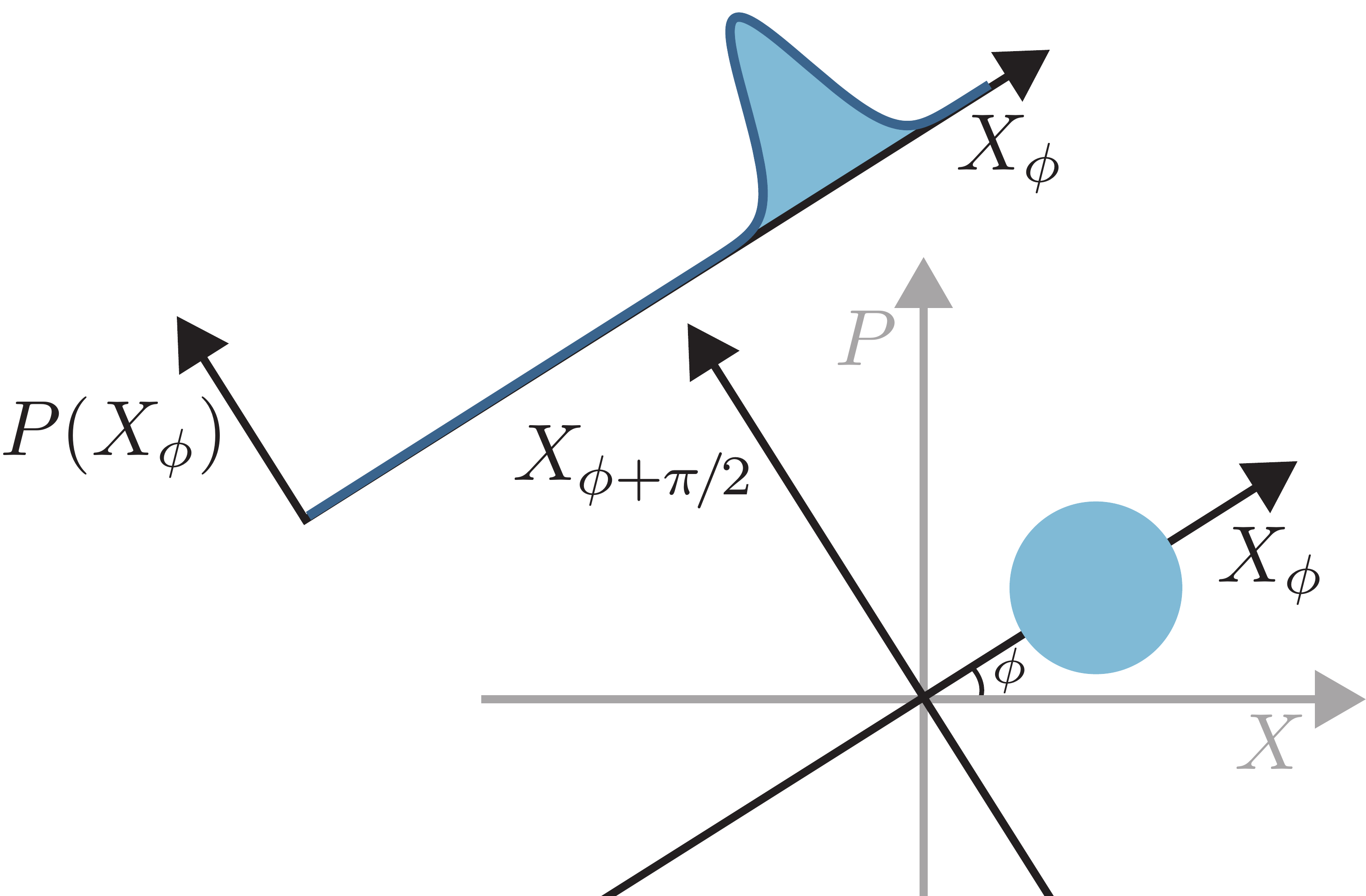}\\
  \caption{
  Pictorial phase-space distribution of a coherent state and its marginal along an axis $X_\phi$ rotated by $\phi$ from $X$.
  }
 \label{fig:PhaseSpaceMarginal}
\end{figure}

Coherent states, already introduced in \cref{eq:CoherentState}, have particularly simple Wigner functions. Indeed, as illustrated schematically in \cref{fig:PhaseSpaceMarginal}, the Wigner function $W_{\ket \beta}(\alpha)$ of the coherent state $\ket\beta$ is simply a Gaussian centered at $\beta$ in phase space:
\begin{equation}\label{eq:WignerCoherentState}
W_{\ket{\beta}}(\alpha) = \frac2\pi e^{-2|\alpha-\beta|^2}.
\end{equation}
The width $1/\sqrt{2}$ of the Gaussian is a signature of quantum noise and implies that coherent states saturate the Heisenberg inequality, $\Delta X \Delta P = 1/2$ with $\Delta O^2 = \langle\hat O^2\rangle-\langle\hat O\rangle^2$. We note that, in contrast to \cref{eq:WignerCoherentState}, Wigner functions take negative values for non-classical states of the field.

In the context of dispersive qubit measurements, the Wigner function is particularly useful because it is related to the probability distribution for the outcome of measurements of the quadratures $\hX$ and $\hP$. Indeed, the marginals  $P(x)$ and $P(p)$, obtained by integrating $W_\rho(x,p)$ along the orthogonal quadrature, are simply given by
\begin{subequations}
  \label{eq:WignerMarginals}
\begin{align}
P(x) = \int_{-\infty}^\infty dp\, W_\rho(x,p) &= \me{x}{\rho}{x},\\
P(p) = \int_{-\infty}^\infty dx\, W_\rho(x,p) &= \me{p}{\rho}{p},
\end{align}
\end{subequations}
where $\ket x$ and $\ket p$ are the eigenstate of $\hX$ and $\hP$, respectively.
This immediately implies that the probability distribution of the outcomes of an ideal homodyne measurement of the quadrature $\hX_\phi$ is given by $P(x_\phi)$ obtained by integrating the Wigner function $W_\rho(\alpha)$ along the orthogonal quadrature $\hX_{\phi+\pi/2}$. This is schematically illustrated for a coherent state in Fig.~\ref{fig:PhaseSpaceMarginal}.

Another useful phase-space function is the Husimi-Q distribution which, for a state $\rho$, takes the simple form
\begin{equation}
Q_\rho(\alpha) = \frac1\pi \me{\alpha}{\rho}{\alpha}.
\end{equation}
This function represents the probability distribution of finding $\rho$ in the coherent state $\ket\alpha$ and, in contrast to $W_\rho(\alpha)$, it is therefore always positive.

Since $Q_\rho(\alpha)$ and $W_\rho(\alpha)$ are both complete descriptions of the state $\rho$, it is not surprising that one can be expressed in terms of the other. For example, in terms of the Wigner function, the Q-function takes the form~\cite{Carmichael2002}
\begin{equation}
Q_\rho(\alpha)
= \frac{2}{\pi}\int_{-\infty}^\infty d^2\beta\, W_\rho(\beta) e^{-2|\alpha-\beta|^2}
= W_\rho(\alpha) * W_{\ket{0}}(\alpha).
\end{equation}
The Husimi-Q distribution $Q_\rho(\alpha)$ is thus obtained by convolution of the Wigner function with a Gaussian, and is therefore smoother than $W_\rho(\alpha)$. As made clear by the second equality, this Gaussian is in fact the Wigner function of the vacuum state, $W_{\ket{0}}(\alpha)$, obtained from \eq{eq:WignerCoherentState} with $\beta=0$.
In other words, the Q-function for $\rho$ is obtained from the Wigner function of the same state after adding vacuum noise. As already illustrated in Fig.~\ref{fig:IQMixer}, heterodyne detection with an IQ mixer  adds (ideally) vaccum noise to the signal before detection.
This leads to the conclusion that the probability distributions for the simultaneous measurement of two orthogonal quadratures in heterodyne detection is given by the marginals of the Husimi-Q distribution rather than of the Wigner function~\cite{Caves2012a}.

\subsection{Dispersive qubit readout}\label{sec:DispersiveQubitReadout}

\subsubsection{Steady-state intra-cavity field}

As discussed in Sec.~\ref{sec:dispersive}, in the dispersive regime the transmon-resonator Hamiltonian is well approximated by
\begin{equation}\label{eq:HQubitDispersiveSimple}
    \hH_\text{disp} \approx \hbar \left(\wc + \chi \sz{} \right) \hat a^\dagger \hat a + \frac{\hbar\wa}2 \hat \sigma_z.
\end{equation}
To simplify the discussion, here we have truncated the transmon Hamiltonian to its first two levels,
absorbed Lamb shifts in the system frequencies, and neglected a transmon-induced nonlinearity of the cavity [$K_a$ appearing in~\cref{eq:HTransmonDispersive}].
As made clear by the first term of the above expression, in the dispersive regime, the resonator frequency becomes qubit-state dependent: If the qubit is in $\ket{g}$ then $\av{\sz{}}=-1$ and the resonator frequency is $\wc-\chi$. On the other hand, if the qubit is in $\ket e$, $\av{\sz{}}=1$ and  $\wc$ is pulled to $\wc+\chi$. In this situation, driving the cavity results in a qubit-state dependent coherent state, $\ket{\alpha_{g,e}}$. Thus, if the qubit is initialized in the superposition $c_g\ket{g} + c_e \ket{e}$, the system evolves to an entangled qubit-resonator state of the form
\begin{equation}\label{eq:entangledQubitPointer}
c_g\ket{g, \alpha_g} + c_e \ket{e, \alpha_e}.
\end{equation}

To interpret this expression, let us recall the paradigm of the Stern-Gerlach experiment.  There, an atom passes through a magnet and the field gradient applies a spin-dependent force to the atom that entangles the spin state of the atom with the momentum state of the atom (which in turn determines where the atom lands on the detector).  The experiment is usually described as measuring the spin of the atom, but in fact it only measures the final position of that atom on the detector.  However, since the spin and position are entangled, we can uniquely infer the spin from the position, provided there is no overlap in the final position distributions for the two spin states.  In this case we have effectively performed a projective measurement of the spin.

By analogy, if the spin-dependent coherent states of the microwave field, $\alpha_{e,g}$, can be resolved by heterodyne detection, then they act as pointer states in the qubit measurement. Moreover, since $\hH_\text{disp}$ commutes with the observable that is measured, $\sz{}$, this is a QND (quantum non-demolition) measurement \cite{Braginsky1980} (in contrast to the Stern-Gerlach measurement which is destructive).  Note that for a system initially in a superposition of eigenstates of the measurement operator, a QND measurement \emph{does} in fact change the state by randomly collapsing it onto one of the measurement eigenstates.  The true test of `QNDness' is that subsequent measurement results are not random but simply reproduce the first measurement result.

\begin{figure}[t]
  \centering
  \includegraphics[width=1\columnwidth]{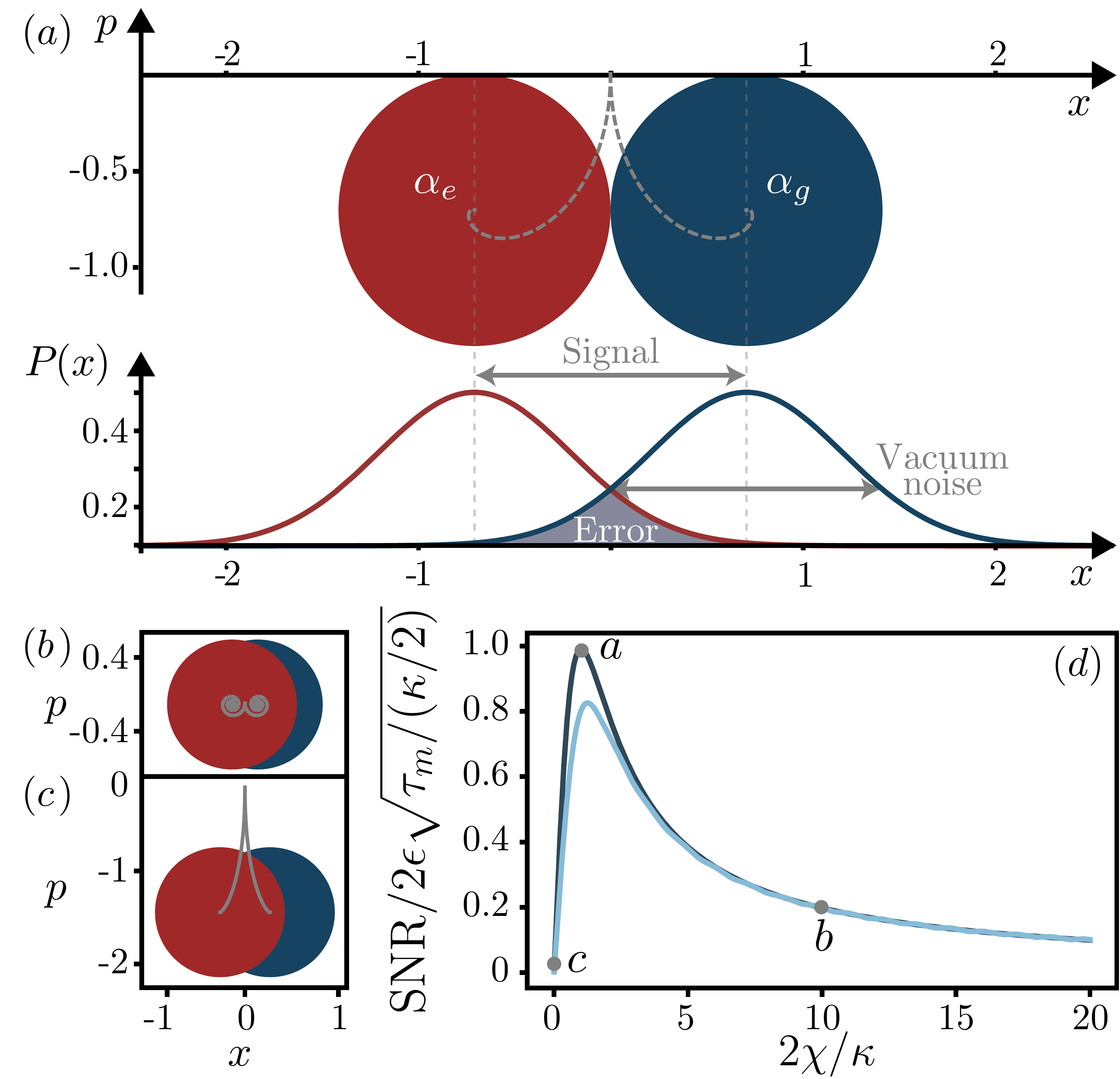}
  \caption{(a) Path in phase space leading up to steady-state of the intra-cavity pointer states $\alpha_g$ and $\alpha_e$ for $2\chi/\kappa = 1$, a measurement drive at the bare cavity frequency with an amplitude leading to one measurement photon at steady-state, and assuming infinite qubit relaxation time. (top). Corresponding marginals along $x$ with the signal, noise and error defined in the text (bottom). The circles of radius $1/\sqrt2$ represent vacuum noise.
  (b) Path in phase space for $2\chi/\kappa = 10$ and (c) $2\chi/\kappa = 0.2$.
  (d) Signal-to-noise ratio as a function of $2\chi/\kappa$ for an integration time $\tau_m/\kappa = 200$ (dark blue) and $\tau_m/\kappa = 10$ (light blue). The maximum of the SNR at short integration time is shifted away from $2\chi/\kappa$. The letters correspond to the ratio $2\chi/\kappa$ of the three previous panels.}
\label{fig:DispersiveQubitReadoutPhaseSpace}
\end{figure}

The objective in a qubit measurement is to maximize the readout fidelity in the shortest possible measurement time. To see how this goal can be reached, it is useful to first evaluate more precisely the evolution of the intra-cavity field under such a measurement.
The intra-cavity field is obtained from the Langevin equation \cref{eq:inouteom} with $\hH_S = \hH_\text{disp}$ and by taking into account the cavity drive as discussed in \secref{sec:drives}. Doing so, we find that the complex field amplitude $\av{\aop}_{\sigma} = \alpha_{\sigma}$ given that the qubit is in state $\sigma = \{g,e\}$ satisfies
\begin{subequations}
\begin{align}
\dot\alpha_e(t)&= -i \varepsilon(t) - (\delta_r + \chi)\alpha_e(t) - \kappa\alpha_e(t)/2,\label{eq:alpha_e}\\
\dot\alpha_g(t)&= -i \varepsilon(t) - (\delta_r - \chi)\alpha_g(t) - \kappa\alpha_g(t)/2\label{eq:alpha_g},
\end{align}
\end{subequations}
with $\delta_r = \wc-\wdrive$ the detuning of the measurement drive to the bare cavity frequency. The time evolution of these two cavity fields in phase space are illustrated for three different values of $2\chi/\kappa$ by dashed gray lines in \cref{fig:DispersiveQubitReadoutPhaseSpace}(a-c). 

Focusing for the moment on the steady-state ($\dot\alpha_\sigma = 0$) response 
\begin{equation}\label{eq:apha_eg_s}
\alpha_{e/g}^\mathrm{s} = \frac{-\varepsilon}{(\delta_r \pm\chi)-i\kappa/2},
\end{equation}
with $+$ for $e$ and $-$ for $g$, results in the steady-state intra-cavity quadratures
\begin{subequations}
  \begin{align}
    \av{\hX}^2_\mathrm{e/g} &= \frac{\varepsilon(\delta_r\pm\chi)}{(\delta_r\pm\chi)^2+(\kappa/2)^2},\\
    \av{\hP}^2_\mathrm{e/g} &= \frac{\varepsilon \kappa/2}{(\delta_r\pm\chi)^2+(\kappa/2)^2}.
  \end{align}
\end{subequations}
When driving the cavity at its bare frequency, $\delta_r=0$, information about the qubit is only contained in the $X$ quadrature, see \cref{fig:DispersiveQubitReadoutPhaseSpace}(a-c).

\begin{figure}[t]
  \centering
  \includegraphics[width=1.0\columnwidth]{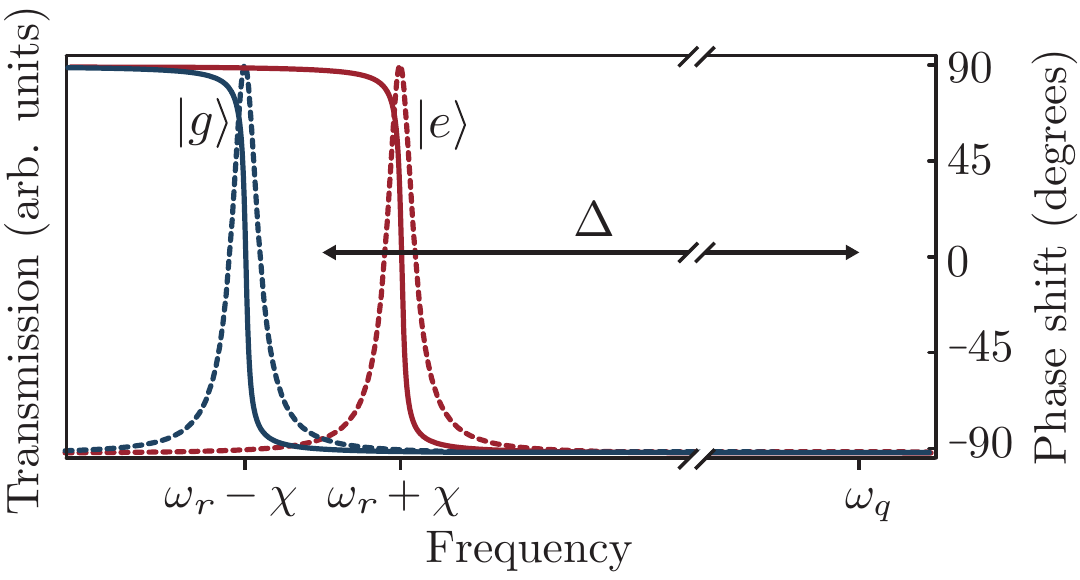}\\
  \caption{Resonator transmission (dashed lines) and corresponding phase shifts (full lines) for the two qubit states (blue:~ground; red:~excited). When driving the resonator close to its pulled frequencies, the resonator response strongly depends on the state of the qubit. Adapted from \cite{Blais2007}.}
  \label{fig:DispersiveCavityPull}
\end{figure}

It is also useful to define the steady-state amplitude
\begin{equation}\label{eq:Amplitude}
A_{e/g}^\mathrm{s} = \sqrt{\av\hX_{e/g}^2+\av\hP_{e/g}^2} = \frac{2\varepsilon}{\sqrt{(\kappa/2)^2 + (\delta_r\pm\chi)^2}}
\end{equation}
and phase
\begin{equation}\label{eq:phiDispersive}
\phi_{e/g}^\mathrm{s} =  \arctan\left(\frac{\av\hX_{e/g}}{\av\hP_{e/g}}\right) = \arctan\left(\frac{\delta_r\pm\chi}{\kappa/2}\right).
\end{equation}
These two quantities are plotted in \cref{fig:DispersiveCavityPull}.  As could already have been expected from the form of $\hH_\text{disp}$, a coherent tone of frequency $\wc\pm\chi$ on the resonator is largely transmitted if the qubit is in the ground (excited) state, and mostly reflected if the qubit is in the excited (ground) state. Alternatively, driving the resonator at its bare frequency $\wc$ leads to a different phase accumulation for the transmitted signal depending on the state of the qubit. In particular, on resonance with the bare resonator, $\delta_r=0$, the phase shift of the signal associated to the two qubit states is simply $\pm\arctan(2\chi/\kappa)$. As a result, in the dispersive regime, measuring the amplitude and/or the phase of the transmitted or reflected signal from the resonator reveals information about the qubit state~\cite{Blais2004}. On the other hand, when driving the resonator at the qubit frequency, for example, to realize a logical gate discussed further in \cref{subsec:SingleQubitGates}, the phase shift of the resonator field only negligibly depends on the qubit state. This results in negligible entanglement between the resonator, and consequently on negligible measurement-induced dephasing on the qubit.

It is very important to note that for purposes of simplification, all of the above discussion has been couched in terms of the amplitude and phase of the oscillating electric field internal to the microwave resonator.  In practice, we can typically only measure the field externally in the transmission line(s) coupled to the resonator.  The relation between the two is the subject of input-output theory discussed in
\cref{sec:environment} and \cref{app:inout}.  The main ideas can be summarized rather simply.  Consider an asymmetric cavity with one port strongly coupled to the environment and one port weakly coupled.  If driven from the weak port, nearly all of the information about the state of the qubit is in the field radiated by the cavity into the strongly coupled port.  The same is true if the cavity is driven from the strongly coupled side, but now the output field is a superposition of the directly reflected drive plus the field radiated by the cavity.  If the drive frequency is swept across the cavity resonance, the signal undergoes a phase shift of $\pi$ in the former case and $2\pi$ in the latter. This affects the sensitivity of the output field to the dispersive shift induced by the qubit.  If the cavity is symmetric, then half the information about the state of the qubit appears at each output port so this configuration is inefficient.  Further details can be found in \cite{Clerk2010}.

\subsubsection{Signal-to-noise ratio and measurement fidelity}
\label{sec:SNR}

Except for the last paragraph, the above discussion concerned the steady-state intra-cavity field from which we can infer the steady-state heterodyne signal. It is, however, also crucial to consider the temporal response of the resonator's output field to the measurement drive since, in the context of quantum computing, qubit readout should be as fast as possible. Moreover, the probability of assigning the correct outcome to a qubit measurement, or more simply put the measurement fidelity, must also be large. As the following discussion hopes to illustrate, simultaneously optimizing these two important quantities requires some care.

As discussed in \cref{sec:FieldDetection}, the quadratures $\hX_f(t)$ and $\hP_f(t)$ are extracted from heterodyne measurement of the resonator output field. Combining these signals and integrating for a time $\tmeas$, the operator corresponding to this measurement takes the form
\begin{equation}
\begin{split}
  \h{M}(\tmeas) & = \int_0^{\tmeas}dt\,
  \left\{
  w_X(t)\left[V_\mathrm{IF}\hX_f(t) + \hat V_\mathrm{noise,X_f}(t) \right]\right.\\
  &\left.\quad\quad\quad\quad\;\; +
  w_P(t)\left[V_\mathrm{IF}\hP_f(t) + \hat V_\mathrm{noise,P_f}(t) \right]   
  \right\},
\end{split}
\end{equation}
where $\hat V_\mathrm{noise,X_f/P_f}(t)$ is the noise in the $X_f/P_f$ quadrature. The weighting functions $w_X(t) = |\av{\hX_f}_e-\av{\hX_f}_g|$ and $w_P(t) = |\av{\hP_f}_e-\av{\hP_f}_g|$ are chosen such as to increase the discrimination of the two qubit states~\cite{Bultink2018,Ryan2015,Magesan2015,Walter2017}. Quite intuitively, because of qubit relaxation, these functions give less weight to the cavity response at long times since it will always reveal the qubit to be in its ground state irrespective of the prepared state~\cite{Gambetta2007}. Moreover, for the situation illustrated in \cref{fig:DispersiveQubitReadoutPhaseSpace}, there is no information on the qubit state in the $P$ quadrature. Reflecting this, $w_P(t)=0$ which prevents the noise in that quadrature from being integrated.

Following \cref{sec:FieldDetection,sec:PhaseSpace}, the probability distribution for the outcome of multiple shots of the measurement of $\h{M}(\tmeas)$ is expected to be Gaussian and characterized by the marginal of the Q-function of the intra-cavity field. Using the above expression, the signal-to-noise ratio ($\SNR$) of this measurement can be defined as illustrated in \cref{fig:DispersiveQubitReadoutPhaseSpace}(a) for the intra-cavity field: it is the separation of the average combined heterodyne signals corresponding to the two qubit states divided by the standard deviation of the signal, an expression which takes the form 
\begin{align}\label{eq:SNR2}
\SNR^2(t)\equiv\frac{|\av{\h{M}(t)}_e-\av{\h{M}(t)}_g|^2}{\av{\h{M}_{N}^2(t)}_e+\av{\h{M}_{N}^2(t)}_g}.
\end{align}
Here, $\av{\h{M}}_\sigma$ is the average integrated heterodyne signal given that the qubit is in state $\sigma$, and $\h{M}_N=\h{M}-\av{\h{M}}$ the noise operator which takes into account the added noise but also the intrinsic vacuum noise of the quantum states.

In addition to the $\SNR$, another important quantity is the measurement fidelity~\cite{Gambetta2007,Walter2017}~\footnote{An alternative definition  known as the assignment fidelity is $1 - \tfrac{1}{2}[P(e|g) - P(g|e)]$~\cite{Magesan2015}. This quantity takes values in $[0,1]$ while formally $\Fmeas \in [-1,1]$. Negative values are, however, not relevant in practice. Indeed, because $\Fmeas = -1$ corresponds to systematically reporting the incorrect value, a fidelity of $1$ is recovered after flipping the measurement outcomes.}
\begin{equation}
\Fmeas = 1 - [P(e|g) + P(g|e)] \equiv 1- \Emeas,
\end{equation}
where $P(\sigma|\sigma')$ is the probability that a qubit in state $\sigma$ is measured to be in state $\sigma'$. In the second equality, we have defined the measurement error $\Emeas$ which, as illustrated in \cref{fig:DispersiveQubitReadoutPhaseSpace}(a), is simply the overlap of the marginals $P_\sigma(x)$ of the Q-functions for the two qubit states. This can be expressed as $\Emeas = \int dx_{\philo+\pi/2}\, \min[P_0(x_{\philo+\pi/2}),P_1(x_{\philo+\pi/2})]$, where the LO phase is chosen to minimize $\Emeas$. Using this expression, the measurement fidelity is found to be related to the $\SNR$ by $\Fmeas =  1 - \mathrm{erfc}(\SNR/2)$, where $\mathrm{erfc}$ is the complementary error function~\cite{Gambetta2007}. It is important to note that this last result is valid only if the marginals are Gaussian. In practice, qubit relaxation and higher-order effects omitted in the dispersive Hamiltonian \cref{eq:HQubitDispersiveSimple} can lead to distortion of the coherent states and therefore to non-Gaussian marginals~\cite{Gambetta2007,Hatridge2013}. Kerr-type nonlinearities that are common in circuit QED tend to create a banana-shaped distortion of the coherent states in phase space, something that is sometimes referred to as bananization \cite{Boutin2017b,Malnou2018,Sivak2019}.

Although we are interested in short measurement times, it is useful to consider the simpler expression for the long-time behavior of the $\SNR$ which suggests different strategies to maximize the measurement fidelity. Assuming $\delta_r = 0$ and ignoring the prefactors related to gain and mixing, we find~\cite{Gambetta2008}
\begin{equation}\label{eq:SNRLongTime}
\SNR(\tmeas\rightarrow\infty) \simeq (2\varepsilon/\kappa) \sqrt{2\kappa\tmeas}\left|\sin{2\phi}\right|,
\end{equation}
where $\phi$ is given by \cref{eq:phiDispersive}; see \cite{Didier2015c} for a detailled derivation of this expression. The reader can easily verify that the choice $\chi/\kappa = 1/2$ maximizes \cref{eq:SNRLongTime}, see \cref{fig:DispersiveQubitReadoutPhaseSpace}(d) \cite{Gambetta2008}. This ratio is consequently often chosen experimentally~\cite{Walter2017}. While leading to a smaller steady-state $\SNR$, other choices of the ratio $\chi/\kappa$ can be more advantageous at finite measurement times.

In the small $\chi$ limit, the factor $2\epsilon/\kappa$ in $\SNR(\tmeas\rightarrow\infty)$ can be interpreted using \cref{eq:alpha_g} as the square-root of the steady-state average intra-cavity measurement photon number. Another approach to improve the $\SNR$ is therefore to work at large measurement photon number $\bar n$. This idea, however, cannot be pushed too far since increasing the measurement photon number leads to a breakdown of the approximations that have been used to derive the dispersive Hamiltonian  \cref{eq:HQubitDispersiveSimple}. Indeed, as discussed in \cref{sec:dispersive} the small parameter in the perturbation theory that leads to the dispersive approximation is not $g/\Delta$ but rather $\bar n/n_\mathrm{crit}$, with $n_\mathrm{crit}$ the critical photon number introduced in \cref{eq:ncrit}. Well before reaching $\bar n/n_\mathrm{crit}\sim 1$, higher-order terms in the dispersive approximation start to play a role and lead to departures from the expected behavior. For example, it is commonly experimentally observed that the dispersive measurement loses its QND character well before $\bar n \sim \ncrit$ and often at measurement photon populations as small as $\bar n\sim 1-10$ \cite{Johnson2011a,minev2019}.
Because of these spurious qubit flips, measurement photon numbers are typically chosen to be well below $\ncrit$~\cite{Walter2017}. While this non-QNDness at $\bar n<\ncrit$ is expected  from the discussion of dressed-dephasing found in \cref{sec:DissipationDispersive}, the predicted measurement-induced qubit flip rates are smaller than often experimentally observed.  We note that qubit transitions at $\bar n>\ncrit$ caused by accidental resonances within the qubit-resonator system have been studied in~\textcite{Sank2016}.

To reach high fidelities, it is also important for the measurement to be fast compared to the qubit relaxation time $T_1$. A strategy to speed-up the measurement is to use a low-Q  oscillator which leads to a faster readout rate simply because the measurement photons leak out more rapidly from the resonator to be detected. However, this should not be done at the price of increasing the Purcell rate $\gamma_\kappa$ to the point where this mechanism dominates qubit decay~\cite{Houck2008}. As discussed in \cref{sec:Purcell}, it is possible to avoid this situation to a large extent by adding a Purcell filter at the output of the resonator~\cite{Reed2010,Jeffrey2014,Bronn2015b}. 

Fixing $\kappa$ so as to avoid Purcell decay and working at the optimal $\chi/\kappa$ ratio, it can be shown that the steady-state response is reached in a time $\propto1/\chi$~\cite{Walter2017}. Large dispersive shifts can therefore help to speed up the measurement. As can be seen from \cref{eq:HQubitDispersiveParametersSW}, $\chi$ can be increased by working at larger qubit anharmonicity or, in other words, larger charging energy $E_C$. Once more, this cannot be pushed too far since the transmon charge dispersion and therefore its dephasing rate increase with $E_C$.

The above discussion shows that QND qubit measurement in circuit QED is a highly constrained problem. The state-of-the-art for such measurements is currently of $\Fmeas\sim98.25\%$ in $\tmeas=48$ ns, when minimizing readout time, and $99.2\%$ in 88 ns, when maximizing the fidelity, in both cases using $\bar n\sim 2.5$ intra-cavity measurement photons~\cite{Walter2017}. These results were obtained by detailed optimization of the system parameters following the concepts introduced above but also given an understanding of the full time response of the measurement signal $|\av{\h{M}(t)}_1-\av{\h{M}(t)}_0|$. The main limitation in these reported fidelity was the relatively short qubit relaxation time of 7.6 $\mu$s. Joint simultanteous dispersive readout of two transmon qubits capacitively coupled to the same resonator has also been realized \cite{Filipp2009b}. 

The very small photon number used in these experiments underscores the importance of quantum-limited amplifiers in the first stage of the measurement chain, see Fig.~\ref{fig:MeasurementChain}. Before the development of these amplifiers, which opened the possibility to perform strong single-shot (i.e.~projective) measurements, the $\SNR$ in dispersive measurements was well below unity, forcing the results of these weak measurements to be averaged over tens of thousands of repetitions of the experiment to extract information about the qubit state~\cite{Wallraff2005}. The advent of near quantum-limited amplifiers (see \cref{sec:QO:AmplificationSqueezing}) has made it possible to resolve the qubit state in a single-shot something which has led, for example, to the observation of quantum jumps of a transmon qubit~\cite{Vijay2011}. 

Finally, we point out that the quantum efficiency, $\eta$, of the whole measurement chain can be extracted from the $\SNR$ using \cite{Bultink2018}
\begin{equation}\label{eq:eta_chain}
  \eta = \frac{\SNR^2}{4\beta_m},
\end{equation}
where $\beta_m = 2\chi \int_0^{\tmeas} dt\,\mathrm{Im}[\alpha_g(t)\alpha_e(t)^*]$ is related to the measurement-induced dephasing discussed further in \cref{sec:AcStark_MeasIndDephasing}. This connection between quantum efficiency, $\SNR$, and measurement-induced dephasing results from the fundamental link between the rate at which information is gained in a quantum measurement and the unavoidable backaction on the measured system \cite{Clerk2010,Korotkov2001}.

\subsubsection{Other approaches}
\label{sec:ReadoutOtherApproaches}

\paragraph{Josephson Bifurcation Amplifier}

While the vast majority of circuit QED experiments rely on the approach described above, several other qubit-readout methods have been  theoretically explored or experimentally implemented. One such alternative is known as the Josephson Bifurcation Amplifier (JBA) and relies on using, for example, a transmission-line resonator that is made nonlinear by incorporating a Josephson junction in its center conductor~\cite{Boaknin2007}. This circuit can be seen as a distributed version of the transmon qubit and is well described by the Kerr-nonlinear Hamiltonian of \cref{eq:HsjPhi4b}~\cite{Bourassa2012}. With a relatively weak Kerr nonlinearity ($\sim -500$ kHz) and under a coherent drive of well chosen amplitude and frequency, this system bifurcates from a low photon-number state to a high photon-number state~\cite{Dykman1980,Manucharyan2007}. By dispersively coupling a qubit to the nonlinear resonator, this bifurcation can be made qubit-state dependent~\cite{Vijay2009}. It is possible to exploit the fact that the low- and high-photon-number states can be easily distinguished to realize high-fildelity single-shot qubit readout~\cite{Mallet2009}.

\paragraph{High-power readout and qubit 'punch out'}

Coming back to linear resonators, while the non-QNDness at moderate measurement photon number mentioned above leads to small measurement fidelity, it was observed by a fearless graduate student that, in the limit of very large measurement power, a fast and high-fidelity single-shot readout is recovered~\cite{Reed2010a}. An intuitive understanding of this observation can be obtained from the Jaynes-Cummings Hamiltonian \eq{eq:HJC}~\cite{Boissonneault2010,Bishop2010a}. Indeed, for $n\gg\sqrt n$, the first term of this Hamiltonian dominates over the qubit-oscillator interaction $\propto g$ such that the cavity responds at its bare frequency $\wc$ despite the presence of the transmon. This is sometimes referred to as `punching out' the qubit and can be understood as a quantum-to-classical transition where, in the correspondence limit, the system behaves classically and therefore responds at the bare cavity frequency $\wc$. Interestingly, with a multi-level system such as the transmon, the power at which this transition occurs depends on the state of the transmon, leading to a high-fidelity measurement. This high-power readout is, however, obtained at the expense of completely losing the QND nature of the dispersive readout~\cite{Boissonneault2010}.

\paragraph{Squeezing}

Finally, the $\sqrt n$ scaling of $\SNR(\tmeas\rightarrow\infty)$ mentioned above can be interpreted as resulting from populating the cavity with a coherent state and is known as the standard quantum limit. It is natural to ask if replacing the coherent measurement tone with squeezed input radiation (see \cref{sec:Squeezing}) can lead to Heisenberg-limited scaling for which the $\SNR$ scales linearly with the measurement photons number~\cite{Giovanetti2004}. To achieve this, one might imagine squeezing a quadrature of the field to reduce the overlap between the two pointer states. In \cref{fig:DispersiveQubitReadoutPhaseSpace}, this corresponds to squeezing along $X$. The situation is not so simple since the large dispersive coupling required for high-fidelity qubit readout leads to a significant rotation of the squeezing angle as the pointer states evolve from the center of phase space to their steady-state. This rotation results in increased  measurement noise due to contributions from the antisqueezed quadrature~\cite{Barzanjeh2014}. Borrowing the idea of quantum-mechanics-free subsystems~\cite{Tsang2012}, it has been shown that Heisenberg-limited scaling can be reached with two-mode squeezing by dispersively coupling the qubit to two rather than one resonator~\cite{Didier2015b}.

\paragraph{Longitudinal readout}

An alternative approach to qubit readout is based on the Hamiltonian $\hH_z$ of \cref{eq:H_Longitudinal} with its longitudinal qubit-oscillator coupling $g_z(\ad + \aop)\sz{}$. In contrast to the dispersive Hamiltonian which leads to a rotation in phase space, longitudinal coupling generates a linear displacement of the resonator field that is conditional on the qubit state. As a result, while under the dispersive evolution there is little information gain about the qubit state at short times [see the poor pointer state separation at short times in \cref{fig:DispersiveQubitReadoutPhaseSpace}(a)], $\hH_z$ rather generates the ideal dynamics for a measurement with a $180^\circ$ out-of-phase displacements of the pointer states $\alpha_g$ and $\alpha_e$. It is therefore expected that this approach can lead to much shorter measurement times than is possible with the dispersive readout \cite{Didier2015c}.

Another advantage is that $\hH_z$ commutes with the measured observable, $[\hH_z,\sz{}]=0$, corresponding to a QND measurement. While the dispersive Hamiltonian $\hH_\text{disp}$ also commutes with $\sz{}$, it is not the case for the full Hamiltonian~\cref{eq:HTransmonJC} from which $\hH_\text{disp}$ is perturbatively derived. As already discussed, this non-QNDness leads to Purcell decay and to a breakdown of the dispersive approximation when the photon populations is not significantly smaller than the critical photon number $\ncrit$. On the other hand, because $\hH_z$ is genuinely QND it does not suffer from these problems and the measurement photon number can, in principle, be made larger under longitudinal than under dispersive coupling. Moreover, given that $\hH_z$ leads to displacement of the pointer states rather than to rotation in phase space, single-mode squeezing can also be used to increase the measurement SNR \cite{Didier2015c}.

Because the longitudinal coupling can be thought of as a cavity drive of amplitude $\pm g_z$ with the sign being conditional on the qubit state, $\hH_z$ leads in steady-state to a pointer state displacement $\pm g_z/(\wc+i\kappa/2)$, see \cref{eq:apha_eg_s}. With $\wc \gg g_z,\, \kappa$
in practice, this displacement is negligible and cannot realistically be used for qubit readout. One approach to increase the pointer state separation is to activate the longitudinal coupling by modulating $g_z$ at the resonator frequency \cite{Didier2015c}. Taking $g_z(t) = \tilde g_z \cos(\wc t)$ leads, in a rotating frame and after dropping rapidly oscillating terms, to the Hamiltonian
\begin{equation}
  \tilde H_z = \frac{\tilde g_z}{2} (\ad+a)\sz{}.
\end{equation}
Under this modulation, the steady-state displacement now becomes $\pm \tilde g_z/\kappa$ and can be significant even for moderate modulation amplitudes $\tilde g_z$. 

Circuits realizing the longitudinal coupling with transmon or flux qubits have been studied \cite{Didier2015c,Kerman2013,Billangeon2015,Billangeon2015a,Richer2016,Richer2017a}. Another approach to realize these ideas is to strongly drive a resonator dispersively coupled to a qubit \cite{Blais2007,Dassonneville2020}. Indeed, the strong drive leads to a large displacement of the cavity field $\aop \rightarrow \aop + \alpha$ which on the dispersive Hamiltonian leads to
\begin{equation}
  \chi\ada\sz{} 
  \rightarrow 
  \chi\ada\sz{} + \alpha\chi (\ad+\aop)\sz{} + \chi\alpha^2\sz{},
\end{equation}
where we have assumed $\alpha$ to be real for simplicity. For $\chi$ small and $\alpha$ large, the second term  dominates therefore realizing a synthetic longitudinal interaction af amplitude $g_z=\alpha\chi$. In other words, longitudinal readout can be realized as a limit of the dispersive readout where $\chi$ approaches zero while $\alpha$ grows such that $\chi\alpha$ is constant. A simple interpretation of this observation is that, for strong drives, the circle on which the pointer states rotate due to the dispersive interaction has a very large radius $\alpha$ such that, for all practical purposes, the motion appears linear.

A variation of this approach which allows for larger longitudinal coupling strengh was experimentally realized by \textcite{Touzard2019} and \textcite{Ikonen2019} and relies on driving the qubit at the frequency of the resonator. This is akin to the cross-resonance gate discussed further in \cref{sec:AllMicrowaveGates} and which leads to the desired longitudinal interaction, see the last term of \cref{eq:CrossResonanceTLS}. A more subtle approach to realize a synthetic longitudinal interaction is to drive a qubit with a Rabi frequency $\Omega_R$ while driving the resonator at the sideband frequencies $\wc\pm\Omega_R$. This idea was implemented by \textcite{Eddins2018} who also showed improvement of qubit readout with single-mode squeezing. Importantly, because these realizations are based on the dispersive Hamiltonian, they suffer from Purcell decay and non-QNDness. Circuits realizing dispersive-like interactions that are not derived from a Jaynes-Cummings interaction have been studied \cite{Dassonneville2020,Didier2015c}.

\section{Qubit-resonator coupling regimes}\label{sec:CouplingRegimes}

We now turn to a discussion of the different coupling regimes that are accessible in circuit QED and how these regimes are probed experimentally. We first consider the resonant regime where the qubit is tuned in resonance with the resonator, before moving on to the dispersive regime characterized by large qubit-resonator detuning. While the situation of most experimental interest is the strong coupling regime where the coupling strength $g$ overwhelms the decay rates, we also touch upon the so-called bad-cavity and bad-qubit limits because of their historical importance and their current relevance to hybrid quantum systems. Finally, we briefly consider the ultrastrong coupling regime where $g$ becomes comparable or is even larger than the system's frequencies. To simplify the discussion, we will treat the artificial atom as a simple two-level system throughout this section.

\subsection{\label{sec:resonant}Resonant regime}

The low-energy physics of the Jaynes-Cummings model is well described by the ground state $\ket{\overline{g,0}}=\ket{g,0}$ and first two excited states
\begin{equation}\label{eq:OnResonanceDoublet}
\begin{split}
\ket{\overline{g, 1}} &= (\ket{g,1}-\ket{e,0})/\sqrt{2},\\
\ket{\overline{e, 0}} &= (\ket{g,1} + \ket{e,0})/\sqrt{2},
\end{split}
\end{equation}
which, as already illustrated in \cref{fig:JCSpectrum}, are split in frequency by $2g$. As discussed in \cref{sec:DispersiveQubitReadout} in the context of the dispersive readout, the coupled qubit-resonator system can be probed by applying a coherent microwave tone to the input of the resonator and measuring the transmitted or reflected signal.

To arrive at an expression for the expected signal in such an experiment, we consider the equations of motion for the field and qubit annihilation operators in the presence of a coherent drive of amplitude $\varepsilon$ and frequency $\wdrive$ on the resonator's input port. In a frame rotating at the drive frequency, these equations take the form
\begin{align}
\av{\dot\aop} &= -\left(\frac{\kappa}{2}+i \delta_r\right)\av\aop - ig\langle\smm{}\rangle - i\varepsilon,
\label{eq:a}\\
\av{\dot{\hat{\sigma}}_-} &= -\left(\gamma_2+i\delta_q\right)\langle \smm{}\rangle + ig\av{\aop\sz{}},
\label{eq:SigmaMinus_Original}
\end{align}
with $\delta_r = \wc-\wdrive$ and $\delta_q = \wa-\wdrive$, and where $\gamma_2$ is defined in \cref{eq:T2_definition}. These expressions are obtained using $\partial_t\av{\h{O}} = \tr{\dot\rho\h{O}}$ and the master equations of \cref{eq:ME_harmonic,eq:ME_transmon} at zero temperature and in the two-level approximation for the transmon. Alternatively, the expression for $\partial_t\av{\aop}$ is simply the average of \cref{eq:inouteom} with $\hat H_S$ the Jayne-Cummings Hamiltonian.

At very low excitation amplitude $\varepsilon$, it is reasonable to truncate the Hilbert space to the first three levels defined above. In this subspace, $\av{\aop\sz{}} = -\av\aop$ since $\aop$ acts nontrivially only if the qubit is in the ground state~\cite{Kimble1994}. It is then simple to compute the steady-state transmitted homodyne power by solving the above expressions with $\partial_t\av{\aop} = \partial_t\av{{\sigma}_-} = 0$ and using \cref{eq:Amplitude} to find 
\begin{equation}\label{eq:CavityTransmission}
|A|^2 = 
\left(\frac{\varepsilon V_\mathrm{IF}}{2}\right)^2
\left|
\frac{\delta_q-i\gamma_2}{(\delta_q-i\gamma_2)(\delta_r-i\kappa/2)-g^2}
\right|^2.
\end{equation}
This expression is exact in the low excitation power limit.  Taking the qubit and the oscillator to be on resonance, $\Delta = \wa-\wc=0$, we now consider the result of cavity transmission measurements in three different regimes of qubit-cavity interaction.

\subsubsection{Bad-cavity limit}

We first consider the bad-cavity limit realized when the cavity decay rate overwhelms the coupling $g$ which is itself larger than the qubit linewidth: $\kappa > g \gg \gamma_2$. This situation corresponds to an overdamped oscillator and, at qubit-oscillator resonance, leads to rapid decay of the qubit. A simple model for this process is obtained using the truncated Hilbert space discussed above where we now drop the cavity drive for simplicity. Because of the very large decay rate $\kappa$, we can asume the oscillator to rapidly reach its steady-state $\partial_t\av{\aop}=0$. Using the resulting expression for $\av{\aop}$ in \cref{eq:SigmaMinus_Original} immediately leads to
\begin{equation}
\av{\dot{\hat{\sigma}}_-}
= -\left(\frac{\gamma_1+\gamma_k'}{2}+\gamma_\varphi\right)\langle \smm{}\rangle,
\end{equation}
where we have defined the Purcell decay rate $\gamma_\kappa' = 4g^2/\kappa$. The expression for this rate has a rather different form than the Purcell rate $\gamma_\kappa = (g/\Delta)^2\kappa$ given in \cref{eq:DispersiveRates}. These two results are, however, not incompatible but have been obtained in very different regimes. An expression for the Purcell rate that interpolates between the two above expressions can be obtained and takes the form $\kappa g^2/[(\kappa/2)^2+\Delta^2]$~\cite{Sete2014}.

\begin{figure}[t]
  \centering
  \includegraphics[width=1\columnwidth]{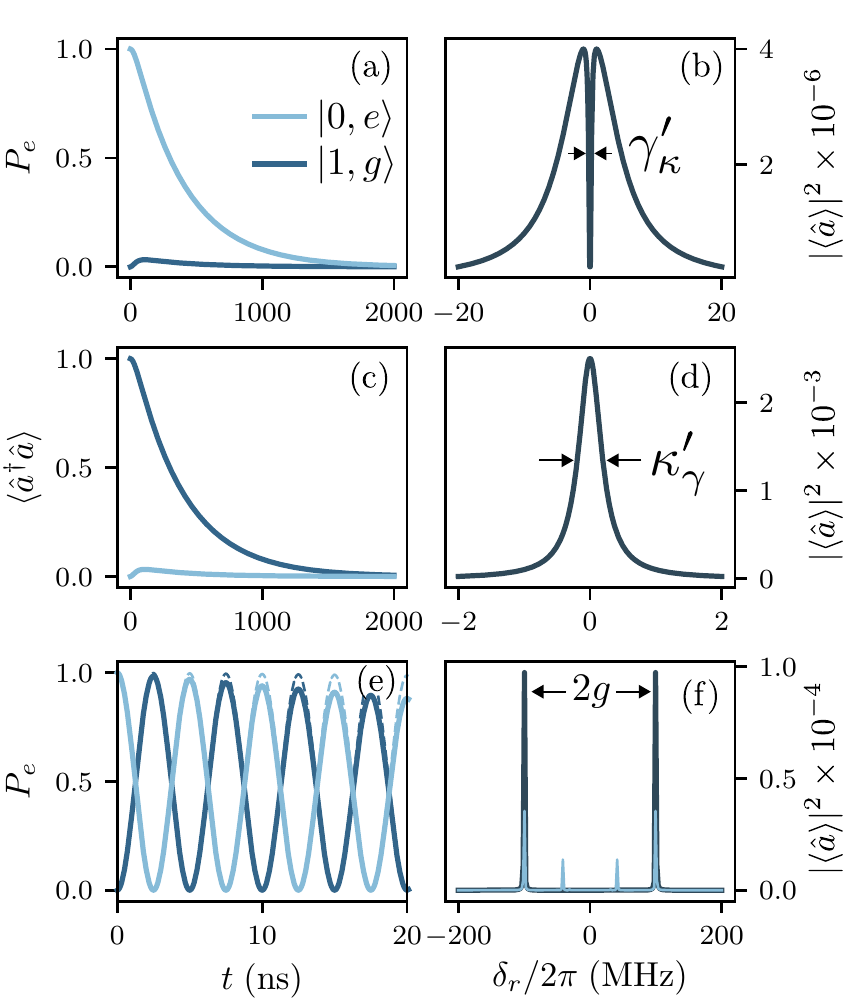}
  \caption{
  Numerical simulations of the qubit-oscillator master equation for (a,c,e) the time evolution starting from the bare state $\ket{0,e}$ (light blue) or $\ket{1,g}$ (blue), and (b,d,f) steady-state response $A^2 = |\av\aop|^2$ as a function of cavity drive frequency (dark blue) for the three coupling regimes. $P_e$: qubit excited state population. (a,b) Bad-cavity limit: $(\kappa, \gamma_1, g)/2\pi = (10,0,1)$ MHz. (c,d) Bad-qubit limit: $(\kappa, \gamma_1, g)/2\pi = (0,10,1)$ MHz. (e,f) Strong coupling: $(\kappa,\gamma_1,g)/2\pi = (0.1, 0.1, 100)$ MHz (dashed lines in panel (e) and solid in (f) and $(\kappa,\gamma_1,g)/2\pi = (1, 1, 100)$ (full lines in e). The light blue line in panel (f) is computed with a thermal photon number of $\bar n_\kappa = 0.35$ rather than $\bar n_\kappa = 0$ for all the other results.}
\label{fig:ResonantRegimes}
\end{figure}

The situation described here is illustrated for $\kappa/g = 10$ and $\gamma_1 = 0$ in \cref{fig:ResonantRegimes}(a) which shows the probability for the qubit to be in its excited state versus time after initializing the qubit in its excited state and the resonator in vacuum. Even in the absence of qubit $T_1$, the qubit is seen to quickly relax to its ground state something which, as discussed in \cref{sec:DissipationDispersive},  is due to qubit-oscillator hybridization.   
\Cref{fig:ResonantRegimes}(b) shows the transmitted power versus drive frequency in the presence of a very weak coherent tone populating the cavity with $\bar n\ll1$ photons. The response shows a broad Lorentzian peak of width $\kappa$ together with a narrow electromagnetically induced transparency (EIT)-like window of width $\gamma_\kappa'$~\cite{rice1996,Mlynek2014b}.  This effect which is due to interference between the intra-cavity field and the probe tone vanishes in the presence of qubit dephasing.

Although not the main regime of interest in circuit QED, the bad-cavity limit offers an opportunity to engineer the dissipation seen by the qubit. For example, this regime has been used to control the lifetime of long-lived donor spins in silicon in a hybrid quantum system~\cite{Bienfait2016a}.

\subsubsection{Bad-qubit limit}

The bad qubit limit corresponds to the situation where a high-Q cavity with large qubit-oscillator coupling is realized, while the qubit dephasing and/or energy relaxation rates is large: $\gamma_2>g\gg\kappa$. Although this situation is not typical of circuit QED with transmon qubits, it is relevant for some hybrid systems that suffer from significant dephasing. This is the case, for example, in early experiments with charge qubits based on semiconductor quantum dots coupled to superconducting resonators \cite{Frey2012,Petersson2012a,Viennot2014a}. 

In analogy to the bad-cavity case, the strong damping of the qubit together with the qubit-resonator coupling leads to the photon decay rate $\kappa_\gamma' = 4g^2/\gamma_1$ which is sometimes known as  the `inverse' Purcell rate. This is illustrated in \cref{fig:ResonantRegimes}(c) which shows the time-evolution of the coupled system starting with a single photon in the resonator and the qubit in the ground state. In this situation, the cavity response is a simple Lorentzian broadened by the inverse Purcell rate, see \cref{fig:ResonantRegimes}(d). If the qubit were to be probed directly rather than indirectly via the cavity, the atomic response would show the EIT-like feature of \cref{fig:ResonantRegimes}(b), now with a dip of width $\kappa_\gamma'$~\cite{rice1996}. The reader should also be aware that qubit-resonator detuning-dependent dispersive shifts of the cavity resonance can be observed in this bad-qubit limit. The observation of such dispersive shifts on its own should not be mistaken for an observation of strong coupling \cite{Wallraff2013}.

\subsubsection{Strong coupling regime}

We now turn to the case where the coupling strength overwhelms the qubit and cavity decay rates, $g > \kappa,\, \gamma_2$. 
In this regime, light-matter interaction is strong enough for a single quantum to be coherently exchanged between the electromagnetic field and the qubit before it is irreversibly lost to the environment. In other words, at resonance $\Delta = 0$ the splitting $2g$ between the two dressed eigenstates $\{\ket{\overline{g, 1}},\ket{\overline{e, 0}}\}$ of \cref{eq:OnResonanceDoublet} is larger than their linewidth $\kappa/2 + \gamma_2$ and can be resolved spectroscopically. We note that, with the eigenstates being half-photon and half-qubit\footnote{According to some authors, these dressed states should therefore be referred to as quton and phobit~\cite{Schuster2007}.}, the above expression for the dressed-state linewidth is simply the average of the cavity and of the qubit linewitdh~\cite{Haroche1992}. \Cref{fig:ResonantRegimes}(f) shows cavity transmission for $(\kappa,\gamma_1,\gamma_\varphi)/g = (0.1, 0.1,0)$ and at low excitation power such that, on average, there is significantly less than one photon in the cavity. The resulting doublet of peaks located at $\wc\pm g$ is the direct signature of the dressed-states $\{\ket{\overline{g, 1}},\ket{\overline{e, 0}}\}$ and is known as the vacuum Rabi splitting. The observation of this doublet is the hallmark of the strong coupling regime. 

\begin{figure}[t]
  \centering
  \includegraphics[width=0.8\columnwidth]{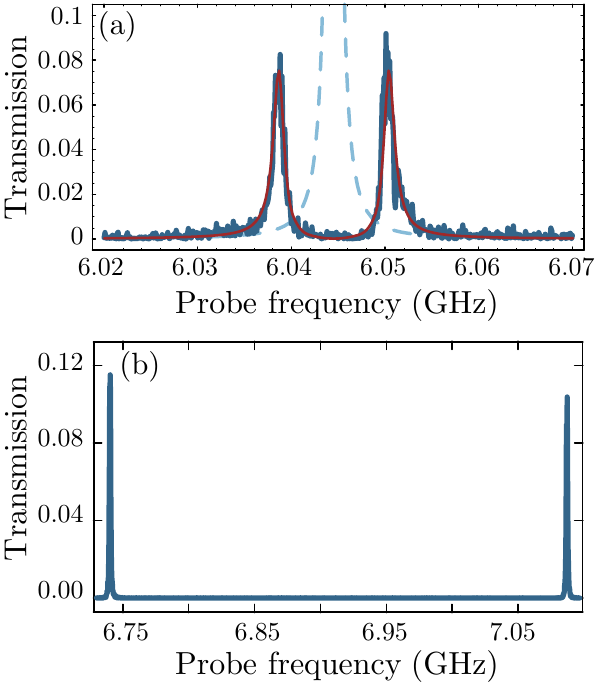}\\
  \caption{
    (a) Transmission-line resonator transmission versus probe frequency in the first observation of vacuum Rabi splitting in circuit QED (full blue line). The qubit is a Cooper pair box qubit with $E_J/h\approx 8\,\rm{GHz}$ and $E_C/h \approx 5.2 \, \rm{GHz}$. The full red line is a calculated spectra with $2g/2\pi \approx \, 11.6$~MHz, $\kappa/2\pi \approx\,0.8$~MHz and $\gamma_2/2\pi\approx0.7$~MHz. As a reference, the dashed light blue line is the measured transmission with the qubit strongly detuned from the resonator. Adapted from \textcite{Wallraff2004}. 
    (b) Resonator transmission with a transmon qubit. The vacuum Rabi splitting is even more resolved with $2g/2\pi = 350$ MHz, $\kappa/2\pi \sim  800$ kHz and $\gamma_2/2\pi \sim 200$ kHz. Notice the change in probe frequency range from panel (a). Adapted from \textcite{Schoelkopf2008}.
  }
\label{fig:VacuumRabiSplitting}
\end{figure}

The first observation of this feature in cavity QED with a single atom and a single photon was reported by~\cite{Thompson1992}. In this experiment, the number of atoms in the cavity was not well controlled and it could only be determined that there was \emph{on average} one atom in interaction with the cavity field. This distinction is important because, in the presence of $N$ atoms, the collective interaction strength is $g\sqrt{N}$ and the observed splitting correspondingly larger~\cite{Tavis1968,Fink2009}. Atom number fluctuation is obviously not a problem in circuit QED and, with the very strong coupling and relatively small linewidths that can routinely be experimentally achieved, reaching the strong coupling regime is not particularly challenging in this system. In fact, the very first circuit QED experiment of \textcite{Wallraff2004} reported the observation of a clear vacuum Rabi splitting with $2g/(\kappa/2 + \gamma_2) \sim 10$, see \cref{fig:VacuumRabiSplitting}(a). This first demonstration used a charge qubit which, by construction, has a much smaller coupling $g$ than typical transmon qubits. As a result, more recent experiments with transmon qubits can display ratios of peak separation to linewidth in the several hundred, see \cref{fig:VacuumRabiSplitting}(b)~\cite{Schoelkopf2008}.

\begin{figure}[t]
  \centering
  \includegraphics[width=0.9\columnwidth]{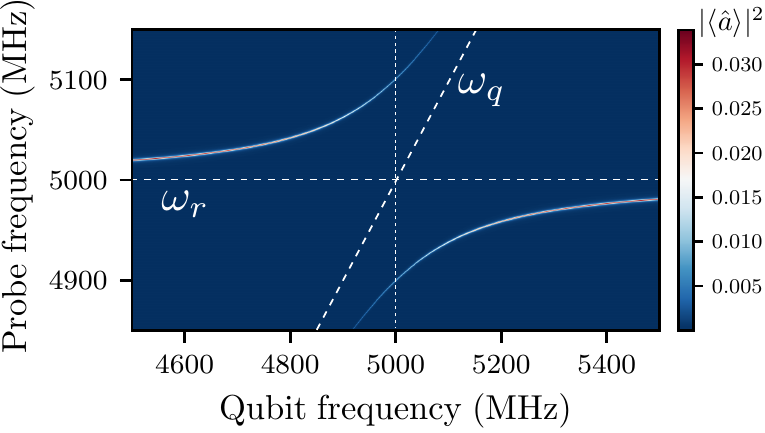}\\
  \caption{
  Vacuum Rabi splitting revealed in numerical simulations of the cavity transmission $A^2 = |\av\aop|^2$ as a function of probe frequency and qubit transition frequency for the same parameters as in \cref{fig:ResonantRegimes}(f) The bare cavity and qubit frequencies are indicated by the horizontal and diagonal dashed lines, respectively. The vacuum Rabi splitting of \cref{fig:ResonantRegimes}(f) is obtained at  resonance ($\wc=\wa$) along the vertical dotted line.
  }
  \label{fig:AvoidedCrossing}
\end{figure}

\Cref{fig:AvoidedCrossing} shows the qubit-oscillator spectrum as a function of probe frequency, as above, but now also as a function of the qubit frequency allowing to see the full qubit-resonator avoided crossing. The horizontal dashed line corresponds to the bare cavity frequency while the diagonal dashed line is the bare qubit frequency. The vacuum Rabi splitting of \cref{fig:ResonantRegimes}(f) is obtained from a linecut (dotted vertical line) at resonance between the bare qubit frequency $\wa$ and the bare cavity frequency $\wc$. Because it is the cavity that is probed here, the response is larger when the dressed-states are mostly cavity-like and disappears away from the cavity frequency where the cavity no longer responds to the probe~\cite{Haroche1992}.

It is interesting to note that the splitting predicted by \cref{eq:CavityTransmission} for the transmitted homodyne signal is in fact smaller than $2g$ in the presence of finite relaxation and dephasing. Although not significant in circuit QED with transmon qubits, this correction can become important in systems such as charge qubits in quantum dots that are not very deep in the strong coupling regime. We also note that the observed splitting can be smaller when measured in reflection rather than in transmission.

Rather than spectroscopic measurements, strong light-matter coupling can also be displayed in time-resolved measurements~\cite{Brune1996}. Starting from the qubit-oscillator ground state, this can be done, for example, by first pulsing the qubit to its first excited state and then bringing it on resonance with the cavity. As illustrated in \cref{fig:ResonantRegimes}(e), this results in oscillations in the qubit and cavity populations at the vacuum Rabi frequency $2g$. Time-resolved vacuum Rabi oscillations in circuit QED were first performed with a flux qubit coupled to a discrete LC oscillator realized in the bias circuitry of the device \cite{Johansson2006}. This experiment was followed by a similar observation with a phase qubit coupled to a coplanar waveguide resonator \cite{Hofheinz2008}.

In the limit of weak excitation power which we have considered so far, the coupled qubit-oscillator system is indistinguishable from two coupled classical linear oscillators. As a result, while the dressed-states that are probed in these experiments are entangled, the observation of an avoided crossing cannot be taken as a conclusive demonstration that the oscillator field is quantized or of qubit-oscillator entanglement. Indeed, a vacuum Rabi splitting can be interpreted as the familiar normal mode splitting of two coupled classical oscillators. 

\begin{figure}[t]
  \centering
  \includegraphics[width=0.5\columnwidth]{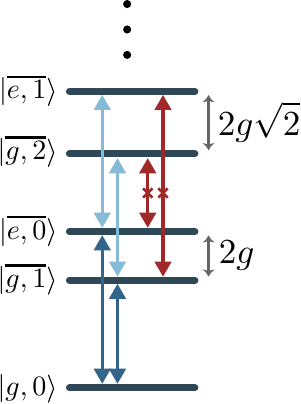}\\
  \caption{
  Ground state and first two doublets of the Jaynes-Cummings ladder. The dark blue arrows correspond to the transitions that are probed in a vacuum Rabi experiment. The transitions illustrated with light blue arrows lead to additional peaks at transition frequencies lying inside of the vacuum Rabi doublet at elevated temperature or increased probe power. On the other hand, the matrix element associated with the red transitions would lead to response at transition frequencies outside of the vacuum Rabi doublet. Those transition are, however, suppressed and are not observed \cite{Rau2004a}.
  } \label{fig:ResonantMultiPhotons}
\end{figure}

A clear signature of the quantum nature of the system can, however, be obtained by probing the $\sqrt n$ dependence of the spacing of the higher excited states of the Jaynes-Cummings ladder already discussed in \cref{sec:JaynesCummings}. This dependence results from the matrix element of the operator $\aop$ and is consequently linked to the quantum nature of the field~\cite{Carmichael1996}. Experimentally, these transitions can be accessed in several ways including by two-tone spectroscopy~\cite{Fink2008}, by increasing the probe tone power~\cite{Bishop2009a}, or by increasing the system temperature~\cite{Fink2010}. The light blue line in \cref{fig:ResonantRegimes}(f) shows cavity transmission with a thermal photon number of $\bar n_\kappa = 0.35$ rather than $\bar n_\kappa = 0$ (dark blue line). At this more elevated temperature, additional pairs of peaks with smaller separation are now observed in addition to the original peaks separated by $2g$. As illustrated in \cref{fig:ResonantMultiPhotons}, these additional structures are due to multi-photon transitions and their $\sqrt n$ scaling reveal the anharmonicity of the Jaynes-Cummings ladder. Interestingly, the matrix elements of transitions that lie outside of the original vacuum Rabi splitting peaks are suppressed and these transitions are therefore not observed, see the red arrow in \cref{fig:ResonantMultiPhotons}~\cite{Rau2004a}. We also note that, at much larger power or at elevated temperature, the system undergoes a quantum-to-classical transition and a single peak at the resonator frequency $\wc$ is observed~\cite{Fink2010}. In short, the impact of the qubit on the system is washed away in the correspondence limit. This is to be expected from the form of the Jaynes-Cummings Hamiltonian \cref{eq:HJC} where the qubit-cavity coupling $\hbar g (\ad\smm{} + \aop\spp{})$ with its $\sqrt{n}$ scaling is overwhelmed by the free cavity Hamiltonian $\hbar\wc \ada$ which scales as $n$. This is the same mechanism that leads to the high-power readout discussed in \cref{sec:ReadoutOtherApproaches}.

Beyond this spectroscopic evidence, the quantum nature of the field and qubit-oscillator entanglement was also demonstrated in a number of experiments directly measuring the joint density matrix of the dressed states. For example, \textcite{Eichler2012b} have achieved this by creating one of the entangled states $\{\ket{\overline{g, 1}},\ket{\overline{e, 0}}\}$ in a time-resolved vacuum Rabi oscillation experiment and, subsequently, measuring the qubit state in a dispersive measurement and the photon state using a linear detection method \cite{Eichler2012}. A range of experiments used the ability to create entanglement between a qubit and a photon through the resonant interaction with a resonator, e.g.~in the context of quantum computation \cite{Mariantoni2011a}, for entangling resonator modes \cite{Mariantoni2011}, and transferring quantum states \cite{Sillanpaa2007}.

\subsection{\label{sec:DispersiveConsequences}Dispersive regime}

For most quantum computing experiments, it is common to work in the dispersive regime where, as already discussed in \cref{sec:dispersive}, 
the qubit is strongly detuned from the oscillator with $|\Delta|\gg g$. There, the dressed eigenstates are only weakly entangled qubit-oscillator states. This is to be contrasted to the resonant regime where these eigenstates are highly entangled resulting in the qubit and the oscillator to completely lose their individual character.

In the two-level system approximation, the dispersive regime is well described by the Hamiltonian $\hH_\text{disp}$ of \cref{eq:HQubitDispersiveSimple}. There, we had interpreted the dispersive coupling as a qubit-state dependent shift of the oscillator frequency. This shift can be clearly seen in \cref{fig:AvoidedCrossing} as the deviation of the oscillator response from the bare oscillator frequency away from resonance (horizontal dashed line). This figure also makes it clear that the qubit frequency, whose bare value is given by the diagonal dashed line, is also modified by the dispersive coupling to the oscillator. To better understand this qubit-frequency shift, it is instructive to rewrite $\hH_\text{disp}$ as
\begin{equation}\label{eq:HQubitDispersiveSimpleACStark}
    \hH_\text{disp} \approx \hbar \wc \hat a^\dagger \hat a + \frac{\hbar}{2} \left[\wa+ 2 \chi \left(\hat a^\dagger \hat a +\frac12 \right)\right] \sz{},
\end{equation}
where it is now clear that the dispersive interaction of amplitude $\chi$ not only leads to a qubit-state dependent frequency pull of the oscillator, but also to a photon-number dependent frequency shift of the qubit given by $2\chi\hat a^\dagger \hat a$. This is known as the ac-Stark shift (or the quantized light shift) and is here accompanied by a Lamb shift corresponding to the factor of $1/2$ in the last term of \cref{eq:HQubitDispersiveSimpleACStark} and which we had dropped in \cref{eq:HQubitDispersiveSimple}. In this section, we explore some consequences of this new point of view on the dispersive interaction, starting by first reviewing some of the basic aspects of qubit spectroscopic measurements.

\subsubsection{Qubit Spectroscopy}

\begin{figure}[t]
  \centering
  \includegraphics[width=1\columnwidth]{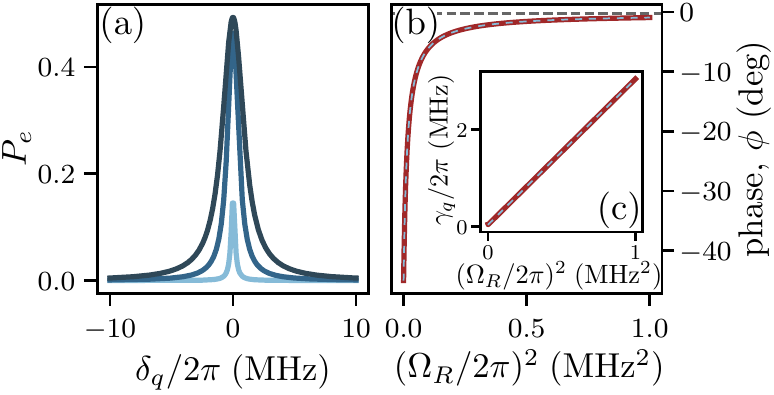}\\
  \caption{ Power broadening of the qubit line. (a) Excited qubit population (left vertical axis) and phase (right vertical axis) as a function of the drive detuning $\delta_q$  for the Rabi amplitudes $\Omega_R/2\pi = 0.1$ MHz (light blue), 0.5 MHz (blue) and 1 MHz (dark blue). The phase is obtained from $\phi = \arctan(2\chi \av{\sz{}}/2)$, with $2\chi/\kappa =1$. 
  (b) Excited qubit population and phase at $\delta_q = 0$ and as a function of $\Omega_R^2$. The horizontal dashed gray line corresponds to qubit saturation, $P_e = 1/2$. 
  (c) Qubit linewidth as a function of $\Omega_R^2$. All three panels have been obtained from numerical simulations of the dispersive qubit master equation with $\gamma_1/2\pi = 0.1$ MHz and $\gamma_\varphi/2\pi$ = 0.1 MHz, to the exception of the dashed blue lines in panels (b) and (c) that correspond to the analytical expressions found in the text.
  } \label{fig:DispersiveQubitSpectroscopy}
\end{figure}

To simplify the discussion, we first consider spectroscopically probing the qubit assuming that the oscillator remains in its vacuum state. This is done by applying a coherent field of amplitude $\alpha_d$
and frequency $\wdrive$ to the qubit, either via a dedicated voltage gate on the qubit or to the input port of the resonator. Ignoring the resonator for the moment, this situation is described by the Hamiltonian $\delta_q \sz{}/2 + \Omega_R \sx{}/2$, where $\delta_q = (\wa+\chi)-\wdrive$ is the detuning between the Lamb-shifted qubit transition frequency and the drive frequency, and $\Omega_R \propto \alpha_d$ is the Rabi frequency. Under this Hamiltonian and using the master equation \cref{eq:ME_transmon} projected on two levels of the qubit, 
 the steady-state probability $P_e=(\av{\sz{}}_\mathrm{s}+1)/2$ for the qubit to be in its excited state (or, equivalently, the probability to be in the ground state, $P_g$) is found to be \cite{Abragam1961}
\begin{equation}\label{eq:lineshape}
        P_{e} = 1 - P_{g} =
        \frac{1}{2}
        \frac
        {\Omega_R^2}
        {\gamma_1\gamma_2+\delta_q^2\gamma_1/\gamma_2+\Omega_R^2}.
\end{equation}
The Lorantzian lineshape of $P_e$ as a function of the drive frequency is illustrated in \cref{fig:DispersiveQubitSpectroscopy}(a). In the limit of strong qubit drive, i.e.~large Rabi frequency $\Omega_R$, the steady-state qubit population reaches saturation with $P_e=P_g=1/2$, see \cref{fig:DispersiveQubitSpectroscopy}(b). Moreover, as the power increases, the full width at half maximum (FWHM) of the qubit lineshape evolves from the bare qubit linewidth given by $\gamma_q = 2 \gamma_2$ to $2\sqrt{{1}/{T_2^2} +\Omega_R^2 {T_1}/{T_2}}$, something that is known as power broadening and which is illustrated in \cref{fig:DispersiveQubitSpectroscopy}(c). In practice, the unbroadenend dephasing rate $\gamma_2$ can be determined from spectroscopic measurements by extrapolating to zero spectroscopy tone power the linear dependence of $\nu_{\rm{HWHM}}^2$. This quantity can also be determined in the time domain from a Ramsey fringe experiment \cite{Vion2002}.\footnote{
Different quantities associated with the dephasing time are used in the literature, the three most common being $T_2$, $T_2^*$ and $T_2^\mathrm{echo}$. While $T_2$ corresponds to the intrinsic or ``natural'' dephasing time of the qubit, $T_2^*\le T_2$ accounts for inhomogeneous broadening. For example, for a flux-tunable transmon, this broadening can be due to random fluctuations of the flux treading the qubit's SQUID loop. A change of the flux over the time of the experiment needed to extract $T_2$ results in a  qubit frequency shifts, something that is measured as a broadening of the qubit's intrinsic linewidth. Notably, the slow frequency fluctuations can be cancelled by applying a $\pi$-pulse midway through a Ramsey fringe experiment. The measured dephasing time is then known as $T_2^\mathrm{echo}$ and is usually longer than $T_2^*$ with its exact value depending on the spectrum of the low-frequency noise affecting the qubit \cite{Martinis2003}. The method of dynamical decoupling which relies on more complex pulse sequences can be used to cancel higher-frequency components of the noise \cite{Bylander2011}.}

In typical optical spectroscopy of atoms in a gas, one directly measures the absorption of photons by the gas as a function of the frequency of the photons.  In circuit QED, one typically performs quantum jump spectroscopy by measuring the probability that an applied microwave drive places the qubit into its excited state.  The variation in qubit population with qubit drive can be measured by monitoring the change in response of the cavity to the spectroscopy drive. As discussed in \cref{sec:DispersiveQubitReadout}, this is realized using two-tone spectroscopy by measuring the cavity transmission, or reflection, of an additional drive of frequency close to $\wc$. In the literature, this second drive is often referred to as the probe or measurement tone, while the spectroscopy drive is also known as the pump tone.  As shown by \cref{eq:phiDispersive}, the phase of the transmitted probe tone is related to the qubit population. In particular, with the probe tone at the bare cavity frequency and in the weak dispersive limit $\chi\ll\kappa$, this phase is simply proportional to the qubit population, $\phi_\mathrm{s} = \arctan(2\chi \av{\sz{}}_\mathrm{s}/2) \approx 2\chi\av{\sz{}}_\mathrm{s}/\kappa$. Monitoring $\phi_\mathrm{s}$ as a function of the spectroscopy tone frequency therefore directly reveals the Lorentzian qubit lineshape~\cite{Schuster2005}.

\subsubsection{AC-Stark shift and measurement-induced broadening}
\label{sec:AcStark_MeasIndDephasing}

\begin{figure}[t]
  \centering
  \includegraphics[width=1.\columnwidth]{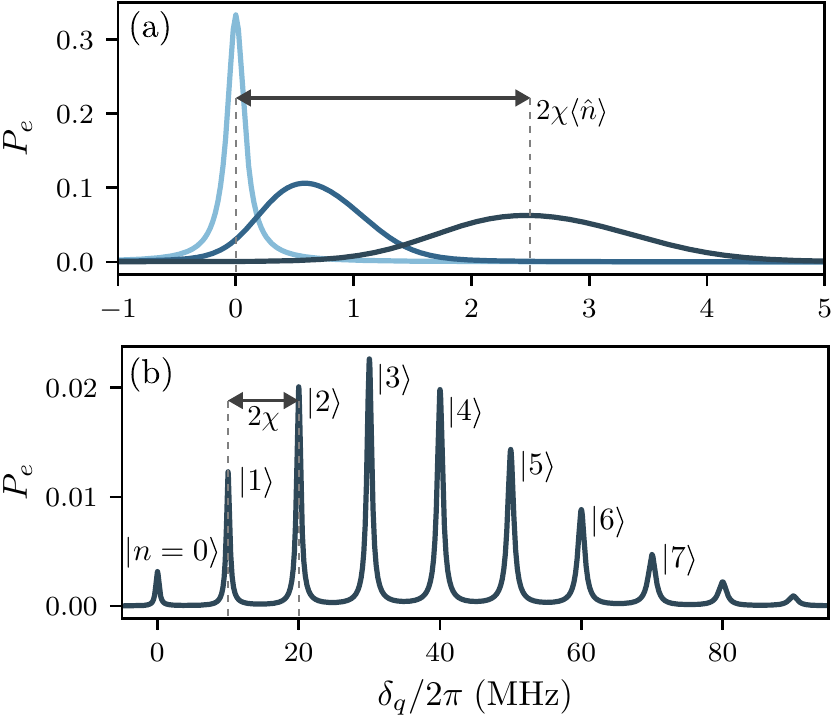}
  \caption{Excited state population as a function of the qubit drive frequency. (a) Dispersive regime with $\chi/2\pi = 0.1$ MHz and (b) strong dispersive limit with $\chi/2\pi = 5$ MHz. The resolved peaks correspond to different cavity photon numbers. 
  The spectroscopy drive amplitude is fixed to $\Omega_R/2\pi = 0.1$ MHz and the damping rates to $\gamma_1/2\pi = \kappa/2\pi = 0.1$ MHz.
  In panel (a) the measurement drive is on resonance with the bare cavity frequency, with amplitude $\epsilon/2\pi =$ (0, 0.2, 0.4) MHz for the light blue, blue and dark blue line, respectively. In panel (b) the measurement drive is at the pulled cavity frequency $\omega_r-\chi$ with amplitude $\epsilon/2\pi = 0.1$ MHz. 
} \label{fig:AcStarkShift}
\end{figure}

In the above discussion, we have implicitly assumed that the amplitude of the measure tone is such that the intra-cavity photon population is vanishingly small, $\av{\ada}\rightarrow 0$. As is made clear by \cref{eq:HQubitDispersiveSimpleACStark}, increase in photon population leads to a qubit frequency shift by an average value of $2 \chi \av{\ada}$. \Cref{fig:AcStarkShift}(a) shows this ac-Stark shift in the steady-state qubit population as a function of spectroscopy frequency for three different probe drive powers. Taking advantage of the dependence of the qubit frequency on measurement power, prior knowledge of the value of $\chi$ allows one to infer the intra-cavity photon number as a function of input pump power from such measurements~\cite{Schuster2005}. However, care must be taken since the linear dependence of the qubit frequency on power predicted in \cref{eq:HQubitDispersiveSimple} is only valid well inside the dispersive regime or, more precisely, at small $\bar n/n_\mathrm{crit}$. We come back to this shortly.

As is apparent from \cref{fig:AcStarkShift}(a), in addition to causing a frequency shift of the qubit, the cavity photon population also causes a broadening of the qubit linewidth. This can be understood simply by considering again the form of $\hH_\text{disp}$ in \cref{eq:HQubitDispersiveSimpleACStark}. Indeed, while in the above discussion we considered only the \emph{average} qubit frequency shift, $2\chi\av{\ad\aop}$, the actual shift is rather given by $2\chi\ad\aop$ such that the full photon-number distribution is important. As a result, when the cavity is prepared in a coherent state by the measurement tone, each Fock state $\ket n$ of the coherent field leads to its own qubit frequency shift $2\chi n$. In the weak dispersive limit corresponding to $\chi/\kappa$ small, the observed qubit lineshape is thus the result of the inhomogeneous broadening due to the Poisson statistics of the coherent state populating the cavity. This effect becomes more apparent as the average measurement photon number $\bar n$ increases and results in a crossover from a Lorentzian qubit lineshape whose linewidth scales with $\bar n$  to a Gaussian lineshape whose linewidth rather scales as $\sqrt{\bar n}$~\cite{Schuster2005,Gambetta2006}. This square-root dependence can be traced to the coherent nature of the cavity field. For a thermal cavity field, a $\bar n (\bar n+1)$ dependence is rather expected and observed~\cite{Bertet2005,Kono2017}.

This change in qubit linewidth due to photon shot noise in the coherent measurement tone populating the cavity can be interpreted as the unavoidable dephasing that a quantum system undergoes during measurement. Using a polaron-type transformation familiar from condensed-matter theory, the cavity can be integrated out of the qubit-cavity master equation and, in this way, the associated measurement-induced dephasing rate can be expressed in the dispersive regime as $\gamma_\mathrm{m}(t) = 2\chi\mathrm{Im}[\alpha_g(t)\alpha_e^*(t)]$, where $\alpha_{g/e}(t)$ are the time-dependent coherent state amplitudes associated with the two qubit states~\cite{Gambetta2008}. In the long time limit, the above rate can be expressed in the more intuitive form $\gamma_\mathrm{m} = \kappa|\alpha_e-\alpha_g|^2/2$, where $\alpha_e-\alpha_g$ is the distance between the two steady-state pointer states~\cite{Gambetta2008}. Unsurprisingly, measurement-induced dephasing is faster when the pointer states are more easily distinguishable and the measurement thus more efficient.  This last expression can also be directly obtained from the entangled qubit-pointer state \cref{eq:entangledQubitPointer} whose coherence decay, at short times, at the rate $\gamma_\mathrm{m}$ under photon loss~\cite{Haroche2006}.

Using the expressions \cref{eq:alpha_g,eq:alpha_e} for the pointer states amplitude, $\gamma_\mathrm{m}$ can be expressed as
\begin{equation}\label{eq:MeasurementInducedDephasingRate}
\gamma_\mathrm{m} = 
\frac
{\kappa \chi ^2 (\bar n_g + \bar n_e)}
{\delta_r^2+\chi^2 + (\kappa/2)^2},
\end{equation}
with $\bar n_\sigma = |\alpha_\sigma|^2$ the average cavity photon number given that the qubit is state $\sigma$. The distinction between $\bar n_g$ and $\bar n_e$ is important if the measurement drive is not symmetrically placed between the two pulled cavity frequencies corresponding to the two qubit states.   Taking $\delta_r = \wc-\wdrive = 0$ and thus $\bar n_g = \bar n_e \equiv \bar n$ for a two-level system, the measurement-induced dephasing rate takes, in the small $\chi/\kappa$ limit, the simple form $\gamma_\mathrm{m} \sim 8 \chi^2 \bar n / \kappa$. Thus as announced above, the qubit linewitdth scales with $\bar n$. With the cautionary remarks that will come below, measuring this linewidth versus the drive power is thus another way to infer $\bar n$ experimentally.

So far, we have been concerned with the small $\chi/\kappa$ limit. However, given the strong coupling and high-quality factor that can be experimentally realized in circuit QED, it is also interesting to consider the opposite limit where $\chi/\kappa$ is large. A first consequence of this strong dispersive regime, illustrated in \cref{fig:AcStarkShift}(b), is that the qubit frequency shift per photon can then be large enough to be resolved spectroscopically~\cite{Gambetta2006,Schuster2007a}. More precisely, this occurs if $2\chi$ is larger than $\gamma_2 + (\bar n + n)\kappa/2$, the width of the $n$th photon peak~\cite{Gambetta2006}. Moreover, the amplitude of each spectroscopic line is a measure of the probability of finding the corresponding photon number in the cavity. Using this idea, it is possible, for example, to experimentally distinguish between coherent and thermal population of the cavity~\cite{Schuster2007a}. This strong dependence of the qubit frequency on the exact photon number also allows for conditional qubit-cavity logical operations where, for example, a microwave pulse is applied such that qubit state is flipped if and only if there are $n$ photons in the cavity~\cite{Johnson2010}. Although challenging, this strong dispersive limit has also been acheived in some cavity QED experiments \cite{Gleyzes2007,Guerlin2007}. This regime has also been acheived in hybrid quantum systems, for example in phonon-number resolving measurements of nanomechanical oscillators \cite{Arrangoiz-Arriola2019,Sletten2019} and magnon-number resolving measurements \cite{Lachance-Quirion2017}.

\begin{figure}[t]
  \centering
  \includegraphics[width=1.0\columnwidth]{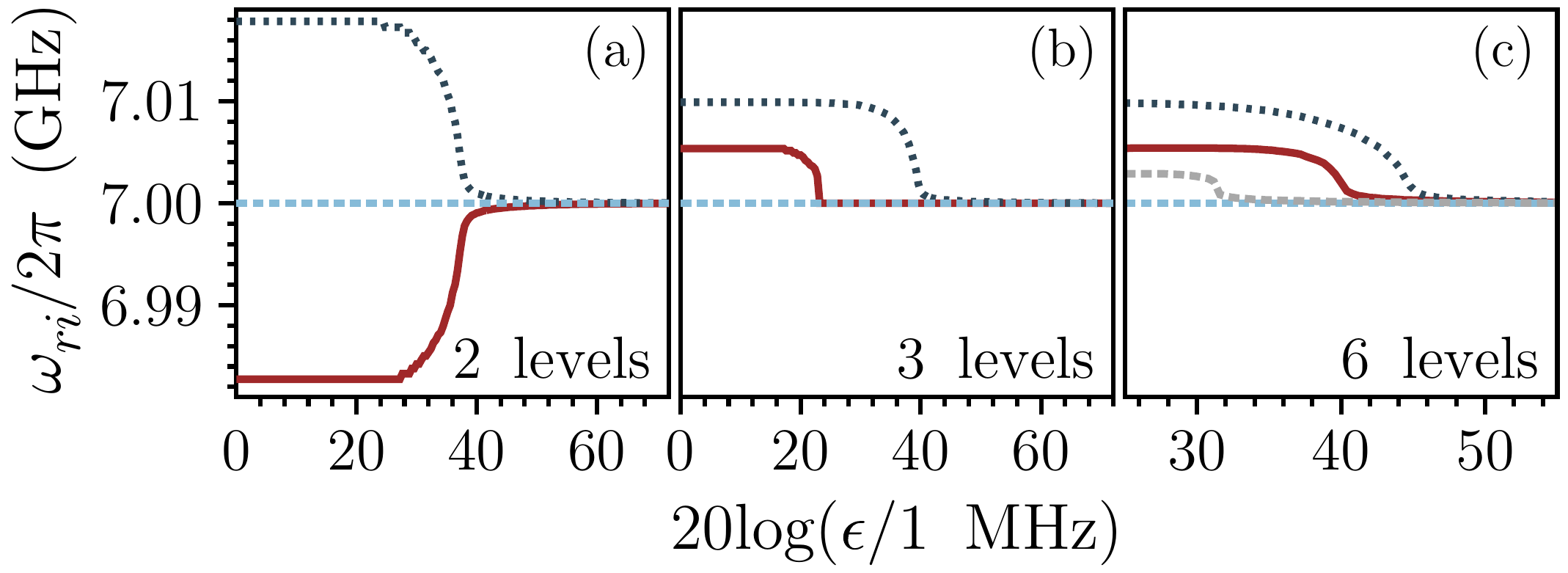}\\
  \caption{Change of effective resonator frequency $\omega_{r\sigma}$ with increasing measurement drive power for the different states $\sigma$ of the transmon qubit (dotted blue line: ground state, full red line: first excited state and dashed gray line: second excited state). The horizontal green dashed line is the bare resonator frequency. (a) Two-level artificial atom, (b) taking into account three levels of the transmon and (c) 6 levels of the transmon. The system parameters are chosen such that $(\omega_{01},\omega_{12},g)/2\pi = (6,5.75,0.1)$~GHz. Adapted from \textcite{Boissonneault2010}.
 } \label{fig:NLCavityPull}
\end{figure}

We now come back to the question of inferring the intra-cavity photon number from ac-Stark shift or qubit linewidth broadening measurements. As mentioned previously, the linear dependence of the ac-Stark shift on the measurement drive power predicted from the dispersive Hamiltonian \cref{eq:HQubitDispersiveSimple} is only valid at small $\bar n/n_\mathrm{crit}$. Indeed, because of higher-order corrections, the cavity pull itself is not constant with $\bar  n$ but rather decreases with increasing $\bar  n$~\cite{Gambetta2006}. This change in cavity pull is illustrated in \cref{fig:NLCavityPull}(a) which shows the effective resonator frequency given that the qubit is in state $\sigma$ as a function of drive amplitude, $\omega_{\mathrm{r}\sigma}(n)= E_{\overline{\sigma,n+1}} - E_{\overline{\sigma,n}}$, with $E_{\overline{\sigma,n+1}}$ the dressed state energies defined in \cref{eq:JCEnergies}~\cite{Boissonneault2010}. At very low drive amplitude, the cavity frequency is pulled to the expected value $\wc\pm\chi$ depending on the state of the qubit. As the drive amplitude increases, and with it the intra-cavity photon number, the pulled cavity frequency goes back to its bare value $\wc$. Panels (b) and (c) show the pulled frequencies taking into account three and six transmon levels, respectively. In contrast to the two-level approximation and as expected from \cref{eq:HTransmonDispersiveSW}, in this many-level situation the symmetry that was present in the two-level case is broken and the pulled frequencies are not symmetrically placed around $\wc$. We note that this change in effective cavity frequency is at the heart of the high-power readout already discussed in \cref{sec:SNR}.

Because of this change in cavity pull, which can be interpreted as $\chi$ itself changing with photon numbers, the ac-Stark shift and the measurement-induced dephasing do not necessarily follow the simple linear dependence expected from $\hH_\text{disp}$. For this reason, it is only possible to safely infer the intra-cavity photon number from measurement of the ac-Stark shift or qubit linewidth broadening at small photon number. It is worth nothing that, in some cases, the reduction in cavity pull can move the cavity frequency closer to the drive frequency, thereby leading to an increase in cavity population. For some system parameters, these two nonlinear effects -- reduction in cavity pull and increase in cavity population -- can partly compensate each other, leading to an \emph{apparent} linear dependence of the qubit ac-Stark with power~\cite{Gambetta2006}. We can only repeat that care must be taken when extracting the intra-cavity photon number in the dispersive regime. 

\subsection{Beyond strong coupling: ultrastrong coupling regime}
\label{sec:ultrastrong}

We have discussed consequences of the strong coupling, $g> \kappa, \gamma_2$, and strong dispersive, $\chi> \kappa, \gamma_2$, regimes which can both easily be realized in circuit QED. Although the effect of light-matter interaction has important consequences, in both these regimes $g$ is small with respect to the system frequencies, $\wc,\wa \gg g$, a fact that allowed us to safely drop counter-rotating terms from~\cref{eq:HTransmonRabi}. In the case of a two-level system this allowed us to work with the Jaynes-Cummings Hamiltonian~\cref{eq:HJC}. The situation where these terms can no longer be neglected is known as the ultrastrong coupling regime.

As discussed in \cref{sec:TransmonOscillator}, the relative smallness of $g$ with respect to the system frequencies can be traced to \cref{eq:g_alpha} where we see that $g/\wc \propto \sqrt\alpha$, with $\alpha\sim1/137$ the fine-structure constant. This is, however, not a fundamental limit and it is possible to take advantage of the flexibility of superconducting quantum circuits to engineer situations where light-matter coupling rather scales as $\propto 1/\sqrt\alpha$. In this case, the smallness of $\alpha$ now helps boost the coupling rather than constraining it. A circuit realizing this idea was first proposed in \cite{Devoret2007} and is commonly known as the in-line transmon. It simply consists of a transmission line resonator whose center conductor is interrupted by a Josephson junction. Coupling strengths has large as $g/\wc \sim 0.15$ can in principle be obtained in this way but increasing this ratio further can be challenging because it is done at the expense of reducing the transmon anharmonicity~\cite{Bourassa2012}.

An alternative approach relies on galvanically coupling a flux qubit to the center conductor of a transmission-line resonator. In this configuration, light-matter coupling can be made very large by increasing the impedance of the center conductor of the resonator in the vicinity of the qubit, something that can be realized by interrupting the center conductor of the resonator by a Josephson junction or a junction array~\cite{Bourassa2009}. In this way, coupling strengths of $g/\wa \sim 1$ or larger can be achieved. These ideas were first realized in \cite{Niemczyk2010,FornDiaz2010} with $g/\wa \sim 0.1$ and more recently with coupling strengths as large as $g/\wa \sim 1.34 $ \cite{Yoshihara2016}. Similar results have also been obtained in the context of waveguide QED where the qubit is coupled to an open transmission line rather than to a localized cavity mode~\cite{Forn-Diaz2016b}.

A first consequence of reaching this ultrastrong coupling regime is that, in addition to a Lamb shift $g^2/\Delta$, the qubit transition frequency is further modified by the so-called Bloch-Siegert shift of magnitude $g^2/(\wa+\wc)$~\cite{bloch:1940a}. Another consequence is that the ground state of the combined system is no longer the factorizable state $\ket{g0}$ but is rather an entangled qubit-resonator state. An immediate implication of this observation is that the master equation \cref{eq:JCMasterEquation}, whose steady-state is $\ket{g0}$, is not an appropriate description of damping in the ultrastrong coupling regime~\cite{Beaudoin2011}. The reader interested in learning more about this regime of light-matter interaction can consult the reviews by \textcite{FornDiaz2019} and \textcite{Kockum2019}.

\section{\label{sec:quantumcomputing}Quantum computing with circuit QED}

\begin{figure*}[t]
  \centering
  \includegraphics[width=1.8\columnwidth]{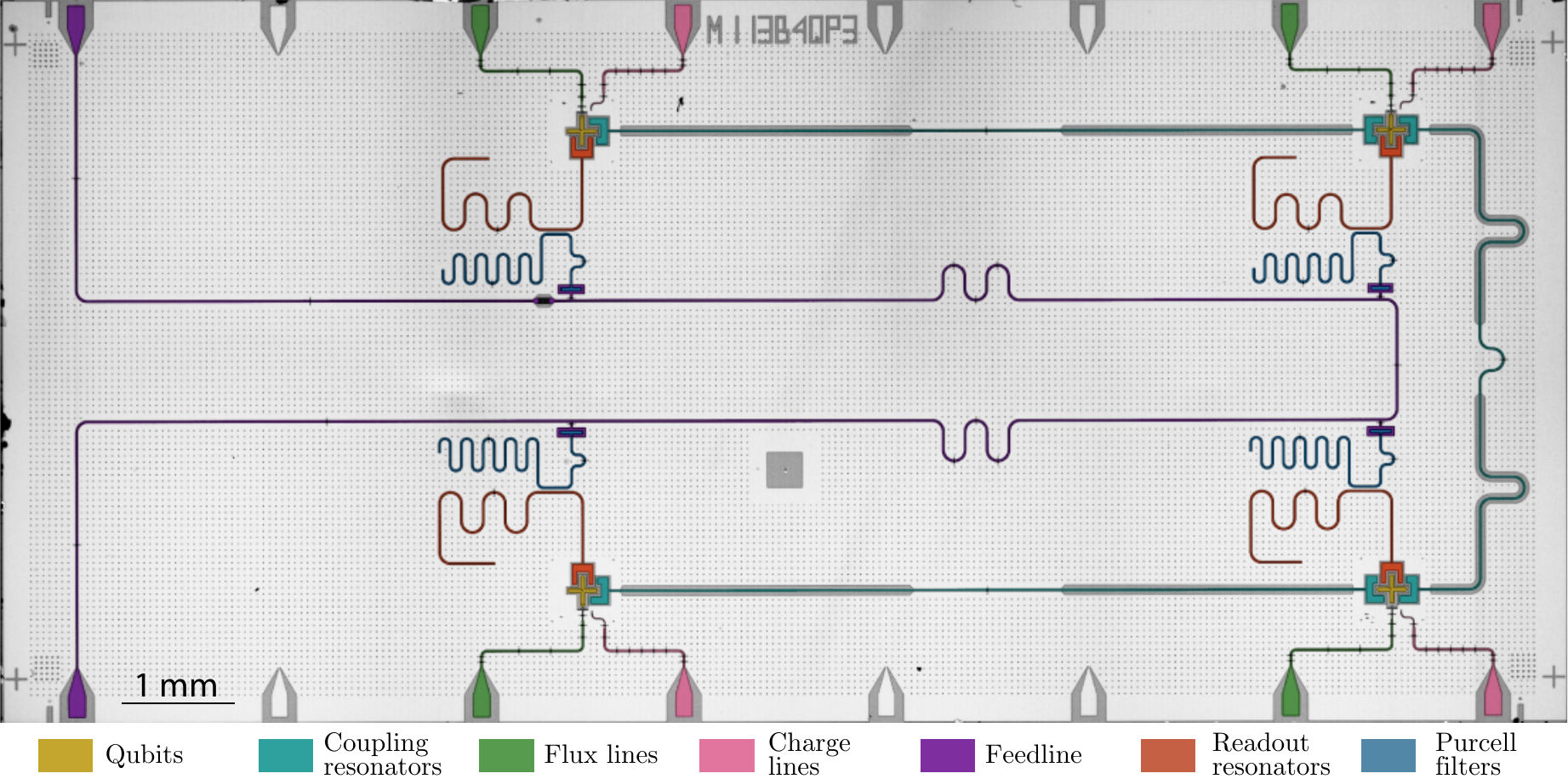}\\
  \caption{
  False colored optical microscope image of a four-transmon device. The transmon qubits are shown in yellow, the coupling resonators in cyan, the flux lines for single qubit tuning in green, the charge lines for single-qubit manipulation in pink, and a common feedline for multiplexed readout in purple, with transmission line resonators for dispersive readout (red) employing Purcell filters (blue). Adapted from \textcite{Andersen2019}.}
 \label{fig:CQED_chip}
\end{figure*}

One of the reasons for the rapid growth of circuit QED as a field of research is its prominent role in quantum computing. The transmon is today the most widely used superconducting qubit, and the dispersive measurement described in~\cref{sec:readout} is the standard approach to qubit readout. Moreover, the capacitive coupling between transmons that are fabricated in close proximity can be used to implement two-qubit gates. Alternatively, the transmon-resonator electric dipole interaction can also be used to implement such gates between qubits that are separated by distances as large as a centimeter, the resonator acting as a quantum bus to mediate qubit-qubit interactions. As illustrated in \cref{fig:CQED_chip}, realizing a quantum computer architecure, even of modest size, requires bringing together in a single working package essentially all of the elements discussed in this review.

In this section, we describe the basic principles behind one- and two-qubit gates in circuit QED. Our objective is not to give a complete overview of the many different gates and gate-optimization techniques that have been developed. We rather focus on the key aspects of how light-matter interaction facilitates coherent quantum operations for superconducting qubits, and describe some of the more commonly used gates to illustrate the basic principles. Unless otherwise noted, in this section we will assume the qubits to be dispersively coupled to the resonator.

\subsection{Single-qubit control}
\label{subsec:SingleQubitGates}

Arbitrary single-qubit rotations can be realized in an NMR-like fashion with voltage drives at the qubit frequency~\cite{Blais2004,Blais2007}. One approach is to drive the qubit via one of the resonator ports~\cite{Wallraff2005}. Because of the large qubit-resonator detuning, a large fraction of the input power is reflected at the resonator, something that can be compensated by increasing the power emitted by the source. The reader will recognize that this approach is very similar to a qubit measurement but, because of the very large detuning, with $\delta_r \gg \chi$  such that $\alpha_e - \alpha_g \sim 0$ according to \cref{eq:apha_eg_s}. As illustrated in \cref{fig:DispersiveCavityPull}, this far off-resonance drive therefore causes negligible measurement-induced dephasing~\cite{Blais2007}. We also note that in the presence of multiple qubits coupled to the same resonator, it is important that the qubits be sufficiently detuned in frequency from each other to avoid the control drive intended for one qubit to inadvertently affect the other qubits. 

Given this last constraint, an often more convenient approach, already illustrated in \cref{fig:dispersivenoise}, is to capacitively couple the qubit to an additional transmission line from which the control drives are applied. Of course, the coupling to this additional control port must be small enough to avoid any impact on the qubit relaxation time. Following~\cref{sec:drives}, the amplitude of the drive as seen by the qubit is given by $\varepsilon = -i\sqrt{\gamma}\beta$, where $\beta$ is the amplitude of the drive at the input port, and $\gamma$ is set by the capacitance between the qubit and the transmission line. A small $\gamma$, corresponding to a long relaxation time, can be compensated by increasing the drive  amplitude $|\beta|$, while making sure that any heating due to power dissipation close to the qubit does not affect qubit coherence. Design guidelines for wiring, an overview of the power dissipation induced by drive fields in qubit drive lines, and their effect on qubit coherence is discussed, for example, in \textcite{Krinner2019}.

Similarly to \cref{eq:DriveOnCavity}, a coherent drive of time-dependent amplitude $\varepsilon(t)$, frequency $\wdrive$ and phase $\phidrive$ on a transmon is then modeled by
\begin{equation}\label{eq:singlequbitdrive}
  \hat H(t) = \hH_\mathrm{q} + \hbar \varepsilon(t)\left(\hat b\dg e^{-i\wdrive t -i\phidrive} + \hat b e^{i\wdrive t +i\phidrive}\right),
\end{equation}
where $\hH_\mathrm{q} = \hbar \omega_q \hat b\dg \hat b - \frac{E_C}{2}(\hat b\dg)^2 \hat b^2$ is the transmon Hamiltonian. Going to a frame rotating at $\wdrive$, $\hat H(t)$ takes the simpler form
\begin{equation}\label{eq:singlequbitdrive_rotatingframe}  
  \hat H' = \hH_\mathrm{q}' + \hbar \varepsilon(t)\left(\hat b\dg e^{-i\phidrive} + \hat b e^{i\phidrive}\right),
\end{equation}
where $\hH_\mathrm{q}' = \hbar \delta_q \hat b\dg \hat b - \frac{E_C}{2}(\hat b\dg)^2 \hat b^2$ with $\delta_q = \omega_q-\omega_d$ the detuning between the qubit and the drive frequencies.

Truncating to two levels of the transmon as in \cref{eq:HJC}, $\hat H'$ takes the form 
\begin{equation}\label{eq:H_qubit_drive_TLS}
  \hat H' = \frac{\hbar \delta_q}{2} \sz{} + \frac{\hbar \Omega_R(t)}{2} \left[\cos(\phidrive)\sx{} + \sin(\phidrive)\sy{}\right],
\end{equation}
where we have introduced the standard notation $\Omega_R = 2\varepsilon$ for the Rabi frequency. This form of $\hat H'$ makes it clear how the phase of the drive, $\phidrive$, controls the axis of rotation on the qubit Bloch sphere. Indeed, for $\delta_q=0$, the choice $\phidrive=0$ leads to rotations around the $X$-axis while  $\phidrive=\pi/2$ to rotations around the $Y$-axis. Since any rotation on the Bloch sphere can be decomposed into $X$ and $Y$ rotations, arbitrary single-qubit control is therefore possible using sequences of on-resonant drives with appropriate phases.

Implementing a desired gate requires turning on and off the drive amplitude. To realize as many logical operations as possible within the qubit coherence time, the gate time should be as short as possible and square pulses are optimal from that point of view. In practice, however, such pulses suffer from important deformation as they propagate down the finite-bandwidth transmission line from the source to the qubit. Moreover, for a weakly anharmonic multi-level system such as a transmon, high-frequency components of the square pulse can cause unwanted transitions to levels outside the two-level computational subspace. This leakage can be avoided by using smooth (e.g.~Gaussian) pulses, but this leads to longer gate times. Another solution is to shape the pulse so as to remove the unwanted frequency components. A widely used approach that achieves this is known as Derivative Removal by Adiabatic Gate (DRAG). It is based on driving the two quadratures of the qubit with the envelope of the second quadrature chosen to be the time-derivative of the envelope of the first quadrature~\cite{Motzoi2009,Gambetta2011a}. More generally, one can cast the problem of finding an optimal drive as a numerical optimization problem which can be tackled with optimal control approaches such as the GRadient Ascent Pulse Engineering (GRAPE)~\cite{Khaneja2005}.

\begin{figure}
  \centering
  \includegraphics{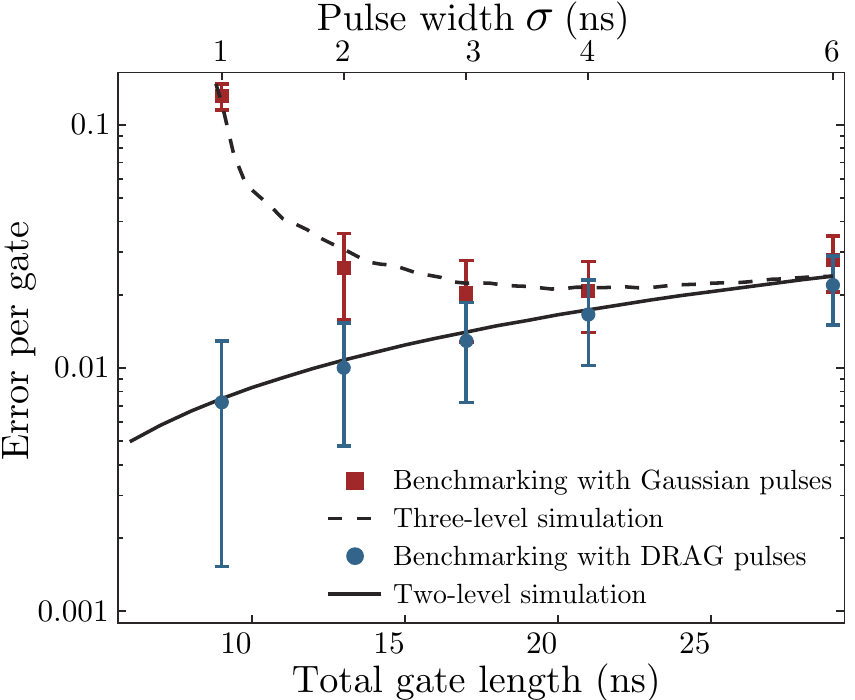}
  \caption{\label{fig:DRAG} Single-qubit gate errors extracted from randomized benchmarking for Gaussian and DRAG pulses as a fonction of total gate time and pulse width $\sigma$ for Gaussian pulses. The experimental results (symbols) are compared to numerical simulations (lines) with two or three transmon levels. Adapted from~\textcite{Chow2010}.}
\end{figure}

Experimental results from~\textcite{Chow2010} comparing the error in single-qubit gates with and without DRAG are shown in~\cref{fig:DRAG}. At long gate times, decoherence is the dominant source of error such that both Gaussian and DRAG pulses initially improve as the gate time is reduced. However, as the pulses get shorter and their frequency bandwidth become comparable to the transmon anharmonicity, leakage leads to large errors for the Gaussian pulses. In contrast, the DRAG results continue to improve as gates are made shorter and are consistent with a two-level system model of the transmon. These observations show that small anharmonicity is not a fundamental obstacle to fast and high-fidelity single-qubit gates. Indeed, thanks to pulse shaping techniques and long coherence times, state of the art single-qubit gate errors are below $10^{-3}$, well below the threshold for topological error correcting codes~\cite{Barends2014,Chen2016}.

While rotations about the $Z$ axis can be realized by concatenating the $X$ and $Y$ rotations described above, several other approaches are used experimentally. Working in a rotating frame as in \cref{eq:H_qubit_drive_TLS} with $\delta_q = 0$, one  alternative method relies on changing the qubit transition frequency such that $\delta_q\neq 0$ for a determined duration. In the absence of drive, $\Omega_R = 0$, this leads to phase accumulation by the qubit state and therefore to a rotation about the $Z$ axis. As discussed in~\cref{sec:tunabletransmon}, fast changes of the qubit transition frequency are possible by, for example, applying a magnetic field to a flux-tunable transmon. However, working with flux-tunable transmons is done at cost of making the qubit susceptible to dephasing due to flux noise. To avoid this, the qubit transition frequency can also be tuned without relying on a flux-tunable device by applying a strongly detuned microwave tone on the qubit. For $\Omega_R/\delta_q \ll 1$, this drive does not lead to Rabi oscillations but induces an ac-Stark shift of the qubit frequency due to virtual transition caused by the drive~\cite{Blais2007}. Indeed, as shown in \cref{sec:AppendixDrivenTransmon}, to second order in $\Omega_R/\delta_q$ and assuming for simplicity a constant drive amplitude, this situation is described by the effective Hamiltonian
\begin{equation}\label{eq:HacStarkDrive}
  \begin{aligned}
    \hat H''
    \simeq{}& \half \left(\hbar\omega_q - \frac{E_C}{2} \frac{\Omega_R^2}{\delta_q^2}\right) \sz{}.
  \end{aligned}
\end{equation}
The last term can be turned on and off with the amplitude of the detuned microwave drive and can therefore be used to realize $Z$ rotations.

Finally, since the $X$ and $Y$ axis in \cref{eq:H_qubit_drive_TLS} are defined by the phase $\phi_d$ of the drive, a particularly simple approach to realize a $Z$ gate is to add the desired phase offset to the drive fields of all subsequent $X$ and $Y$ rotations and two-qubit gates. This so-called virtual $Z$-gate can be especially useful if the computation is optimized to use a large number of $Z$-rotations~\cite{McKay2017}.

\subsection{Two-qubit gates}

\begin{figure*}
  \centering
  \includegraphics[width=1.5\columnwidth]{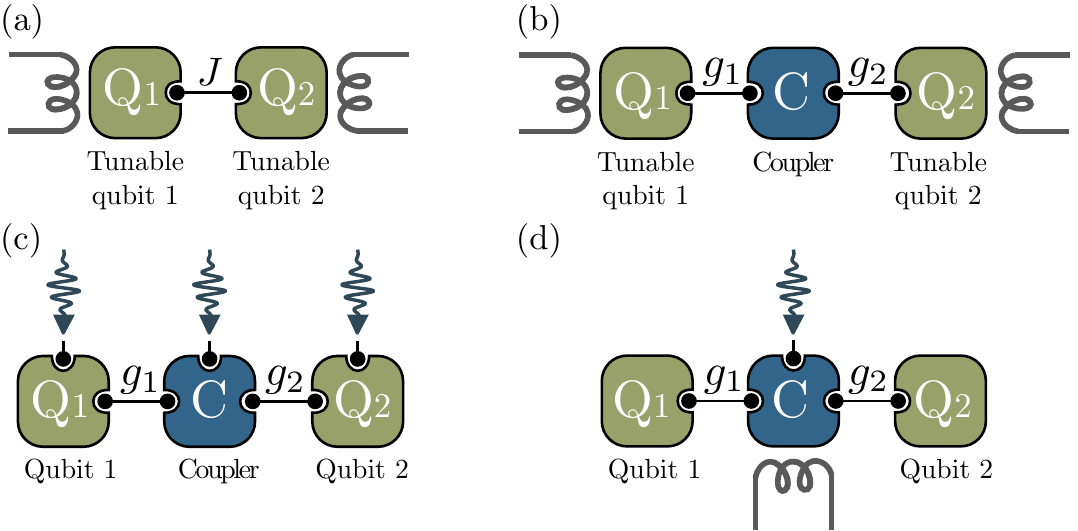}
  \caption{\label{fig:2QubitGates} Schematic illustration of some of the two-qubit gate schemes discussed in the text. Exchange interaction between two qubits (a) from direct capacitive coupling or (b) mediated by a coupler such as a bus resonator. The qubits are tuned in and out of resonance with each other to activate and deactivate the interaction, respectively. (c) All-microwave gates activated by microwave drives on the qubits and/or a coupler such as a bus resonator. In this scheme, the qubits can have a fixed frequency. (d) Parametric gates involving modulating a system parameter, such a tunable coupler. Inspired from \textcite{Yan2018c}.
  }
\end{figure*}

Two-qubit gates are generally more challenging to realize than single-qubit gates. Error rates for current two-qubit gates are typically around one to a few percent, which is an order of magnitude higher than those of single-qubit gates. Recent experiments are, however, closing this gap~\cite{Foxen2020}. Improving two-qubit gate fidelities at short gate times is a very active area of research, and a wide variety of approaches have been developed. A key challenge in realizing two-qubit gates is the ability to rapidly turn interactions on and off. While for single-qubit gates this is done by simply turning on and off a microwave drive, two-qubit gates require turning on a coherent qubit-qubit interaction for a fixed time. Achieving large on/off ratios is far more challenging in this situation.  

Broadly speaking, one can divide two-qubit gates into different categories depending on how the qubit-qubit interaction is activated. The main approaches discussed in the following are illustrated schematically in~\cref{fig:2QubitGates}. An important distinction between these different schemes is whether they rely on frequency-tunable qubits or not. Frequency tunability is convenient because it can be used to controllably tune qubits into resonance with one another qubit or with a resonator. Using flux-tunable transmons has led to some of the fastest and highest fidelity two-qubit gates to date, see \cref{fig:2QubitGates}(a,b) \cite{Barends2014,Chen2014m,Arute2019}. However, as mentioned previously this leads to additional qubit dephasing due to flux noise. An alternative are all-microwave gates which only use microwave drives, either on the qubits or on a coupler bus such as a resonator, to activate an effective qubit-qubit interaction, see \cref{fig:2QubitGates}(c). 
Finally, yet another category of gates are parametric gates where a system parameter is modulated in time at a frequency which bridges an energy gap between the states of  two qubits. Parametric gates can be all-microwave but, in some instances, involve modulating system frequencies using external magnetic flux, see \cref{fig:2QubitGates}(d).

\subsubsection{Qubit-qubit exchange interaction}

\paragraph{Direct capacitive coupling}\label{sec:DirectCoupling}

One of the conceptually simplest ways to realize two-qubit gates is through direct capacitive coupling between the qubits, see \cref{fig:2QubitGates}(a). In analogy with~\cref{eq:HTransmonJC}, the Hamiltonian describing this situation reads
\begin{equation}\label{eq:gates:H_swap}
  \hH = \hat H_{q1} + \hat H_{q2} + \hbar J (\bd_1\bop_2+\bop_1\bd_2),
\end{equation}
where $\hat H_{qi} = \hbar \omega_{qi}\hat b_i\dg \hat b_i - E_{C_i} (\hat b_i\dg)^2\hat b_i^2/2$ is the Hamiltonian of the $i$th transmon and $\bop_i$ the corresponding annihilation operator. The interaction amplitude $J$ takes the form
\begin{equation}\label{eq:gates:twoqubitg}
\hbar J = \frac{2 E_{C1}E_{C2}}{E_{C_c}}\left(\frac{E_{J1}}{2E_{C1}}
\times
\frac{E_{J2}}{2E_{C2}}\right)^{1/4},
\end{equation}
with $E_{Ji}$ and $E_{Ci}$ the transmon Josephson and charging energies, and $E_{C_c} = e^2/2C_c$ the charging energy of the coupling capacitance labelled $C_c$. This beam-splitter Hamiltonian describes the coherent exchange of an excitation between the two qubits. In the two-level approximation, assuming tuned to resonance qubits, $\omega_{q1} = \omega_{q2}$, and moving to a frame rotating at the qubit frequency, \cref{eq:gates:H_swap} takes the familiar form
\begin{equation}
  \hH' = \hbar J (\hat \sigma_{+1}\hat \sigma_{-2} + \hat \sigma_{-1}\hat \sigma_{+2}).
\end{equation}
Evolution under this Hamiltonian for a time $\pi/(4J)$ leads to an entangling $\sqrt{i\text{SWAP}}$ gate which, up to single-qubit rotations, is equivalent to a controlled NOT gate (CNOT)~\cite{Loss1998}.

As already mentioned, to precisely control the evolution under $ \hH'$ and therefore the gate time, it is essential to be able to vary the qubit-qubit interaction with a large on/off ratio. 
There are essentially two approaches to realizing this. The most straightforward way is to tune the qubits in resonance to perform a two-qubit gate, and to strongly detune them to stop the coherent exchange induced by $\hH'$ \cite{Blais2003a,Dewes2012,Bialczak2010}. Indeed, for $J/\Delta_{12} \ll 1$ where $\Delta_{12} = \omega_{q1}-\omega_{q2}$ is the detuning between the two qubits, the coherent exchange $J$ is suppressed and can be dropped from \cref{eq:gates:H_swap} under the RWA. A more careful analysis following the same arguments and approach used to describe the dispersive regime (cf. \cref{sec:dispersive}) shows that, to second order in $J/\Delta_{12}$, a residual qubit-qubit interaction of the form  $(J^2/\Delta_{12}) \sz{1}\sz{2}$ remains. This unwanted interaction in the off state of the gate leads to a conditional phase accumulation on the qubits. As a result, the on-off ratio of this direct coupling gate is estimated to be $\sim \Delta_{12}/J$. In practice, this ratio cannot be made arbitrarily small because increasing the detuning of one pair of qubits in a multi-qubit architecture might lead to accidental resonance with a third qubit. This direct coupling approach was implemented by \textcite{Barends2014} using frequency tunable transmons with a coupling $J/2\pi = 30$ MHz and an on/off ratio of 100. We note that the unwanted phase accumulation due to the residual $\sz{1}\sz{2}$ can in principle be eliminated using refocusing techniques borrowed from nuclear magnetic resonance~\cite{Slichter95}.

Another approach to turn on and off the swap interaction is to make the $J$ coupling itself tunable in time. This is conceptually simple, but requires more complex coupling circuitry typically involving flux-tunable elements that can open additional decoherence channels for the qubits. One advantage is, however, that tuning a coupler rather than qubit transition frequencies helps in reducing the frequency crowding problem. This approach is used, for example, by~\cite{Chen2014m} where two transmon qubits are coupled via a flux-tunable inductive coupler. In this way, it was possible to realize an on/off ratio of 1000, with a maximum coupling of 100 MHz corresponding to a $\sqrt{i\text{SWAP}}$ gate in 2.5 ns. A simpler approach based on a frequency tunable transmon qubit acting as coupler, as suggested in \textcite{Yan2018c}, was also used to tune qubit-qubit coupling from 5 MHz to $-40$ MHz going through zero coupling with a gate time of $\sim$ 12 ns and a gate infidelity of $\sim 0.5\%$ \cite{Arute2019}.

\paragraph{Resonator mediated coupling}

An alternative to the above approach is to use a resonator as a quantum bus mediating interactions between two qubits, see \cref{fig:2QubitGates}(b)~\cite{Blais2004,Blais2007,Majer2007}. An advantage compared to direct coupling is that the qubits do not have to be fabricated in close proximity to each other. With the qubits coupled to the same resonator, and in the absence of any direct coupling between the qubits, the Hamiltonian describing this situation is
\begin{equation}\label{eq:gates:H_resonatormediated}
  \hH = \hat H_{q1} + \hat H_{q2} + \hbar \wc \ad\aop + \sum_{i=1}^2\hbar g_i (\ad\bop_i+\aop\bd_i).
\end{equation}
One way to make use of this pairwise interaction is, assuming the resonator to be in the vacuum state, to first tune one of the two qubits in resonance with the resonator for half a vacuum Rabi oscillation cycle, swapping an excitation from the qubit to the resonator, before tuning it back out of resonance. The second qubit is then tuned in resonance mapping the excitation from the resonator to the second qubit \cite{Sillanpaa2007}. While this sequence of operations can swap the quantum state of the first qubit to the second, clearly demonstrating the role of the resonator as a quantum bus, it does not correspond to an entangling two-qubit gate. 

\begin{figure}
  \centering
  \includegraphics{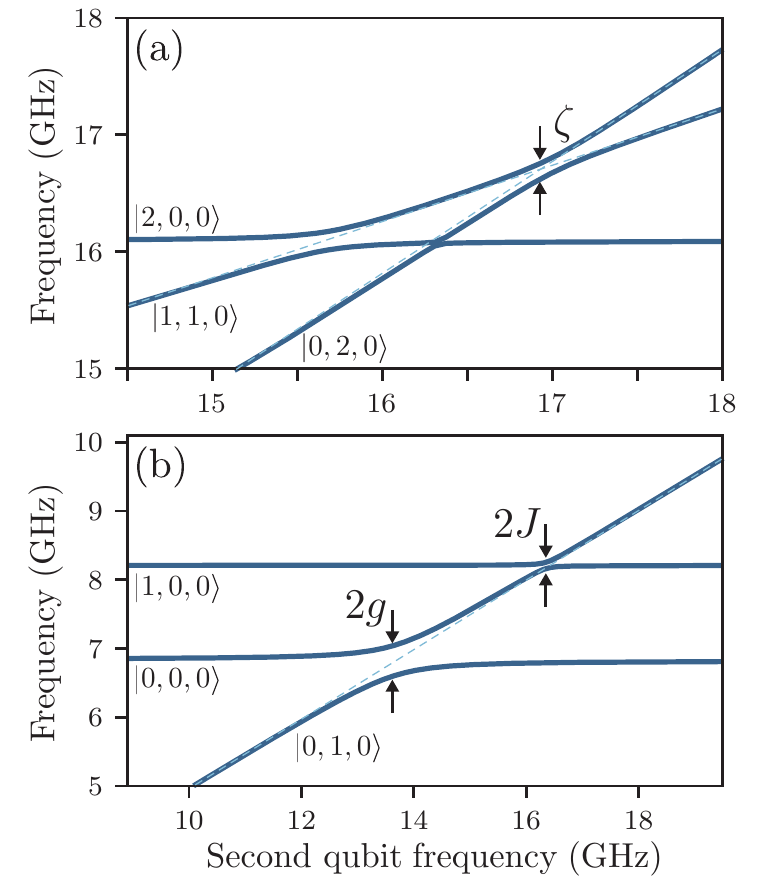}
  \caption{\label{fig:1102}
Spectrum of two transmon qubits coupled to a common resonator as a function of the frequency of the second qubit in the (a) 2-excitation and (b) 1-excitation manifold. The full lines are obtained by numerical diagonalization of \cref{eq:gates:H_resonatormediated} in the charge basis with 5 transmon levels and 5 resonator levels, and with parameters adapted from \textcite{DiCarlo2009}: $E_{J1(2)}/h= 28.48 (42.34)$ GHz, $E_{C1(2)}/h = 317 (297)$ MHz and $g_{1(2)}/2\pi = 199 (190)$ MHz. In the one-excitation manifold, both the $2g$ anticrossing of the first qubit with the resonator and the $2J$ anticrossing of the two qubits are visible. In the two-excitation manifold, the 11-02 anticrossing of magnitude $\zeta$ can be seen. Notice the change in horizontal scale between the two panels.  The states are labelled as $\ket{\text{1st qubit, 2nd qubit, resonator}}$. The dashed light blue lines are guides to the eye following the bare frequency of the first qubit.
}
\end{figure}

Alternatively, a two-qubit gate can be performed by only virtually populating the resonator mode by working in the dispersive regime where both qubits are far detuned from the resonator~\cite{Blais2004,Blais2007,Majer2007}. Building on the results of \cref{sec:dispersive}, in this situation the effective qubit-qubit interaction is revealed by using the approximate dispersive transformation $\hat U = \exp\left[\sum_i \frac{g_i}{\Delta_i}\left( \ad\bop_i - \aop\bd_i\right) \right]$ on  \cref{eq:gates:H_resonatormediated}.
Making use of the  Baker-Campbell-Hausdorff expansion \cref{eq:BCH} to second order in $g_{i}/\Delta_i$, we find
\begin{equation}\label{eq:gates:H_swap2}
  \begin{aligned}
    \hH' ={}& \hat H_{q1}' + \hat H_{q2}' + \hbar J (\bd_1\bop_2+\bop_1\bd_2)\\
    & + \hbar \tilde{\omega}_r \hat a\dg \hat a + \sum_{i=1}^2 \hbar \chi_{ab_i} \hat a\dg \hat a \hat b_i\dg \hat b_i \\
    & + \sum_{i\neq j}\hbar \Xi_{ij}\bd_i\bop_i\left( \bd_j \bop_i + \bd_i \bop_j \right),
  \end{aligned}
\end{equation}
with $H_{qi}' \simeq \hbar\tilde \omega_{qi}\hat b_i\dg \hat b_i - \frac{E_{Ci}}{2} (\hat b_i\dg)^2\hat b_i^2$ the transmon Hamiltonians, $\chi_{ab_i} \simeq - 2 E_{Ci} g_i^2/\Delta_i^2$ a cross-Kerr coupling between the resonator and the $i$th qubit. The frequencies $\tilde \omega_{qi}$ and $\tilde \omega_{r}$ include the Lamb shift. The last line can be understood as an excitation number dependent exchange interaction with $\Xi_{ij} = E_{Ci} g_ig_j/(2\Delta_i\Delta_j$). Since this term is much smaller than the $J$-coupling it can typically be neglected. Note that we have not included a self-Kerr term of order $\chi_{ab_i}$ on the resonator. This term is of no practical consequences in the dispersive regime where the resonator is only virtually populated. The resonator-induced $J$ coupling in $\hH'$ takes the form
\begin{equation}\label{eq:gates:J_mediated}
  J = \frac{g_1g_2}{2}\left(\frac1\Delta_1 + \frac1\Delta_2\right)
\end{equation}
and reveals itself in the frequency domain by an anticrossing of size $2J$ between the qubit states $\ket{01}$ and $\ket{10}$. This is illustrated in \cref{fig:1102}(b) which shows the eigenenergies of the Hamiltonian \cref{eq:gates:H_resonatormediated} in the 1-excitation manifold. In this figure, the frequency of qubit 1 is swept while that of  qubit 2 is kept constant at $\sim 8$ GHz with the resonator at $\sim 7$ GHz. From left to right, we first see the vacuum Rabi splitting of size $2g$ at $\omega_{q1} = \wc$, followed by a smaller anticrossing of size $2J$ at the qubit-qubit resonance. It is worth mentioning that the above expression for $J$ is only valid for single-mode oscillators and is renormalized in the presence of multiple modes~\cite{Filipp2011a,Solgun2019}.

To understand the consequence of the $J$-coupling in the time domain, it is useful to note that, if the resonator is initially in the vacuum state, it will remain in that state under the influence of  $\hH'$. In other words, the resonator is only virtually populated by its dispersive interaction with the qubits. For this reason, with the resonator initialized in the vacuum state, the second line of~\cref{eq:gates:H_swap2} can for all practical purposes be ignored and we are back to the form of the direct coupling Hamiltonian of \cref{eq:gates:H_swap}. Consequently, when both qubits are tuned in resonance with each other, but still dispersive with respect to the resonator, the latter acts as a quantum bus mediating interactions between the qubits. An entangling gate can thus be performed in the same way as with direct capacitive coupling, either by tuning the qubits in and out of resonance with each other \cite{Majer2007} or by making the couplings $g_i$ tunable \cite{Gambetta2011,Srinivasan2011}.

\subsubsection{\label{sec:gates:1102} Flux-tuned 11-02 phase gate}

The $11$-$02$ phase gate is a controlled-phase gate that is well suited to weakly anharmonic qubits such as transmons~\cite{Strauch2003,DiCarlo2009}. It is obtained from the exchange interaction of~\cref{eq:gates:H_swap} and can thus be realized through direct (static or tunable) qubit-qubit coupling or  indirect coupling via a resonator bus.

In contrast to the $\sqrt{i\text{SWAP}}$ gate, the $11$-$02$ phase gate is not based on tuning the qubit transition frequencies between the computational states into resonance with each other, but rather exploits the third energy level of the transmon. The $11$-$02$ gate relies on tuning the qubits to a point where the states $\ket{11}$ and $\ket{02}$ are degenerate in the absence of $J$ coupling. As illustrated in \cref{fig:1102}(a), the qubit-qubit coupling lifts this degeneracy by an energy $\zeta$  whose value can be found perturbatively \cite{DiCarlo2009}. Because of this repulsion caused by coupling to the state $\ket{02}$, the energy $E_{11}$ of the state $\ket{11}$ is smaller than $E_{01}+E_{10}$ by $\zeta$. Adiabatically flux tuning the qubits in and out of the $11-02$ anticrossing therefore leads to a conditional phase accumulation which is equivalent to a controlled-phase gate.

To see this more clearly, it is useful to write the unitary corresponding to this adiabatic time evolution as
\begin{equation}\label{eq:gates:CZ_general}
  \hat C_Z(\phi_{01},\phi_{10},\phi_{11}) = \left(
    \begin{array}{cccc}
      1 & 0 & 0 & 0 \\
      0 & e^{i\phi_{01}} & 0 & 0 \\
      0 & 0 & e^{i\phi_{10}} & 0 \\
      0 & 0 & 0 & e^{i\phi_{11}}
    \end{array}
  \right),
\end{equation}
where $\phi_{ab} = \int dt\, E_{ab}(t)/\hbar$ is the dynamical phase accumulated over total flux excursion. Up to single-qubit rotations, this is equivalent to a standard controlled-phase gate since
\begin{equation}
\begin{split}
\hat C_Z(\phi) 
&= \text{diag}(1,1,1,e^{i\phi})\\ 
&= \hat R_Z^1(-\phi_{10})\hat R^2_Z(-\phi_{01}) \hat C_Z(\phi_{01},\phi_{10},\phi_{11}),
\end{split}
\end{equation}
with $\phi=\phi_{11}-\phi_{01}-\phi_{10} = \int dt \, \zeta(t)$ and where $\hat R^i_Z(\theta) = \text{diag}(1,e^{i\theta})$ is a single qubit phase gate acting on qubit $i$. For $\phi \neq 0$ this is an entangling two-qubit gate and, in particular, for $\phi=\pi$ it is a controlled-$Z$ gate (CPHASE). 

Rather than adiabatically tuning the flux in and out of the $11-02$ resonance, an alternative is to non-adiabatically  pulse to this anti-crossing \cite{Strauch2003,DiCarlo2010,Yamamoto2010}. In this sudden approximation, leaving the system there for a time $t$, the state $\ket{11}$ evolves into $\cos(\zeta t /2\hbar)\ket{11}+\sin(\zeta t /2\hbar)\ket{02}$. For $t = h/\zeta$, $\ket{11}$ is mapped back into itself but acquires a minus sign in the process. On the other hand, since they are far from any resonance, the other logical states evolve trivially. This therefore again results in a CPHASE gate. In this way, fast controlled-$Z$ gates are possible. For direct qubit-qubit coupling in particular, some of the fastest and highest fidelity two-qubit gates have been achieved this way with error rates below the percent level and gate times of a few tens of ns \cite{Barends2014,Chen2014m}.

Despite its advantages, a challenge associated with this gate is the distortions in the flux pulses due to the finite bandwidth of the control electronics and line. In addition to modifying the waveform experienced by the qubit, this can lead to long time scale distortions where the flux at the qubit at a given time depends on the previous flux excursions. This situation can be partially solved by pre-distorting the pulses taking into account the known distortion, but also by adapting the applied flux pulses to take advantage of the symmetry around the transmon sweet-spot to cancel out unwanted contributions~\cite{Rol2019a}.

\subsubsection{All-microwave gates}
\label{sec:AllMicrowaveGates}

Because the on/off ratio of the gates discussed above is controlled by the detuning between the qubits, it is necessary to tune the qubit frequencies over relatively large frequency ranges or, alternatively, to have tunable coupling elements. In both cases, having a handle on the qubit frequency or qubit-qubit coupling opens the system to additional dephasing. Moreover, changing the qubit frequency over large ranges can lead to accidental resonance with other qubits or uncontrolled environmental modes, resulting in energy loss. For these reasons, it can be advantageous to control two-qubit gates in very much the same way as single-qubit gates: by simply turning on and off a microwave drive. In this section, we describe two so-called all-microwave gates: the resonator-induced phase (RIP) gate and the cross-resonance (CR) gate. Both are based on fixed-frequency far off-resonance qubits with an always-on qubit-resonator coupling. The RIP gate is activated by driving a common resonator and the CR gate by driving one of the qubits. Other all-microwave gates which will not be discussed further here include the sideband-based iSWAP~\cite{Leek2009}, the bSWAP~\cite{Poletto2012}, the microwave-activated CPHASE~\cite{Chow2013} and the fg-ge gate \cite{Zeytinoglu2015,Egger2019}.

\paragraph{Resonator-induced phase gate}

The RIP gate relies on two strongly detuned qubits that are dispersively coupled to a common resonator mode. The starting point is thus \cref{eq:gates:H_swap2} where we now neglect the $J$ coupling by taking $|\omega_{q1}-\omega_{q2}| \gg J$. In the two-level approximation and accounting for a drive on the resonator, this situation is described by the Hamiltonian
\begin{equation}\label{eq:gates:H_RIP}
  \begin{aligned}
    \hH'/\hbar ={}& \frac{\tilde \omega_{q1}}{2} \hat \sigma_{z1} + \frac{\tilde \omega_{q2}}{2} \hat \sigma_{z2} + \tilde \wc \hat a\dg \hat a \\
    +& \sum_{i=1}^2 \chi_{i} \hat a\dg \hat a \hat \sigma_{zi}
    + \varepsilon(t)(\hat a\dg e^{-i\wdrive t} + \hat a e^{i\wdrive t}),
  \end{aligned}
\end{equation}
where $\varepsilon(t)$ is the time-dependent amplitude of the resonator drive and $\wdrive$ its frequency. Note that we also neglect the resonator self-Kerr nonlinearity.

The gate is realized by adiabatically ramping on and off the drive $\varepsilon(t)$, such that the resonator starts and ends in the vacuum state. Crucially, this means that the resonator is unentangled from the qubits at the start and end of the gate. Moreover, to avoid measurement-induced dephasing, the drive frequency is chosen to be far from the cavity mode, $\tilde\delta_r = \twc-\wdrive \gg \kappa$. Despite this strong detuning, the dispersive shift causes the resonator frequency to depend on the state of the two qubits and, as a result, the resonator field evolves in a closed path in phase space that is qubit-state dependent. This leads to a different phase accumulation for the different qubit states, and therefore to a controlled-phase gate of the form of \cref{eq:gates:CZ_general}.

This conditional phase accumulation can be made more apparent by moving \cref{eq:gates:H_RIP} to a frame rotating at the drive frequency and by applying the polaron transformation $\hat U = \exp[\hat \alpha'(t) \hat a\dg - \hat \alpha^{*\prime}(t)\hat a]$ with $\alpha'(t) = \alpha(t) - \sum_i \chi_i \hat \sigma_{zi}/\tilde\delta_r$ on the resulting Hamiltonian. This leads to the approximate effective Hamiltonian~\cite{Puri2016c}
\begin{equation}
  \begin{split}
    \hat H'' \simeq{}& \sum_i \hbar\left[\frac{\tilde\delta_{qi}}{2} + \chi|\alpha(t)|^2\right]\hat\sigma_{zi} + \hbar\delta_r \hat a\dg \hat a \\
  &+ \sum_{i=1}^2 \hbar \chi_{i} \hat a\dg \hat a \hat \sigma_{zi}
  - \hbar\frac{2\chi_1\chi_2|\alpha(t)|^2}{\delta_r}\hat \sigma_{z1}\hat \sigma_{z2},
  \end{split}
\end{equation}
with $\tilde\delta_x = \tilde\omega_x-\wdrive$ and where the field amplitude $\alpha(t)$ satisfies $\dot\alpha = -i\tilde\delta_r\alpha - i\epsilon(t) $. In this frame, it is clear how the resonator mediates a $\sz{1}\sz{2}$ interaction between the two qubits and therefore leads to a conditional phase gate. This expression also makes it clear that the need to avoid measurement-induced dephasing with $\tilde\delta_r\gg\kappa$ limits the effective interaction strength and therefore leads to relatively long gate times. This can, however, be mitigated by taking advantage of pulse shaping techniques~\cite{Cross2015} or by using squeezed radiation to erase the which-qubit information in the output field of the resonator~\cite{Puri2016c}. Similarly to the longitudinal readout protocol discussed in~\cref{sec:ReadoutOtherApproaches}, longitudinal coupling also offers a way to overcome many of the limitations of the conventional RIP gate~\cite{Royer2017}.

Some of the advantages of this two-qubit gate are that it can couple qubits that are far detuned from each other and that it does not introduce significant leakage errors~\cite{Paik2016}. 
This gate was demonstrated by~\textcite{Paik2016} with multiple transmons coupled to a 3D resonator, achieving error rates of a few percent and gate times of several hundred nanoseconds.

\paragraph{Cross-resonance gate}

The cross-resonance gate is based on qubits that are detuned from each other and coupled by an exchange term $J$ of the form of \cref{eq:gates:H_swap} or \cref{eq:gates:H_swap2}~\cite{Rigetti2010,Chow2011}. While the RIP gate relies on off-resonant driving of a common oscillator mode, this gate is based on directly driving one of the qubits at the frequency of the other. Moreover, since the resonator is not directly used and, in fact, ideally remains in its vacuum throughout the gate, the $J$ coupling can be mediated by a resonator or by direct capactitive coupling.

In the two-level approximation and in the absence of the drive, this interaction takes the form
\begin{equation}\label{eq:H_ExchangeTLS}
  \hH = 
  \frac{\hbar\omega_{q1}}{2}\sz{1}
  +\frac{\hbar\omega_{q2}}{2}\sz{2}
  +\hbar J (\spp{1}\smm{2}+\smm{1}\spp{2}).
\end{equation}
To see how this gate operates, it is useful to diagonalize $\hH$ using the two-level system version of the transformation \cref{eq:UDispersive}. The result takes the same general form as \cref{eq:HDiagonalLinear} and \cref{eq:DressedFrequenciesDispersive}, after projecting to two levels. In this frame, the presence of the $J$ coupling leads to a renormalization of the qubit frequencies which for strongly detuned qubits, $|\Delta_{12}| = |\omega_{q1}-\omega_{q2}|\gg |J|$, take the values $\tilde\omega_{q1} \approx \omega_{q1} + J^2/\Delta_{12}$ and $\tilde\omega_{q2} \approx \omega_{q2} - J^2/\Delta_{12}$ to second order in $J/\Delta_{12}$. In the same frame, a drive on the first qubit, $\hbar\Omega_R(t)\cos (\wdrive t) \sx{1}$, takes the form \cite{Chow2011}
\begin{equation}\label{eq:CrossResonanceTLS}
\begin{split}
  &\hbar\Omega_R(t)\cos (\wdrive t) 
  \left( \cos\theta \sx{1} + \sin\theta\sz{1}\sx{2} \right)\\
  &\approx
  \hbar\Omega_R(t)\cos (\wdrive t) 
  \left(
    \sx{1}     
    +
    \frac{J}{\Delta_{12}} \sz{1}\sx{2}
  \right),
\end{split}
\end{equation}
with $\theta = \arctan(2J/\Delta_{12})/2$ and where the second line is valid to first order in $J/\Delta_{12}$. As a result, driving the first qubit at the frequency of the second qubit, $\wdrive = \tilde\omega_{q2}$, activates the term $\sz{1}\sx{2}$ which can be used to realize a CNOT gate.
 
More accurate expressions for the amplitude of the CR term $\sz{1}\sx{2}$ can be obtained by taking into account more levels of the transmons. In this case, the starting point is the Hamiltonian \cref{eq:gates:H_swap} with, as above, a drive term on the first qubit
\begin{equation}\label{eq:gates:H_CR}
  \begin{aligned}
    \hH ={}& \hat H_{q1} + \hat H_{q2} + \hbar J (\bd_1\bop_2+\bop_1\bd_2)\\
    &+ \hbar \varepsilon(t)(\hat b_1\dg e^{-i\wdrive t} + \hat b_1 e^{i\wdrive t}),
  \end{aligned}
\end{equation}
where $\wdrive \sim \omega_{q2}$. Similarly to the previous two-level system example, it is useful to eliminate the $J$-coupling. We do this by moving to a rotating frame at the drive frequency for both qubits, followed by a Schireffer-Wolff transformation to diagonalize the first line of~\cref{eq:gates:H_CR} to second order in $J$, see \cref{sec:SW}. The drive term is modified under the same transformation by using the explicit expression for the Schrieffer-Wolff generator $\hat S = \hat S^{(1)} + \dots$ given in~\cref{eq:SW:explicit_expansions_generator}, and the Baker-Campbell-Hausdorff formula~\cref{eq:BCH} to first order: $e^{\hat S} \bop_1 e^{-\hat S} \simeq \bop_1 + [\hat S^{(1)},\bop_1]$. The full calculation is fairly involved and here we only quote the final result after truncating to the two lowest levels of the transmon qubits~\cite{Magesan2018,Tripathi2019a}
\begin{equation}\label{eq:gates:H_CR_approx}
  \begin{aligned}
    \hH' \simeq{}& \frac{\hbar\tilde\delta_{q1}}{2}\hat \sigma_{z1} + \frac{\hbar\tilde\delta_{q2}}{2}\hat \sigma_{z2}
    + \frac{\hbar\chi_{12}}{2}\hat\sigma_{z1}\hat\sigma_{z2} \\
    +& \hbar \varepsilon(t)\left( \hat \sigma_{x1} - J' \hat \sigma_{x2} -\frac{E_{C_1}}{\hbar}\frac{J'}{\Delta_{12}}  \hat \sigma_{z1} \hat \sigma_{x2} \right).
  \end{aligned}
\end{equation}
In this expression, the detunings include frequency shifts due to the $J$ coupling with
$\tilde\delta_{q1} = \omega_{q1} + J^2/\Delta_{12} + \chi_{12} - \wdrive$ and 
$\tilde\delta_{q2} = \omega_{q2} - J^2/\Delta_{12} + \chi_{12} - \wdrive$.
The parameters $\chi_{12}$ and $J'$ are given by
\begin{subequations}
  \begin{align}
    \chi_{12} &= \frac{J^2}{\Delta_{12} + \frac{E_{C_2}}{\hbar}}  - \frac{J^2}{\Delta_{12} - \frac{E_{C_1}}{\hbar}},\\
    J' &
    = \frac{J}{\Delta_{12}-\frac{E_{C_1}}{\hbar}}.
  \end{align}
\end{subequations}

\Cref{eq:CrossResonanceTLS,eq:gates:H_CR_approx} agree in the limit of large anharmonicity $E_{C_{1,2}}$ and we again find that a drive on the first qubit at the frequency of the second qubit activates the CR term $\sz{1}\sx{2}$. However, there are important differences at finite $E_{C_{1/2}}$, something which highlights the importance of taking into account the multilevel nature of the transmon. Indeed, the amplitude of the CR term is smaller here than in \cref{eq:CrossResonanceTLS} with a two-level system. Moreover, in contrast to the latter case, when taking into account multiple levels of the transmon qubits we find a spurious interaction $\sz{1}\sz{2}$ of amplitude $\chi_{12}$ between the two qubits, as well as a drive on the second qubit of amplitude $J'\varepsilon(t)$. This unwanted drive can be echoed away with additional single-qubit gates~\cite{Corcoles2013,Sheldon2016}. The $\sz{1}\sz{2}$ interaction is detrimental to the gate fidelity as it effectively makes the frequency of the second qubit dependent on the logical state of the first qubit.
Because of this, the effective dressed frequency of the second qubit cannot be known in general, such that it is not possible to choose the drive frequency $\wdrive$ to be on resonance with the second qubit, irrespective of the state of the first. As a consequence, the CR term $\sz{1}\sx{2}$ in~\cref{eq:gates:H_CR_approx} will rotate at an unknown qubit-state dependent frequency, leading to a gate error.
The $\sz{1}\sz{2}$ term should therefore be made small, which ultimately limits the gate speed. Interestingly, for a pair of qubits with equal and opposite anharmonicity, $\chi_{12}=0$ and this unwanted effect is absent. This cannot be realized with two conventional transmons, but is possible with other types of qubits~\cite{Winik2020,Ku2020}.

Since $J'$ is small, another caveat of the CR gate is that large microwave amplitudes $\varepsilon$ are required for fast gates. For the typical low-anharmonicity of transmon qubits, this can lead to leakages and to effects that are not captured by the second-order perturbative results of \cref{eq:CrossResonanceTLS,eq:gates:H_CR_approx}. More detailed modeling based on the Hamiltonian of \cref{eq:gates:H_CR} suggests that classical crosstalk induced on the second qubit from driving the first qubit can be important and is a source of discrepancy between the simple two-level system model and experiments~\cite{Magesan2018,Tripathi2019a,Ware2019}. Because of these spurious effects, CR gate times have typically been relatively long, of the order of 300 to 400 ns with gate fidelities $\sim$ 94--96\%~\cite{Corcoles2013}. However, with careful calibration and modeling beyond \cref{eq:gates:H_CR_approx}, it has been possible to push gate times down to the 100--200 ns range with error per gates at the percent level~\cite{Sheldon2016}.

Similarlrly to the RIP gate, advantages of the CR gate include the fact that realizing this gate can be realized using the same drive lines that are used for single-qubit gates. Moreover, it works with fixed frequency qubits which often have longer phase coherence times than their flux-tunable counterparts. However, both the RIP and the CR gate are slower than what can now be achieved with additional flux control of the qubit frequency or of the coupler. We also note that, due to the factor $E_{C1}/\hbar\Delta_{12}$ in the amplitude of the $\hat \sigma_{z1}\hat \sigma_{x2}$ term, the detuning of the two qubits cannot be too large compared to the anharmonicity, putting further constraints on the choice of the qubit frequencies. This may lead to frequency crowding issues when working with large numbers of qubits.

\subsubsection{Parametric gates}

As we have already discussed, a challenge in realizing two-qubit gates is activating a coherent interaction between two qubits with a large on/off ratio. The gates discussed so far have aimed to achieve this in different ways. The $\sqrt{i\text{SWAP}}$ and the $11$-$02$ gates are based on flux-tuning qubits into a resonance condition or on a tunable coupling element. The RIP gate is based on activating an effective qubit-qubit coupling by driving a resonator and the CR gate by driving one of the qubits. Another approach is to activate an off-resonant interaction by modulating a qubit frequency, a resonator frequency, or the coupling parameter itself at an appropriate frequency. This parametric modulation provides the energy necessary to bridge the energy gap between the far detuned qubit states. Several such schemes, known as parametric gates, have been theoretically developed and experimentally realized, see for example~\textcite{Bertet2006,Niskanen2006,Liu2007,Niskanen2007,Beaudoin2012a,Strand2013,Kapit2015a,McKay2016,Naik2017,Sirois2015a,Reagor2018,Caldwell2018,Didier2017}.

The key idea behind parametric gates is that modulation of a system parameter can induce transitions between energy levels that would otherwise be too far off-resonance to give any appreciable coupling. We illustrate the idea first with two directly coupled qubits described by the Hamiltonian
\begin{equation}
  \begin{aligned}
    \hH ={}& \frac{\hbar \omega_{q1}}{2}\hat \sigma_{z1} + \frac{\hbar \omega_{q2}}{2}\hat \sigma_{z2}
    + J(t)\hat\sigma_{x1}\hat\sigma_{x2},
  \end{aligned}
\end{equation}
where we assume that the coupling is periodically modulated at the frequency $\wmod$, $J(t) = J_0 + \tilde J\cos(\wmod t)$. Moving to a rotating frame at the qubit frequencies, the above Hamiltonian takes the form 
\begin{equation}
  \begin{aligned}
    \hH' = J(t)\bigg(&
    e^{i(\omega_{q1}-\omega_{q2})t}\hat\sigma_{+1}\hat\sigma_{-2} \\
    &+e^{i(\omega_{q1}+\omega_{q2})t}\hat\sigma_{+1}\hat\sigma_{+2}
    + \hc
    \bigg).
  \end{aligned}
\end{equation}
Just as in \cref{sec:DirectCoupling}, if the coupling is constant $J(t) = J_0$, and $J_0/(\omega_{q1}-\omega_{q2}),\,J_0/(\omega_{q1}+\omega_{q2})\ll1$, then all the terms of $\hH'$ are fast-rotating and can be  neglected. In this situation, the gate is in the off state. On the other hand, by appropriately choosing the modulation frequency $\wmod$, it is possible to selectively activate some of these terms. Indeed, for $\wmod = \omega_{q1}-\omega_{q2}$, the terms $\hat \sigma_{+1}\hat\sigma_{-2} + \hc$ are no longer rotating and are effectively resonant. Dropping the rapidly rotating terms, this leads to 
\begin{equation}
  \begin{aligned}
    \hH' \simeq \frac{\tilde J}{2}\left(
    \hat\sigma_{+1}\hat\sigma_{-2}
    +\hat\sigma_{-1}\hat\sigma_{+2}
    \right).
  \end{aligned}
\end{equation}
As already discussed, this interaction can be used to generate entangling gates such as the $\sqrt{i\text{SWAP}}$. If rather $\wmod = \omega_1+\omega_2$ then $\hat\sigma_{+1}\hat\sigma_{+2} +\hc$ is instead selected. 

In practice, it can sometimes be easier to modulate a qubit or resonator frequency rather than a coupling strength. To see how this leads to a similar result, consider the Hamiltonian
\begin{equation}
  \begin{aligned}
    \hH ={}& \frac{\hbar \omega_{q1}(t)}{2}\hat \sigma_{z1} + \frac{\hbar \omega_{q2}}{2}\hat \sigma_{z2}
    + J\hat\sigma_{x1}\hat\sigma_{x2}.
  \end{aligned}
\end{equation}
Taking $\omega_{q1}(t) = \omega_{q1} + \varepsilon\sin(\wmod t)$, the transition frequency of the first qubit develops frequency modulation (FM) sidebands. The two qubits can then be effectively brought into resonance by choosing the modulation to align one of the FM sidebands with $\omega_{q2}$, thereby rendering the $J$ effectively coupling resonant. This can be seen more clearly by moving to a rotating frame defined by the unitary
\begin{equation}
\hat U = 
e^{
      -\frac{i}{2}\int_0^t dt'\, \omega_{q1}(t')\hat\sigma_{z1}
    }
e^{
    - i\omega_{q2}t \hat\sigma_{z2}/2 
    },
\end{equation}
where the Hamiltonian takes the form \cite{Beaudoin2012a,Strand2013}
\begin{equation}
  \begin{aligned}
    \hH' ={}& J
    \sum_{n=-\infty}^\infty J_n\left(\frac{\varepsilon}{\wmod}\right)
    \bigg(
    i^n e^{i(\Delta_{12} - n\wmod)t}\hat\sigma_{+1}\hat\sigma_{-2} \\
    &+i^n e^{i(\omega_{q1}+\omega_{q2} - n\wmod t)t}\hat\sigma_{+1}\hat\sigma_{+2}
    + \hc
    \bigg).
  \end{aligned}
\end{equation}
To arrive at the above expression, we have used the Jacobi-Anger expansion 
$e^{iz\cos\theta} = \sum_{n=-\infty}^\infty i^n J_n(z) e^{in\theta}$, with $J_n(z)$ Bessel functions of the first kind. Choosing the modulation frequency such that $n\omega_m = \Delta_{12}$ aligns the $n$th sideband with the resonator frequency such that a resonant qubit-resonator interaction is recovered. The largest contribution comes from the first sideband with $J_{1}$ which has a maximum around $J_1(1.84)\simeq 0.58$, thus corresponding to an effective coupling that is a large fraction of the bare $J$ coupling. Note that the assumption of having a simple sinusoidal modulation of the frequency neglects the fact that the qubit frequency has a nonlinear dependence on external flux for tunable transmons. This behavior can still be approximated by appropriately varying $\Phi_x(t)$~\cite{Beaudoin2012a}.

Parametric gates can also be mediated by modulating the frequency of a resonator bus to which qubits are dispersively coupled \cite{McKay2016}. Much as with flux-tunable transmons, the resonator is made tunable by inserting a SQUID loop in the center conductor of the resonator \cite{Sandberg2008a,Castellanos2007}. Changing the flux threading the SQUID loop changes the SQUID's inductance and therefore the effective length of the resonator. As in a trombone, this leads to a change of the resonator frequency. An advantage of modulating the resonator bus over modulating the qubit frequency is that the latter can have a fixed frequency, thus reducing its susceptibility to flux noise.

Finally, it is worth pointing out that while the speed of the cross-resonance gate is reduced when the qubit-qubit detuning is larger than the transmon anharmonicity, parametric gates do not suffer from this problem. As a result, there is more freedom in the choice of the qubit frequencies with parametric gates, which is advantageous to avoid frequency crowding related issues such as addressability errors and crosstalk. We also note that the modulation frequencies required to activate parametric gates can be a few hundred MHz, in contrast to the RIP gate or the CR gate which require microwave drives. Removing the need for additional microwave generators simplifies  the control electronics and \emph{may} help make the process more scalable. 
A counterpoint is that fast parametric gates often require large modulation amplitudes, which can be challenging.

\subsection{Encoding a qubit in an oscillator}
\label{sec:CatCodes}

\begin{table*}[t]\label{Table:AmpDampComparison}
  \begin{tabular}{ ccc }
   &4-qubit code  & Simplest binomial code\\
     \hline
     \hline
   Code word $|0_\mathrm{L}\rangle$ & $\frac{1}{\sqrt{2}}(|0000\rangle+|1111\rangle)$ &$\frac{1}{\sqrt{2}}(|0\rangle+|4\rangle)$  \\
    \hline
    Code word $|1_\mathrm{L}\rangle$& $\frac{1}{\sqrt{2}}(|1100\rangle+|0011\rangle) $ &$|2\rangle$ \\
    \hline
    Mean excitation number $\bar n$& 2 & 2 \\
   \hline
   Hilbert space dimension & $2^4=16$ & $\{0,1,2,3,4\}=5$ \\
   \hline
   Number of correctable errors & $\{\hat I,\sigma_1^-,\sigma_2^-,\sigma_3^-,\sigma_4^-\}=5$ & $\{\hat I,a\}=2$ \\
   \hline
   Stabilizers &$\hat S_1=\hat Z_1\hat Z_2,\, \hat S_2=\hat Z_3\hat Z_4,\, \hat S_3=\hat X_1\hat X_2\hat X_3\hat X_4$ & $\hat P=(-1)^{\hat n}$\\
   \hline
   Number of Stabilizers & 3&1\\
   \hline
   Approximate QEC? & Yes, 1st order in $\gamma t$&Yes, 1st order in $\kappa t$\\
  \end{tabular}
  \caption{Comparison of qubit and bosonic codes for amplitude damping. $\gamma$ and $\kappa$ are respectively the qubit and oscillator energy relaxation rates.}
\end{table*}

So far we have discussed encoding of quantum information into the first two energy levels of an artificial atom, the cavity being used for readout and two-qubit gates. However, cavity modes often have superior coherence properties than superconducting artificial atoms, something that is especially true for the 3D cavities discussed in \cref{sec:3D} \cite{Reagor2016}. This suggests that encoding quantum information in the oscillator mode can be advantageous. Using oscillator modes to store and manipulate quantum information can also be favorable for quantum error correction which is an essential aspect of scalable quantum computer architectures \cite{Nielsen2000}. 

Indeed, in addition to their long coherence time, oscillators have a simple and relatively well-understood error model: to a  large extent, the dominant error is single-photon loss. Taking advantage of this, it is possible to design quantum error correction codes that specifically correct for this most likely error. This is to be contrasted to more standard codes, such as the surface code, which aim at detecting and correcting both amplitude and phase errors~\cite{Fowler2012}. Moreover, as will become clear below, the infinite dimensional Hilbert space of a single oscillator can be exploited to provide the redundancy which is necessary for error correction thereby, in principle, allowing using less physical resources to protect quantum information than when using two-level systems. Finally, qubits encoded in oscillators can be concatenated with conventional error correcting codes, where the latter should be optimized to exploit the noise resilience provided by the oscillator encoding~\cite{Tuckett2018,Tuckett2019b,Tuckett2019,Puri2019,Guillaud2019,Grimsmo2020}.

Of course, as we have already argued, nonlinearity remains essential to prepare and manipulate quantum states of the oscillator. When encoding quantum information in a cavity mode, a dispersively coupled artificial atom (or other Josephson junction-based circuit element) remains present but only to provide nonlinearity to the oscillator ideally without playing much of an active role.

Oscillator encodings of qubits investigated in the context of quantum optics and circuit QED include
cat codes \cite{CochMilbMunr99,Ralph2003,Gilchrist2004,Mirrahimi2014,Ofek2016,Puri2017,Grimm2019,Lescanne2019},
the related binomial codes~\cite{Michael2016a,Hu2019},
and Gottesman-Kitaev-Preskill (GKP) codes~\cite{Gottesman2001b,HOME-GKP2019,Campagne2019}, as well as a two-mode amplitude damping code described in \cite{Chuang97}.

To understand the basic idea behind this approach, we first consider the simplest instance of the binomial code in which a qubit is encoded in the following two states of a resonator mode~\cite{Michael2016a}
\begin{align}\label{eq:catcodes:kitten}
  \ket{0_L} = \frac{1}{\sqrt 2}\left(\ket 0 + \ket 4\right),\qquad \ket{1_L} = \ket 2,
\end{align}
with Fock states $\ket n$. The first aspect to notice is that for both logical states, the average photon number is $\bar n = 2$ and, as a result, the likelihood of a photon loss event is the same for both states. An observer detecting a loss event will therefore not gain any information allowing her to distinguish whether the loss came from $\ket{0_L}$ or from $\ket{1_L}$. This is a necessary condition for a quantum state encoded using the logical states \cref{eq:catcodes:kitten} to not be `deformed' by a photon loss event. Moreover, under the action of $\aop$, the arbitrary superposition $c_0\ket{0_L}+c_1\ket{1_L}$ becomes $c_0\ket{3}+c_1\ket{1}$ after normalization. The coefficients $c_0$ and $c_1$ encoding the quantum information are intact and the original state can in principle be recovered with a unitary transformation. By noting that while the original state only has support on even photon numbers, the state after a photon loss only has support on odd photon numbers, we see that the photon loss event can be detected by measuring photon number parity $\hat P = (-1)^{\hat n}$. The parity operator thus plays the role of a stabilizer for this code~\cite{Nielsen2000,Michael2016}.

This simple encoding should be compared to directly using the Fock states $\{\ket0,\ket1\}$ to store quantum information. Clearly, in this case, a single photon loss on $c_0\ket{0}+c_1\ket{1}$ leads to $\ket{0}$ and the quantum information has been irreversibly lost. Of course, this disadvantage should be contrasted to the fact that the rate at which  photons are lost, which scales with $\bar n$, is (averaged over the code words) four times as large when using the encoding \cref{eq:catcodes:kitten}, compared to using the Fock states $\{\ket0,\ket1\}$. This observation reflects the usual conundrum of quantum error correction: using more resources (here more photons) to protect quantum information actually increases the natural error rate.  The protocol for detecting and correcting errors must be fast enough and accurate enough to counteract this increase. The challenge for experimental implementations of quantum error correction is thus to reach and go beyond the break-even point where the encoded qubit, here \cref{eq:catcodes:kitten}, has a coherence time exceeding the coherence time of the unencoded constituent physical components, here the Fock states $\{\ket0,\ket1\}$. Near break-even performance with the above binomial code has been experimentally reported  by \textcite{Hu2019}.

The simplest binomial code introduced above is able to correct a single amplitude-damping error (photon loss).  Thus if the correction protocol is applied after a time interval $\delta t$, the probability of an uncorrectable error is reduced from ${\mathcal O}(\kappa\,\delta t)$ to ${\mathcal O}((\kappa\,\delta t)^2)$, where $\kappa$ is the cavity energy decay rate.

To better understand the simplicity and efficiency advantages of bosonic QEC codes, it is instructive to do a head-to-head comparison of the simplest binomial code to the simplest qubit code for amplitude damping.  The smallest qubit code able to protect logical information against a general single-qubit error requires five qubits \cite{Bennett1996,Laflamme96,Knill2001}.  However, the specific case of the qubit amplitude-damping channel can be corrected to first order against single-qubit errors using a 4-qubit code \cite{Leung97} that, like the binomial code, satisfies the Knill-Laflamme conditions \cite{Knill1997} to lowest order and whose two logical codewords are
\begin{subequations}\label{eq:4qubitcode}
  \begin{align}
    \ket{0_\mathrm{L}} &= \frac{1}{\sqrt{2}}\left(|0000\rangle+|1111\rangle\right),\\
    \ket{1_\mathrm{L}} &= \frac{1}{\sqrt{2}}\left(|1100\rangle+|0011\rangle\right).
      \end{align}
\end{subequations}

This four-qubit amplitude damping code and the single-mode binomial bosonic code for amplitude damping are compared in Table~\ref{Table:AmpDampComparison}.  Note that, just as in the binomial code, both codewords have mean excitation number equal to two and so are equally likely to suffer an excitation loss. The logical qubit of \cref{eq:4qubitcode} lives in a Hilbert space of dimension $2^4=16$ and has four different physical sites at which the damping error can occur.  Counting the case of no errors, there are a total of five different error states which requires measurement of three distinct error syndromes $\hat Z_1\hat Z_2$, $\hat Z_3\hat Z_4$, and $\hat X_1\hat X_2\hat X_3\hat X_4$ to diagnose (where $\hat P_i$ refers to Pauli operator $\hat P$ acing on qubit $i$). The required weight-two and weight-four operators have to date not been easy to measure in a highly QND manner and with high fidelity, but some progress has been made towards this goal \cite{Chow2014,Chow2015,Corcoles2015,Riste2015,Takita2016}. In contrast, the simple bosonic code in \cref{eq:catcodes:kitten} requires only the lowest five states out of the (formally infinite) oscillator Hilbert space. Moreover, since there is only a single mode, there is only a single error, namely photon loss (or no loss), and it can be detected by measuring a single stabilizer, the photon number parity. It turns out that, unlike in ordinary quantum optics, photon number parity is relatively easy to measure in circuit QED with high fidelity and minimal state demolition \cite{Sun2014,Ofek2016}.  It is for all these reasons that, unlike the four-qubit code, the bosonic code \cref{eq:catcodes:kitten} has already been demonstrated experimentally to (very nearly) reach the break-even point for QEC \cite{Hu2019,LuyanSun2020}. Generalizations of this code to protect against more than a single photon loss event, as well as photon gain and dephasing, are described in \textcite{Michael2016a}.

\begin{figure}
    \centering
    \includegraphics[width=1.0\columnwidth]{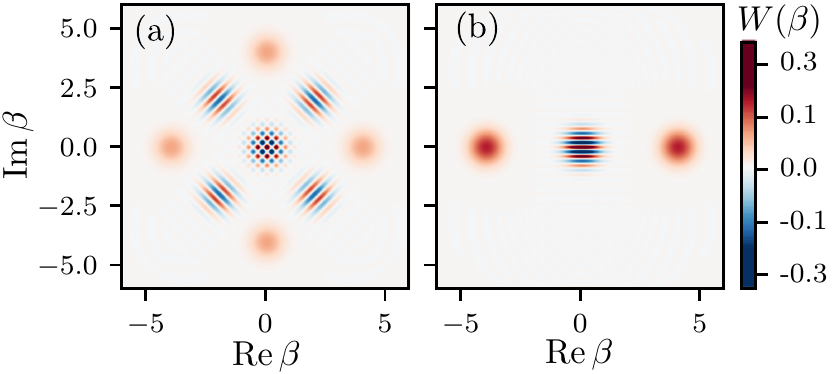}
    \caption{\label{fig:catcodes:fourlegs}Plot of Wigner function $W(\beta)$ obtained numerically for four-component (a) and two-component (b) cat states with $\alpha=4$. Red is positive and blue is negative.}
\end{figure}

Operation slightly exceeding break-even has been reported by \cite{Ofek2016} with cat-state bosonic encoding which we  describe now. In the encoding used in that experiment, each logical code word is a superposition of four coherent states referred to as a four-component cat code~\cite{Mirrahimi2014}:
\begin{subequations}\label{eq:catcodes:fourlegs}
\begin{align}
  \ket{0_L} &= \mathcal N_0\left( \ket{\alpha} + \ket{i\alpha} + \ket{-\alpha} + \ket{-i\alpha} \right),\\
  \ket{1_L} &= \mathcal N_1\left( \ket{\alpha} - \ket{i\alpha} + \ket{-\alpha} - \ket{-i\alpha} \right),
\end{align}
\end{subequations}
where $\mathcal N_i$ are normalization constants, with $\mathcal N_0\simeq \mathcal N_1$ for large $|\alpha|$. The Wigner functions for the $\ket{0_L}$ codeword is shown in~\cref{fig:catcodes:fourlegs}(a) for $\alpha=4$. The relationship between this encoding and the simple code in~\cref{eq:catcodes:kitten} can be seen by writing~\cref{eq:catcodes:fourlegs} using the expression \cref{eq:CoherentState} for $\ket\alpha$ in terms of Fock states. One immediately finds that $\ket{0_L}$ only has support on Fock states $\ket{4n}$ with $n=0,1,\dots$, while $\ket{1_L}$ has support on Fock states $\ket{4n+2}$, again for $n=0,1,\dots$. It follows that the two codewords are mapped onto orthogonal states under the action of $\aop$, just as the binomial code of~\cref{eq:catcodes:kitten}.
Moreover, the average photon number $\bar n$ is approximately equal for the two logical states in the limit of large $|\alpha|$. The protection offered by this encoding is thus similar to that of the binomial code in~\cref{eq:catcodes:kitten}. In fact, these two encodings belong to a larger class of codes characterized by rotation symmetries in phase space~\cite{Grimsmo2020}. 

We end this section by discussing an encoding that is even simpler than~\cref{eq:catcodes:fourlegs}, sometimes referred to as a two-component cat code. In this case, the codewords are defined simply as $\ket{+_L} = \mathcal N_0(\ket{\alpha} + \ket{-\alpha})$ and $\ket{-_L} = \mathcal N_1(\ket{\alpha} - \ket{-\alpha}$~\cite{CochMilbMunr99,Ralph2003,Gilchrist2004,Mirrahimi2014,Puri2017}. The Wigner function for $\ket{+_L}$ is shown in~\cref{fig:catcodes:fourlegs}(b).
The choice to define the above codewords in the logical $\hat X_L$ basis instead of the $\hat Z_L$ basis is, of course, just a convention, but turns out to be convenient for this particular cat code.
In contrast to \cref{eq:catcodes:kitten,eq:catcodes:fourlegs}, these two states are \emph{not} mapped to two orthogonal states under the action of $\aop$.
To understand this encoding, it is useful to consider the logical $\hat Z_L$ basis states in the limit of large $|\alpha|$
\begin{subequations}\label{eq:catcodes:twolegs}
\begin{align}
  \ket{0_L} &= \frac{1}{\sqrt2}(\ket{+_L}+\ket{-_L}) = \ket{\alpha} + \mathcal{O}(e^{-2|\alpha|^2}),\\
  \ket{1_L} &= \frac{1}{\sqrt2}(\ket{+_L}-\ket{-_L}) = \ket{-\alpha} + \mathcal{O}(e^{-2|\alpha|^2}).
\end{align}
\end{subequations}
As is made clear by the second equality, for large enough $|\alpha|$ these logical states are very close to coherent states of the same amplitude but opposite phase. The action of $\aop$ is thus, to a very good approximation, a phase flip since $\aop\ket{0_L/1_L} \sim \pm \ket{0_L/1_L}$.

The advantage of this encoding is that, while photon loss leads to phase flips, the bit-flip rate is exponentially small with $|\alpha|$. This can be immediately understood from the golden rule whose relevant matrix element for bit flips is $\me{1_L}{\aop}{0_L} \sim \me{-\alpha}{\aop}{\alpha} = \alpha e^{-2|\alpha|^2}$.
In other words, if the qubit is encoded in a coherent state with many photons, losing one simply does not do much. This is akin to the redundancy required for quantum error correction. As a result, the bit-flip rate ($1/T_1$) decreases \emph{exponentially} with $|\alpha|^2$ while the phase flip rate increases only \emph{linearly} with $|\alpha|^2$. The crucial point is that the bias between bit and phase flip error rates increases exponentially with $\alpha$, which has been verified experimentally \cite{Grimm2019,Lescanne2019}. While the logical states \cref{eq:catcodes:twolegs} do not allow for recovery from photon-loss errors, the strong asymmetry between different types of errors can be exploited to significantly reduce the qubit overhead necessary for fault-tolerant quantum computation \cite{Puri2019,Guillaud2019}. The basic intuition behind this statement is that the qubit defined by \cref{eq:catcodes:twolegs} can be used in an error correcting code tailored to predominantly correct the most likely error (here, phase flips) rather than devoting ressources to correcting both amplitude and phase errors \cite{Tuckett2018,Tuckett2019b,Tuckett2019}. 

Another bosonic encoding that was recently demonstrated in circuit QED is the Gottesman-Kitaev-Preskill (GKP) code~\cite{Campagne2019}.  This demonstration is the first QEC experiment able to correct all logical errors and it came close to reaching the break-even point. While all the bosonic codes described above are based on codewords that obey rotation symmetry in phase space, the GKP code is instead based on translation symmetry. We will not describe the GKP encoding in more detail here, but refer the reader to the review by \textcite{Terhal2020}.

\section{Quantum optics on a chip}
\label{sec:QuantumOptics}

\begin{figure*}
  \centering
  \includegraphics{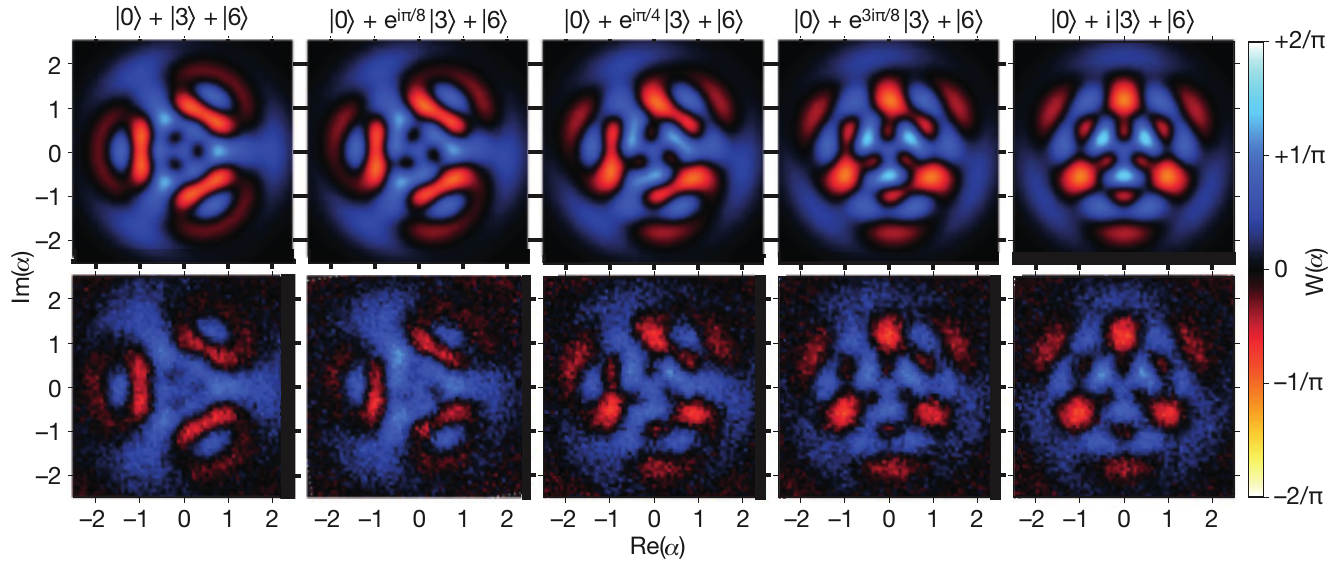}
  \caption{\label{fig:Hofheinz2009}
  Wigner function of the intra-cavity field for Fock state superpositions $\ket{0}+e^{i\varphi}\ket{3}+\ket{6}$ for five values of the phase $\varphi$, see panel titles. Top row: theory, Bottom row: experimental data. Figure adapted from \textcite{Hofheinz2009}.}
\end{figure*}

The strong light-matter interaction realized in circuit QED together with the flexibility allowed in designing and operating superconducting quantum circuits has opened the possibility to explore the rich physics of quantum optics at microwave frequencies in circuits. As discussed previously it has, for example, made possible the clear observation of vacuum Rabi splitting, of photon-number splitting in the strong-dispersive regime, as well as of signatures of ultrastrong light-matter coupling. The new parameter regimes that can be achieved in circuit QED have also made it possible to test some of the theoretical predictions from the early days of quantum optics and to explore new research avenues. A first indication that circuit QED is an ideal playground for these ideas is the strong Kerr nonlinearity relative to the decay rate, $K/\kappa,$ that can readily be achieved in circuits. Indeed, from the point of view of quantum optics, a transmon is a Kerr nonlinear oscillator that is so nonlinear that it exhibits photon blockade. Given the very high Q factors that can be achieved in 3D superconducting cavities, such levels of nonlinearity can also readily be obtained in microwave resonators by using transmons or other Josephson junction-based circuits to induce nonlinearity in electromagnetic modes. 

Many of the links between circuit QED and quantum optics have already been highlighted in this review. In this section, we continue this discussion by presenting some further examples. The reader interested in learning more about quantum optics at microwave frequencies can consult the review article by \textcite{Gu2017b}.

\subsection{Intra-cavity fields}

Because superconducting qubits can rapidly be tuned over a wide frequency range, it is possible to bring them in and out of resonance with a cavity mode on a time scale which is fast with respect to $1/g$, the inverse of the qubit-cavity coupling strength. For all practical purposes, this is equivalent to the thought experiment of moving an atom in and out of the cavity in cavity QED. An experiment by \textcite{Hofheinz2008} took advantage of this possibility to prepare the cavity in Fock states up to $\ket{n=6}$. With the qubit and the cavity in their respective ground states and the two systems largely detuned, their approach is to first $\pi$-pulse the qubit to its excited state. The qubit frequency is then suddenly brought in resonance with the cavity for a time $1/2g$ such as to swap the qubit excitation to a cavity photon as the system evolves under the Jaynes-Cummings Hamiltonian \cref{eq:HJC}. The interaction is then effectively stopped by moving the qubits to its original frequency, after which the cycle is repeated until $n$ excitations have been swapped in this way. Crucially, because the swap frequency between the states $\ket{e,n-1}$ and $\ket{g,n}$ is proportional to $\sqrt{n}$, the time during which qubit and cavity are kept in resonance must be adjusted accordingly at each cycle. The same $\sqrt{n}$ dependence is then used to probe the cavity state using the qubit as a probe~\cite{Hofheinz2008,Brune1996}.

Building on this technique and using a protocol proposed by \textcite{Law96} for cavity QED, the same authors have demonstrated the preparation of arbitrary states of the cavity field and characterized these states by measuring the cavity Wigner function \cite{Hofheinz2009}. \Cref{fig:Hofheinz2009} shows the result of this Wigner tomography for superpositions involving up to six cavity photons (top row: theory, bottom row: data). As noted in \textcite{Hofheinz2008}, a downside of this sequential method is that the preparation time rapidly becomes comparable to the Fock state lifetime, limiting the Fock states which can be reached and the fidelity of the resulting states. 

Taking advantage of the very large $\chi/\kappa$ which can be reached in 3D cavities, an alternative to create such states is to cause qubit transitions conditioned on the Fock state of the cavity. Together with cavity displacements, these photon-number dependent  qubit transitions can be used to prepare arbitrary cavity states \cite{Krastanov2015,Heeres2015}. Combining these ideas with numerical optimal control has allowed \textcite{Heeres2017} to synthesize cavity states with high fidelity such as Fock states up to $\ket{n=6}$ and four-legged cat states. 

The long photon lifetime that is possible in 3D superconducting cavities together with the possibility to realize a single-photon Kerr nonlinearity which overwhelms the cavity decay, $K/\kappa > 1$, has enabled a number of similar experiments such as the observation of collapse and revival of a coherent state in a Kerr medium \cite{Kirchmair2013} and the preparation of cat states with nearly 30 photons \cite{Vlastakis2013}. Another striking example is the experimental encoding of qubits in oscillator states already discussed in \cref{sec:CatCodes}.

\subsection{Quantum-limited amplification}
\label{sec:QO:AmplificationSqueezing}

Driven by the need for fast, high-fidelity single-shot readout of superconducting qubits, superconducting low-noise linear microwave amplifiers are a subject of intense research. There are two broad classes of linear amplifiers. First, phase-preserving amplifiers that amplify equivalently both quadratures of the signal. Quantum mechanics imposes that these amplifiers add a minimum of half a photon of noise to the input signal~\cite{Caves1982,Caves2012a,Clerk2010}. Second, phase-sensitive amplifiers which amplify one quadrature of the signal while squeezing the orthogonal quadrature. This type of amplifier can in principle operate without adding noise~\cite{Caves1982,Clerk2010}. Amplifiers adding the minimum amount of noise allowed by quantum mechanics, phase preserving or not, are referred to as quantum-limited amplifiers. We note that, in practice, phase sensitive amplifiers are useful if the quadrature containing the relevant information is known in advance, a condition that is realized when trying to distinguish between two coherent states in the dispersive qubit readout discussed in \cref{sec:DispersiveQubitReadout}.

While much of the development of near-quantum-limited amplifiers has been motivated by the need to improve qubit readout, Josephson junction based amplifiers have been theoretically investigated~\cite{Yurke1987} and experimentally demonstrated as early as the late 80`s~\cite{Yurke1988,Yurke1989}. These amplifiers have now found applications in a broad range of contexts. In their simplest form, such an amplifier is realized as a driven oscillator mode rendered weakly nonlinear by incorporating a Josephson junction and are generically known as a Josephson parametric amplifier (JPA). 

For weak nonlinearity, the Hamiltonian of a driven nonlinear oscillator is well approximated by 
\begin{equation}\label{eq:DPA1}
H = \omega_0 \ada + \frac{K}{2}\hat{a}^{\dag 2}{\aop}^2	+ \epsilon_\mathrm{p} (\ad e^{-i\omega_\mathrm{p} t}+ \aop e^{i\omega_\mathrm{p} t}),
\end{equation}
where $\omega_0$ is the system frequency, $K$ the negative Kerr nonlinearity, and $\epsilon_\mathrm{p}$ and $\omega_\mathrm{p}$ are the pump amplitude and frequency, respectively. The physics of the JPA is best revealed by applying a displacement transformation $\Dop^\dag(\alpha)\aop \Dop(\alpha) = a + \alpha$ to $H$ with $\alpha$ chosen to cancel the pump term. Doing so leads to the transformed Hamiltonian
\begin{equation}\label{eq:DPA}
	H_\mathrm{JPA} = \delta \ada + \frac{1}{2}\left(\epsilon_2 \hat{a}^{\dag 2} + \epsilon_2^*{\aop}^2\right) + H_\mathrm{corr},
\end{equation}
where $\delta = \omega_0 + 2|\alpha|^2 K - \omega_\mathrm{p}$ is the effective detuning, $\epsilon_2 = \alpha^2 K$, and are $H_\mathrm{corr}$ correction terms which can be dropped for weak enough pump amplitude and Kerr nonlinearity, i.e.~when $\kappa$ is large in comparison to $K$ and thus the drive does not populate the mode enough for higher-order nonlinearity to become important~\cite{Boutin2017b}. The second term, of amplitude $\epsilon_2$, is a two-photon pump which is the generator of quadrature squeezing. Depending on the size of the measurement bandwidth, this leads to phase preserving or sensitive amplification when operating the device close to but under the parametric threshold $\epsilon_2 < \sqrt{\delta^2 + (\kappa/2)^2}$, with $\kappa$ the device's single-photon loss rate \cite{Wustmann2013}. Rather than driving the nonlinear oscillator as in \cref{eq:DPA1}, an alternative approach to arrive at $H_\mathrm{JPA}$ is to replace the junction by a SQUID and to apply a flux modulation at $2\omega_0$~\cite{Yamamoto2008}.

\Cref{eq:DPA} is the Hamiltonian for a parametric amplifier working with a single physical oscillator mode. Using appropriate filtering in the frequency domain, single-mode parametric amplifiers can be operated in a phase-sensitive mode, when detecting the emitted radiation over the full bandwidth of the physical mode, see e.g. \textcite{Eichler2011a}. This is also called the degenerate mode of operation. Alternatively, the same single-oscillator-mode amplifier can be operated in the phase-preserving mode, when separating frequency components above and below the pump in the experiment, e.g.~by using appropriately chosen narrow-band filters, see for example \textcite{Eichler2011a}. Parametric amplifiers with two or multiple physical modes are also frequently put to use \cite{Roy2016c} and can be operated both in the phase-sensitive and phase-preserving modes, e.g.~in degenerate or non-degenerate mode of operation, as for example demonstrated in \textcite{Eichler2014a}.

Important parameters which different approaches for implementing JPAs aim at optimizing include  amplifier gain, bandwidth and dynamic range. The latter refers to the range of power over which the amplifier acts linearly, i.e. powers at which the amplifier output is linearly related to its input. Above a certain input power level, the correction terms in \cref{eq:DPA} resulting from the junction nonlinearity can no longer be ignored and lead to saturation of the gain~\cite{Abdo2012,Kochetov2015,liu2017,Boutin2017b,Planat2019a}. For this reason, while transmon qubits are operated in a regime where the single-photon Kerr nonlinearity is large and overwhelms the decay rate, JPAs are operated in a very different regime with $|K|/\kappa \sim 10^{-2}$ or smaller. 

An approach to increase the dynamic range of  JPAs is to replace the Josephson junction of energy $E_J$ by an array of $M$ junctions, each of energy $ME_J$~\cite{Castellanos2007,Castellanos2008,Eichler2014}. Because the voltage drop is now distributed over the array, the bias on any single junction is $M$ times smaller and therefore the effective Kerr nonlinearity of the device is reduced from $K$ to $K/M^2$. As a result, nonlinear effects kick-in only at increased input signal powers leading to an increased dynamic range. Importantly, this can be done without degrading the amplifier's bandwidth~\cite{Eichler2014}. Typical values are $\sim$ 50 MHz bandwidth with $\sim -117$~dBm saturation power for $\sim$ 20 dB gain~\cite{Planat2019a}. Impedance engineering can be used to improve these numbers further~\cite{Roy2015c}. 

Because the JPA is based on a localized oscillator mode, the product of its gain and bandwidth is approximately constant. Therefore, increase in one must be done at the expense of the other~\cite{Clerk2010,Eichler2014}. As a result, it has proven difficult to design JPAs with enough bandwidth and dynamic range to simultaneously measure more than a few transmons~\cite{Jeffrey2014}. 

To avoid the constant gain-bandwidth product which results from relying on a resonant mode, a drastically different strategy, known as the Josephson traveling-wave parametric amplifier (JTWPA), is to use an open nonlinear medium in which the signal co-propagates with the pump tone. While in a JPA the signal interacts with the nonlinearity for a long time due to the finite Q of the circuit, in the JTWPA the long interaction time is rather a result of the long propagation length of the signal through the nonlinear medium~\cite{O'Brien2014a}. In practice, JTWPA are realized with a metamaterial transmission line whose center conductor is made from thousands of Josephson junctions in series~\cite{Macklin2015}. This device does not have a fixed gain-bandwidth product and has been demonstrated to have 20 dB over as much as 3 GHz bandwidth while operating close to the quantum limit~\cite{Macklin2015,Planat2019b,White2015}. Because every junction in the array can be made only very weakly nonlinear, the JTWPA also offers large enough dynamic range for rapid multiplexed simultaneously readout of multiple qubits~\cite{Heinsoo2018a}. 

\subsection{Propagating fields and their characterization}

\subsubsection{Itinerant single and multi-photon states}

In addition to using qubits to prepare and characterize quantum states of intra-cavity fields, it is also possible to take advantage of the strong nonlinearity provided by a qubit to prepare states of propagating fields at the output of a cavity. This can be done, for example, in a cavity with relatively large decay rate $\kappa$ by tuning a qubit into and out of resonance with the cavity~\cite{Bozyigit2011} or by applying appropriatly chosen drive fields~\cite{Houck2007}. Alternatively, it is also possible to change the cavity decay rate in time to create single-photon states~\cite{Sete2013,Yin2013b}.

The first on-chip single-photon source in the microwave regime was realized with a dispersively coupled qubit engineered such that the Purcell decay rate $\gamma_\kappa$ dominates the qubit's intrinsic non-radiative decay rate $\gamma_1$~\cite{Houck2007}. In this situation, exciting the qubit leads to rapid qubit decay by photon emission. In the absence of single-photon detectors working at microwave frequencies, the presence of a photon was observed by using a nonlinear circuit element (a diode) whose output signal is proportional to the square of the electric field, $\propto (\ad+\aop)^2$, and therefore indicative of the average photon number, $\av{\ada}$, in repeated measurements. 

Rather than relying on direct power measurements, techniques have also been developed to reconstruct arbitrary correlation functions of the cavity output field from the measurement records of the field qudratures \cite{daSilva2010,Menzel2010}. These approaches rely on multiple detection channels with uncorrelated noise to quantify and subtract from the data the noise introduced by the measurement chain. In this way, it is possible to extract, for example, first- and second-order coherence functions of the microwave field. Remarkably, with enough averaging, this approach does not require quantum-limited amplifiers, although the number of required measurement runs is drastically reduced when such amplifiers are used when compared to HEMT amplifiers.

This approach was used to measure second-order coherence functions, $G^2(t,t+\tau) = \av{\ad(t)\ad(t+\tau)\aop(t+\tau)\aop(t)}$, in the first demonstration of antibunching of a pulsed single microwave-frequency photon source \cite{Bozyigit2011}. The same technique also enabled the observation of resonant photon blockade at microwave frequencies \cite{Lang2011} and, using two  single photon sources at the input of a microwave beam splitter, the indistinguishability of microwave photons was demonstrated in a Hong-Ou-Mandel correlation function measurement \cite{Lang2013}. Moreover, a similar approach was used to characterize the blackbody radiation emitted by a 50 $\Omega$ load resistor~\cite{Mariantoni2010}.

Building on these results, it is also possible to reconstruct the quantum state of itinerant microwave fields from measurement of the fields moments. This technique relies on interleaving two types of measurements: measurements on the state of interest and ones in which the field is left in the vacuum as a reference to subtract away the measurement chain noise~\cite{Eichler2011}. In this way, the Wigner function of arbitrary superpositions of vacuum and one-photon Fock states have been reconstructed~\cite{Eichler2011,Kono2018}. This technique was extended to propagating modes containing multiple photons~\cite{Eichler2012}. Similarly, entanglement between a (stationary) qubit and a propagating mode was quantified in this approach with joint state tomography~\cite{Eichler2012,Eichler2012b}. Quadrature-histogram analysis also enabled, for example, the measurement of correlations between radiation fields \cite{Flurin2015}, and the observation of entanglement of itinerant photon pairs in waveguide QED \cite{Kannan2020}.

\subsubsection{Squeezed microwave fields}
\label{sec:Squeezing}

\begin{figure}
  \centering\includegraphics[width=0.9\columnwidth]{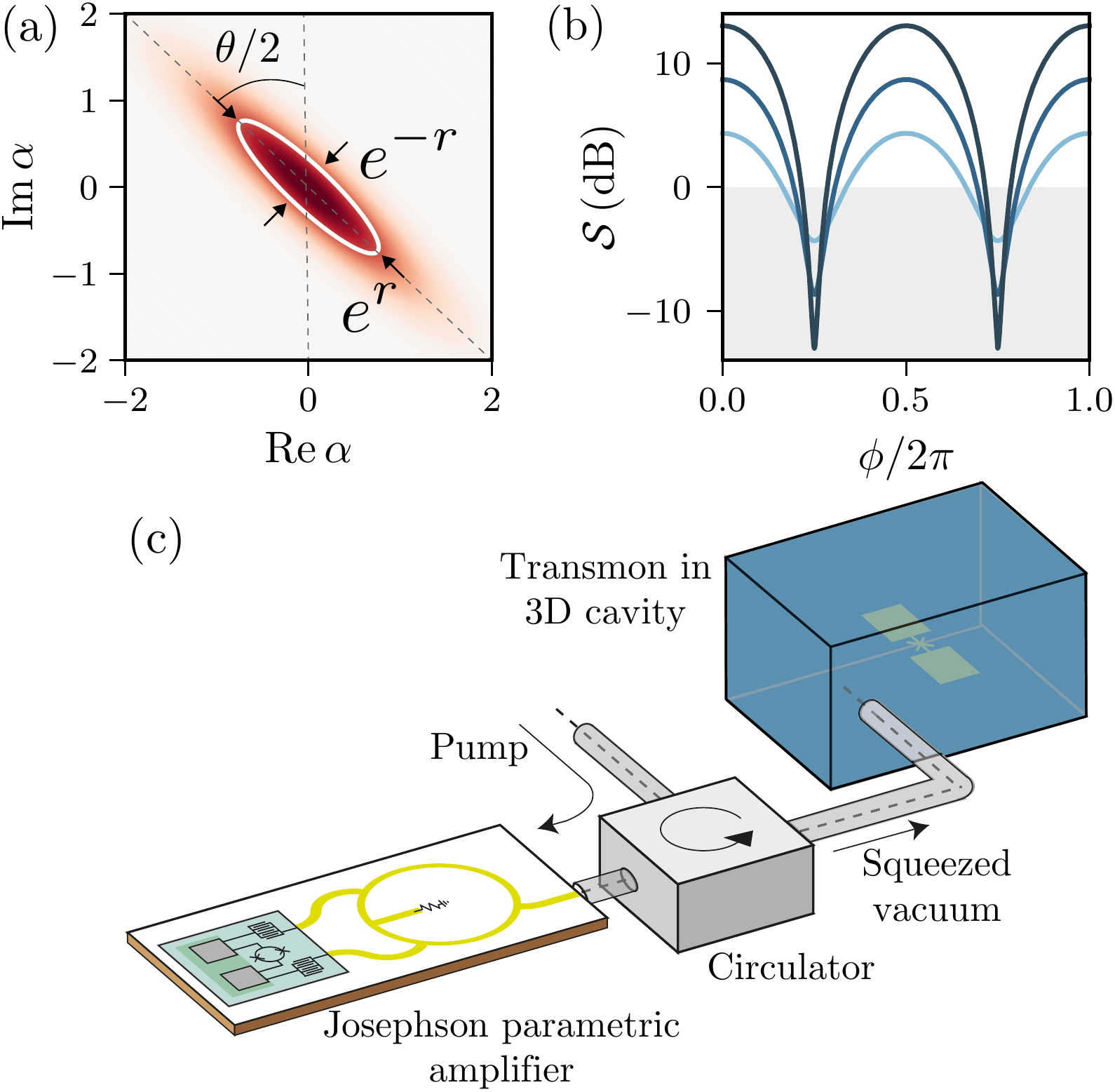}
  \caption{\label{fig:squeezing}
  (a) Wigner function of a squeezed vacuum state $S(\zeta)\ket0$ with $r=0.75$ and $\theta = \pi/2$. The white countour line is an ellipse of semi-minor axis $e^{-r}/2$ and semi-major axis $e^{r}/2$.  (b) Squeezing level versus $\phi$ for $r=$ 0.5, 1.0, 1.5 and $\theta = \pi$. The horizontal line corresponds to vacuum noise $\Delta X^2_\mathrm{vac} = 1/2$. (c) Experimental setup used by \textcite{Murch2013a} to prepare a squeezed vacuum state using a Josephson parametric amplifier and to send, via a circulator (gray box), this state to a transmon qubit in a 3D cavity (blue box). Panel c) is adapted from \textcite{Murch2013a}.}
\end{figure}

Operated in the phase-sensitive mode, quantum-limited amplifiers are sources of squeezed radiation. Indeed, for $\delta = 0$ and ignoring the correction terms, the JPA Hamiltonian of \cref{eq:DPA} is the generator of the squeezing transformation
\begin{equation}
	S(\zeta) = e^{\frac{1}{2} \zeta^* \aop^2 - \frac{1}{2} \zeta \hat{a}^{\dag 2}},
\end{equation}
which takes vacuum to squeezed vacuum, $\ket\zeta = S(\zeta)\ket0$. In this expression, $\zeta  = r e^{i\theta}$ with $r$ the squeezing parameter and $\theta$ the squeezing angle. As illustrated in \cref{fig:squeezing}(a), the action of $S(\zeta)$ on vacuum is to `squeeze' one quadrature of the field at the expense of `anti-squeezing' the orthogonal quadrature while leaving the total area in phase space unchanged. As a result, squeezed states, like coherent states, saturate the Heisenberg inequality. 

This can be seen more clearly from the variance of the quadrature operator $\hX_\phi$ which takes the form
\begin{equation}
	\Delta X_\phi^2 = \frac{1}{2}\left(e^{2r}\sin^2\tilde\phi + e^{-2r}\cos^2\tilde\phi\right),
\end{equation}
where we have defined $\tilde\phi = \phi - \theta/2$. In experiments, the squeezing level is often reported in dB computed using the expression
\begin{equation}
  \mathcal{S} = 10 \log_{10} \frac{\Delta X^2_\phi}{\Delta X^2_\mathrm{vac}}.
\end{equation}
\Cref{fig:squeezing}(b) shows this quantity as a function of $\phi$. It reaches its minimal value $e^{- 2r}/2$ at $\phi =[\theta + (2n+1)\pi]/2$ where the variance $\Delta X^2_\phi$ dips below the vacuum noise level $\Delta X^2_\mathrm{vac} = 1/2$ (horizontal line).

Squeezing in Josephson devices was observed already in the late 80's~\cite{Yurke1988,Yurke1989,Movshovich1990}, experiments that have been revisited with the development of near quantum-limited amplifiers~\cite{Castellanos2008,Zhong2013}. Quantum state tomography of an itinerant squeezed state at the output of a JPA was reported by \textcite{Mallet2011}. There, homodyne detection with different LO phases on multiple preparations of the same squeezed state, together with maximum likelihood techniques, was used to reconstruct the Wigner function of the propagating microwave field. Moreover, the photon number distribution of a squeezed field was measured using a qubit in the strong dispersive regime~\cite{Kono2017}. As is clear from the form of the squeezing transformation $S(\zeta)$, squeezed vacuum is composed of a superposition of only even photon numbers \cite{Schleich1987}, something which \textcite{Kono2017} confirmed in experiments. 

Thanks to the new parameter regimes that can be achieved in circuit QED, it is possible to experimentally test some long-standing theoretical predictions of quantum optics involving squeezed radiation. For example, in the mid-80's theorists predicted how dephasing and resonance fluorescence of an atom would be modified in the presence of squeezed radiation~\cite{Gardiner1986,Carmichael1987}. Experimentally testing these ideas in the context of traditional quantum optics with atomic systems, however, represents a formidable challenge~\cite{Turchette1998,Carmichael2016}. The situation is different in circuits where squeezed radiation can easily be guided from the source of squeezing to the qubit playing the role of artificial atom. Moreover, the reduced dimensionality in circuits compared to free-space atomic experiments limits the number of modes that are involved, such that the artificial atom can be engineered so as to preferentially interact with a squeezed radiation field.

Taking advantage of the possibilities offered by circuit QED, \textcite{Murch2013} confirmed the prediction that squeezed radiation can inhibit phase decay of an (artifical) atom~\cite{Gardiner1986}. In this experiment, the role of the two-level atom was played by the hybridized cavity-qubit state $\{\ket{\overline{g,0}},\ket{\overline{e,0}}\}$. Moreover, squeezing was produced by a JPA over a bandwidth much larger than the natural linewitdh of the two-level system, see \cref{fig:squeezing}(c). According to theory, quantum noise below the vacuum level along the squeezed quadrature leads to a reduction of dephasing. Conversely, along the anti-squeezed quadrature, the enhanced fluctuations lead to increased dephasing. For the artificial atom, this results in different time scales for dephasing along orthogonal axis of the Bloch sphere. In the experiment, phase decay inhibition along the squeezed quadrature was such that the associated dephasing time increased beyond the usual vacuum limit of $2T_1$. By measuring the dynamics of the two-level atom, it was moreover possible to reconstruct the Wigner distribution of the itinerant squeezed state produced by the JPA. Using a similar setup, \textcite{Toyli2016a} studied resonance fluorescence in the presence of squeezed vacuum and found excellent agreement with theoretical predictions~\cite{Carmichael1987}. In this way, it was possible to infer the level of squeezing (3.1 dB below vacuum) at the input of the cavity.

The discussion has so far been limited to squeezing of a single mode. It is also possible to squeeze a pair of modes, which is often  referred to as two-mode squeezing. Labeling the modes as $\aop_1$ and $\aop_2$, the corresponding squeezing transformation reads
\begin{equation}
	S_{12}(\zeta) = e^{\frac{1}{2} \zeta^* \aop_1\aop_2 - \frac{1}{2} \zeta \ad_1\ad_2}.
\end{equation}
Acting on vacuum, $S_{12}$ generates a two-mode squeezed state which is an entangled state of modes $\aop_1$ and $\aop_2$. As a result, in isolation, the state of one of the two entangled modes appears to be in a thermal state where the role of the Boltzmann factor $\exp(-\beta\hbar\omega_i)$, with $\omega_{i=1,2}$ the mode frequency, is played by $\tanh^2r$~\cite{Barnett2002}. 
In this case, correlations and therefore squeezing is revealed when considering joint quadratures of the form $\hX_1 \pm \hX_2$ and $\hP_1 \pm \hP_2$, rather than the quadratures of a single mode as in~\cref{fig:squeezing}(a).
In Josephson-based devices, two-mode squeezing can be produced using nondegenerate parametric amplifiers of different types~\cite{Roy2016c}. Over 12 dB of squeezing below vacuum level between modes of different frequencies, often referred to as signal and idler in this context, has been reported~\cite{Eichler2014a}. Other experiments have demonstrated two-mode squeezing in two different spatial modes, i.e.~entangled signals propagating along different transmission lines~\cite{Bergeal2012,Flurin2012}.

\subsection{Remote Entanglement Generation}

Several approaches to entangle nearby qubits have been discussed in \cref{sec:quantumcomputing}. In some instances it can, however, be useful to prepare entangled states of qubits separated by larger distances. Together with protocols such as quantum teleportation, entanglement between distant quantum nodes can be the basis of a `quantum internet' \cite{Kimble2008}. Because optical photons can travel for relatively long distances in room temperature optical fiber while preserving their quantum coherence, this vision appears easier to realize at optical than at microwave frequencies. Nevertheless, given that superconducting cables at millikelvin temperatures have similar losses per meter as optical fibers \cite{Kurpiers2017}, there is no reason to believe that complex networks of superconductor-based quantum nodes cannot be realized. One application of this type of network is a modular quantum computer architecture where the nodes are relatively small-scale error-corrected quantum computers connected by quantum links \cite{Monroe2014,Chou2018a}.

One approach to entangle qubits fabricated in distant cavities relies on entanglement by measurement, which is easy to understand in the  case of two qubits coupled to the same cavity. Assuming the qubits to have the same dispersive shift $\chi$ due to coupling to the cavity, the dispersive Hamiltonian in a doubly rotating frame takes the form 
\begin{equation}
	H = \chi (\sz{1}+\sz{2})\ada.
\end{equation}
Crucially, the cavity pull associated with odd-parity states $\{\ket{01},\ket{10}\}$ is zero while it is $\pm2\chi$ for the even-parity states $\{\ket{00},\ket{11}\}$. As a result, for $\chi \gg \kappa$, a tone at the bare cavity frequency leads to a large cavity field displacement for the even-parity subspace. On the other hand, the displacement is small or negligible for the odd-parity subspace. Starting with a uniform unentangled superposition of the states of the qubits, homodyne detection therefore stochastically collapses the system to one of these subspaces thereby preparing an entangled state of the two qubits \cite{Lalumiere2010}, an idea that was realized experimentally \cite{Riste2013}. 

The same concept was used by \textcite{Roch2014} to entangle two transmon qubits coupled to two 3D cavities separated by more than a meter of coaxial cable. There, the measurement tone transmitted through the first cavity is sent to the second cavity, only after which it is measured by homodyne detection. In this experiment, losses between the two cavities -- mainly due to the presence of a circulator preventing any reflection from the second cavity back to the first cavity -- as well as finite detection efficiency was the main limit to the acheivable concurrence, a measure of entanglement, to 0.35.

While the above protocol probabilistically entangles a pair of qubits, a more powerful but also more experimentally challenging approach allows, in principle, to realize this in a fully deterministic fashion \cite{Cirac1997}. Developed in the context of cavity QED, this scheme relies on mapping the state of an atom strongly coupled to a cavity to a propagating photon. By choosing its wave packet to be time-symmetric, the photon is absorbed with unit probability by a second cavity also containing an atom. In this way, it is possible to exchange a quantum state between the two cavities. Importantly, this protocol relies on having a unidirectional channel between the cavities such that no signal can propagate from the second to the first cavity. At microwave frequencies, this is achieved by inserting a circulator between the cavities. By first entangling the emitter qubit to a partner qubit located in the same cavity, the quantum-state transfer protocol can be used to entangle the two nodes.

Variations on this more direct approach to entangle remote nodes have been implemented in circuit QED \cite{Kurpiers2018,Campagne-Ibarcq2018,Axline2018}. All three experiments rely on producing time-symmetric propagating photons by using the interaction between a transmon qubit and cavity mode. Multiple approaches to shape and catch propagating photons have been developed in circuit QED. For example, \textcite{Wenner2014} used a transmission-line resonator with a tunable input port to catch a shaped microwave pulse with over 99\% probability. Time-reversal-symmetric photons have  been created by \textcite{Srinivasan2014} using 3-island transmon qubits \cite{Srinivasan2011,Gambetta2011} in which the coupling to a microwave resonator is controlled in time so as to shape the mode function of spontaneously emitted photons. In a similar fashion, shaped single photons can be generated by modulating the boundary condition of a semi-infinite transmission line using a SQUID \cite{Forn-Diaz2017} which effectively controls the spontaneous emission rate of a qubit coupled to the line and emitting the photon.

Alternatively, the remote entanglement generation experiment of \textcite{Kurpiers2018} rather relies on a microwave-induced amplitude- and phase-tunable coupling between the qubit-resonator $\ket{f0}$ and $\ket{g1}$ states, akin to the fg-ge gate already mentioned in \cref{sec:AllMicrowaveGates} \cite{Zeytinoglu2015}. Exciting the qubit to its $\ket{f}$ state followed by a $\pi$-pulse on the $f0-g1$ transition transfers the qubit excitation to a single resonator photon which is emitted as a propagating photon.
This single-photon wave packet can be shaped to be time-symmetric by tailoring the envelope of the $f0-g1$ pulse \cite{Pechal2014}. By inducing the reverse process with a time-reversed pulse on a second resonator also containing a transmon, the itinerant photon is absorbed by this second transmon. In this way, an arbitrary quantum state can be transferred with a probability of 98.1\% between the two cavities separated by 0.9 m of coaxial line bisected by a circulator \cite{Kurpiers2018}. By rather preparing the emitter qubit in a $(\ket{e}+\ket{f})/\sqrt{2}$ superposition, the same protocol deterministically prepares an entangled state of the two transmons with a fidelity of 78.9\% at a rate of 50 kHz \cite{Kurpiers2018}. The experiments of \textcite{Axline2018} and \textcite{Campagne-Ibarcq2018} reported similar Bell-state fidelities using different approaches to prepare time-symmetric propagating photons \cite{Pfaff2017}. The fidelity reported by the three experiments suffered from the presence of a circulator bisecting the nearly one meter-long coaxial cable separating the two nodes.
Replacing the lossy commercial circulator by an on-chip quantum-limited version could improve the fidelity \cite{Chapman2017,Kamal2011,Metelmann2015}. By taking advantage of the multimode nature of a meter long transmission line, it was also possible to deterministically entangle remote qubits without the need of a circulator. In this way, a bidirectional communication channel between the nodes is established and deterministic Bell pair production with 79.3\% fidelity has been reported \cite{Leung2019}.

\subsection{\label{sec:WaveguideQED}Waveguide QED}

The bulk of this review is concerned with the strong coupling of artificial atoms to the confined electromagnetic field of a cavity. Strong light-matter coupling is also possible in free space with an atom or large dipole-moment molecule by tightly confining an optical field in the vicinity of the atom or molecule \cite{Schuller2010}. A signature of strong coupling in this setting is the extinction of the transmitted light by the single atom or molecule acting as a scatterer. This extinction results from destructive interference of the light beam with the collinearly emitted radiation from the scatterer. Ideally, this results in 100\% reflection. In practice, because the scatterer emits in all directions, there is poor mode matching with the focused beam and reflection of $\sim 10$\% is observed with a single atom \cite{Tey2008} and $\sim 30\%$ with a single molecule \cite{Maser2016}.

Mode matching can, however, be made to be close to ideal with electromagnetic fields in 1D superconducting transmission lines and superconducting artificial atoms where the artificial atoms can be engineered to essentially only emit in the forward and backward directions along the line \cite{Shen2005}. In the first realization of this idea in superconducting quantum circuits, \textcite{Astafiev2010} observed extinction of the transmitted signal by as much as 94\% by coupling a single flux qubit to a superconducting transmission line. Experiments with a transmon qubit have seen extinction as large as 99.6\% \cite{Hoi2011}. Pure dephasing and non-radiative decay into other modes than the transmission line are the cause of the small departure from ideal behavior in these experiments. Nevertheless, the large observed extinction is a clear signature that radiative decay in the transmission line $\gamma_\mathrm{r}$ (i.e.~Purcell decay) overwhelms non-radiative decay $\gamma_\mathrm{nr}$. In short, in this cavity-free system referred to as waveguide QED, $\gamma_\mathrm{r}/\gamma_\mathrm{nr} \gg 1$ is the appropriate definition of strong coupling and is associated with a clear experimental signature: the extinction of transmission by a single scatterer.

Despite its apparent simplicity, waveguide QED is a rich toolbox with which a number of physical phenomena have been investigated \cite{Roy2017a}. This includes Autler-Townes splitting \cite{Abdumalikov2010}, single-photon routing \cite{Hoi2011}, the generation of propagating nonclassical microwave states \cite{Hoi2012b}, as well as large cross-Kerr phase shifts at the single-photon level \cite{Hoi2013b}.

In another experiment, \textcite{Hoi2015} studied the radiative decay of an artificial atom placed in front of a mirror, here formed by a short to ground of the waveguide's center conductor. In the presence of a weak drive field applied to the waveguide, the atom relaxes by  emitting a photon in both directions of the waveguide. The radiation emitted towards the mirror, assumed here to be on the left of the atom, is reflected back to interact again with the atom after having acquired a phase shift $\theta = 2\times 2\pi l/\lambda + \pi$, where $l$ is the atom-mirror distance and $\lambda$ the wavelength of the emitted radiation. The additional phase factor of $\pi$ accounts for the hard reflection at the mirror. Taking into account the resulting multiple round trips, this modifies the atomic radiative decay rate which takes the form $\gamma(\theta) = 2\gamma_\mathrm{r} \cos^2(\theta/2)$ \cite{Hoi2015,Koshino2012,Glaetzle2010a}. 

For $l/\lambda = 1/2$, the radiative decay rate vanishes corresponding to destructive interference of the right-moving field and the left-moving field after multiple reflections on the mirror. In contrast, for $l/\lambda = 1/4$, these fields interfere constructively leading to enhanced radiative relaxation with $\gamma(\theta) = 2\gamma_\mathrm{r}$. The ratio $l/\lambda$ can be modified by shorting the waveguide's center conductor with a SQUID. In this case, the flux threading the SQUID can be used to change the boundary condition seen by the qubit, effectively changing the distance $l$ \cite{Sandberg2008a}. The experiment of  \textcite{Hoi2015} rather relied on flux-tuning of the qubit transition frequency, thereby changing $\lambda$. In this way, a modulation of the qubit decay rate by a factor close to 10 was observed. A similar experiment has been reported with a trapped ion in front of a movable mirror \cite{Eschner2001}.

Engineering vacuum fluctuations in this system has been pushed even further by creating microwave photonic bandgaps in waveguides to which transmon qubits are coupled \cite{Liu2016a,Mirhosseini2018a}. For example, \textcite{Mirhosseini2018a} have coupled a transmon qubit to a metamaterial formed by periodically loading the waveguide with lumped-element microwave resonators. By tuning the transmon frequency in the band gap where there is zero or only little density of states to accept photons emitted by the qubit, an increase by a factor of 24 of the qubit lifetime was observed.

An interpretation of the `atom in front of a mirror' experiments is that the atom interacts with its mirror image. Rather than using a boundary condition (i.e. a mirror) to study the resulting constructive and destructing interferences and change in the radiative decay rate, it is also possible to couple a second atom to the same waveguide \cite{Lalumiere2013,vanLoo2013}. In this case, photons (real or virtual) emitted by one atom can be absorbed by the second atom leading to interactions between the atoms separated by a distance $2l$. Similar to the case of a single atom in front of a mirror, when the separation between the atoms is such that $2l/\lambda = 1/2$, correlated decay of the pair of atoms at the enhanced rate $2\gamma_1$ is expected \cite{Lalumiere2013,Chang2012a} and experimentally observed \cite{vanLoo2013}. On the other hand, at a separation of $2l/\lambda = 3/4$, correlated decay is replaced by coherent energy exchange between the two atoms mediated by virtual photons \cite{Lalumiere2013,Chang2012a,vanLoo2013}. We note that the experiments of \textcite{vanLoo2013} with transmon qubits agree with a Markovian model of the interaction of the qubits with the waveguide \cite{Lalumiere2013,Chang2012a,Lehmberg1970}. Deviations from these predictions are expected as the distance between the atoms increases \cite{Zheng2013a}.

Finally, following a proposal by \textcite{Chang2012a}, an experiment by \textcite{Mirhosseini2019} used a pair of transmon qubits to act as an effective cavity for a third transmon qubit, all qubits being coupled to the same waveguide. In this way, vacuum Rabi oscillations between the dark state of the effective cavity and the qubit playing the role of atom were observed, confirming that the strong-coupling regime of cavity QED was achieved.

\subsection{Single microwave photon detection}
\label{sec:SinglePhotonDetector}

The development of single-photon detectors at infrared, optical and ultraviolet frequencies has been crucial to the field of quantum optics and in fundamental tests of quantum physics \cite{Hadfield2009,Eisaman2011}. High-efficiency photon detectors are, for example, one of the elements that allowed the loophole-free violation of Bell's inequality \cite{Hanson2015,Shalm2015a,Giustina2015a}. Because microwave photons have orders of magnitude less energy than infrared, optical or ultraviolet photons, the realization of a photon detector at microwave frequencies is more challenging. Yet, photon detectors in that frequency range would find a number of applications, including in quantum information processing \cite{Kimble2008,Narla2016}, for quantum radars \cite{Barzanjeh2015,Chang2019,Barzanjeh2019}, and for the detection of dark matter axions \cite{Lamoreaux2013}.

Non-destructive counting of microwave photons localized in a cavity has already been demonstrated experimentally by using an (artificial) atom as a probe in the strong dispersive regime \cite{Gleyzes2007,Schuster2007a}. Similar measurements have also been done using a transmon qubit mediating interactions between two cavities, one containing the photons to be measured and a second acting as a probe \cite{Johnson2010}. The detection of itinerant microwave photons remains, however, more challenging. A number of theoretical proposals have appeared \cite{Helmer2009b,Romero2009,Wong2017,Kyriienko2016,Koshino2013b,Koshino2016a,Sathyamoorthy2014,Fan2014a,Leppakangas2018,Royer2018}. One common challenge for these approaches based on absorbing itinerant photons in a localized mode before detecting them can be linked to the quantum Zeno effect. Indeed, continuous monitoring of the probe mode will prevent the photon from being absorbed in the first place. Approaches to mitigate this problem have been introduced, including using an engineered, impedance matched $\Lambda$-system used to deterministically capture the incoming photon \cite{Koshino2016a}, and using the bright and dark states of an ensemble of absorbers \cite{Royer2018}.

Despite these challenges, first itinerant microwave photon detectors have been achieved in the laboratory \cite{Narla2016,Chen2011a,Oelsner2017,Inomata2016}, in some cases achieving photon detection without destroying the photon in the process~\cite{Kono2018,Besse2017,Lescanne2019a}. Notably, a microwave photon counter was used to measure a superconducting qubit with a fidelity of 92\% without using a linear amplifier between the source and the detector \cite{Opremcak2018a}. Despite these advances, the realization of a high-efficiency, large-bandwith, QND single microwave photon detector remains a challenge for the field.

\section{Outlook}

Fifteen years after its introduction~\cite{Blais2004,Wallraff2004}, circuit QED is a leading architecture for quantum computing and an exceptional platform to explore the rich physics of quantum optics in new parameter regimes. Circuit QED has, moreover, found applications in numerous other fields of research as discussed in the body of the review and in the following. In closing this review, we turn to some of these recent developments. 

Although there remain formidable challenges before large-scale quantum computation becomes a reality, the increasing number of qubits that can be wired up, as well as the improvements in coherence time and gate fidelity, suggests that it will eventually be possible to perform computations on circuit QED-based quantum processors that are out of reach of current classical computers. Quantum supremacy on a 53-qubit device has already been claimed \cite{Arute2019}, albeit on a problem of no immediate practical interest. There is, however, much effort deployed in finding useful computational tasks which can be performed on Noisy Intermediate-Scale Quantum (NISQ) devices \cite{Preskill2018}. First steps in this direction include the determination of  molecular energies with variational quantum eigensolvers \cite{OMalley2016,Kandala2017,Colless2018} or boson sampling approachs \cite{Wang2019a} and machine learning with quantum-enhanced features \cite{Havlicek2019}.

Engineered circuit QED-based devices also present an exciting avenue toward performing analog quantum simulations. In contrast to quantum computing architectures, quantum simulators are usually tailored to explore a single specific problem. An example are arrays of resonators which are capacitively coupled to allow photons to hop from resonator to resonator. Taking advantage of the flexibility of superconducting quantum circuits, it is possible to create exotic networks of resonators such as lattices in an effective hyperbolic space with constant negative curvature \cite{Kollar2019}. Coupling qubits to each resonator realizes a Jaynes-Cummings lattice which exhibits a quantum phase transition similar to the superfluid-Mott insulator transition in  Bose–Hubbard lattices \cite{Houck2012}. Moreover, the nonlinearity provided by capactitively coupled qubits, or of Josephson junctions embedded in the center conductor of the resonators, creates photon-photon interactions. This leads to effects such as photon blockade bearing some similarities to Coulomb blockade in mesoscopic systems \cite{Schmidt2013}. Few resonator- and qubit-devices are also promising for analog quantum simulations. Examples are the exploration of a simple model of the light harvesting process in photosynthetic complexes in a circuit QED device under the influence of both coherent and incoherent drives \cite{Potocnik2018}, and the analog simulation of dissipatively stabilized strongly correlated quantum matter in a small photon Bose–Hubbard lattice \cite{Ma2019}.

Because it is a versatile platform to interface quantum devices with transition frequencies in the microwave domain to photons stored in superconducting resonators at similar frequencies, the ideas of circuit QED are also now used to couple to a wide variety of physical systems. An example of such hybrid quantum systems are semiconducter-based double quantum dots coupled to superconducting microwave resonators. Here, the position of an electron in a double dot leads to a dipole moment to which the resonator electric field couples. First experiments with gate-defined double quantum dots in nanotubes \cite{Delbecq2011}, GaAs \cite{Frey2012,Toida2013,Wallraff2013}, and InAs nanowires \cite{Petersson2012a} have demonstrated dispersive coupling and its use for characterizing charge states of quantum dots \cite{Burkard2020}. These first experiments were, however, limited by the very large dephasing rate of the quantum dot's charge states, but subsequent experiments have been able to reach the strong coupling regime \cite{Mi2017,Stockklauser2017,Bruhat2018}. Building on these results and by engineering an effective spin-orbit interaction \cite{PioroLadriere2008,Beaudoin2016a}, it has been possible to reach the strong coupling regime with single spins \cite{Mi2018,Samkharadze2018,Landig2018}.

When the coupling to a single spin cannot be made large enough to reach the strong coupling regime, it can be possible to rely on an ensemble of spins to boost the effective coupling. Indeed, in the presence of an ensemble of $N$ emitters, the coupling strength to the ensemble is enhanced by $\sqrt N$ \cite{Imamoglu2009,Fink2009}, such that for large enough $g\sqrt N$ the strong coupling regime can be reached. First realization of these ideas used ensembles of $\sim 10^{12}$ spins to bring the coupling from a few Hz to $\sim 10$ MHz with NV centers in diamond \cite{Kubo2010} and Cr$^{3+}$ spins in ruby \cite{Schuster2010a}. One objective of these explorations is to increase the sensitivity of electron paramagnetic resonance (EPR) or electron spin resonance (ESR) spectroscopy for spin detection with the ultimate goal of achieving the single-spin limit. A challenge in reaching this goal is the long lifetime of single spins in these systems which limits the repetition rate of the experiment. By engineering the coupling between the spins and an LC oscillator fabricated in close proximity, it has been possible to take advantage of the Purcell effect to reduce the relaxation time from $10^3$s to $1$s \cite{Bienfait2016a}. This faster time scale allows for faster repetition rates thereby boosting the sensitivity, which could lead to spin sensitivities on the order of $0.1~\mathrm{spin}/\sqrt{\mathrm{Hz}}$ \cite{Haikka2017}.

Mechanical systems operated in the quantum regime also benefited from the ideas of circuit QED  \cite{Aspelmeyer2013}. An example is a suspended aluminium membrane that plays the role of a vacuum gap capacitor in a microwave LC oscillator. The frequency of this oscillator depends on the separation between the plates of the capacitor leading to a coupling between the oscillator and the flexural mode of the membrane. Strong coupling between mechanical motion and the LC oscillator has been demonstrated \cite{Teufel2011a}, which allowed to sideband cool the motion of the mechanical oscillator to phonon occupation number as small as $n_\mathrm{phonon}\sim0.34$. \cite{Teufel2011}. Squeezed radiation generated by a Josephson parametric amplifier was also used to cool beyond the quantum backaction limit to $n_\mathrm{phonon}\sim0.19$ \cite{Clark2017}. Building on these ideas, entanglement of the mechanical motion and the microwave fields was demonstrated \cite{Palomaki2013a} as well as coherent state transfer between itinerant microwave fields and a mechanical oscillator \cite{Palomaki2013}.

Hybrid systems are also important in the context of microwave to optical frequency transduction in the quantum regime. This is a very desirable primitive for quantum networks, as it would allow quantum processors based on circuit QED to be linked optically over large distances. A variety of hybrid systems are currently being investigated for this purpose, including electro-optomechanical, electro-optic and magneto-optic ones~\cite{Higginbotham2018,Lambert2019,Lauk2020}. Two other hybrid quantum systems that have recently emerged are quantum surface acoustic waves interacting with superconducting qubits \cite{Gustafsson2014}, and quantum magnonics where quanta of excitation of spin-wave modes known as magnon are strongly coupled to the field of a 3D  microwave cavity \cite{Lachance-Quirion2019}.

\section*{Acknowledgments}
We thank Alexandre Choquette-Poitevin, Agustin Di Paolo, Christopher Eichler, Jens Koch, Dany Lachance-Quirion, Yasunobu Nakamura, William Oliver, Alexandru Petrescu, Baptiste Royer, Gerd Sch\"on and Irfan Siddiqi for discussions and comments on the manuscript.
This work was undertaken thanks to funding from NSERC, the Canada First Research Excellence Fund, the U.S. Army Research Office Grant No.~W911NF-18-1-0411 and W911NF-18-1-0212, and the Australian Research Council (ARC) via Discovery Early Career Research Award (DECRA) DE190100380.

\appendix

\section{Hamiltonian of a voltage biased transmon}
\label{sec:AppendixTRcoupling}

An excellent introduction to the quantization of electromagnetic circuits can be found in~\textcite{Vool2017}. Here, we only give a brief introduction to this topic by means of two examples that are used throughout this review: a transmon qubit biased by an external voltage source, and a transmon coupled to an LC oscillator. 

\subsection{Classical gate voltage}

\begin{figure}[t]
  \centering
  \includegraphics{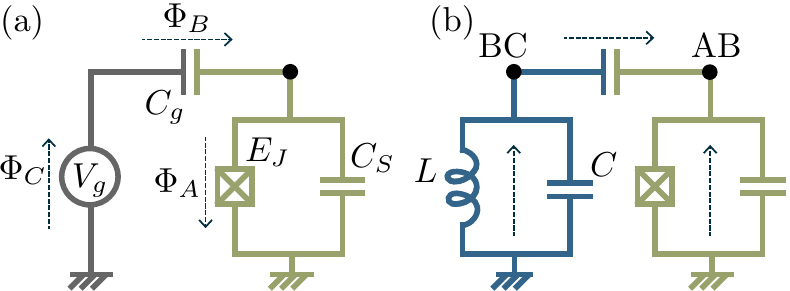}
  \caption{(a) Voltage-biased transmon qubit with the three relevant flux branches. (b) Replacing the classical voltage source by a LC oscillator. The dashed arrows indicate the sign convention.
  \label{fig:app:transmon}}
  \end{figure}

Consider first the circuit shown in~\cref{fig:app:transmon}(a), illustrating a transmon biased by an external voltage $V_g$. Following \textcite{Vool2017}, we start by associating a branch flux $\Phi_i(t) = \int_{-\infty}^t dt'\, V_i(t')$ to each branch of the circuit, with $V_i$ the voltage across branch $i=A,B,C$ indicated in~\cref{fig:app:transmon}(a). Because Kirchoff's laws impose constraints between the branch fluxes, these fluxes are not independent variables and are therefore not independent degrees of freedom of the circuit. Indeed, Kirchoff's voltage law dictates that $V_C + V_B + V_A = V_g + \dot \Phi_B + \dot \Phi_A = 0$ where we have used the sign convention dictated by the (arbitrarily chosen) orientation of the arrows in~\cref{fig:app:transmon}(a). This constraint allows us to eliminate $\Phi_B$ in favor of $\Phi_A$. Moreover, following Kirchoff's current law, the currents $I_A$ and $I_B$ flowing into and out of the node indicated by the black dot in~\cref{fig:app:transmon}(a) obey $I_A = I_B$. This constraint can be expressed in terms of the branch fluxes using the constitutive relations for the capacitances $C_g$ and $C_\Sigma = C_S + C_J$
\begin{equation}
Q_A = C_\Sigma\dot\Phi_A,\quad
Q_B = C_g\dot\Phi_B,
\end{equation}
as well as the Josephson current relation
\begin{equation}
I_J = I_c\sin\varphi_A,
\end{equation}
where $\varphi_A = (2\pi/\Phi_0)\Phi_A$ and $I_c$ is the critical current. We can thus write $I_A = \dot Q_A + I_J = C_\Sigma\ddot \Phi_A + I_c\sin\varphi_A$ and $I_B = \dot Q_B = C_g \ddot \Phi_B$. Combining the above expressions, we arrive at 
\begin{equation}\label{eq:app:cq:Phi_A}
    C_\Sigma \ddot \Phi_A + I_c\sin\varphi_A 
    = - C_g(\ddot \Phi_A + \ddot \Phi_C).
\end{equation}
 Here, $\dot\Phi_C = V_g$ is the applied bias voltage and the only dynamical variable in the above equation is thus $\Phi_A$. As can easily be verified, this equation of motion for $\Phi_A$ can equivalently be derived from the Euler-Lagrange equation for $\Phi_A$ with the Lagrangian
\begin{equation}\label{eq:app:cq:L_T}
    \mathcal L_T = \frac{C_\Sigma}{2}\dot\Phi_A^2 + \frac{C_g}{2}(\dot\Phi_A + \dot\Phi_C)^2
    + E_J\cos\varphi_A,
\end{equation}
where $E_J = (\Phi_0/2\pi)I_c$.

The corresponding Hamiltonian can be found by first identifying the canonical momentum associated to the coordinate $\Phi_A$, $Q_A = \partial \mathcal L_T/\partial \dot\Phi_A = (C_\Sigma+C_g)\dot\Phi_A + C_g\dot\Phi_C$, and performing a Legendre transform to obtain~\cite{Goldstein2001}
\begin{equation}
    H_T = Q_A\dot\Phi_A - \mathcal L_T = \frac{(Q_A - C_gV_g)^2}{2(C_\Sigma+C_g)} - E_J\cos\varphi_A,
\end{equation}
where we have made the replacement $\dot\Phi_C = V_g$ and dropped the term $C_gV_g^2/2$ which only leads to an overall shift of the energies. Promoting the conjugate variables to non-commuting operators $[\hat\Phi_A, \hat Q_A]=i\hbar$, we arrive at~\cref{eq:Hsj} where we have assumed that $C_g \ll C_\Sigma$ to simplify the notation.

\subsection{Coupling to an LC oscillator}

As a model for the simplest realization of circuit QED, we now replace the voltage source by an LC oscillator, see \cref{fig:app:transmon}(b). The derivation follows the same steps as before, now with $V_g+\dot\Phi_B-\dot\Phi_A = 0$ and $I_A+I_B=0$ because of the different choice of orientation for branch $A$. Moreover, at the node labelled BC we have $I_B=I_C$. Eliminating $\Phi_B$ as before and using the constitutive relations for the capacitance $C$ and inductance $L$ of the LC oscillator to express the current through the oscillator branch as $I_C = C\ddot \Phi_C + \Phi_C/L$, we find
\begin{equation}\label{eq:app:cq:Phi_C}
    C\ddot\Phi_C + \frac{\Phi_C}{L} = C_g(\ddot\Phi_A - \ddot \Phi_C).
\end{equation}
In contrast to the above example, $\Phi_C$ is now a dynamical variable in it's own right rather than being simply set by a voltage source. Together with \Cref{eq:app:cq:Phi_A} which still holds, \cref{eq:app:cq:Phi_C} can equivalently be derived using the Euler-Lagrange equations with the Lagrangian
\begin{equation}\label{eq:L}
    \mathcal L = \mathcal L_T + \mathcal L_{LC},
\end{equation}
where $\mathcal L_T$ is given in~\cref{eq:app:cq:L_T} and $\mathcal L_{LC} = \frac{C}{2}\dot\Phi_C^2 - \frac{1}{2L}\Phi_C$. 

It is convenient to write \cref{eq:L} as $\mathcal L = T - V$ with $T = \half \*\Phi^T \*C \* \Phi$ and $V=\Phi_C/2L - E_J\cos\varphi_A$ where we have defined the vector $\*\Phi = (\Phi_A, \Phi_C)^T$ and the capacitance matrix
\begin{equation}
    \*C =
    \left(\begin{array}{cc}
    C_\Sigma + C_g & - C_g \\
    -C_g & C + C_g
    \end{array}\right).
\end{equation}
Defining the vector of conjugate momenta $\*Q = (Q_A, Q_C)^T$, the Hamiltonian is then~\cite{Goldstein2001}
\begin{equation}\label{eq:app:cq:HTransmonResonatorNoApprox}
    \begin{aligned}
    H ={}& 
    \half \*Q^T \*C^{-1} \*Q + V\\
    ={}&\frac{(C+C_g)}{2\bar C^2}Q_A^2 + \frac{C_g}{\bar C^2}Q_A Q_C - E_J\cos\varphi_A\\
    &+ \frac{(C_\Sigma + C_g)}{2\bar C^2}Q_C^2 + \frac{\Phi_C}{2L},
    \end{aligned}
\end{equation}
where we have defined $\bar C^2 = C_gC_\Sigma + C_gC + C_\Sigma C$.
The limit $C_g \ll C_\Sigma,\, C$ results in the simplified expression
\begin{equation}\label{eq:app:cq:HTransmonResonator}
    \begin{aligned}
    H \simeq{}& \frac{\left(Q_A + \frac{C_g}{C} Q_C\right)^2}{2C_\Sigma} - E_J\cos\varphi_A + H_{LC},
    \end{aligned}
\end{equation}
with $H_{LC} = \frac{Q_C^2}{2C} + \frac{\Phi_C}{2L}$ the Hamiltonian of the LC circuit.
By promoting the flux and charge variables to operators, and defining $\hat n = \hat Q_A/2e$, $\hat n_r = (C_g/C)\hat Q_C/2e$ and diagonalizing $\hat H_{LC}$ as in~\cref{sec:HO}, we arrive at~\cref{eq:HTransmonResonator} for a single mode $m=r$.

\Cref{eq:app:cq:HTransmonResonator} can easily be generalized to capacitive coupling between other types of circuits, such as resonator-resonator, transmon-transmon or transmon-transmission line coupling by simply replacing the potential energy terms $-E_J\cos\varphi_A$ and $\Phi_C^2/2L$ to describe the type of circuits in question. This leads, for example, to~\cref{eq:gates:H_swap} for two capacitively coupled transmons after introducing ladder operators as in~\cref{eq:phiTransmon,eq:nTransmon}.

\section{\label{sec:unitarytransforms}Unitary transformations}

We introduce a number of unitary transformations often employed in the field of circuit QED. The starting point is the usual transformation 
\begin{equation}
  \hat H_U = \hat U^\dagger \hat H \hat U - i\hbar \hat U^\dagger \dot{\hat U},
\end{equation}
of a Hamiltonian under a time-dependent unitary $\hat U$ with the corresponding transformation for the states $\ket{\psi_U} = \hat U^\dagger \ket{\psi}$.
Since the unitary can be written as $\hat U = \exp(-\hat S)$ with $\hat S$ an anti-Hermitian operator, a very useful result in this context is the Baker-Campbell-Hausdorff (BCH) formula, which holds for any two operators $\hat S$ and $\hat H$
\begin{equation}\label{eq:BCH}
  \begin{aligned}
    e^{\hat S} \hat H e^{-\hat S} ={}& \hat H + [\hat S, H] + \frac{1}{2!} [\hat S,[\hat S, \hat H]] + \dots \\
    ={}& \sum_{n=0}^\infty \frac{1}{n!} \cC_{\hat S}^n[\hat H],
  \end{aligned}
\end{equation}
where in the last line we have introduced the short-hand notation $\cC_{\hat S}^n[\hat H] = \stackrel{n \text{ times}}{[\hat S, [\hat S, [\hat S}, \dots, \hat H]]]$ and $\cC_{\hat S}^0[\hat H] = \hat H$~\cite{Boissonneault2009}.

\subsection{\label{sec:SW}Schrieffer-Wolff perturbation theory}

We often seek unitary transformation that diagonalize the Hamiltonian of an interacting system.
Exact diagonalization can, however, be impractical, and we must resort to finding an effective Hamiltonian which describes the physics at low energies using perturbation theory. A general approach to perturbation theory which we follow here is known as a Schrieffer–Wolff transformation~\cite{Schrieffer1966,Bravyi2011schrieffer}. The starting point is a generic Hamiltonian of the form
\begin{equation}
    \hat H = \hat H_0 + \hat V,
\end{equation}
with typically $\hat H_0$ some free Hamiltonian and $\hat V$ a perturbation.
We divide the total Hilbert space of our system into different subspaces such that $\hat H_0$ does not couple states in different subspaces while $\hat V$ does. The goal of the Schrieffer-Wolff transformation is to arrive at an effective Hamiltonian for which the different subspaces are completely decoupled.

The different subspaces, which we label by an index $\mu$, can conveniently be defined by a set of projection operators~\cite{Zhu2012,Cohen-Tannoudji1998} 
\begin{equation}
    \hat P_\mu = \sum_n \ket{\mu,n}\bra{\mu,n},
\end{equation}
where $\ket{\mu,n}$, $n=0,1,\dots$, is an orthonormal basis for the subspace labeled $\mu$. 
For the Schrieffer-Wolff transformation to be valid, we must assume that $\hat V$ is a small perturbation. Formally, the operator norm $||\hat V|| = \max_{\ket\psi}||\hat O \ket\psi||$ should be smaller than half the energy gap between the subspaces we intend to decouple; see Eq.~(3.1) of \textcite{Bravyi2011schrieffer}. While $\hat V$ is often formally unbounded in circuit QED applications, the operator is always bounded when restricting the problem to physically relevant states. 

The Schrieffer-Wolff transformation is based on finding a unitary transformation $\hat U=e^{-\hat S}$ which approximately decouples the different subspaces $\mu$ by truncating the Baker-Campbell-Hausdorff formula~\cref{eq:BCH} at a desired order. We first expand both $\hat H$ and $\hat S$ in formal power series
\begin{subequations}\label{eq:SW:formal_expansions}
\begin{align}
    \hat H ={}& \hat H^{(0)} + \varepsilon \hat H^{(1)} + \varepsilon^2 \hat H^{(2)} + \dots,\\
    \hat S ={}& \varepsilon \hat S^{(1)} + \varepsilon^2 \hat S^{(2)} + \dots,
\end{align}
\end{subequations}
where $\varepsilon$ is a fiducial parameter introduced to simplify order counting and which we can ultimately set to $\varepsilon \to 1$. The Schrieffer-Wolff transformation is found by inserting~\cref{eq:SW:formal_expansions} back into~\cref{eq:BCH}, and collecting terms at each order $\varepsilon^k$. We can then iteratively solve for $S^{(k)}$ and $\hat H^{(k)}$ by requiring that the resulting Hamiltonian $\hat H_U$ is block-diagonal (i.e.~it does not couple different subspaces $\mu$) at each order, and the additional requirement that $\hat S$ is itself block off-diagonal~\cite{Bravyi2011schrieffer}.

For the reader's convenience, the explicit results up to $k=2$ are for the generator (with $\varepsilon = 1$)
\begin{subequations}\label{eq:SW:explicit_expansions_generator}
\begin{align}
&\braket{\mu,n|\hat S^{(1)}|\nu,l} = \frac{\braket{\mu,n|\hat V|\nu,l}}{E_{\mu,n}-E_{\nu,l}},\\
&\begin{aligned}
\braket{\mu,n|\hat S^{(2)}|\nu,l} ={}& \sum_{k}
\Bigg( \frac{\braket{\mu,n|\hat V|\nu,k}}{E_{\mu,n}-E_{\nu,l}} \frac{\braket{\nu,k|\hat V|\nu,l}}{E_{\mu,n}-E_{\nu,k}}\\
&- \frac{\braket{\mu,n|\hat V|\mu,k}}{E_{\mu,n}-E_{\nu,l}} \frac{\braket{\mu,k|\hat V|\nu,l}}{E_{\mu,k}-E_{\nu,l}}\Bigg)
\end{aligned},
\end{align}
\end{subequations}
for $\nu\neq\mu$, while block-diagonal matrix element where $\mu=\nu$ are zero,
and
\begin{subequations}\label{eq:SW:explicit_expansions}
\begin{align}
&\hat H^{(0)} = \hat H_0,\\
&\hat H^{(1)} = \sum_\mu \hat P_\mu \hat V \hat P_\mu,\\
&\begin{aligned}
\braket{\mu,n|\hat H^{(2)}|\mu,m} &= \sum_{\nu\neq\mu,l} \braket{\mu,n|\hat V|\nu,l}\braket{\nu,l|\hat V|\mu,m}\\
\times \half \bigg( &\frac{1}{E_{\mu,n}-E_{\nu,l}} + \frac{1}{E_{\mu,m}-E_{\nu,l}}  \bigg),
\end{aligned}
\end{align}
\end{subequations}
for the transformed Hamiltonian (block off-diagonal matrix element are zero, i.e., $\braket{\mu,n|\hat H^{(2)}|\nu,m}=0$ for $\mu\neq\nu$). In these expressions, $E_{\mu,n}$ refers to the bare energy of $\ket{\mu,n}$ under the unperturbed Hamiltonian $\hat H_0$. An explicit formulae for $\hat H^{(k)}$ up $k=4$ can be found, e.g. in~\textcite{Winkler2003}.

\subsection{\label{sec:SW:multilevel}Schrieffer-Wolff for a multilevel system coupled to an oscillator in the dispersive regime}

As an application of the general result of \cref{eq:SW:explicit_expansions} we consider in this section a situation that is commonly encoutered in circuit QED: An arbitrary artificial atom coupled to a single mode oscillator in the dispersive regime.
Both the transmon artificial atom and the two-level system discussed in~\cref{sec:dispersive} are special cases of this more general example. The artificial atom, here taken to be a generic multilevel system, is described in its eigenbasis with the Hamiltonian $\hat H_\text{atom} = \sum_j \hbar\omega_j \ket j \bra j$. The full Hamiltonian is therefore given by
\begin{equation}
    \hat H = \hbar \wc \ada + \sum_j \hbar\omega_j \ket j \bra j + \left( \hat B \ad + \hat B\dg \aop\right),
\end{equation}
where $\hat B$ is an arbitrary operator of the artificial atom which couples to the oscillator. For example, in the case of capacitive coupling, it is proportional to the charge operator with $\hat B \sim i\hat n$, cf.~\cref{eq:HTransmonResonator}.

By inserting resolutions of the identity $\hat I = \sum_j \ket j \bra j$, the interaction term can be re-expressed in the atomic eigenbasis as \cite{Koch2007}
\begin{equation}\label{eq:SW:multilevel_hamiltonian}
    \begin{aligned}
    \hat H ={}& \hbar \wc \ada + \sum_j \hbar\omega_j \ket j \bra j \\
    &+ \sum_{ij} \hbar \left( g_{ij} \ket i \bra j \ad + g_{ij}^* \ket j \bra i \aop\right),
    \end{aligned}
\end{equation}
where $\hbar g_{ij} = \braket{i|\hat B|j}$, and with $g_{ij} = g_{ji}$ if $\hat B = \hat B\dg$.

To use~\cref{eq:SW:explicit_expansions}, we identify the first line of~\cref{eq:SW:multilevel_hamiltonian} as $\hat H_0$ and the second line as the perturbation $\hat V$. The subspaces labeled by $\mu$ are in this situation one-dimensional, $\hat P_\mu = \ket\mu\bra\mu$, with $\ket{\mu} = \ket{n,j} = \ket{n}\otimes \ket{j}$, $\ket{n}$ an oscillator number state and $\ket j$ an artificial atom eigenstate.
A straightforward calculation yields the second order result~\cite{Zhu2012}
\begin{equation}\label{eq:SW:multilevel}
  \begin{aligned}
  \hH_\mathrm{disp} = e^{\hat S} \hH e^{-\hat S} \simeq{}& \hbar \wc \ada + \sum_j \hbar (\omega_j + \Lambda_j) \ket j \bra j \\
  &+ \sum_j \hbar \chi_j \ad \aop \ket j \bra j,
  \end{aligned}
\end{equation}
where
\begin{equation}
\Lambda_j = \sum_i \chi_{ij},\quad
\chi_j = \sum_i \left(\chi_{ij} - \chi_{ji} \right),
\end{equation}
with
\begin{equation}
    \chi_{ij} = \frac{|g_{ji}|^2}{\omega_j-\omega_i-\omega_r}.
\end{equation}
Note that we are following here the convention of \textcite{Koch2007} rather than of \textcite{Zhu2012} for the definition of $\chi_{ij}$.

Projecting~\cref{eq:SW:multilevel} on the first two-atomic levels $j=0,1$ with the convention $\hat\sigma_z = \ket 1\bra 1 - \ket 0\bra 0$ we obtain
\begin{equation}\label{eq:SW:twolevel}
  \begin{aligned}
  \hH_\mathrm{disp} \simeq{}& \hbar \wc' \ada + \frac{\hbar\omega_q'}{2}\sz{}
  + \hbar\chi \ad \aop \sz{},
  \end{aligned}
\end{equation}
where we have dropped a constant term and defined $\wc' = \wc + (\chi_0 + \chi_1)/2$, $\wa'=\omega_1-\omega_0 + \Lambda_1-\Lambda_0$ and $\chi= (\chi_1 - \chi_0)/2$.

\subsubsection{\label{sec:SW:transmon}The transmon}

The transmon capacitively coupled to an oscillator is one example of the above result. From \cref{eq:HTransmonJC}, we identify the free Hamiltonian as 
\begin{equation}
  \hH_0 = \hbar\wc \ada + \hbar\omega_q \bdb - \frac{E_C}{2}\bd\bd\bop\bop,
\end{equation}
and the perturbation as
\begin{equation}
    \hat V = \hbar g (\bd\aop+\bop\ad).
\end{equation}
In this nonlinear oscillator approximation for the transmon, the transmon eigenstates are number states $\bd\bop \ket{j} = j\ket j$, with $j=0,1,\dots,\infty$. Moreover, the coupling operator is $\hat B = \hbar g \bop$, and thus
\begin{equation}
    g_{j,j+1} = g \braket{j|\bop|j+1} = g\sqrt{j+1} = g_{j,j+1}^*,
\end{equation}
while all other matrix elements $g_{ij}$ are zero.
We consequently find
\begin{subequations}
\begin{align}
    &\Lambda_{j} = \chi_{j-1,j} = \frac{j g^2}{\wa - E_C/\hbar(j - 1) - \omega_r},\\
    &\begin{aligned}
    \chi_{j} ={}& \chi_{j-1,j} - \chi_{j,j+1} \\
    ={}& g^2\left(\frac{j}{\omega_j-\omega_{j-1}-\omega_r} - \frac{j+1}{\omega_{j+1}-\omega_j-\omega_r}\right),
    \end{aligned}
\end{align}
\end{subequations}
for $j>0$, while for $j=0$ we have $\Lambda_0 = 0$ and $\chi_0 = -\chi_{01} = -g^2/\Delta$ where 
$\Delta \equiv \omega_q-\omega_r$.
In the two-level approximation of \cref{eq:SW:twolevel}, this becomes~\cite{Koch2007}
\begin{subequations}\label{eq:SW:transmonshifts}
\begin{align}
\wc' ={}& \wc - \frac{\chi_{12}}{2} = \wc - \frac{g^2}{\Delta-E_C/\hbar},\\
\wa' ={}& \omega_1-\omega_0 + \chi_{01} = \wa + \frac{g^2}{\Delta},\\
\chi ={}& \chi_{01} - \frac{\chi_{12}}{2} = - \frac{g^2 E_C/\hbar}{\Delta\left(\Delta-E_C/\hbar\right)},
\end{align}
\end{subequations}
which are the results quoted in~\cref{eq:HQubitDispersiveParametersSW}.

Recall that this Schrieffer-Wolff perturbation theory is only valid if the perturbation $\hat V$ is sufficiently small. Following \textcite{Bravyi2011schrieffer}, a more precise statement is that we require $2||\hat V|| < \Delta_\text{min}$, where $\Delta_\text{min}$ is the smallest energy gap between any of the bare energy eigenstates $\ket{n} \otimes \ket j$, where $\ket n$ is a number state for the oscillator. Here, $\hat V = g(\bd\aop + \bop\ad)$ is formally unbounded but physical states have finite excitation numbers. Therefore, replacing the operator norm by $\braket{n,j|\hat V\dg \hat V|n,j}^{1/2}$ and using $\Delta_\text{min} = |\Delta- j E_C/\hbar|$ corresponding to the minimum energy gap between neighboring states $\ket{n,j}$ and $\ket{n\pm 1,j\mp 1}$, we find that a more precise criterion for the validity of the above perturbative results is
\begin{equation}
    n \ll n_\text{crit} \equiv \frac{1}{2j + 1} \left(\frac{|\Delta-j E_C/\hbar|^2}{4 g^2} - j\right).
\end{equation}
Setting $j=0$, this gives the familiar expression $n_\text{crit} = (\Delta/2g)^2$, while setting $j=1$ gives a more conservative estimate.
As quoted in the main text, the appropriate small parameter is therefore $\bar n/n_\mathrm{crit}$, with $\bar n$ the average oscillator photon number. For the second order effective Hamiltonian $\hH_\text{disp}$ to be an accurate description of the system requires $\bar n/n_\mathrm{crit}$ to be significantly smaller than unity (it is difficult to make a precise statement but the criteria $\bar n/n_\mathrm{crit}\lesssim 0.1$ is often used).

\subsubsection{The Jaynes-Cummings model}
\label{sec:AppendixDispersiveTLS}

It is interesting to contrast the above result in which the transmon is treated as a multilevel system with the result obtained if the artificial atom is truncated to a two-level system \emph{before} performing the Schireffer-Wolff transformation. That is, we start with the Jaynes-Cummings Hamiltonian
\begin{equation}
\begin{split}
    \hH_\mathrm{JC}
    = \hbar\wc \ada + \frac{\hbar\wa}{2}\sz{} + \hbar g(\ad\smm{} + \aop\spp{}).
\end{split}
\end{equation}
Identifying the first two terms as the unperturbed Hamiltonian $\hat H_0$ and the last term as the interaction $\hat V$, we can again apply~\cref{eq:SW:explicit_expansions}. Alternatively, the result can be found more directly from~\cref{eq:SW:twolevel} by taking $g_{01} = g_{01}^* = g$ and all other $g_{ij} = 0$. The result is
\begin{equation}
\wc' = \wc,\quad
\wa' =  \wa + \frac{g^2}{\Delta},\quad
\chi = \frac{g^2}{\Delta},
\end{equation}
with $\Delta = \omega_q - \omega_r$ as before.
We see that the results agree with~\cref{eq:SW:transmonshifts} only in the limit $E_C/\hbar \gg \Delta,\, g$. Importantly, since $E_C$ is relatively small compared to the detuning $\Delta$ in most transmon experiments, the value for $\chi$ predicted from the Jaynes-Cummings model is far from the multi-level case. Moreover, following the same argument as above, we find that the Schrieffer-Wolff transformation is valid for photon numbers $\bar n<n_\text{crit}$ with $n_\text{crit} = (\Delta/2g)^2-j$ with $j=0,1$ for the ground and excited qubit states, respectively.

It is interesting to note that the transformation used here to approximately diagonalize the Jaynes-Cummings Hamiltonian can be obtained by Taylor expanding the generator $\Lambda(\hat N_T)$ of the unitary transformation \cref{eqn:JCdiag_transform} which exactly diagonalizes $\hH_\mathrm{JC}$. This exercise also leads to the conclusion that $\bar n/n_\mathrm{crit}$, with $n_\mathrm{crit} = (\Delta/2g)^2$, is the appropriate small parameter. Alternatively,  $\hH_\text{disp}$ can also be obtained simply by Taylor expanding the diagonal form \cref{eq:HJCdiagonal} of $\hH_\mathrm{JC}$ \cite{Boissonneault2010}.

\subsection{Bogoliubov approach to the dispersive regime}\label{sec:AppendixDispersiveTransformation}

We derive the results presented in~\cref{sec:dispersive:bb}. Our starting point is thus the transmon-resonator Hamiltonian~\cref{eq:HTransmonJC} and our final result the dispersive Hamiltonian of \cref{eq:HTransmonDispersive}.

It is first useful to express \cref{eq:HTransmonJC} as a sum of a linear and a nonlinear part, $\hH = \hHL + \hHNL$ where
\begin{align}
  \hHL ={}& \hbar\wc \ada + \hbar\omega_q \bdb + \hbar g (\bd\aop+\bop\ad),\label{eq:app:HTransmonDispersiveL}\\
  \hHNL ={}& -\frac{E_C}{2} \bd\bd\bop\bop.\label{eq:app:HTransmonDispersiveNL}
\end{align}
The linear Hamiltonian $\hHL$ can be diagonalized exactly with the Bogoliubov transformation
\begin{equation}
  \hat U = \exp\left[\Lambda (\hat a^\dagger \hat b - \hat a \hat b^\dagger) \right].
\end{equation}
Under this unitary transformation, the annihilation operators transform  as $\hat U^\dagger \hat a \hat U = \cos(\Lambda)\hat a + \sin(\Lambda)\hat b$, $\hat U^\dagger \hat b \hat U = \cos(\Lambda)\hat b - \sin(\Lambda)\hat a$, leading to
\begin{equation}
\begin{aligned}
  \hHL' ={}& \hat U^\dagger \hHL \hat U
  = \tilde \wc \hat a^\dagger \hat a + \tilde \omega_q \hat b^\dagger \hat b \\
  &+\left[g\cos(2\Lambda) -\frac{\Delta}{2}\sin(2\Lambda)\right](\hat a^\dagger \hat b + \hat a \hat b^\dagger),
\end{aligned}
\end{equation}
where 
\begin{align}
  \twc  ={}& \cos^2(\Lambda)\wc + \sin^2(\Lambda)\omega_q - g\sin(2\Lambda),\\
  \tilde\omega_q ={}& \cos^2(\Lambda)\omega_q + \sin^2(\Lambda)\wc + g\sin(2\Lambda).
\end{align}
To cancel the last term of $\hHL'$, we take $\Lambda = \half \arctan(2\lambda)$ with $\lambda = g/\Delta$ and $\Delta = \omega_q-\wc$ to obtain the diagonal form
\begin{equation}
\begin{aligned}
  \hHL' ={}& \hbar\twc  \hat a^\dagger \hat a + \hbar\tilde\omega_q \hat b^\dagger \hat b,
\end{aligned}
\end{equation}
with the mode frequencies
\begin{align}
  \twc  ={}& \half\left(\wc + \omega_q - \sqrt{\Delta^2+4g^2}\right),\\
  \tilde\omega_q ={}& \half\left(\wc + \omega_q + \sqrt{\Delta^2+4g^2}\right).
\end{align}
The same transformation on $\hHNL$ gives
\begin{equation}
  \begin{aligned}
    \hHNL' ={}& \hat U^\dagger \hHNL \hat U \\
    ={}& -\frac{E_C}{2}\cos^4(\Lambda)(\hat b^\dagger)^2 \hat b^2 -\frac{E_C}{2} \sin^4(\Lambda) (\hat a^\dagger)^2 \hat a^2\\
    &-2E_C\cos^2(\Lambda)\sin^2(\Lambda)\hat a^\dagger \hat a \hat b^\dagger \hat b\\
    &+ E_C\cos^3(\Lambda)\sin(\Lambda)\left(\hat b^\dagger \hat b\,\hat a^\dagger b + \hc \right) \\
    &+ E_C \cos(\Lambda)\sin^3(\Lambda)\left(\hat a^\dagger \hat a\, \hat a \hat b^\dagger + \hc\right)\\
    &-\frac{E_C}{2} \cos(\Lambda)^2\sin(\Lambda)^2[(\hat a^\dagger)^2 \hat b^2 + \hc].
  \end{aligned}
\end{equation}

Note that, at this stage, the transformation is exact. In the dispersive regime, we expand the mode frequencies and $\hHNL'$ in powers of $\lambda$. For the nonlinear part of the Hamiltonian, this yields
\begin{equation}\label{eq:app:dispersive:H_NL_lambda}
  \begin{aligned}
    \hHNL'
    ={}& -\frac{E_C}{2}(\hat b^\dagger)^2 \hat b^2 -\lambda^4\frac{E_C}{2} (\hat a^\dagger)^2 \hat a^2\\
    &-2\lambda^2E_C\hat a^\dagger \hat a \hat b^\dagger \hat b\\
    &+ \lambda E_C (\hat b^\dagger \hat b\,\hat a^\dagger \hat b + \hc)\\
    &+ \lambda^3E_C (\hat a^\dagger \hat a \,\hat a \hat b^\dagger + \hc)\\
    &-\lambda^2\frac{E_C}{2}[(\hat a^\dagger)^2 \hat b^2 + \hc] + \mathcal{O}(\lambda^5).
  \end{aligned}
\end{equation}
The magnitude $\lambda^2 E_C$ of the cross-Kerr term $\hat a^\dagger \hat a \hat b^\dagger \hat b$ in this expression does not coincide with Eq.~(3.12) of \textcite{Koch2007}. To correct this situation, we apply an additional transformation to eliminate the third line of \cref{eq:app:dispersive:H_NL_lambda}. This term is important because it corresponds, roughly, to an exchange interaction $\hat a^\dagger \hat b + \hat b^\dagger \hat a$ with an additional number operator $\hat b^\dagger \hat b$ which distinghuishes the different transmon levels. To eliminate this term, we apply a Schrieffer-Wolff transformation to second order with the generator $S = \lambda' (\hat b^\dagger \hat b\,\hat a^\dagger \hat b - \hc)$ where $\lambda' = \lambda E_C/[\Delta + E_C(1-2\lambda^2)]$. Neglecting the last two lines of \cref{eq:app:dispersive:H_NL_lambda} and ommiting a correction of order $\lambda^2$, we arrive at~\cref{eq:HTransmonDispersive} which agrees with \textcite{Koch2007}. 

\subsection{\label{sec:AppendixDrivenTransmon}Off-resonantly driven transmon}

We derive \cref{eq:HacStarkDrive} describing the ac-Stark shift resulting from an off-resonant drive on a transmon qubit. Our starting point is \cref{eq:singlequbitdrive} which takes the form
\begin{equation}
  \hat H(t) = \hbar \omega_q \hat b\dg \hat b - \frac{E_C}{2}(\hat b\dg)^2 \hat b^2 + \hbar \epsilon(t)\hat b\dg  + \hbar \epsilon^*(t) \hat b,
\end{equation}
where we have defined $\epsilon(t) = \varepsilon(t) e^{-i\wdrive t -i\phidrive}$. To account for a possible time-dependence of the drive envelope $\varepsilon(t)$ it is useful to apply the time-dependent displacement transformation
\begin{equation}
  \hat U(t) = e^{\alpha^*(t) \hat b - \alpha(t) \hat b\dg}.
\end{equation}
Under $\hat U(t)$, $\hat b$ transforms to $\hat U\dg \hat b \hat U = \hat b - \alpha(t)$, while
\begin{equation}
  \hat U\dg \dot{\hat U} = \dot \alpha^*(t) \hat b - \dot \alpha(t) \hat b\dg.
\end{equation}
Using these expressions, the transformed Hamiltonian becomes
\begin{equation}
  \begin{aligned}
    \hat H' ={}& \hat U\dg H \hat U - i\hat U\dg \dot{\hat U} \\
    \simeq{}& \hbar\omega_q (\hat b\dg \hat b - \alpha^* \hat b - \alpha \hat b\dg)\\
    &- \frac{E_C}{2} [(\hat b\dg)^2 \hat b^2 + 4|\alpha|^2 \hat b\dg \hat b)\\
    &+ \hbar\epsilon\hat b\dg + \hbar \epsilon^* \hat b
    -i\hbar(\dot \alpha^* \hat b - \dot \alpha \hat b\dg),
  \end{aligned}
\end{equation}
where we have dropped fast-rotating terms and a scalar. The choice
\begin{equation}
    \dot \alpha(t) = -i\omega_q \alpha(t) + i\epsilon(t),
\end{equation}
cancels the linear drive term leaving
\begin{equation}\label{eq:gates:Stark}
  \begin{aligned}
    \hat H'(t)
    \simeq{}& [\hbar\omega_q- 2E_C |\alpha(t)|^2] \hat b\dg \hat b
    - \frac{E_C}{2} (\hat b\dg)^2 \hat b^2.
  \end{aligned}
\end{equation}
Taking a constant envelope $\varepsilon(t)=\varepsilon$ for simplicity such that $|\alpha(t)|^2 = (\varepsilon/\delta_q)^2$, the above expression takes the compact form
\begin{equation}
  \begin{aligned}
    \hat H''(t)
    \simeq{}& \half \left(\hbar\omega_q - E_C \frac{\Omega_R^2}{2 \delta_q^2}\right) \sz{},
  \end{aligned}
\end{equation}
in the two-level approximation which is \cref{eq:HacStarkDrive} of the main text.

It is instructive to obtain the same result now using the Schrieffer-Wolff approach. Assuming a constant envelope $\varepsilon(t)=\varepsilon$ and with $\phidrive=0$ for simplicity, our starting point is
\begin{equation}
  \hat H = \hbar \delta_q \hat b\dg \hat b - \frac{E_C}{2}(\hat b\dg)^2 \hat b^2 + \hbar \varepsilon(\hat b\dg  + \hat b),
\end{equation}
in a frame rotating at $\wdrive$ and where $\delta_q = \wa-\wdrive$. We treat the drive as a perturbation and apply the second order formula~\cref{eq:SW:explicit_expansions} to obtain
\begin{equation}
  \begin{aligned}
  \hat H_U \simeq{}& \hbar \delta_q \hat b\dg \hat b - \frac{E_C}{2}(\hat b\dg)^2 \hat b^2 \\
  + |\varepsilon|^2 \sum_j &\left( \frac{j}{\delta_q - \frac{E_C(j-1)}{\hbar}} - \frac{j+1}{\delta_q - \frac{E_C j}{\hbar}} \right) \ket j \bra j\\
  \simeq{} \hbar \delta_q& \hat b\dg \hat b - \frac{E_C}{2}(\hat b\dg)^2 \hat b^2 
  - 2E_C \frac{|\varepsilon|^2}{\delta_q^2} \bd\bop - \frac{|\varepsilon|^2}{\delta_q}, 
  \end{aligned}
\end{equation}
where $\ket j$ is used to label transmon states, as before. In the last approximation we have kept only terms to $\mathcal{O}(j E_C /\delta_q)$. This agrees with~\cref{eq:gates:Stark} for $|\alpha|^2 = |\varepsilon/\delta_q|^2$. More accurate expressions can be obtained by going to higher order in perturbation theory \cite{Schneider2018b}.

\section{\label{app:inout}Input-output theory}

Following closely \textcite{Yurke1984,Yurke2004}, we derive the input-output equations of \cref{sec:inout}. As illustrated in \cref{fig:inout}, we consider an LC oscillator located at $x=0$ and which is capacitively coupled to a semi-infinite transmission line extending from $x=0$ to $\infty$. In analogy with~\cref{eq:HamiltonianResonatorContinous}, the Hamiltonian for the transmission line is
\begin{equation}
    \hat H_\tml = \int_{-\infty}^{\infty} dx\,
    \theta(x) \bigg\{
    \frac{\hat Q_\tml(x)^2}{2 c} 
    + \frac{\left[\partial_x \hat \Phi_\tml(x)\right]^2}{2l}
    \bigg\},
\end{equation}
where $c$ and $l$ are, respectively, the capacitance and inductance per unit length, and $\theta(x)$ the Heaviside step function. The flux and charge operators satisfy the canonical commutation relation $[\hat \Phi_\tml(x),\hat Q_\tml(x')] = i\hbar\delta(x-x')$. 

On the other hand, the Hamiltonian of the LC oscillator of frequency $\wc=1/\sqrt{L_\mathrm{r}C_\mathrm{r}}$ is $\hat H_S = \hat Q_\mathrm{r}^2/(2C_\mathrm{r}) +\hat \Phi_\mathrm{r}^2/(2L_\mathrm{r})$ and the interaction Hamiltonian takes the form
\begin{equation}\label{eq:inout:H_int}
  \hat H_\text{int} = \int_{-\infty}^\infty dx\, \delta(x) \frac{C_\kappa}{c C_\mathrm{r}} \hat Q_\mathrm{r} \hat Q_\tml(x),
\end{equation}
where $C_\kappa$ is the coupling capacitance between the oscillator and the line. In deriving~\cref{eq:inout:H_int}, we have neglected renormalizations of $c$ and $C_\mathrm{r}$ due to $C_\kappa$ (c.f~\cref{sec:AppendixTRcoupling}). The total Hamiltonian is thus $\hat H = \hat H_S + \hat H_\tml + \hat H_\text{int} = \int_{-\infty}^\infty dx \, \mathcal H$, where we have introduced the Hamiltonian density $\mathcal H$ in the obvious way. 

Using these results, Hamilton's equations for the field in the transmission line  take the form
\begin{align}
  &\begin{aligned}
    \dot{\hat \Phi}_\tml(x) 
    ={}& \theta(x) \frac{\hat Q_\tml(x)}{c}
    + \delta(x) \frac{C_\kappa}{C_\mathrm{r} c}\hat Q_\mathrm{r},
  \end{aligned}\\
  &\begin{aligned}
  \dot{\hat Q}_\tml(x) 
  ={}& \partial_x \left[\theta(x) \frac{\partial_x\hat \Phi_\tml(x)}{l}\right].
  \end{aligned}
\end{align}
These two equations can be combined into a wave 
equation for $\hat \Phi_\tml$ which, for $x>0$, reads
\begin{equation}\label{eq:inout:waveeq}
  \ddot{\hat \Phi}_\tml(x) = v^2\partial_x^2 \hat \Phi_\tml(x),
\end{equation}
and where $v = 1/\sqrt{lc}$ is the speed of light in the line. At the location $x=0$ of the oscillator, we instead find
\begin{equation}
  \begin{aligned}
    \ddot{\hat \Phi}_\tml(x) ={}& \theta(x)v^2\left[\delta(x) \partial_x \hat \Phi_\tml(x) + \partial_x^2 \hat \Phi_\tml(x) \right]\\
    &+ \delta(x) \frac{C_\kappa}{C_\mathrm{r}c}\dot{\hat Q}_\mathrm{r},
  \end{aligned}
\end{equation}
where we have used $\partial_x\theta(x) = \delta(x)$. We integrate the last equation over $-\varepsilon < x < \varepsilon$ and subsequently take $\varepsilon\to 0$ to find the boundary condition
\begin{equation}\label{eq:inout:boundary}
  \begin{aligned}
    v^2 \partial_x \hat \Phi_\tml(x=0) = - \frac{C_\kappa}{C_\mathrm{r}c}\dot{\hat Q}_\mathrm{r}.
  \end{aligned}
\end{equation}

From~\cref{eq:inout:waveeq}, we find that  the general solution for the flux and charge fields, defined as $\hat Q_\tml(x,t) = c\partial_t\hat \Phi_\tml(x,t)$, can be written for $x>0$ as $\hat \Phi_\tml(x,t) = \hat \Phi_\text{L}(x,t) + \hat \Phi_\text{R}(x,t)$ and $\hat Q_\tml(x,t) = \hat Q_\text{L}(x,t) + \hat Q_\text{R}(x,t)$, with the subscript L/R denoting left- and right-moving fields
\begin{subequations}\label{eq:inout:fields}
\begin{align}
  &\begin{aligned}
    \hat\Phi_\text{L/R}(x,t) ={}& \int_0^\infty d\omega\, \sqrt{\frac{\hbar}{4 \pi\omega cv}} e^{\pm i\omega x/v + i\omega t} \hat b_{\text{L/R}\omega}\dg \\
  & + \hc,
  \end{aligned}\\
  &\begin{aligned}
    \hat Q_\text{L/R}(x,t) ={}& i \int_0^\infty d\omega\, \sqrt{\frac{\hbar\omega c}{4\pi v}} e^{\pm i\omega x/v + i\omega t} \hat b_{\text{L/R}\omega}\dg \\
  &- \hc
  \end{aligned}
\end{align}
\end{subequations}
In this expression, we introduced the operators $\hat b_{\nu\omega}$ satisfying $[\hat b_{\nu\omega}, \hat b_{\mu\omega'}] = \delta_{\nu\mu}\delta(\omega-\omega')$ for $\nu=$L, R.

Because of the boundary condition at $x=0$, the left- and right-moving fields are not independent. To see this, we first note that, following from the form of $\hat \Phi_\tml(x,t)$,
\begin{equation}\label{eq:inout:Ohmslaw}
  \begin{aligned}
    Z_\tml \frac{\partial_x \hat \Phi_\tml(x,t)}{l} ={}& \dot{\hat \Phi}_\text{L}(x,t) - \dot{\hat \Phi}_\text{R}(x,t),
  \end{aligned}
\end{equation}
with $Z_\tml = \sqrt{l/c}$ the characteristic impedance of the transmission line. Noting that $\hat I(x) = \partial_x \hat \Phi_\tml(x)$ is the current and defining voltages $\hat V_\text{L/R}(x) = \dot{\hat \Phi}_\text{L/R}(x)$, we can recognize~\cref{eq:inout:Ohmslaw} as Ohm's law. Using~\cref{eq:inout:boundary}, we finally arrive at the boundary condition of \cref{eq:voltage_inout} at $x=0$
\begin{equation}\label{eq:inout:inoutrel0}
  \begin{aligned}
    \hat V_\text{out}(t) - \hat V_\text{in}(t) = Z_\tml \frac{C_\kappa}{C_\mathrm{r}} \dot{\hat Q}_\mathrm{r},
  \end{aligned}
\end{equation}
where we have introduced the standard notation $\hat V_\text{in/out}(t) = \hat V_\text{L/R}(x=0,t)$. 

Using the mode expansion of the fields in~\cref{eq:inout:fields} together with \cref{eq:HatPhiQ} for the LC oscillator charge operator in terms of the ladder operator $\aop$, \cref{eq:inout:inoutrel0} can be expressed as
\begin{equation}\label{eq:inout:inoutrel1}
  \begin{aligned}
    & \int_0^\infty d\omega\, \sqrt{\frac{\omega }{4 \pi cv}} e^{- i(\omega-\wc) t} \left(\hat b_{\text{R}\omega} - \hat b_{\text{L}\omega} \right) \\
    ={}& -i \wc Z_\tml \frac{C_\kappa}{C_\mathrm{r}} \sqrt{\frac{\wc C_\mathrm{r}}{2}} \hat a,
  \end{aligned}
\end{equation}
where we have neglected terms rotating at $\omega+\wc$. After some re-arrangements this can be written in the form of the standard input-output boundary condition~\cite{Collett1984,Gardiner1985}
\begin{equation}\label{eq:inout:inoutrel}
   \bout(t) - \bin(t) = \sqrt{\kappa} \hat a(t),
\end{equation}
with input and output fields defined as
\begin{subequations}
  \begin{align}
    \bin(t) ={}& \frac{-i}{\sqrt{2\pi}} \int_{-\infty}^\infty d\omega\, \hat b_{L\omega}e^{-i(\omega-\wc) t},\label{eq:bin_appendix}\\
    \bout(t) ={}& \frac{-i}{\sqrt{2\pi}} \int_{-\infty}^\infty d\omega\, \hat b_{R\omega}e^{-i(\omega-\wc) t}.\label{eq:bout_appendix}
  \end{align}
\end{subequations}
and the photon loss rate $\kappa$ is given by
\begin{equation}
  \kappa =  \frac{Z_\tml C_\kappa^2 \wc^2}{C_\mathrm{r}}.
\end{equation}
There are two further approximations which are made when going from~\cref{eq:inout:inoutrel1} to~\cref{eq:inout:inoutrel}: We have extended the range of integration over frequency from $[0,\infty)$ to $(-\infty,\infty)$, and we have replaced the factor $\sqrt{\omega}$ by $\sqrt{\wc}$ inside the integrand. Both approximations are made based on the assumptions that only terms with $\omega\simeq\wc$ contribute significantly to the integral in~\cref{eq:inout:inoutrel1}.

Moreover, we rewrite~\cref{eq:inout:Ohmslaw} as
\begin{equation}
  \begin{aligned}
    \partial_x \hat \Phi_\tml(x,t) ={}& Z_\tml \left[ \hat Q_\text{L}(x,t) - \hat Q_\text{R}(x,t) \right]\\
    ={}& Z_\tml \left[ 2\hat Q_\text{L}(x,t) - \hat Q_\tml(x,t) \right],
  \end{aligned}
\end{equation}
where in the last equality we have used $\hat Q_\tml(x,t) = \hat Q_\text{L}(x,t) + \hat Q_\text{R}(x,t)$. At $x=0$, this gives 
\begin{equation}
  \begin{aligned}
    \hat Q_\tml(x=0,t) = 2\hat Q_\text{L}(x=0,t) + \frac{1}{v} \frac{C_\kappa}{C_\mathrm{r}} \hat Q_\mathrm{r}(t).
  \end{aligned}
\end{equation}
Using this result in the Heisenberg representation equations of motion for the LC oscillator,
\begin{align}
  &\dot{\hat \Phi}_\mathrm{r} = \frac{i}{\hbar}[\hat H,\hat \Phi_\mathrm{r}] = \frac{\hat Q_\mathrm{r}}{C_\mathrm{r}} + \frac{C_\kappa}{C_\mathrm{r} c}\hat Q_\tml(x=0),\\
  &\dot{\hat Q}_\mathrm{r} = \frac{i}{\hbar}[\hat H,\hat Q_\mathrm{r}] = -\frac{\hat \Phi_\mathrm{r}}{L_\mathrm{r}},
\end{align}
we arrive at a single equation of motion for the oscillator charge
\begin{equation}\label{eq:inout:inouteom0}
  \ddot{\hat Q}_\mathrm{r} = - \wc^2\left[ \hat Q_\mathrm{r} + \frac{C_\kappa}{c} \left( \frac{1}{v} \frac{C_\kappa}{C_\mathrm{r}} \hat Q_\mathrm{r} + 2\hat Q_\text{in} \right) \right].
\end{equation}
Again writing $\hat Q_\mathrm{r}$ in terms of bosonic creation and annihilation operators, it is possible to express \cref{eq:inout:inouteom0} in the form of the familiar Langevin equation \cref{eq:inouteom} for the mode operator $\aop(t)$. This standard expression is obtained after neglecting fast rotating terms and  making the following ``slowly varying envelope'' approximations \cite{Yurke2004}
\begin{subequations}
\begin{align}
  \frac{d^2}{d t^2} \hat ae^{-i\wc t} \simeq{}& -\wc^2 \hat ae^{-i\wc t} - 2i\wc \dot{\hat a}e^{-i\wc t},\\
  \frac{d}{d t} \hat ae^{-i\wc t} \simeq{}& -i\wc \hat ae^{-i\wc t},\\
  \frac{d}{d t} \hat b_{-\omega}e^{-i\omega t} \simeq{}& -i\wc \hat b_{-\omega} e^{-i\omega t}.
\end{align}
\end{subequations}
\Cref{eq:inouteom} can be viewed as a Heisenberg picture analog to the Markovian master equation~\cref{eq:ME_harmonic}.

\bibliographystyle{apsrmp}
\bibliography{QudevRefDB}

\end{document}